\documentclass[galaxies,review,accept,pdftex,moreauthors]{Definitions/mdpi}

\firstpage{1} 
\pubvolume{1}
\issuenum{1}
\articlenumber{0}
\pubyear{2024}
\copyrightyear{2024}
\externaleditor{Academic Editor: Firstname Lastname}
\datereceived{6 May 2025} 
\daterevised{6 August 2025} 
\dateaccepted{7 August 2025} 
\datepublished{ } 
\hreflink{https://doi.org/} 

\newcommand{\ergs}{\mbox{ erg s}^{-1}}
   
\newcommand{\cc}{\mbox{ cm}^{-3}}
\newcommand{\kms}{\mbox{ km s}^{-1}}
\newcommand{\kpc}{\mbox{ kpc}}

 \newcolumntype{A}{>{\arraybackslash\hsize=0.05\hsize}X}
\newcolumntype{B}{>{\RaggedRight\arraybackslash\hsize=1\hsize}X}
\newcolumntype{C}{>{\RaggedRight\arraybackslash\hsize=0.8\hsize}X}
 \newcolumntype{D}{>{\RaggedRight\arraybackslash\hsize=2.15\hsize}X}

\Title{
 Jet Feedback on kpc Scales: A Review
}

\TitleCitation{Jet Feedback on kpc Scales: A Review}
 
\Author{Dipanjan Mukherjee 
 \orcidA{} }

\AuthorNames{Dipanjan Mukherjee}

\AuthorCitation{Mukherjee, D.}

\address[1]{Inter-University Centre for Astronomy and Astrophysics, Post Bag-4, Pune University, Ganeshkhind, \mbox{Pune 411007, India}; dipanjan@iucaa.in}

\abstract{Relativistic jets from AGN are an important driver of feedback in galaxies. They 
interact with their environments over a wide range of physical scales during their lifetime, 
and an understanding of these interactions is crucial for unraveling the role of 
supermassive black holes in shaping galaxy evolution. The impact of such jets has been 
traditionally  considered in the context of heating large-scale environments. However, in 
the last few decades, there has been additional focus on the immediate impact of jet 
feedback on the host galaxy itself. In this review, we outline the development of various 
numerical simulations from the onset of research on jets to the present day, where 
sophisticated numerical techniques have been employed to study  jet feedback, including a 
range of physical processes. The jets can act as important agents of energy injection into
a host's ISM, as confirmed in both observations of multi-phase gas as well as in 
simulations. Such interactions have the potential to impact the kinematics of the gas as 
well as star formation. We summarize recent results from simulations of  jet 
feedback on kpc scales and outline the broader implications for observations and galaxy 
evolution. }


\begin{document}


\section{Introduction}
\subsection{A Brief Overview of Classical AGN Feedback}\label{sec:overview}
Feedback from supermassive black holes (SMBH) in large early-type galaxies has been 
 established as a major influencer of galaxy evolution 
\citep{fabian12a,harrison24a}. However, the exact mechanism by which active galactic 
nuclei (AGN) affect the galaxy and its environment---and its implications on the galaxy's properties---is still not settled. From a historical perspective, since the 
advent of X-ray observations of galaxy clusters, cooling flows of gas cooled via thermal 
Bremsstrahlung from the cluster environment \cite{lea73a} have been both postulated 
from theoretical modeling \cite{cowie77a} and observationally \mbox{confirmed 
\cite{fabian77a,fabian94a}}. However, the fate of the  gas cooled below X-ray-emitting 
temperatures ($\lesssim$1--2 keV) was left uncertain due to lack of distinct observational 
signatures \citep{peterson01a,tamura01a}. This prompted considerations of re-heating of 
the gas by some mechanism with feedback from the AGN being a viable 
source~\citep{croton06a,fabian12a}. The early concept of AGN feedback was
primarily proposed to investigate two major implications: {(}
a) the well-known $M-\sigma$ relation due to the co-evolution of the SMBH and 
the galaxy core \cite{silk98a} and {(}b) a heating mechanism to offset the over-cooling of 
the cluster cores \cite{croton06a,bower06a}. These two different tracks eventually led to 
the 
evolution of the concept of dual-mode feedback by AGN: {(}a)  \emph{{Quasar}}  or 
\emph{{Establishment}} mode---related to the local impact of AGN-driven outflows 
and the
co-evolution of the SMBH and galaxy mass---and {(}b) \emph{{Radio} 
}  or \emph{{Maintenance}} mode---catering to the large-scale heating of gas reservoirs 
external to the galaxy {and regulating galaxy growth by preventing cooling flows}.   In this 
dual-mode scenario, the role of relativistic jets has been largely confined to their impact 
on extra-galactic gas in the \emph{{Radio/Maintenance}} mode, whereas 
non-relativistic winds in high-Eddington ratio systems have been considered to be the 
primary driver of \emph{{Quasar/Establishment}} mode feedback. However, in recent 
decades, a large body of literature has demonstrated, from both theory and observations, 
that jets can have a significant impact on the ISM of the host galaxy. This makes the earlier 
dual-mode distinction ambiguous in some cases, requiring rethinking of the traditional 
definitions (see Harrison et al. 2024 \citep{harrison24a} for a discussion). 

\subsection{Scope of the Current Review}\label{sec.scope}
 Over the last few decades, there have been some excellent reviews on different aspects of 
 the topic of relativistic jets and their feedback by various authors. However, their scopes
 and focuses have been different and often non-overlapping. {Some of the recent 
 comprehensive reviews in this domain can be placed in the following broad groups:
 \begin{itemize}
     \item Advances in the physics of relativistic jets themselves (e.g., Blandford et al. 2019 
     \cite{blandford19a}) or their simulations (e.g., Marti 2019 \cite{marti19a}, Komissarov 
     and Porth 2021 \cite{komissarov21a}, and Perucho 2023 \cite{perucho23a}).
     \item Astrophysical implications of jets and outflows in general (e.g., Veilleux et al. 
     2020 \cite{veilleux20a} and Laha et al. 2021 \cite{laha21a}).
     \item AGN feedback and its implications (e.g., Fabian 2012 \cite{fabian12a}, Harrison 
     2017 \cite{harrison17a}, Morganti 2017 \cite{morganti17a}, Eckert et al. 2021 
     \cite{eckert21a}, Combes 2021 \cite{combes21a}, Bourne and Yang 2023 \cite{yang23a}, 
     and Harrison and Ramos Almeida 2024 \cite{harrison24a}).
     \item The varied nature of AGN sources and their radio-loud counterparts (such as 
     Tadhunter 2016 \cite{tadhunter16a}, O'dea and Saikia 2021 \cite{odea21a}, Hardcastle 
     and Croston 2020 \cite{hardcastle20a}, and Baldi \mbox{2023 \cite{baldi23a}).}
     \item Gas in and around AGN host galaxies leading to feeding and feedback (e.g., 
     Morganti and Oosterloo 2018 \citet{morganti18a}, Storchi-Bergmann 2019 
     \cite{storchiBergmann19a}, Gaspari et al. 2020 \cite{gaspari20a}, and Combes 2023 
     \cite{combes23a}).
     \item  The episodic nature of AGN outbursts and duty cycles. (e.g., Morganti 2017 
     \cite{morganti17a}).     
 \end{itemize}}
 The above works provide broad overviews of the various complex astrophysical processes 
 related to the topics of feedback and galaxy evolution.  However,  
 only a few detailed reviews  have discussed the complex issues regarding the 
 interactions of 
 such outflows, specifically jets, with the host galaxy itself (such as Wagner et al. 2016 
 \cite{wagner16a}, Mukherjee et al. 2021 \cite{mukherjee21b}, Morganti et al. 2023 
 \cite{morganti23a}, and Krause et al. 2023~\cite{krause23a}). In this review, we discuss the simulation techniques developed over the past few decades    for studying AGN jets in general and jet--ISM interaction in particular. We also summarize the status of 
 observational studies of jet--ISM interaction and their  implications for galaxy evolution. The review is not meant to be a 
 comprehensive summary of all accumulated results to date. Rather, it highlights  the major achievements in this field and their historical 
 developments, to place them in the broader context  of AGN feedback and  galaxy evolution.

\section{Modeling Jet-Driven Feedback at Galactic Scales}
\label{sec:small_scales}
\subsection{Jets in Homogeneous Medium}\label{sec.homog}

\begin{figure}[H]
	
	\includegraphics[width=0.9\linewidth]{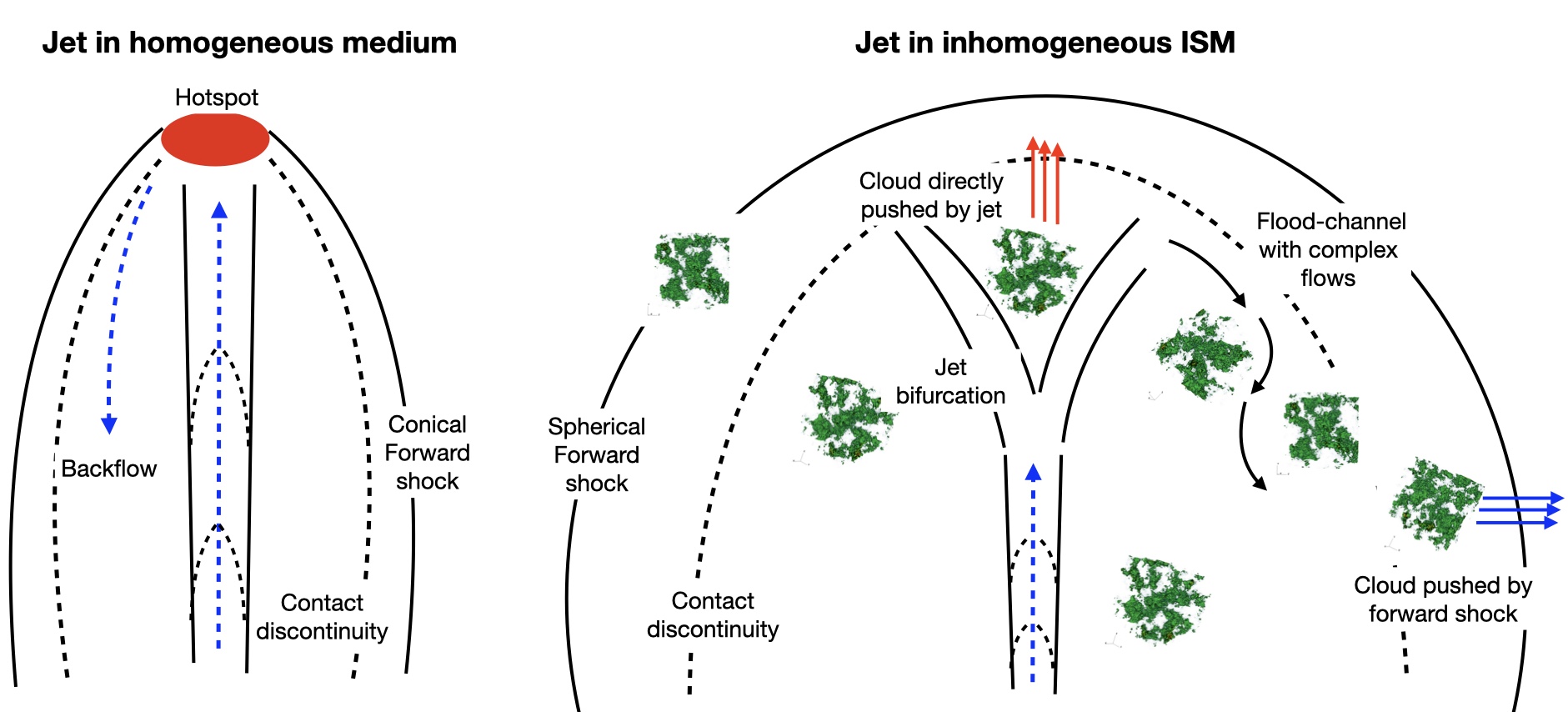}
	\caption{{A} 
		cartoon of a jet and its cocoon evolving  in a homogeneous medium (\textbf{left}) and 
		clumpy ISM (\textbf{right}). The jet in a smooth homogeneous medium has a 
		collimated beam with recollimation shocks, a conical forward shock, followed by 
		contact discontinuity corresponding to the density jump between the cocoon filled by 
		the non-thermal jet material and the swept-up gas from the external medium. A jet in 
		an inhomogeneous ISM results in a more spherical-shaped forward shock as the jet 
		beam is trapped by intervening clouds. The jet material is channeled through gaps 
		between clouds  (`flood-channel' phase \citep{sutherland07a}). Clouds directly in the 
		path of jet beam are more strongly impacted. Clouds embedded in the evolving 
		forward shock on the sides face lower shock velocities. See Section~\ref{sec.evol} for 
		more details on the confined phase.}
	\label{fig.cartoon}
\end{figure}

In the mid and late 1970s, there were several seminal  theoretical models 
to explain the dynamics and emission from extra-galactic relativistic jets, such as the 
`twin-exhaust' \cite{blandford74a} and beam models \cite{scheuer74a}, the Blandford--Znajek (B-Z) jet launch 
mechanism \cite{blandford77a} and diffusive shock acceleration  \cite{blandford78a}, which helped shape the future study of jets and non-thermal emission. 
Attempts at simulating such jet beams were made even at early stages as well, although 
with limited resolution \citep{rayburn77a}. The first detailed 2D simulations exploring the structures of
 hypersonic jet beams  were published nearly simultaneously in 1982 by
\citet{yokosawa82a} and \citet{norman82a},
with the latter paper being more widely recognized in the literature. 
The \citet{yokosawa82a} paper showed that the nature of the jet beam (ballistic vs. 
turbulent) and formation of well-defined backflows depend on the jet and the  
contrast of density between the jet and ambient media. \citet{norman82a} presented a more 
detailed description of the structure of jet beams with features such as a 
working surface and backflow, as proposed in Blandford and Rees 1974 
\cite{blandford74a} (see Figure~\ref{fig.cartoon}). These papers spawned several other
numerical works that probed different aspects of the dynamics of supersonic jet beams, such as 
beamed synchrotron emission \citep{wilson83a}, 3D generalization 
\citep{williams84a,arnold86a,hardee92,norman93a},  stability of slab jets 
\citep{norman88a}, and MHD simulations \citep{clarke86a,clarke89a}. Future works have 
built on the early success of such  numerical simulations with larger domains, grid sizes 
and resolution, although true convergence of the cocoon and beam structures remain 
elusive \citep{koessl88a,mignone10a} due to the small-scale structures generated with an 
increase in resolution.

The relativistic nature of jet flows had been inferred in the 1970s from observations of 
superluminal motion of jet knots  in radio studies  (e.g., 
\citep{cohen77a,cohen79a,O'dell78a,blandford77b,blandford79b}). Numerical simulations 
of steady-state jets with relativistic solvers had been presented as early as  1987 
by~\citet{wilson87b}. Full-fledged dynamic simulations with relativistic solvers followed in the next decade 
\citep[e.g.][]{vanputten93a,marti94a,marti95a,marti97a,duncan94a,koide96a,rosen99a}. A 
key focus of these simulations was to reconfirm the proposed model of the jet 
structure in the relativistic limit and explore the dependence of the jet's dynamics on various properties of the jet. 
They established that the internal structures of relativistic jets show significant 
dependence on the Mach number of the jet beam, bulk Lorentz factor and internal pressure. 
Highly relativistic flows are more stable due to longer growth times of instabilities \cite{bodo13a}. On the other hand, jets with progressively lower Mach 
numbers show markedly different behavior, varying from more 
internal structures in warm jets ($\mathcal{M} \sim 2$)  to  stable cocoons for  
hotter jets ($\mathcal{M} \lesssim 1.6$) \cite{rosen99a}. This results  either because small-scale perturbations 
are ill-resolved by the numerical grid or because KH instabilities couple poorly with the 
jet flow. Attempts at simulating magnetized relativistic jets have been made in 
tandem as well (e.g., \citep{vanputten96a,koide96a,nishikawa97a,nishikawa98a}). 
However, such simulations received more momentum with the development of 
efficient high-resolution shock-capturing schemes in the late 1990s and early 2000s (see, for 
example, 
\citep{komissarov99a,komissarov99b,koldoba02a,delzanna02a,delzanna03a,leismann05a,mignone05a,mignone06a}
 and references~therein). 

In later years, advances in computational power and high-order numerical schemes 
{(see \citet{marti03a} for a review)} have led to a wide range of AGN jet simulations. Such works have 
probed a diverse range of topics, such as  the impact of fluid instabilities on the jet and cocoon 
structures 
{(for some recent examples see} 
\cite{mignone10a,mukherjee20a,meenakshi23a,mattia23a,perucho23a,rossi24a,upreti24a,costa24a,costa25a}),
 the origin of turbulent structures and dynamics of lower-power FR-I jets 
(e.g., 
\citep{perucho07a,rossi08a,perucho14b,massaglia16a,massaglia19a,rossi20a,massaglia22a,bhattacharjee24a}),
and the formation of their FR-II counterparts (such as 
\citep{perucho19a,seo21a,perucho22a}), to name a few. Another major focus has been to 
compare the jet dynamics with predictions from semi-analytical models of jet evolution (e.g., 
\cite{begelman89a,kaiser97a}). Several papers have proposed that jets undergo a self-similar expansion \citep{falle91a,kaiser97a,perucho19a}, although such an assumption may not hold for the entire life span of the jet \citep{carvalho02a,o'neill05a}, especially 
for the early phase of evolution \citep{mukherjee20a}. In many cases, however, the general scaling laws 
predicted by \citet{begelman89a}, duly modified for a power-law ambient density profile, show a good agreement with simulations \citep{perucho19a,mukherjee20a}.

Simulations of large-scale jets have been further driven by efforts to understand the impact of 
jets on cluster scale environment and the resultant non-thermal emission 
(e.g., 
\citep{perucho11a,perucho14a,hardcastle13b,hardcastle14a,english16a,english19a,chen19a,jerrim24a,giri25a}). Such studies have given detailed results of the dynamics of large-scale jets, 
energy transfer to the environment, evolution of synchrotron surface brightness and 
polarization characteristics as a function of jet length. These have further motivated 
simulation-based scaling laws relating the synchrotron power to the jet's mechanical 
power \citep{hardcastle18a}. In recent years, simulations of jets have 
also been utilized to address other science goals such as production of ultra-high-energy 
cosmic rays from shocks driven by jets \citep{matthews19a,seo23a,seo24a}, 
impact of multi-species fluid on jet dynamics and emission (e.g., 
\citep{ohmura23a,ohmura23b,joshi23a}), jet precession and production of X-shaped 
structures 
(\citep{rossi17a,nawaz14a,nawaz16a,horton20a,giri22a,giri23a,giri24a}),
impact of in situ particle acceleration of non-thermal electrons 
(e.g., 
\citep{tregillis01a,tregillis04a,vaidya18a,mukherjee21a,meenakshi24a,chen23a,dubey23a,dubey24a}
), etc., demonstrating the diverse areas of interest on this topic.

\subsection{Jets in Inhomogeneous Medium}
\subsubsection{Non-Relativistic Simulations}\label{sec.NRsim}
\emph{\underline{{The early phase:}
}}  
The earliest suggestion of jets interacting with intervening gas clouds was 
proposed  {as early as 1979, to explain the observed variability of jet 
emission~\citep{blandford79a} or the knots in the jet of M87 
\citep{blandford79b}}
. {Although future works revealed such knots in M87 \cite{schreier82a,biretta83a,falle85a,biretta95a,perlman99a} and in others, 
 		such as Cen A, to arise from mechanisms related to 
 		hydrodynamics of the jet itself 		\cite{,kraft02a,hardcastle03b,worrall09a,Bogensberger24a}, static shocks in 
 		Cen A have still been attributed to possible obstructions in the jet's 
 		path (e.g., from dense clouds) \cite{hardcastle03a}.} Later, in the 1990s, studies of 
Gigahertz-Peaked Spectrum (GPS), Compact Steep Spectrum (CSS) or Compact Symmetric 
object (CSO)  
\citep{begelman96a,bicknell97a,odea98a,odea21a}, further renewed interest in the jet's impact on its dense environment. Free-free absorption by intervening 
ionized gas, either as swept-up matter in the forward shock or pre-existing clouds engulfed 
by the evolving bubble of a radio jet~\citep{bicknell97a,begelman99a}, was  proposed  to explain the turnover in the radio spectrum of such sources. This motivated 
several theoretical simulations to probe the evolutionary stages of a jet through the host's 
ISM, as further outlined below.

\emph{\underline{{Jet-single cloud interactions:}}}
Some of the earliest 2D simulations of  jets drilling through an inhomogeneous ISM were by DeYoung 1993
\cite{deyoung93a} and Steffen et al. 1997 \cite{steffen97b}, who  considered a random distribution of spherical (or 
point-like) dense structures to mimic an inhomogeneous ISM. However, several key 
aspects of the physics of jet--ISM interaction were first elucidated by  simpler  
configurations of jets piercing an oblique density discontinuity 
\citep{hooda96a,hooda98a,hughes02a}.  These papers highlighted that such interactions can tilt the jet's Mach disc 
 and disrupt the regular jet-cocoon structure into turbulent vortices. These results gave an early hint of disruption of the jet beam, which was further demonstrated by more complex simulations later on. 

Similarly, other works have 
explored the impact of jets on individual spherical clouds, 
(e.g., \citep{higgins99a,wang2000a,witta04a,choi07a,tortora09a,antonuccio08a}), or on multiple randomly distributed 
clouds (\mbox{e.g., \citep{tortora09a,dutta24a}}),  an extension of the~\mbox{\citet{deyoung93a}} setup. These so called `cloud-crushing' simulations\endnote{Simulations 
of gas clouds in a wind are often informally referred to as `cloud-crushing' experiments \cite{seidl25a}. The nomenclature arises as they investigate the survivability of clouds embedded inside inside gas flows. They have been performed 
more widely  in the context of more gentler star formation-driven outflows (e.g., see
\citep{klein94a,cooper09a,pittard10a, 
scannapieco15a,bandabarragan16a,pittard16a,bandabarragan18a,gronke18a,cottle20a}, and 
references therein). The basic physics and results from such simulations also holds true  for 
AGN-driven winds, which however are hotter, and have higher velocities than star formation driven outflows.} were an important first step 
in understanding how jets propagate through an inhomogeneous ISM, besides addressing other related questions, such as the  origin of bent  radio jets  (wide-angle tailed sources) or  
asymmetric hybrid jet morphologies, etc. Subsequent studies  included increasingly complex physical processes such 
as atomic and molecular \mbox{cooling \citep{fragile04a,fragile17a}}, idealized set-ups of 
shear layers and mixing \citep{krause07a}, self-gravity and star formation 
\citep{dugan17a,gardner17a}, and realistic morphological models of inhomogeneous 
molecular clouds~\citep{mandal24a,lauzikas24a}. 

\begin{figure}[H]
	
	\includegraphics[width=0.95\linewidth]{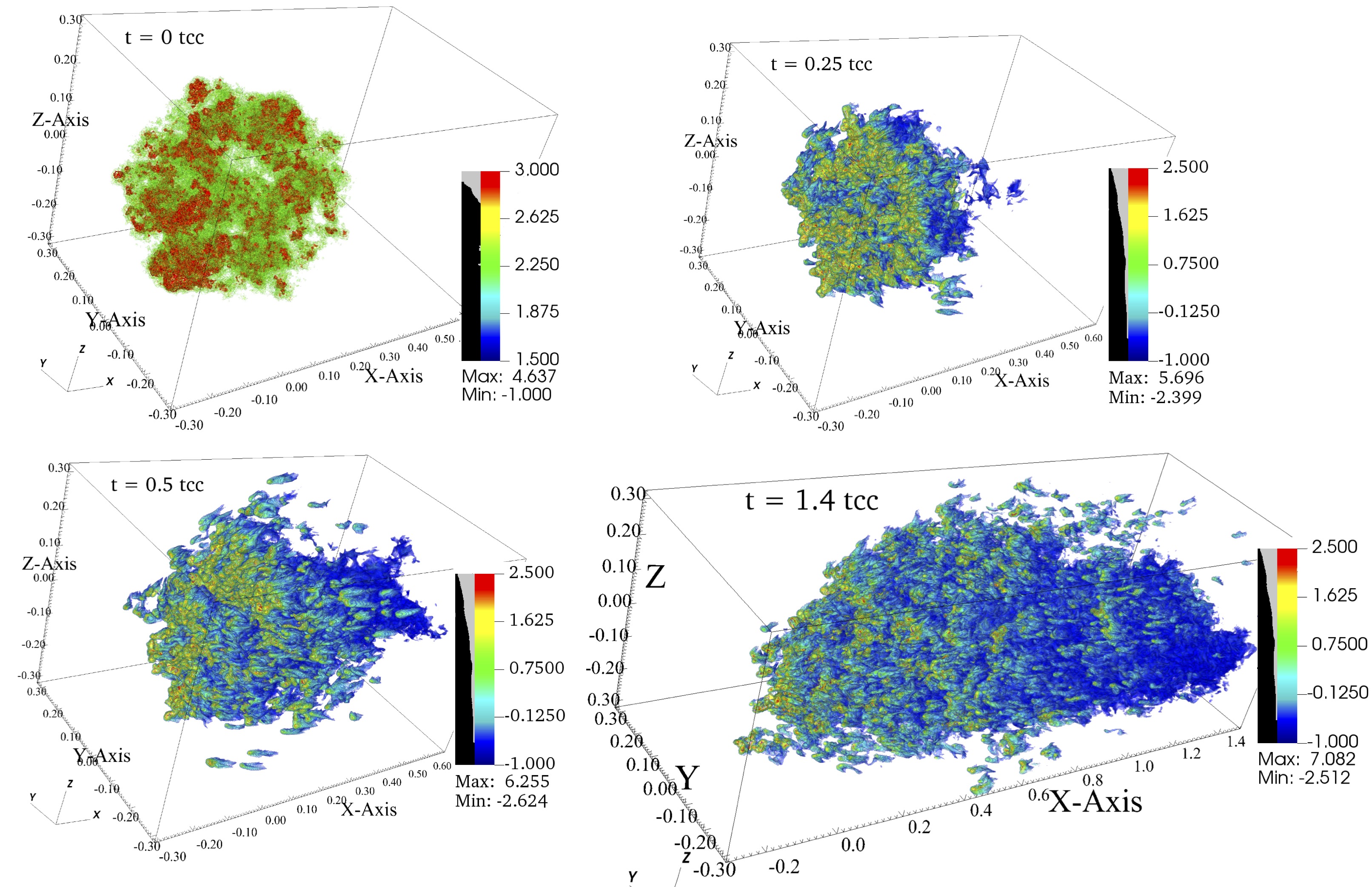}
	\caption{{Three-dimensional} 
		visualizations of density distribution of a fractal cloud being impacted by an AGN-driven wind, from the simulation GC45\_K3 of \citet{mandal24a}. The top right panel 
		corresponds to the initial compression phase, which is followed by the onset of 
		ablation due to Kelvin--Helmholtz instabilities and shear flows (lower left panel). 
		The cloud is seen to eventually disperse several mini-cloudlets, which are swept up with the flow, forming extended cometary tails. Such detailed interactions and micro-structures are 
		usually missed in global simulations of jet--ISM interaction due to inadequate 
		resolution.}
	\label{fig.ankushfig}
\end{figure}

Resolved simulations of jet--cloud interaction
are insightful in providing 
the details of how the jets/outflows are affected by the presence of a cloud 
\citep{hughes02a,wang2000a,jeyakumar09a,nolting22a}, and more importantly, the various 
evolutionary stages of the clouds themselves (e.g., see Figure~\ref{fig.ankushfig}). { Since they focus on a single cloud, such simulations often have sufficient resolution (at least $\gtrsim$120 volume 
elements required across a cloud for convergence \citep{klein94a,bandabarragan16a}) to capture the }
different fluid instabilities operating at jet-cloud interfaces~\citep{fragile04a, mandal24a, lauzikas24a}, which otherwise become difficult 
to follow on global scales. Recent works have also included upgraded models of  star 
formation  \citep{mandal24a,lauzikas24a} to quantify the positive/negative feedback that may result from the  radiative shocks driven inside such clouds by the AGN outflows. However, a limitation of 
such individual jet--cloud simulations is that they do not probe the global 
impact on the ISM at larger scales, and the evolutionary stages of the jet through the inhomogeneous medium.

\emph{\underline{{The first studies of Jets in fractal inhomogeneous medium:}}} A 
separate line of simulations probed the passage of jets through a large-scale inhomogeneous ISM,  a more realistic 
depiction than the early works of \citet{deyoung93a,steffen97b}. Although such 
simulations have relatively moderate resolution ($\sim$10--20 cells across a cloud  
\citep{wagner11a,mukherjee18b}) than the previously mentioned single jet--cloud works, 
they  probe the global impact of the outflow on the turbulent structures in the central few 
kpc of the galaxy. Hence, such studies are a bridge between the highly resolved 
jet--ISM interaction of single clouds and large-scale cosmological simulations with much 
poorer resolutions [$\sim$100 pc], which cannot capture the internal structures of 
molecular clouds in any detail. 

A key new detail of such simulations was the use of 
a fractal density distribution as a realistic model of the ISM. Two-dimensional simulations of jets through 
a fractal ISM were first introduced in the early 2000s \citep{bicknell03a,saxton05a}.  The 
first detailed 3D simulations  were presented in Sutherland and Bicknell 2007
(\cite{sutherland07a}, hereafter SB07), where a non-relativistic jet was injected through a 
two-phase ISM. 
This was a pioneering paper in many aspects. It laid  the technical foundation for 
several future publications and also elucidated the basic evolutionary stages of a jet 
breaking out through an inhomogeneous ISM. These studies were later improved upon by 
relativistic simulations of jet--ISM interaction, summarized in the next section 
(Section~\ref{sec.rel}).

\emph{\underline{{Other non-relativistic jet--ISM simulations with improved 
physics:}}} Besides the SB07 study, several other papers  have explored 
different aspects of jet--ISM interaction through non-relativistic simulations. 
In one such series,  Gaibler et al. (2011,2012), \cite{gaibler11a,gaibler12a} and  Dugan et al. (2014,2017)  \cite{dugan14a,dugan17a}, simulated jet feedback on a large inhomogeneous gas disk \mbox{($\sim$30 $\kpc$}), in contrast to the smaller disks ($\sim1-2$ kpc)  
considered in other studies  
\citep{sutherland07a,mukherjee16a,mukherjee18b,tanner22a}. However, the larger scales 
necessitated a more modest resolution of $\Delta x \sim$67 pc, which was more than 10 times poorer than that of  
other similar simulations. Nonetheless, these simulations truly probed the 
global impact of an AGN jet on a large-scale disk. 
They outlined several important results, such as (a) asymmetric jet morphologies due 
to local density inhomogeneities~\citep{gaibler11a},  (b) large-scale compression-driven positive 
feedback~\citep{gaibler12a}, (c) a ring-shaped region with  enhanced 
SFR surrounding the central cavity, (d) appearance of hyper-velocity stars with strong 
non-circular velocities \citep{dugan14a}, (e) comparison of jet vs. wind-driven feedback, and (f)
impact of jet orientation \citep{dugan17a} on feedback efficiency,  a precursor to the later study by 
\citet{mukherjee18b}. The simulations provided observational predictions of active jet--ISM interaction, and provided diagnostics to identify past such activities in the form of disturbed 
stellar kinematics. While the above papers considered a preexisting fractal density 
 as an initial condition, other recent works have also explored the generation of a self-consistent 
inhomogeneous ISM by stellar feedback, before jet launch (e.g.,\cite{clavijo24a}). However, most of the 
above studies have only performed hydrodynamic simulations. 
Only a handful of papers \cite{asahina17a,clavijo24a} have explored the impact of 
magnetic fields on jet--cloud interaction, which remains an area to be explored in the future.

In recent years, a more  self-consistent treatment of  evolution jets within their environment 
has been carried out in another series of publications (e.g., 
\citet{fiacconi18a}, \citet{talbot21a,talbot22a,talbot24a}). These works presented a novel 
sub-grid prescription for black hole accretion and ejection,  based on a thin accretion disk model, duly 
accounting for variation of the accretion disk's mass and the angular momentum 
exchange between the in-falling gas, the black hole and the accretion disk. The jet power is determined 
by the Blandford--Znajek mechanism, with the efficiency parameterized from GRMHD 
results \citep{fiacconi18a,talbot21a}. The innovative model has been employed to study  
the mutual evolution of jets and a sub-kpc circum-nuclear disk (of radius $\sim 70$ pc and 
height $\sim 9$ pc) \citep{talbot21a,talbot22a}, as well as larger kpc scale gas disks \cite{talbot24a}. The \citet{talbot21a,talbot22a} studies probed outflows from 
lower-mass black holes ($\sim$$10^6 M_\odot$), with jets of low kinetic 
power ($\sim$$10^{42}\ergs$), whereas higher-power jets were explored in the subsequent paper \citep{talbot24a}.

Although the parameter space explored in these simulations is more representative of Seyfert galaxies than typical massive radio-loud AGN, the qualitative results are similar to other studies of jet--ISM interaction (see Section~\ref{sec.results} 
for a general summary). However, one of the key outcomes of the \citet{talbot22a} work, not well explored in earlier simulations, is 
the self-consistent evolution of the jet angular momentum, including reorientation of an inclined jet, driven by the Bardeen--Peterson effect. The simulations also predicted significant 
cold ($T<10^4$ K) outflows from the circum-nuclear disk, with  increased rates for 
inclined jets, in good agreement with observations \citep{garcia14a,garcia21a}. 
Although the authors did not find significant evolution of the black hole spin during a single 
outburst, the generality of the method makes it suitable for implementation 
in large-scale cosmological simulations. This  has been demonstrated in Talbot et al. \cite{talbot24a}, where the framework was used to trace black hole growth and feedback over cosmic time.

\subsubsection{Relativistic Simulations}\label{sec.rel}
\emph{\underline{{Why relativistic hydrodynamics?}}} A drawback of using a non-relativistic 
framework for simulating AGN jets that are inherently relativistic in their bulk flows is 
the difference in the momentum exchange with the external environment. The momentum 
conservation equation in relativistic hydrodynamics {is} 
\begin{equation}
	\frac{\partial }{\partial t}\left(\gamma^2 \rho h \mathbf{v} \right) + 
	\boldsymbol{\nabla}\cdot\left(\gamma^2 \rho h \mathbf{v} \mathbf{v} + \mathbf{I} 
	p\right) = 0,
\end{equation}
where $\rho h = \rho c^2 + \rho \epsilon + p$ is the relativistic enthalpy, $\rho \epsilon$  
the internal energy, and $\mathbf{I}$ an identity tensor. The Lorentz factor of the flow 
velocity is  $\gamma$. For a jet with a given rest-frame density, pressure and bulk 
velocity, a relativistic formulation implies higher momentum  imparted by the jet beam, at 
least by a factor of $\gamma^2$. For example, the difference ($\gamma^2-1$) becomes 
$\sim$10\% even for mildly relativistic flows of $\beta \sim 0.3c$ ($\gamma \sim 
1.05$).  Thus, non-relativistic fluid dynamics will undervalue the 
momentum imparted by the jet. Of course, the total energy flux of a relativistic 
and non-relativistic jet, with identical fluid parameters, is not the same. Hence, 
non-relativistic simulations of jets have often employed an equivalent jet beam by 
fixing the jet pressure, velocity and injection radius to values suited for a non-relativistic simulation, but deriving the jet
density to match the total energy flux. The relations between the densities for such an 
equivalent non-relativistic  ($\rho_{\rm nr}$) and its relativistic counterpart ($\rho_r$) 
were derived by \citet{komissarov96a} as  
\begin{equation}
	\rho_{\rm nr} = 2\rho_r \gamma^2\left(\frac{\gamma}{\gamma + 1} + \frac{1}{\chi}   
	\right) \;\;,\;\; \chi = \frac{\rho c^2}{\rho \epsilon + p} = \frac{(\Gamma - 
	1)}{\Gamma }\frac{\rho c^2}{p}, \label{eq.norelrho}
\end{equation}
where $\chi$ is the ratio of the rest mass energy and the non-relativistic part of the 
rest-frame enthalpy ($\rho h - \rho c^2$). Equation~(\ref{eq.norelrho}) shows that a flux-matched  non-relativistic jet has a higher density, and hence a heavier jet. This results in 
narrower jet cocoons, faster jet propagation, higher mach numbers 
\citep{komissarov96a,rosen99a} and lower cavity pressures \citep{perucho17a} than that 
of relativistic jets. Thus, irrespective of the choice of initial jet parameters, whether done 
by strictly ignoring relativistic effects, or by deriving effective jet parameters by matching 
fluxes, the momentum balance is strongly affected by the neglect of relativistic solvers 
while evolving AGN jets \citep{komissarov96a,rosen99a,komissarov21a}. However, the 
accuracy of the momentum exchange is crucial for the physics of jet--ISM interaction and 
the implications for local-scale AGN feedback effects by the jets. This necessitates the usage 
of relativistic solvers in simulations of jet feedback. 

\emph{\underline{{Relativistic Jets in a static fractal ISM:}}} The first such simulations were 
presented in Wagner and Bicknell 2011~\cite{wagner11a} and later expanded with a larger set of simulations in Wagner et al. 2012~\cite{wagner12a}, which probed different volume filling factors of the dense gas. These 
simulations modeled a relativistic jet ploughing through a static fractal ISM 
\citep{sutherland07a}, immersed in a constant-density background halo. The simulations 
did not have an external gravitational field of the galaxy. The dense ISM was assumed to 
be distributed spherically, unlike SB07 \citep{sutherland07a}, who considered a disk. The 
suite of simulations probed several different parameters of the simulations relevant for 
studying jet--ISM interaction, such as {(}i) jet power: $10^{43}$--$10^{46} \ergs$,  {(}ii) 
mean cloud density: $10^2$--$10^3 \cc$, {(}iii) volume filling factor\endnote{The volume filling factor is 
defined as $f_V = \int_{\rho_{\rm crit}}^\infty p(\rho) d\rho$, with $\rho_{\rm 
crit}/\mu = n_h T_h/(n_{w0} T_{\rm crit})$. Here $p(\rho)$ is the density probability 
distribution function (PDF). $T_{\rm crit}$ is the critical temperature of the dense clouds, 
beyond which the fractal density is replaced by the halo gas in the simulation. Since the 
dense clouds are considered to be in pressure equilibrium with the halo gas, the $T_{\rm 
crit}$ essentially implies a lower cut-off of the lognormal density PDF ($\rho_{\rm 
crit}$).}: $f_V \sim$ 0.027--0.4 and {(}iv) cloud sizes: 10--50 pc (see Table 2 of 
\citet{wagner12a}). All the simulations were run for a jet of $\gamma = 10$ and $\chi = 
1.6$ (see Equation~(\ref{eq.norelrho})). 

These simulations, along with SB07 \citep{sutherland07a}, were the first to explicitly 
identify the various  stages of a jet's evolution, as it channels through an inhomogeneous 
ISM (see Section~\ref{sec.evol} for more details). They examined the effect of the 
jet on ablation and acceleration of ISM clouds, the dynamics of the jet-driven bubble, and the resulting impact 
on ISM energetics, placing these processes in the broader context of AGN feedback in 
galaxies. These results provided definitive proof that jets can significantly impact the host's ISM, which has been further expanded upon and confirmed in future 
simulations. Additionally, these papers 
highlighted two other  impacts of jets, not often highlighted in other studies: 
\begin{itemize}
	\item  Moderately powerful jets ($P_{\rm jet} \gtrsim 10^{43}\ergs$) can potentially clear an ISM with small-sized clouds ($\lambda \lesssim$ 10--20 pc). Such jets are capable of driving outflows with mean radial velocities higher than the stellar velocity dispersion, even for modest values of Eddington ratios ($\eta = P_{\rm jet}/L_{\rm Edd} 
	\sim 10^{-4}$--$10^{-3}$), thereby indicating successful negative feedback. In contrast, sufficiently accelerating larger clouds, such as 
	 Giant Molecular Clouds (GMCs hereafter)  with  sizes of $\gtrsim$50 pc 
	\cite{hughes10a,hughes13a,faesi18a},  require substantially higher Eddington ratio 
	(\mbox{$\eta \gtrsim$ 0.01--0.1)} and jet powers ($P_{\rm jet} \gtrsim 10^{45} 
	\ergs$).  This suggests that such cases require more efficient outbursts from a larger mass SMBH to power global outflows. Thus, larger clouds enhance jet confinement and are also more resilient to ablation, as also confirmed in later works \citep{mukherjee16a}. 
	
	\item The jets were found to provide a {strong  mechanical  advantage (i.e., a value greater than unity), which is  defined as the ratio of total outward momentum of the clouds to the momentum imparted by the jet until a given time (e.g., $P_{\rm jet}t/c$). This enhancement arises because the high-pressure bubble created by the jet accelerates the clouds in addition to the direct ram pressure of the flow. This is similar to the high-momentum boost conjectured for the energy conserving phase of a general AGN-driven outflow \cite{zubovas12a,faucher12a}. 
	} The temporal evolution of the mechanical advantage  correlates with the fraction of kinetic energy transferred to the ISM,  peaking at $\sim $ 20--30\%, which was later refined to slightly lower values by future simulations \citep{mukherjee16a,mukherjee17a}. Overall, these results indicate strong coupling of the jet with the ISM.
\end{itemize}

\emph{\underline{{Jet--ISM interaction in dynamic environments:}}} The earlier studies were extended by four subsequent papers---Mukherjee et al. 2016 \cite{mukherjee16a}, Bicknell et al. 2018 \cite{bicknell18a}, Mukherjee et al. 2018a 
\cite{mukherjee18a} and Mukherjee et al. 2018b \cite{mukherjee18b}---which had several new developments: {(}i) an external gravitational potential 
(double isothermal) with a hydrostatic atmosphere, which enabled more accurate  estimates of gas kinematics (see Section~\ref{sec.kinematics}), {(}ii) initialization of the ISM  with a  turbulent velocity dispersion along with the  fractal density, which added further realism 
to the setups and  enabled successful comparison with  observations
(see Section~\ref{sec.obs}). These simulations have formed the primary benchmark for studies of jet--ISM interactions in recent years. 

The papers listed above mainly considered high-power jets  ($P_{\rm jet} \sim 
10^{44}$--$10^{46}\ergs$). Tanner and Weaver 2022 \mbox{\cite{tanner22a}} extended such simulations to a broader range of powers, including simulations with very low jet powers ($P_{\rm jet} \lesssim 10^{42}\ergs$), compared to what was previously explored. While these simulations broadly
recovered the earlier results \citep{mukherjee18b}, one key distinction was that the higher-power jets ($P_{\rm jet} \gtrsim 10^{44}\ergs$) were not efficiently confined and found to eventually drill through 
the ISM. In contrast, lower-power jets were more prone to disruption and breakup. Such confinement of low-power jets  likely results from the weaker jet momentum flux due to the lower jet density and pressure, which in turn 
causes longer cloud ablation and jet confinement time scales  (see Appendix~\ref{sec.confine} for a discussion on jet confinement).  Similar results have also been reported 
in more recent non-relativistic simulations \citep{borodina25a}. 

\section{Summary of Key Results}\label{sec.results}
Although  results of different simulations may vary due to different choices of  jet parameters, the ambient medium or micro-physics, a common set of general outcomes can be ascertained. In the following sections, we outline the broad summary of some general conclusions and the key results of the simulations of jet--ISM interaction.
\subsection{Evolutionary Stages of the Jet Through an Inhomogeneous 
Medium}\label{sec.evol}
\begin{figure}[H]
	
	\includegraphics[width=1\linewidth]{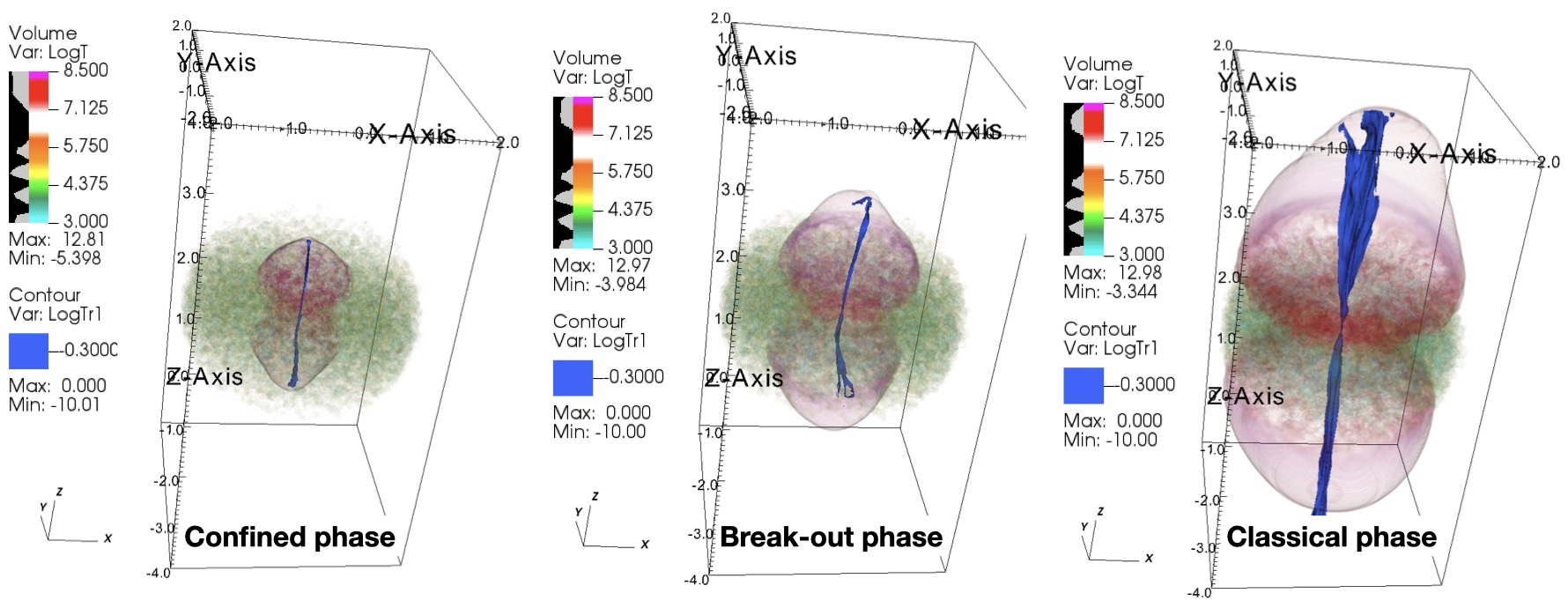}
	\caption{{Evolution} 
		of a jet through a dense kpc scale gas disk, depicting the three phases of evolution 
		outlined in Section~\ref{sec.evol}. The 3D visualizations  show the gas temperature 
		($\log(T)$)  and the jet tracer in blue at different times. The results are from simulation B of 
		Mukherjee et al. 2018b \cite{mukherjee18b}, where a jet of power $P_j=10^{45}\ergs$ is launched perpendicular 
		to the disk plane. The red-colored contours  trace the cocoon of hot gas expanding into 
		the ISM. Post-break-out, the hot pressurized cocoon spreads over the disk and engulfs 
		it from the upper and lower regions. See Sections~\ref{sec.global} and \ref{sec.disp} 
		for broader discussion.  }
	\label{fig.jetphase}
\end{figure}

In general, one can identify a common set of evolutionary phases of a jet moving through as gas-rich ISM (first presented in \citep{sutherland07a}). This is illustrated in 
Figure~\ref{fig.jetphase}.

\begin{itemize}
	\item \emph{{The Confined phase}}: The jet remains confined within the clumpy 
	ISM ($\sim$0.5--1 kpc), resulting in the formation of a \emph{flood-channel} scenario 
	(see right panel of Figure~\ref{fig.cartoon}). The jet plasma is diverted to low density 
	channels through the clouds, through which it percolates into the ISM. The jet beam's forward progress 
	is temporarily halted. However, the backflows from the stalled jet disperse its energy  over a quasi-spherical volume. This creates a highly pressurized energy-driven bubble, enclosed by a forward shock, sweeping through the ambient medium. 
	Simulations find that the timescale  of such a confined phase can last from a few hundred kilo-years 
	to  $\sim$2 Myr, and depends on various factors, such as the jet power, the density of the ambient 
	medium and the  spatial extent of the dense gas. {In fact,  low-power jets may remain  confined 
	 for a long time, without ever evolving to the later stages~\cite{wagner12a,mukherjee16a,tanner22a,borodina25a}.}  An approximate 
	analytical treatment of the duration of confinement is presented in 
	Appendix~\ref{sec.confine}.  Since these conditions can vary significantly between 
	different galaxies, the impact of the jet and the efficiency of coupling with the ISM can have 
	a wide variation as well. 
	
	\item \emph{{Jet breakout phase}}: In this phase, the jet and its resultant hemispherical bubble break free from the dense ISM and evolve further. Although free from the confines of the ISM, the  jet-driven bubble can still indirectly impact the dense gas in the galaxy. 
     The bubble remains 
	 over-pressurized and eventually engulfs the ISM. 
    The combined impact of the bubble and backflows from the tip of the jet drives shocks into clouds away from the jet 
	axis. This enhances turbulence~\citep{mukherjee18b} and impacts star formation in	the inner few kpc of the \mbox{galaxy~\citep{mukherjee18b,bieri16a,mandal21a}}. 
	{Eventually, at late times, as the bubble's pressure decreases due to its expansion, the impact on the dense ISM is weakened.}    
	
	\item \emph{{The classical phase}}: Beyond the breakout phase, the jet carves a 
	clear path through the ISM. Subsequent energy flows have less impact on the ISM. The 
	jet-head proceeds into the low-density stratified  halo gas. Beyond this point, the 
	dynamics of the jet are similar to the conventional models of jet propagation into a static 
	homogeneous medium. The dynamics of the ISM and perturbed velocity dispersion of 
	the clouds start to decay back to the pre-jet levels \citep{mandal21a}.  
\end{itemize}

Of the above stages, the confined phase is of primary interest in the context of AGN 
feedback. During the confined phase, there can be strong coupling of the jet with the dense 
gas, which would result in a {significant} transfer of  the jet's energy flux into the ISM (e.g., 
$\sim$10--20\% for the simulations presented in 
\citep{wagner12a,mukherjee16a,mukherjee17a}) in the form of kinetic energy, which creates 
local outflows  and also  heats the gas via radiative shocks. The efficiency of such an interaction depends  on the following {parameters}:  
\begin{enumerate}
	\item The volume filling factor of the dense gas ($f_V$).
	\item Jet's orientation with respect to the ISM morphology, with jets inclined to a gas disk 
	showing more coupling with the ISM ($\theta_j$).  
	\item Jet power ($P_j$). 
	\item Mean density of the clouds in the ISM ($n_c$). 
\end{enumerate}

{Thus,} 
the efficiency of jet-driven feedback on kpc scales depends on a four-dimensional parameter 
space. The maximal impact, of course, is for a high-power jet, directly oriented into a dense 
ISM (e.g., jets pointed into a gas disk \citep{cielo18a,mukherjee18a,mukherjee18b}), having clouds with high mean density, which results in longer  confinement of the jet. More detailed 
discussion on the impact of parameters 1, 3 and 4 on jet confinement is discussed later in 
Appendix~\ref{sec.confine}. However, the interactions can also be gentle if one of the 
parameters is weak, even though others are prominent. For example, although NGC 3100 
\citep{ruffa19a} hosts a moderately 
powerful jet ($P_j \sim 10^{44}\ergs$) along with a  dense gas disk observed in CO 1-0, the impact of the jet on the disk's kinematics is 
minimal. This is likely due to weak jet--disk coupling,  arising from the  relative orientation of the jet 
away from the disk's plane. However, on the other hand, several detailed spatially 
resolved observations have uncovered more telltale smoking gun signatures of the strong 
impact of the jet on the confining ISM, as discussed later in Section~\ref{sec.obs}. 

\subsection{Global Impact on the ISM}\label{sec.global}
Simulations of jets through the inhomogeneous ISM have strongly supported that jets can 
cause a large-scale effect on the central few kpcs of the galaxy, contrary to earlier beliefs 
that such thin collimated structures are less important in the global context 
\cite{ostriker10a}.  There are again three distinct types of impact of the jet, which has 
varied effects on the ISM.
\begin{enumerate}
	\item \emph{{Direct impact of jet-beam}} ($\lesssim$1 kpc): Clouds directly along 
	the path of the jet are strongly affected by the flow and are eventually destroyed. Such an 
	interaction influences the evolution of both the clouds and the jet. For large clouds  that may nearly cover the jet's width (e.g., a GMC of size 
	$\gtrsim 50$ pc), the 
	jet is  strongly decelerated until the cloud moves away from its path, or is completely 
	disintegrated. The region of such impact is usually confined to $\lesssim$1 kpc, where 
	the jet beam and its ensuing backflow directly interact with the ISM. This region 
	experiences  much higher turbulent velocity dispersion and density enhancement 
	\citep{mandal21a} due to stronger ram pressure-driven shocks. In addition, the stronger 
	interaction in the central region also results in mass removal and formation of a cavity 
	\citep{gaibler12a,mukherjee18b,mandal21a}. However, simulations that better resolve 
	the cloud structures show that such cavities are not completely devoid of dense gas 
	\citep{mukherjee18a,cielo18a}. Strands of dense cloud cores compressed to high densities by radiative shocks 
	remain embedded inside such cavities and are slowly 
	ablated by the jet-driven flows \citep{mukherjee18a,mukherjee18b}.
	
	\item \emph{{Indirect impact by energy bubble}} ($\gtrsim$1 kpc): As discussed 
	 in Section~\ref{sec.evol}, the energy of the confined jet  spreads out in the form of an 
	expanding energy bubble and sweeps through the ISM. The nature of this indirect coupling of the jet and the ambient gas	depends on the evolutionary phase of the jet (see Section~\ref{sec.evol}). During the 
	\emph{{jet-confinement}} phase, the forward shock of the energy bubble sweeps through the ISM. The 
	embedded clouds face a steady  radial outflow of the jet plasma, which is re-directed 
	in lateral directions, away from the jet axis, through the flood-channel mechanism. This results 
	in outward radial flows inside the ISM. In the \emph{{jet-breakout}} phase and beyond, the jet expands beyond the immediate confines along its 
	path and the over-pressured cocoon engulfs the ISM. This is more prominent for gas 
	disks, as shown in the right panel of Figure~\ref{fig.jetphase}. Such indirect interactions 
	are responsible for more large-scale impact of the jet, beyond the central 1 kpc. 
	This raises the velocity dispersion of the gas and also shock heats a large volume of the ISM \citep{meenakshi22a,meenakshi22b}. Inclined jets that remain	strongly confined within the ISM  
	\citep{dugan17a,mukherjee18a,mukherjee18b,tanner22a} are more agents of feedback, as the decelerated jet-head allows plasma to spread and create widespread radial 
	flows.
\end{enumerate}

{The }
fact that jets can, in principle, affect a volume of the ISM larger than their apparent width 
near their launch axis has strong implications for AGN feedback. This demonstrates that 
jets can create strong global outflows, impact gas kinematics and star formation rate of the galaxy, as further
discussed below.

\subsection{Impact on ISM Kinematics}\label{sec.kinematics}
As discussed above, the jets can affect a significant fraction of the ISM during the confined 
and breakout phase, which results in fast multiphase outflows 
\citep{sutherland07a,wagner12a,mukherjee16a,mukherjee17a,mukherjee18a,dugan17a,talbot22a,tanner22a}.
 The extent of the interaction and nature of the outflows depend on the four key factors listed in Section~\ref{sec.evol}. 
We list below some of the broad major inferences that can be drawn from these theoretical 
simulations.

\subsubsection{Multi-phase Outflow:}\label{sec.outflow}

In a realistic system, one would expect a wide range of gas phases to co-exist: (a) dilute hot 
gas in the halo of galaxies, (b) dense gas collisionally ionized by shocks or photoionized by radiation from a 
central source like AGN \citep{meenakshi22a} or shock precursors \citep{sutherland17a}, (c)
warm molecular gas likely representing cooling fronts of shocks or turbulence that have 
penetrated the dense \mbox{gas \citep{nesvadba11a,zovaro19a}}, (d) cold dense molecular 
gas \citep{morganti15a,salome23a}, and (e) neutral (e.g., HI) gas \citep{morganti23a}. However, 
most simulations consider a single fluid system and do not explicitly track the chemical 
evolution of the constituents of the gas, due to the computational complexities. Although 
some recent works have started some preliminary investigations to model differences in 
\mbox{composition \citep{perucho21a,perucho23a,perucho24a}} in jet-gas simulations, such efforts are still in their infancy. 

\begin{figure}[H]
	
	\includegraphics[width=0.8\linewidth,keepaspectratio]{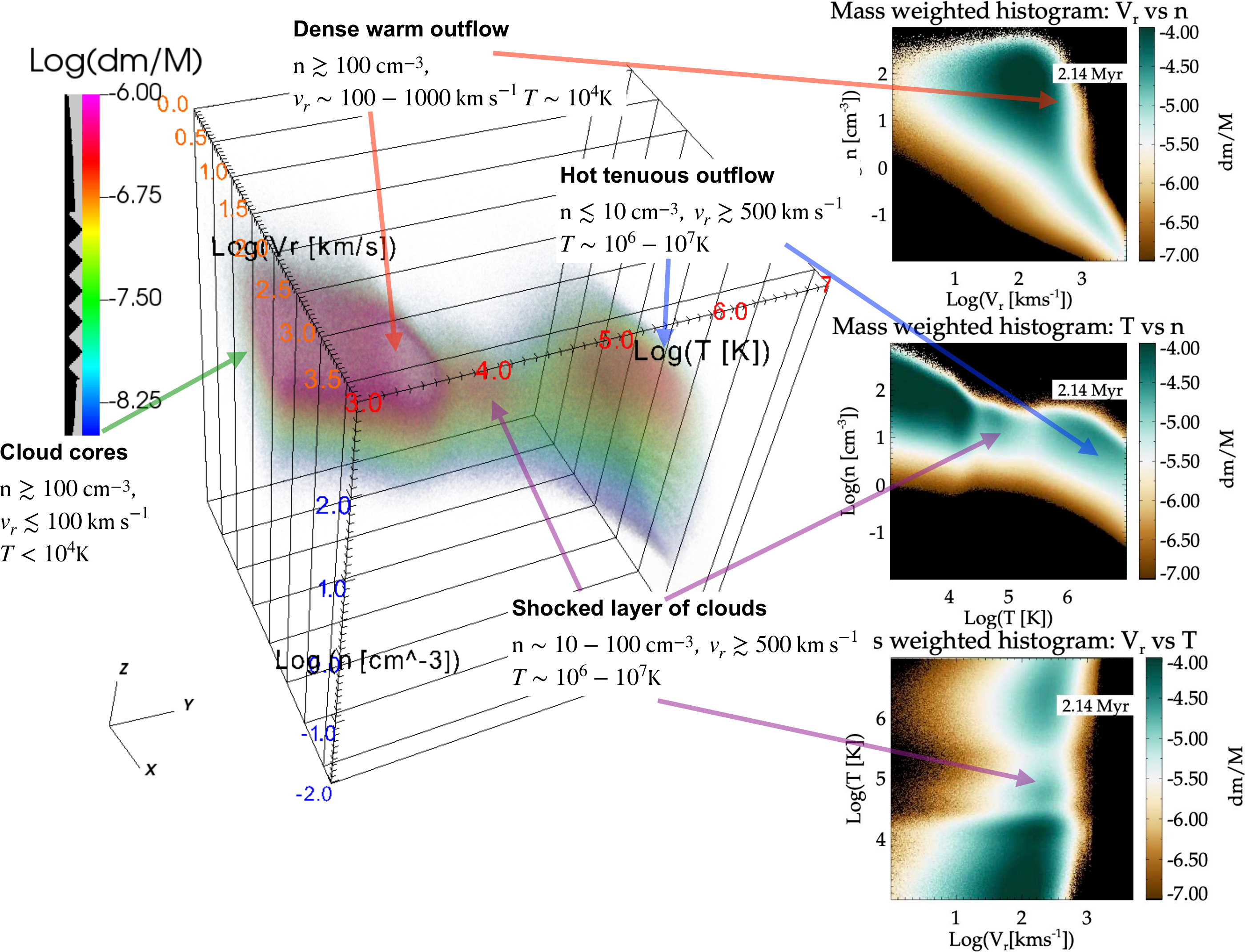}
	\caption{{A} 
		3D visualization of mass distribution as a function of positive radial velocity ($v_r$), 
		density ($n$) and temperature ($T$), to depict the \emph{{multi-phase}} nature of 
		the jet-impacted ISM. The results are from the data corresponding to the last panel of 
		Figure 14 of simulation D from \citep{mukherjee18b}, at 2.14 Myr. Right panels show 
		the corresponding 2D mass distributions obtained by summing the 3D histogram 
		along a chosen axis. Several distinct phases have been be identified. See text in 
		Section~\ref{sec.kinematics} for more details.}
	\label{fig.phasespace}
\end{figure}

Nonetheless, simulations that can resolve the density sub-structures inside an 
inhomogeneous ISM can  track the variation of density, temperature and velocity 
structures of the fluid, which act as a proxy for the different phases 
\citep{mukherjee18a,meenakshi22a,murthy22a}. Some theoretical papers have attempted 
to understand the multi-phase nature of the ISM and disentangle the relative contributions 
of different gas properties in the outflows by evaluating 2D histograms 
\citep{mukherjee16a,mukherjee17a,mandal24a} or analyzing the simulations based on 
temperature thresholds \citep{talbot22a}. A better representation of the multi-dimensional 
nature of the phase space is shown in Figure~\ref{fig.phasespace}, where the mass 
distribution is represented in terms of the three primary variables of interest, viz. density 
($n$), temperature ($T$) and the outward radial velocity ($v_r$) representing the outflowing 
gas. The corresponding 2D distributions  obtained by summing along each axis of the 3D 
distribution are plotted on the right. However, the summed 2D distributions often fail to 
capture the variation in the phase space visible in the 3D image. The plot is for the last panel of 
Figure 14 of simulation D from \citet{mukherjee18b}, which corresponds to a jet of power 
$P_j = 10^{45}\ergs$, launched at $45^\circ$ to the axis of the disk.

Several distinct identifiable regions have been highlighted in the 3D figure 
(Figure~\ref{fig.phasespace}). 
\begin{itemize}
	\item \emph{{Cloud cores:}} There is collection of mass at $T \sim 1000$ K, with 
	high densities near the left face of the 3D figure. This corresponds to the cores of the 
	clouds, with a temperature near the cooling floor of the simulation ($T=10^3$ K). The 
	clouds have some positive radial velocity ($v_r \lesssim 100 \kms$), which likely 
	is a mixture of the turbulent bulk velocity of the clouds injected at the beginning of the simulation and also 	mild acceleration after jet--ISM interaction.
    
	\item \emph{{Dense warm outflow:}} There is a  collection of mass in 
	Figure~\ref{fig.phasespace} that is shifted from the cloud cores,  extending from $T\sim 10^3$ 
	K to $T\sim 10^4$ K in temperature,  \mbox{$v_r \sim$ 100--1000 $\kms$} in velocity 
	and density of $n \gtrsim 100 \cc$. This phase corresponds to the dense shock-heated gas accelerated to high velocities that has now cooled. This phase has the highest mass amongst all of 
	the outflowing gas and is the dominant contributor to the kinetic 
	energy budget of the outflows. In observational studies, this phase would correspond to 
	the warm molecular gas \citep{nesvadba10a,nesvadba11a,collet16a,nesvadba17a} or 
	 the cold gas outflows \citep{morganti15a,murthy22a,murthy24a}, as 
	modeled in \mbox{\citet{mukherjee18a}}. 
	
	{However, one must note that the lack of explicit chemical evolution and molecular cooling 
	(however, see \cite{perucho21a,perucho24a} for recent updates) in these simulations limits  quantitative comparison with such observed 
	phases. Nonetheless, the distinct feature in the above phase space diagram qualitatively 
	indicates the multi-phase nature of the ISM shocked by jets and the probable location of 
	the dense molecular phase in the 3D phase space of the simulated gas distribution.}
	
	\item \emph{{Shocked cloud layers:}} Beyond the dense warm phase, there is 
	another distinct, but small, collection of mass, peaking between $T\sim10^4$--$10^5$ K, 
	and at a lower density (\mbox{$n\sim $ 10--100$ \cc$}) than the dense warm phase. 
	The temperature range  above corresponds to the peak of the cooling curve. This phase is composed of 
	 the outskirts of the clouds being shocked by the enveloping pressure bubble or 
	shocked dense cloud-lets ablated from large clouds 
	\citep{mukherjee18b,meenakshi22b}. It accounts for the majority of the 
	observed emission in optical lines used as diagnostics of shock ionization, such as [OII], 
	[OIII], [SII], etc. \citep{mukherjee18a,meenakshi22b}. It should be noted that the mass 
	represented in this phase is small compared to the dense phase. Hence, masses of
	ionized gas inferred from  such shocked gas are often lower limits to the total ISM mass, which is often difficult to estimate due to the lack of multi-wavelength coverage.
	\item \emph{{Hot tenuous outflow:}}  The jet-driven outflows push out the 
	 gas ablated from the clouds in a tenuous hot form ($n \lesssim 10 \cc$, $T> 10^6$ K). This manifests as an elongated  
	tail in the phase-space distributions of Fig.~\ref{fig.jetphase}, which extends to very low densities and high velocities. Such a hot, 
	tenuous gas can potentially be observed in X-ray wavebands \citep{sutherland07a}. However,
	detecting the soft thermal X-rays from shocked regions, with sufficient spatial resolution  to
	distinguish them from the central nucleus, is challenging, owing to photoionizing radiation from the 
	AGN. Nonetheless, such X-rays from shocked gas have been tentatively confirmed in several sources, {e.g., 3C 171 
	\cite{hardcastle10a}, 3C 305 \cite{hardcastle12a}, PKS 2152-69 \cite{worrall12a}, B2 
	0258+35 \cite{fabbiano22a}. A broader review of such 
	cases, including both jetted and non-jetted AGN, is presented in \citet{fabbiano22b}.  }
\end{itemize}

\subsubsection{Galactic Fountain}
Although the jets can launch strong local outflows, {a blow-out of the major fraction of the 
    ISM, as often required by semi-analytical models of AGN 
    feedback~\cite{croton06a,bower06a}, is not obtained in many of simulations discussed above. Studies investigating resolution dependence of cosmological simulations   
 have shown that simulations with higher resolution retain more gas in the galaxy
\cite{bourne15a}. Hence, simulations that better resolve denser 
ISM structures \cite{mukherjee16a,mukherjee18b}, including those with 
non-relativistic AGN winds \cite{gabor14a,costa20a}, find it difficult to clear out the galaxy. The total mass weighted mean 
velocities are, in fact, negative in some cases \citep{mukherjee16a,mukherjee17a}. Although some recent large-scale simulations have shown 
strong outflows (sometimes reaching up to $\sim$100 kpc \cite{talbot24a}), a fraction 
of the uplifted gas falls back at later stages.  }
{However, it must be noted that the above results depend on various conditions of the ISM, 
such as the mean density, cloud sizes, efficiency of ablation 
\cite{mukherjee18b,wagner16a}, etc., and the power of the outflow. Low-density clouds with smaller sizes may be 
pushed out with higher efficiency. Nonetheless, denser ISM in general provides stronger 
resistance to large-scale ejection. } 

The discussion above indicates that although jet feedback simulations can cause strong localized outflows, a significant 
fraction of the ISM may remain within the gravitational potential of the host galaxy. This is best 
demonstrated by the escape fraction plot in Figure 20 of \citet{mukherjee16a}. It shows 
that only $\lesssim$10\% of the ISM  moves beyond the central few kpc. Such outflows, 
without escape, will give rise to  galactic fountains, where the expelled gas is expected to be 
recycled within the galaxy's confines \citep{mukherjee16a}.

\subsubsection{Turbulent Velocity Dispersion:}\label{sec.disp} 
A key focus of local 
AGN feedback studies in both observational and theoretical domains has been to 
understand the influence of AGN-driven outflows on the turbulence of the ISM. Resolved 
simulations of jet--ISM interactions have shown that as the jet-driven bubble sweeps over 
the central few kpc of the ISM, it can significantly raise the velocity dispersion by an 
order of magnitude from its initial value. However, this seems to depend on the phase of 
the gas. For example, \citet{mukherjee18b} show that hotter shock ionized gas ($T>10^4$ 
K) will have higher-velocity dispersion ($\sim$400--600 $\kms$) than the colder 
component ($\lesssim$100 $\kms$). This is because the shocks progress very slowly 
within the dense cores. The primary impact of the jet-driven bubble is on the ablated 
cloudlets stripped from the larger clouds. Random bulk motions of such clouds add to 
the velocity dispersion. However, the dense gas is not completely undisturbed. Detailed 
comparisons of the kinematics  of dense gas in galaxies such as IC~5063 \citep{morganti15a} 
and B2-0258 \citep{murthy22a,murthy25a} with  simulations 
\citep{mukherjee18a,mukherjee18b,meenakshi22b} have shown excellent correspondence 
with observed gas kinematics. An interesting new feature identified in both 
simulations \citep{mukherjee18a,meenakshi22b} and observations 
\citep{venturi21a,riffel14a,girdhar22a,ulivi24a,ruschelDutra21a} is the appearance of 
high-velocity dispersion in directions perpendicular to the jet. Such features are 
conjectured to arise from deceleration of a jet strongly inclined into the gas disk, resulting 
in outflows of plasma both along the minor axis following the path of least resistance 
\citep{mukherjee18a,meenakshi22b}, as well as into the plane of the gas disk 
\cite{audibert23a}. A combined effect of both types of motions is predicted to result in 
such apparent enhanced widths perpendicular to the jet \cite{meenakshi22b,audibert23a}.

\subsection{Impact on Star Formation Rate}
One of the primary motivations of AGN feedback studies has been to understand the 
impact of AGN-driven outflows on the instantaneous and long-term star formation rate (SFR) of galaxies. In simulations, star formation rates have been predominantly 
estimated as
\begin{equation}
	\mbox{SFR} = \epsilon \frac{\rho}{t_{\rm ff}} \;\, ; \;\; t_{\rm ff} = \left(\frac{3\pi}{32 
	G \rho}\right)^{1/2},
\end{equation}
where $\epsilon$ is often assumed to be a constant efficiency factor ($\lesssim$0.01) and 
$t_{\rm ff}$ is the local free-fall time.  {Both negative and positive feedback scenarios have been 
predicted by simulations.}  Since estimates of star formation rates only depend on the gas density, a reduction in SFR  can be achieved either by  {removing}  gas from a galaxy's potential {or by ablating dense 
 clouds} to lower densities. Similarly, SFR can be enhanced by shock-driven compression of gas 
\cite{wagner16a}. {Some large-scale simulations that probe maintenance mode feedback using 
self-consistent modeling of feedback cycles and star formation have demonstrated that AGN outflows can suppress star formation (e.g.,  for some recent works, see 
\citep{su21a,weinberger23a,yang23a}, and references therein). However, many of these 
studies often are unable resolve a preexisting dense inhomogeneous ISM of the host galaxy. Gas 
removal, and hence reduction in SFR via mass loss, is also more effective for the lower 
resolutions \cite{bourne15a} in such studies, as compared to simulations discussed earlier. Properly testing the impact of outflows on a dense 
inhomogeneous ISM, thus, necessitates resolving the structures on the scales of molecular clouds 
($\sim$50--100 pc).}

Spatially resolved simulations of jet-ISM interaction have predicted a central cavity (hence 
negative feedback) surrounded by a ring of SFR-enhanced region due to compression of gas in 
the immediate rim of the cavity \citep{gaibler12a,dugan17a,mukherjee18b,clavijo24a}. 
Strong compression from the ensuing pressure bubble can potentially  
promote star formation in extended regions of the galaxy  as well \citep{gaibler12a,bieri16a}. Such theoretical 
predictions of positive feedback compliment some observed sources, especially  where star forming streams are found to align with outflows~\citep{salome15a,lacy17a,nesvadba20a,duggal24a}. {Additionally}, several 
radio-loud galaxies are known to have significantly reduced  SFR 
\citep{nesvadba10a,nesvadba21a}. {
The nature of the density structures also determine how the outflows impact the SFR.
 Low-density, smaller clouds are easily ablated, resulting in negative feedback 
\cite{wagner16a,talbot24a}. On the other hand, simulations with higher gas densities, similar 
to the dense cores of molecular clouds ($n \gtrsim 100\cc$), are more resilient  
\cite{gabor14a,costa20a} and often show enhancement of SFR \cite{gaibler12a}.} {Such diverse results  should be explained by a single self-consistent model 
of star formation.}


In a recent study, \citet{mandal21a} proposed a new  method  to 
estimate the SFR in simulations that resolve turbulent gas structures. The work applies the well known
theoretical framework of \emph{turbulence-regulated star formation} in molecular clouds 
\citep{krumholz05a,federrath12a} to estimate the SFR in AGN feedback simulations. The model duly accounts for the variation of local free-fall time as a 
function of the gas density, the virial parameter ($\alpha_{\rm vir} = 2 E_{\rm kin}/E_{\rm 
grav}$) and the Mach number of the gas ($\mathcal{M}$). 
Only those regions whose Jean's length  is lower than the sonic scale ($\lambda_J = \left(\pi c_s^2/(G 
\rho)\right)^{1/2} \lesssim \lambda_s$) are considered to be gravitationally unstable to form stars.  The sonic scale is defined as the length 
scale at which the turbulent velocity dispersion is lower than the local sound speed. At scales 
higher than the sonic scale, the turbulent pressure offsets the gravitational collapse. The above criteria yield a density threshold for star formation that depends on the properties of the local turbulence, which is a  significant upgrade from the simplistic gas-density based prescription outlined above.

The turbulence-regulated method has been used in post-processing to estimate the SFR of jet--ISM interaction simulations by \citet{mandal21a}. In a departure from 
either pure positive or negative feedback, the above approach with improved micro-physics  reveals some new 
aspects. {(}i) There is a mild global reduction in SFR during the onset of the jet--ISM 
interaction, {(}ii) inefficient positive feedback occurs in the inner regions directly impacted 
by the jet, and hence, they encounter both jet-driven compression as well as enhanced turbulence, {(}iii) 
the ISM goes through a sequence of evolutionary phases in the Kennicutt--Schmidt (KS) 
plot, until the jet breakout, beyond which the turbulent velocities return to pre-jet levels. 
Although in its infancy, the inclusion of a \emph{turbulence-regulated star formation} model  in large-scale 
simulations shows promise in addressing several existing issues related to the impact of AGN 
feedback processes on SFR.

{In addition to turbulence, other factors such as photo-ionization from the central AGN 
have also been considered as a potential source of preventing star formation. Post-process 
analysis of some  jet-feedback simulations has found that AGN-driven photoionization  does not significantly affect the SFR, for the chosen parameters of their studies  \cite{roos15a,meenakshi22a}. 
However, this is a largely unexplored domain which requires more work, with detailed inputs of gas 
chemistry \cite{richings17a} for more definitive results. }

\section{Observational Implications}\label{sec.obs}
As outlined  in Section~\ref{sec.scope}, several reviews have summarized the 
observational evidence and implications of AGN feedback. We refer the reader to 
\citet{harrison24a}, which is a more recent addition to the series. However, in the following 
sections, we briefly summarize the observational results related to jet--ISM 
interaction in particular, which is a narrower focus than the broad discussions presented earlier.

\subsection{Observations of Jet--ISM Interactions}
Evidence of jet--ISM interactions was presented even in the early days of studies of relativistic jets in galaxies. One of the first such reports  was in 1972 \cite{kruit72a}, which  conjectured that a ``nuclear explosion'' had expelled  ionized gas 
co-located with radio-emitting ridges in NGC 4258. The authors at that time had discounted the idea that 
relativistic electrons were streamed directly from the nucleus, as the concept of non-thermal plasma moving at bulk relativistic speeds 
was still not well-developed. Future studies of this well-studied galaxy have now established that it
harbors  radio jets which are co-planar with the gas disk and strongly interact with the ISM \citep{cecil2000a,ogle14a,appleton18a}. Later, 
in the early 1980s, several  galaxies were found to have radio  emission co-spatial with ionized 
gas emission (e.g., 3C 66B, 3C 31, NGC 315, 
3C 449 \cite{butcher80a}, 3C 277.3 \cite{miley81a}, 3C 305 \cite{heckman82a},  3C 171 
\cite{heckman84a}, 3C 293 \cite{VanBreugel84b}, 3C 310 with a proposed radio bubble 
from a trapped jet \cite{vanBreugel84c}, etc.). The spectra of such sources often revealed elevated kinematics of the ionized gas and  depolarization of radio emission, all of 
which was indicative of jet--ISM interaction. 

Later on, several other papers reported alignment of radio and optical/UV emission in multiple sources \cite{chambers87a,mccarthy87a,devries97a,devries99a}. Such observations  presented evidence  of
  widespread prevalence of jet--ISM interaction in general and also possible signatures of positive feedback, if the UV emission was from newly formed stars {(see reviews by 
\citet{mccarthy93a} and \citet{odea21a}).}{
	However,  further  analysis revealed possible contamination of the observed UV emission by two sources: (i) radiation from the central AGN, either through scattering from dust or direct 
	photoionization, (ii) nebular continuum emission \cite{fabian89a,tadhunter92a,dickson95a,tadhunter02a,tadhunter16a}. Only in a small fraction of such sources did young stars contribute to the UV budget \cite{tadhunter16a}. Thus, it remains 
	unclear whether such alignments can be considered to be strong signatures of jet-induced star formation. More recent results, such as 
	\citet{duggal24a}, have expanded on this, with  observations of spatially extended UV knots, a few kpc away from the central nucleus, confirming the
	possibility of the presence of young stars. However, more work is needed to ascertain the 
	extent of AGN contribution in such systems. }


In recent years, better access to spatially resolved sensitive observations has resulted in a large collection of  multi-wavelength and multi-phase observations of jet--ISM interaction (see Figure~\ref{fig.ObsSim} for an example). A collection of 
sources has been presented in Table~\ref{tab.sourceList1} of 
Appendix~\ref{sec.sourceList}, along with comments on the nature of the interaction 
where applicable. These sources  show signs of interaction of a jet with the ambient gas, such as  spatially resolved outflows and/or enhanced velocity dispersion of 
the gas. The list is not meant to be a complete collection of all observed cases, but a 
representative sample showing the diverse nature of observations of jet--ISM interaction. In 
many cases, however, the central AGN is also energetically capable of driving the observed outflows. However, several factors, especially resolved spatial 
morphology of the gas kinematics and jet, support the jet-driven 
feedback scenario.  

\begin{figure}[H]
	
	\includegraphics[width=0.95\linewidth]{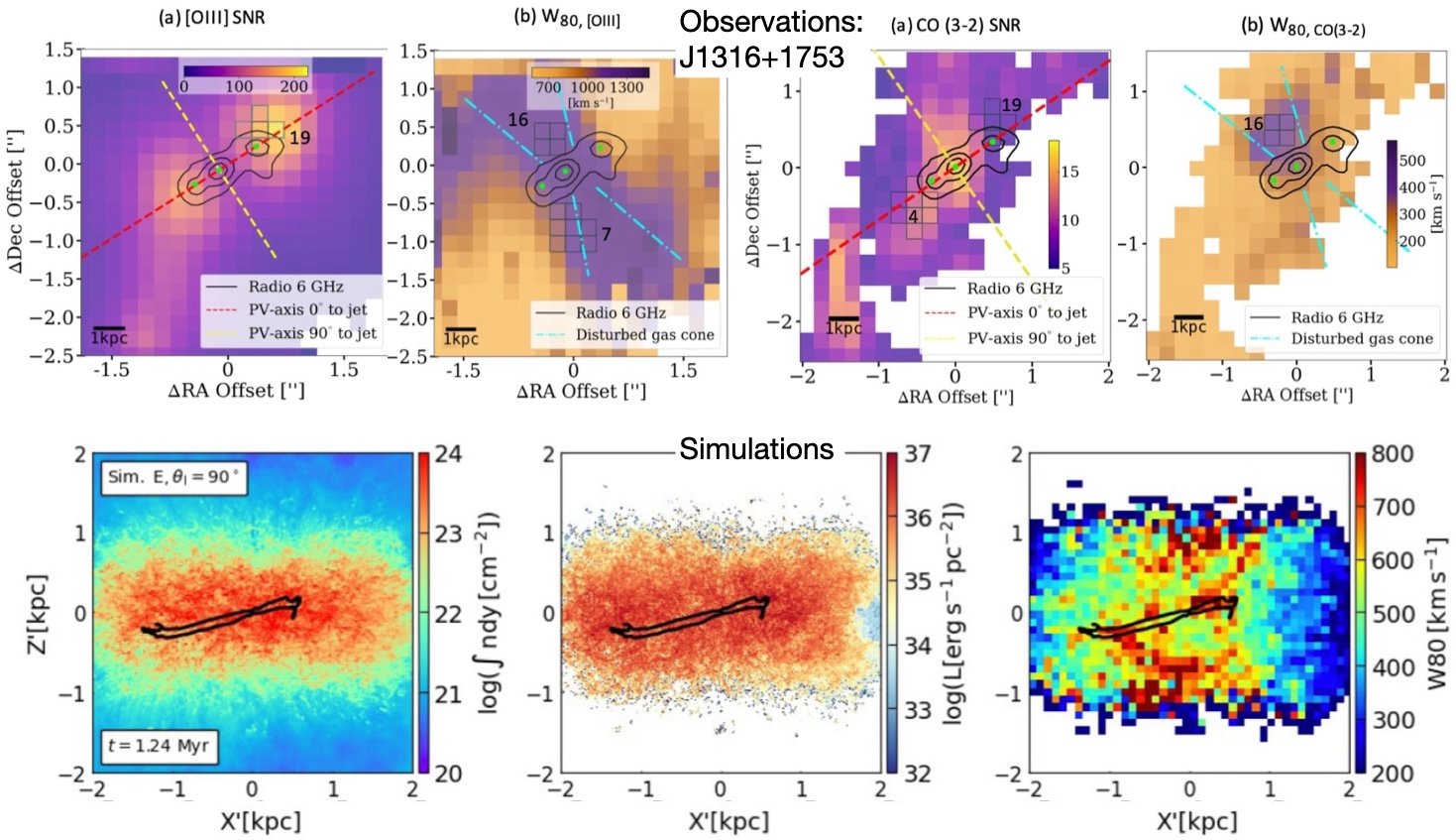}
	\caption{{\textbf{Top}:} 
		Representation of the top two panels of Figures~5 and 6 from \citet{girdhar22a} 
		showing enhanced kinematics in ionized and molecular gas of J1316+1753, a prototype 
		of multi-phase observation of jet--ISM interaction. \textbf{Bottom}: Representation of 
		the middle panel of Figure~8 of \citet{meenakshi22b}, showing predicted [OIII] 
		emission and line widths (W80) from simulations of jet--ISM interaction, with 
		enhanced widths perpendicular to the jet, as also observed in multi-phase 
		observations, such as top panel. Credits: \textbf{Top}: Girdhar et al. 2022 
		\cite{girdhar22a}, reproduced with permission, \textcopyright MNRAS. 
		\textbf{Bottom}: Meenakshi et al. 2022 \cite{meenakshi22b}, reproduced with 
		permission. \textcopyright MNRAS. }
	\label{fig.ObsSim}
\end{figure}

Besides  spatially resolved studies of individual galaxies, several  papers have 
explored observations of jets driving outflows in larger samples of radio-loud galaxies (e.g., \cite{kukreti23a,kukreti25a,rivera24a,nandi25a}, etc.). Such studies confirm the widespread prevalence of  
jet-driven outflows influencing their host. {This is further extended by the discovery of ionized outflows 
in the  ``red geyser'' galaxies, which comprise up to $\sim$10\% of the local quiescent 
population of massive galaxies \cite{cheung16a}. Star formation has been ruled out as the source of radio emission in such systems \cite{roy18a}. The favored interpretation is that the outflows are powered by jets \cite{roy21a}, as the AGN luminosity is inadequate. Additionally, several  well resolved sources have jet-like morphologies \cite{roy21a}. If jets are indeed powering such outflows, it would imply that low-power AGN jets play an important role 
in galaxy evolution, as predicted in some earlier works 
\cite{mukherjee16a,murthy22a}.}

To summarize, the observations discussed above  strongly support the results of 
the spatially resolved simulations outlined in the earlier part of this review. Such observations are  
further augmented by predictions of observable signatures of jet--ISM interaction 
from simulations, such as models for ionized gas kinematics by Meenakshi et al. 
2022b~\cite{meenakshi22b}.  Direct comparison of observations  with 
simulations has been very fruitful for several sources in recent years, e.g., IC 5063 \cite{mukherjee18a}, 4C 
31.04 \cite{zovaro19a}, B2 0258+35 \cite{murthy19a,fabbiano22a,murthy22a,murthy25a},  
2MASSX J23453269-044925 \cite{nesvadba21a,mulard23a}, Tea Cup galaxy 
\cite{audibert23a}, etc. This highlights the growing synergy between studies exploring spatially resolved 
simulations and their observational counterparts.

\subsection{Implications for Compact and Peaked Spectrum Sources (CSS/GPS/CSO)} Observational studies of systems with 
compact jets such as Compact Symmetric Objects (CSO) \cite{Orienti25a}, High Frequency Peakers (HFP) \cite{dallacasa2000a}, GHz-Peaked Spectrum (GPS) \cite{orienti16a}, Compact Steep Spectrum (CSS) \cite{odea21a} 
are directly related to the topic confined, young radio jets  
discussed in this review. The origin and nature of such compact radio sources, which are often characterized by a  peaked radio spectrum, are still debated in the literature. It remains unclear if
 they are (a) young evolving sources or (b) jets trapped by the host's ISM   \citep{odea98a,odea21a}. Source ages inferred from  synchrotron spectral studies \citep{murgia03a} or kinematic age estimates from VLBI 
studies \cite{an12a} favor the former, \emph{young}-jet scenario. However, such 
 short confinement times can also be due to the  powerful ($P_{1.4} \mbox{GHz} 
\gtrsim 10^{25}~\mbox{W Hz}^{-1}$) nature of these jets \citep{an12a,patil20a} and do not rule out jet--ISM interaction. There is evidence to suggest that such galaxies are gas-rich, e.g., (i) high rotation 
measures \citep{odea98a,rossetti08a,mantovani13a,orienti16a,odea21a}, (ii) high X-ray column 
depths  \citep{guiainazzi06a,Siemiginowska08a,Siemiginowska16a} 
that are well correlated with HI gas distribution  \citep{ostorero10a,ostorero17a}, implying
 a common origin of both phases, (iii) enhanced IR emission  indicating presence of
reprocessed dust emission~\citep{patil20a,patil22a,nascimento22a}, etc. There are also direct 
detections of atomic and molecular gas in many such systems, as described in reviews by 
\citet{odea98a}, \citet{fanti09a} and \citet{odea21a}. Recent observations 
\citep{miranda25a} have also uncovered strong ionized outflows in a large number of 
such  systems with compact jets, indicating strong coupling of the jet with the gas, as predicted by
simulations (see Section~\ref{sec.outflow}). 

Hence, given the abundance of 
possibilities of jet--ISM interaction in these systems, such galaxies may indeed follow the 
evolutionary sequence outlined in Section~\ref{sec.evol} and remain confined for some 
duration while within the host's potential.  Approximate analytical estimates of jet confinement timescales are presented in Appendix~\ref{sec.confine}. 
Besides the traditional peaked spectrum sources, recent surveys have uncovered a large 
sample of compact FR0 sources  \citep{baldi18a,baldi23a}, where the jet remains compact. 
Compact jet-like systems have also been found in resolved radio images of  radio-quiet systems as well \citep{jarvis21a,njeri25a}. Long confinement times 
($\tau \gtrsim 5$ Myr) of low-power jets ($P_j \lesssim 10^{41} \ergs$),  even for modest 
mean densities of the ISM clouds  ($n_c \sim 10^2$--$10^3$), can explain the compactness of 
such sources. Indeed, molecular gas has been detected in several 
such galaxies with compact radio emission \citep{jarvis20a,molyneux24a}. This makes 
such sources ideal test beds for investigating the physical processes of jet--ISM interaction discussed in this review.

\section{Concluding Perspectives}
This review outlines the development of numerical  simulations of relativistic jets 
propagating through their environment, with a particular focus on  jet--ISM interaction. In 
addition, the observational implications of such processes have also been discussed. The last 
few decades have seen a very prominent growth of studies in this domain. This is in 
contrast to earlier general skepticism (e.g., see the arguments in Section~1 
of~\mbox{\citet{ostriker10a}}) regarding the impact of jets in the 
\emph{{Establishment phase}} of AGN feedback. However, as discussed in this review, 
such perceptions are starting to change. Some of the major points that have emerged in 
this context are summarized below.

\subsection{Are Radio-Loud AGNs Gas-Rich?} One of the major concerns regarding the role of jets 
in their host galaxies was whether radio-loud galaxies have sufficient gas  to 
be affected by jets in the first place. The traditional view has been that in the nearby universe, powerful 
radio jets are usually found in early-type galaxies (ETGs), which were considered to be gas-poor. However, systematic surveys of such systems have uncovered a significant fraction 
($\sim$25\% \cite{davis19a}) to host dense gas, with higher fractions for radio-loud AGN 
($\gtrsim$34\% \cite{tadhunter24a}). A  summary of the various surveys can be found in 
Table~4 of \citet{tadhunter24a}. The recent review by Ruffa and Davis 2024 \mbox{\cite{ruffa24a}} gives 
more details on the properties of molecular gas in the local ETG. Interestingly, the fraction of radio-loud galaxies containing molecular gas and the estimated $H_2$ masses ($10^7$--$10^{10} M_{H2}$) have  been found to increase 
with redshift \cite{audibert22a}. Thus, 
dense gas is present in a significant fraction of radio galaxies, raising the possibility for 
jet-driven  feedback if jets couple significantly with the ISM.

\subsection{Radio-Detected Fraction of AGN?} Earlier studies of AGN populations had 
demonstrated that the radio-detected fraction of AGN reaches up to $\sim$30\% for high-mass galaxies \citep{best05a,mauch07a}. Though small, this is non-negligible. More recent 
sensitive radio surveys \cite{sabater19a} have extended these studies to lower radio luminosities 
Although the higher fractions result from inclusion of weaker AGN, whose radio powers are lower by an order of magnitude, such results support the widespread presence of radio 
activity in galaxies.
Furthermore, it is important to note that the traditional definition of 
``radio-loudness'', inferred from correlations of [OIII] and 1.4 GHz radio luminosity,  does
not imply radio ``silence''. As demonstrated in recent surveys \cite{jarvis21a,njeri25a},   
a significant fraction of traditional ``radio-quiet'' sources may harbor nuclear radio 
emission driven by an AGN, and more specifically, a jet. They may also demonstrate jet--ISM interaction, as 
shown in \mbox{Figure~\ref{fig.ObsSim}~\cite{girdhar22a}}.

\subsection{Large-Scale impact on the host galaxy?} Although the apparent 
beams of radio jets are often found to be thin, collimated structures, their large-scale influence has been well demonstrated by both simulations and observations, as 
outlined in this review. The observed size of a jet in radio wavelengths may often 
under-represent its wider impact (e.g., in 4C 31.04 \cite{zovaro19a}, and several other 
sources in {Appendix}
~\ref{sec.sourceList}), as the jet plasma is broken into low-density streams during the flood-channel 
phase of its evolution. However, even though the jet's impact may extend beyond its immediate confines, in most cases, the disturbances are restricted to the central few kpc of the galaxy. Though 
non-negligible, there needs to be better proof for a wider-scale impact, to 
confirm/discard the predictions from simulations. 

Observations have now firmly established that jets can drive multi-phase outflows in the 
central few kpc of galaxies, in line with predictions from simulations (see 
Section~\ref{sec.obs}). The broad line widths of such outflowing gas, due to  
turbulent gas motions, indicate the jet's ability to strongly affect the kinematics of the ambient gas. However, the 
 long-term implications of such activity for galaxy evolution, 
particularly in relation to star formation, remain an open question. Although several 
prominent radio-loud sources show a deficiency of star formation rate 
\cite{nesvadba10a,nesvadba20a}, the ubiquitousness of such jet-driven negative feedback has 
been questioned in other recent studies. For example, \citet{molyneux24a} find that jets do not significantly disturb the molecular gas at larger scales in their observed sample of galaxies. Even though recent theoretical studies, such as the \emph{turbulence-regulated star formation} framework, predict that jets can regulate SFR over large scales, such models are still in their infancy. Furthermore, the following two points should be noted while assessing the impact of AGN on the star formation rate and galaxy's mass assembly: (a) a given jet/AGN feedback episode may not have an instantaneous impact on the SFR, as the star formation time scales may differ from the dynamical times, (b) impacts from repeated jet/AGN activity will  accumulate over time and jointly affect the galaxy's growth. For more detailed discussions, see sections 2.1, 5 and 6 of the
reviews \cite{harrison24a}.
Hence, more systematic studies of the long-term impact of jets in 
particular, and AGN in general, are needed in the future to answer these questions.

\subsection{AGN Winds and Jets} 

{The current review primarily discusses the feedback from relativistic jets  on their host galaxies. However, the major 
agent of direct feedback on the host, so far, has been considered to be driven
by non-relativistic winds. This is primarily because of the smaller fraction of powerful 
radio-loud galaxies as compared to AGN \cite{sabater19a}. The possibility of the central 
AGN powering large-scale outflows was proposed in the early 1980s by analytical 
works such as Weymann et al. 1982 \cite{weymann82a}, Schiano {1985, 1986} 
	\cite{schiano85a,schiano86a}, etc. Such ideas were brought  to the focus of cosmological 
	models by the later seminal paper of \citet{silk98a}, and other subsequent recent works 	\cite{king03a,king05a,zubovas12a,faucher12a,zubovas14a}.  Although the  origin of such 
	a wind can be due to radiation pressure from the AGN or an accretion disk-driven MHD 
	wind (see Section 4.2 of \citet{faucher12a} for a discussion), at large scales ($\gtrsim$1 
	kpc), the wind can be treated as a mechanical outflow. Earlier studies assumed that  efficient inverse-Compton 
	cooling from the AGN radiation field will render the 
	wind to be momentum-driven close to the AGN, while at larger distances ($\gtrsim$1 kpc), it was expected to be  energy conserving  
	 \cite{king03a,zubovas12a}. However, \citet{faucher12a} have shown 
	that inefficient cooling of the protons is expected to create  a two-temperature plasma 
	and result in an energy-conserving flow at all scales. The authors showed that this explained the observed 
	momentum boosts in many systems. }

\textls[-10]{Such analytical models have mostly focused on an idealized, spherically symmetric 
ISM being cleared by the AGN-driven wind. Realistic simulations of AGN winds clearing 
an inhomogeneous galaxy have been less explored so far, with only a few recent papers 
highlighting such interactions (e.g., Wagner et al. 2013 \cite{wagner13a}, Gabor and 
Bournard 2014 \cite{gabor14a}, Hopkins et al. 2016 \cite{hopkins16a},  Bieri et al. 2017 
\cite{bieri17b}, Costa et al. 2018 \cite{costa18a}, Costa et al. 2020 \cite{costa20a}, Ward et 
al. 2024 \cite{ward24a}, etc.)}. Some of the general conclusions of such simulations are 
similar to the multi-phase jet--ISM interaction simulations discussed above. 
{(}a) The nuclear outflows can have large-scale impact, generating shocks and 
multi-phase outflows. {(}b) They generate a nuclear cavity and can potentially deplete a 
fraction of the dense gas cores of star-forming fuel. {(}c)  A complete blow-out of the 
dense cloud cores in the ISM is often not tenable and  depends on the nature of the ISM. 
These conclusions support the current proposition that the interaction of a jet with the dense ISM resembles 
the \emph{Quasar/Establishment mode} of feedback, which has hitherto been primarily attributed to AGN winds. 
However, direct quantitative comparison of such jetted vs. non-jetted outflows of similar 
powers has not been widely carried out (see \cite{cielo18a,weinberger23a} for some 
exceptions). The difference between jets and winds as feedback agents needs to be better 
quantified through systematic comparisons in the future.

{From the observational perspective, powerful winds and jets have distinctly different 
signatures. AGN winds are characterized by fast outflowing ionized \mbox{gas  
\cite{tombesi10a,tombesi14a,rankine20a}}, whereas powerful jets have collimated radio 
emission. However, given the discussion in the earlier part of the review on how jets can 
also  drive multi-phase outflows, and also since AGN winds can also result in synchrotron 
emission via shocks \cite{zakamska14a,nims15a}, the distinctions are often blurred in 
radio-detected AGN. Identifying the physical origin of such outflows and the origin of the 
radio emission itself is a rapidly evolving topic of research \cite{panessa19a}.
}

{
	A decade back, synchrotron emission from traditional radio-quiet quasars was 
	considered to arise from shocks driven by the AGN winds 
	\cite{zakamska14a,wylezalek16a}. However, several recent results of large-scale radio 
	surveys have also leaned towards a jet origin of the radio \mbox{emission  
	\cite{morabito19a,rivera24a,alban24a,fawcett25a}}. This has been further supported with 
	clear identification of collimated jet-like morphologies in resolved radio observations of 
	compact radio-quiet \mbox{galaxies \cite{jarvis21a,njeri25a}}. More detailed 
	observational campaigns as well as theoretical developments are required to disentangle 
	such degeneracies. Recent simulations of \cite{meenakshi24a} have attempted to 
	address this to some extent by modeling properties of the synchrotron emission from both jets and 
	winds. However, emission morphologies, especially at coarser resolutions, do not provide very clear distinctions. Polarization signatures were found to be 
	different between the two processes, but more work needs to be carried out to provide more 
	concrete predictions. 
}

\subsection{Need for Theoretical Improvements}  As outlined in the review, recent 
simulation efforts have reached high levels of sophistication and realism in modeling 
jet--ISM interaction and its impact on galaxy evolution. However, there do remain several 
lacunae that need improvement. 
\begin{enumerate}
	\item \textit{\underline{{Higher resolution and longer simulations:}}}
	A primary drawback of the kpc-scale simulations of jet--ISM interaction is the inability to 
	resolve cooling length scales at the outer surface of dense clouds. For example, as 
	outlined in Appendix A of \citet{meenakshi22b}, the typical cooling 
	length\endnote{Cooling length of a shock can be approximately defined as the distance traversed by the shock with a velocity $V_{\rm sh}$ during a typical cooling time: $L_{\rm cool} = V_{\rm sh} \times t_{\rm 
	cool}$. The cooling time scale, $t_{\rm cool}$, is obtained by dividing the internal energy per unit volume ($\gamma$ is $p/(\gamma -1)$) for an ideal gas with adiabatic index) by the cooling rate ($n^2\lambda$) \cite{sutherland93c}. For a more accurate temperature depedent defintion, see section 10 of Sutherland and Dopita 2017 \cite{sutherland17a}. } in multi-phase simulations of \mbox{\citet{mukherjee18b}} ranges from 
	$\sim$0.014 to 1 pc, well below the resolution of the simulations. Achieving such resolutions 
	will require an order of magnitude increase in current resources, which remains a 
	challenging task. Such resolutions are also necessary to better understand the 
	shock--cloud interaction, as demonstrated in Figure~\ref{fig.ankushfig}. Such intricate 
	substructures of the cloudlets are not resolved in current simulations. 
	
	Besides the need for better resolutions, most of the simulations in this domain have been 
	carried out for only a few Myr, due to the limitations of computational time requirements. 
	However, this explores only a very short phase of the jet and galaxy's lifetime. Larger-scale studies exploring the jet-driven heating--cooling feedback cycles  \citep{gaspari11a,gaspari12a,yang16a,yang23a} have explored longer run 
	times of up to a Gyr. However, they do not resolve the multi-phase gas structures internal 
	to the ISM. Future efforts have to explore at least a few tens of Myr of run time, with 
	self-consistent injection of AGN power, to account for at least one duty cycle of the AGN. Such simulations 
	 would require larger computational resources, which are expected to 
	become available in the near future.
	
	\item \textit{\underline{{Better chemistry of gas phases:}}}
    Existing simulations can be further updated with models that more accurately capture the micro-physics of these systems.
    One such area is the treatment of the chemistry of ionized and molecular gas phases and other species, such as dust.
    In most numerical codes, a single fluid prescription is adopted, where the
	cooling of collisionally ionized gas is derived from pre-computed tables that depend on the total gas densities and 
	temperatures. Only a few works  have included more sophisticated treatments of multi-species fluids \cite{richings17a,perucho24a}, an area that requires significant improvement in the future. In addition to this, the impact of 
	photoionizing radiation from the central AGN has been largely unexplored in large-scale simulations of AGN feedback, barring a few works 
	\citep{roos15a,bieri17b,cielo18a,meenakshi22a}. Although well explored for studying 
	cloud dynamics in broad line regions or close to wind launch zones (e.g., see 
	\citep{proga04a,proga14a,dyda25a}, and references therein), their effect on larger kpc-scale simulations is yet to be fully explored. 
	
	\item \textit{\underline{{Magnetic fields:}}}
	Another ill-explored parameter is the effect of magnetic fields on shock--cloud dynamics 
	and star formation. Very few simulations of jet--ISM interaction have included the 
	evolution of magnetic fields \citep{asahina17a,clavijo24a}. Magnetic fields can 
	potentially change the nature of shock--cloud interaction by affecting Kelvin--Helmholtz 
	growth rates. They will also affect the estimates of turbulence-regulated star formation rates 
	\cite{federrath12a}, and should be explored in more detail. 
	
	\item\textit{\underline{{Cosmic ray feedback:}}}
	Another key ingredient overlooked in the current literature is the effect of cosmic rays 
	on the fluid dynamics of jet--ISM interaction, in particular, and AGN feedback in general. 
    Jet--cloud interfaces undergoing diffusive shock acceleration are expected to be active sites for the production of high-energy cosmic rays.
	Such cosmic rays  will provide additional momentum and pressure to the fluid, 
	which would in turn affect the local dynamics of the gas. This has been tentatively 
	explored in some cases, for example, in IC 5063 \cite{dasyra22a}. Inclusion of cosmic ray 
	diffusion and heating in MHD simulations {in general 
	\cite{snodin06a,dubois16a,thomas19a,thomas21a,chan19a}   and   studies of galaxy 
	evolution and jet simulations in particular 
	\cite{fulai11a,yang12a,ruszkowski17a,ehlert18a,farcy22a,su24a} is}  being actively 
	pursued by several groups. However, their impact is yet to be investigated in the context of 
	multi-phase AGN feedback.
	
	\item { \textit{\underline{{Jet composition, plasma processes and instabilities:}}} 
	Most of the simulations of large-scale relativistic jets 
		do not explicitly account for plasma composition, which can influence  the nature of the solution but is numerically challenging to implement (see 
		Section 8.2.1 of \citet{marti03a}). Only a handful of works have 
		explored these issues. Some studies have used a modified EOS incorporating fixed ratios of 
		leptonic and hadronic components in an otherwise single fluid \mbox{description 
		\cite{scheck02a,chattopadhyay2000a,joshi23a}}. Other recent works have used a more sophisticated 
		two-temperature fluid treatment  including electron--ion interaction, which is modeled in a sub-grid  
		framework \cite{ohmura23a,ohmura23b}. 
		However, such attempts are still in their infancy for jet simulations, although active 
		development is on-going in other related domains (e.g., \cite{chael25a,salas25a}).}

	{ 
		Besides jet microphysics, another overlooked domain is the impact of small-scale 
		plasma processes in general and MHD instabilities in particular. Accounting for 
		plasma effects in large-scale jets is understandably difficult, due to the large 
		separation of scales. Nonetheless, particle-in-cell studies  of idealized jets or  shear 
		\mbox{layers 
		\cite{nishikawa03a,liang13a,nishikawa16a,rieger19a,chand24a,dutan25a}} have 
		demonstrated the importance of considering fundamental plasma effects such as 
		Weibel, two-stream instabilities, etc.  (see Meli and Nishikawa 2021 
		\cite{meli21a} for a detailed review). Although    sophisticated simulations have been performed to investigate particle acceleration of non-thermal electrons by relativistic 
		shocks \cite{sironi15a} or reconnection processes     \cite{sironi25a}, the broader 
		  impact of such plasma processes on jet dynamics and emission requires further scrutiny.
	}

	{
		Regarding large-scale fluid instabilities, it is well known that Kelvin--Helmholtz \cite{bodo94a,hardee2000a,perucho04a,perucho05a}, 
		current-driven instabilities (CDIs) 
		\cite{giannios06a,moll08a,mignone10a,mizuno14a,mizuno09a,bromberg19a}, or a 
		mixture of the two,   can operate in relativistic jets 		\cite{mizuno07a,mckinney09a,rossi20a,rossi24a,musso24a}. Linear stability analyses 
		of such instabilities find the growth rates to depend on various jet parameters 
		such as the bulk Lorentz factor \cite{perucho10a}, jet magnetization and magnetic pitch 
		parameter \cite{bodo13a,chow23a}, jet's rotation \cite{bodo16a,bodo19a}, jet opening 
		angle \cite{bromberg16a,tchekhovskoy16a}, etc. Such instabilities can have strong 
		implications for jet collimation and turbulence, which would in turn impact the 
		various morphological classifications of jets (see \mbox{\citet{costa24a}} for an 
		example). Spatially resolved observations of helical structures in jets or their 
		ridge-lines have also hinted towards the presence of such MHD processes in a few 
		sources, such as 3C 273 \cite{lobanov01a,perucho06a,nikonov23a}, M 87 
		\cite{lobanov03a,hardee11a,martens16a}, S5 0836+710 
		\cite{perucho07a,perucho12a,VegaGarcia19a}, NGC 315 \cite{worrall07a,park24a}, 3C 
		279 \cite{fuentes23a}, 3C 84 \cite{paraschos25a}, etc. Although several large-scale 
		simulations have well demonstrated the onset and impact of such instabilities (e.g., 
		\cite{bromberg16a,massaglia19a,perucho20a,mukherjee20a}), more work is needed to 
		unravel how  such MHD processes  affect jet dynamics and their non-thermal emission. 
	}
\end{enumerate}

\vspace{6pt} 



\funding{This research received no external funding.}

\dataavailability{No new data were created or analyzed in this study. Data sharing is not 
applicable to this article. The original contributions presented in this study are included in 
the article. 
	Further inquiries can be directed to the corresponding author.}

\acknowledgments{Several ideas expressed here have benefited from conferences and 
workshops in the recent past. I would like to thank the organizers of the following 
meetings: {(}i)  ``AGN on the Beach'' conference, held at Tropea, from 10 to 15 September 
2023, where this review was first presented, {(}ii)  ``The importance of jet induced 
feedback on galaxy scales'', held at the Lorentz Centre, from 23 to 28 October 2023 and 
{(}iii)  ``Jets on the rocks: On the trail of radio activity in and around galaxies'' held at 
Bad Moos, Sesto, from 14 to 18 July 2025. 
	I would like to thank Isabella Prandoni and Illaria Ruffa for their encouragement in 
	completing the review. I thank the following colleagues: {(}i) Raffaella Morganti and 
	Clive Tadhunter for a thorough read and sharing several fruitful suggestions to improve 
	the draft, {(}ii) Martin Bourne for discussions and suggestions during the early phase 
	of compilation of the review, {(}iii) Gianluigi Bodo and Paola Rossi for helpful 
	discussions in general and inputs on impact of MHD instabilities on jets, {(}iv) 
	Ankush Mandal for making available the data used to create Figure~\ref{fig.ankushfig}, 
	{(}v)  M. Meenakshi and Christopher Harrison for their consent in reproducing 
	figures used in Figure~\ref{fig.ObsSim} and helpful suggestions. I also thank the four 
	anonymous referees for their helpful suggestions and detailed comments, which helped 
	improve clarity of certain sections.  }

\conflictsofinterest{{~}}


\appendixtitles{yes} 
\appendixstart
\appendix
\section{Duration of the Confined Phase of the Jet in the ISM}\label{sec.confine}

The duration of the confined phase depends on the extent of the dense gas, its volume 
filling factor and density, along the path of the jet as well as the jet's properties, ranging 
from  $\sim$100 kyr \cite{sutherland07a,wagner11a,wagner12a} to $\sim$1--2 Myr 
\cite{mukherjee16a,mukherjee17a,mukherjee18b}. An approximate estimate of the jet 
confinement may be obtained by computing the advance speed of the jet through the 
dense ambient medium, which is derived below. 
{In the following equations  the jet Lorentz factor is represented as $\gamma_j$, which is 
different from the adiabatic index $\Gamma$ used for an ideal equation of state (EOS) viz. 
$\rho \epsilon = p/(\Gamma -1)$. An ideal EOS has not been explicitly assumed in the 
derivations below. The variables $\rho$, $h$, $\epsilon$ and $p$ have the usual 
definitions of density,  specific enthalpy, specific internal energy and pressure, respectively. 
Variables with subscript $j$ refer to the jet and those with subscript $a$ refer to the 
ambient medium. 
}

The relevant equations are as follows:
\begin{itemize}

\item \textit{{Jet power} ($P_j$)}: This is defined as rest mass subtracted relativistic 
energy flux \cite{wagner11a,mukherjee16a,mukherjee20a} through a cylindrical surface of 
radius $r_j$ (jet-radius), assuming the jet velocity to be perpendicular to the surface. 
{
	\begin{align}
		P_j &= \left(\gamma_j^2 \rho_j h_j - \gamma_j \rho_jc^2\right)v_j \pi r_j^2 
		=  \pi r^2_j v_j \frac{\gamma_j^2 \rho_j c^2}{\chi}\left(1 + \frac{\gamma_j 
		-1}{\gamma_j} \chi\right) \label{eq.power}\\
		& \mbox{where } \chi = \frac{\rho_j c^2}{\rho_j h_j -\rho_j c^2} = \frac{\rho_j 
		c^2}{\rho \epsilon + p}
	\end{align}
}
The non-dimensional parameter $\chi$, also defined earlier in 
Equation~(\ref{eq.norelrho}), was first introduced in \citet{bicknell94a} and is an useful 
indicator of the nature and composition of the jet plasma (see discussion in Appendix A of 
\citep{mukherjee20a}).

\item \textit{{Jet-head velocity} ($v_h$):} This is obtained by equating the momentum 
flux of the jet to the external medium (e.g., clouds), in the frame of the jet's working surface (see Sections 3 and 3.3 of 
\citep{marti97a,mukherjee20a}, respectively). Using the expression of the jet power ($P_j$) 
as presented in Equation~(\ref{eq.power}), one can replace the jet density ($\rho_j$) to 
express the jet advance speed in terms of the jet power and ambient density 
(Equations~(\ref{eq.vh2}) and (\ref{eq.vh3})). 

\begin{align}
	v_h &= \frac{\gamma_j \sqrt{\eta_R}}{1+ \gamma_j \sqrt{\eta_R}} v_j\simeq 
	\gamma_j v_j \left(\frac{\rho_j}{\rho_a}\right)^{1/2} \left(1 + 
	\frac{1}{\chi}\right)^{1/2} \label{eq.vh1} \\
	& \mbox{ where }  \eta_R = \frac{\rho_j h_j}{\rho_a h_a} \simeq 
	\left(\frac{\rho_j}{\rho_a} \right)\left( 1 + \frac{1}{\chi}\right), \, \mbox{ for } \rho_a 
	h_a \simeq \rho_a c^2 \nonumber. \\
	v_h &= \left(\frac{P_j v_j}{\rho_a c^2 \pi r_j^2}\right)^{1/2} \left(\frac{1 + \chi}{1 + 
	\frac{\gamma_j -1}{\gamma_j} \chi}\right)^{1/2} \label{eq.vh2} \\
	&\simeq 1.75\times10^3 \mbox{km s}^{-1} \left(\frac{P_j}{10^{43}\mbox{erg 
	s}^{-1}}\right)^{1/2} \left(\frac{v_j}{0.98 
	c}\right)^{1/2}\left(\frac{n_a}{1\mbox{cm}^{-3}}\right)^{-1/2} \nonumber\\
	&\times\left(\frac{r_j}{20 \mbox{pc}}\right)^{-1} 
	\left(\frac{g(\chi,\gamma_j)}{1.111}\right)^{1/2} \,\, ; \mbox{ for } (\gamma_j = 5, 
	\chi = 1), \label{eq.vh3} \\ 
	& \mbox{ where } g(\chi,\gamma_j) =\frac{1 + \chi}{1 + \frac{\gamma_j 
	-1}{\gamma_j} \chi}.  \nonumber
\end{align}
The approximate form of $v_h$ in Equation~(\ref{eq.vh1}) arises from the assumption 
that jet density is usually very small when compared to the mean ambient value 
($\rho_j/\rho_a \ll 1$), such that $\gamma_j\sqrt{\eta_R} < 1$. This  is expected for 
physically motivated  parameters of typical jets (see Equations (20)--(22) of Mukherjee et 
al. 2020 \cite{mukherjee20a}).  

\item \textit{{Jet confinement timescale:}} It is apparent that dense clouds along the 
path of the jet can strongly decelerate jets. Typical densities of molecular clouds can range 
from \mbox{$n_c \sim 10^2$ to $10^5~\mbox{cm}^{-3}$}, resulting in a decrease in advance 
speed by several orders of magnitude. An approximate time scale of confinement can be 
assessed by computing the travel time of the jet-head for a scale height $L$, e.g., $L \sim 
500$ pc, which is typical of core radii of bulges in elliptical galaxies. Assuming the volume filling factor 
of the dense clouds to be $fv$, the jet confinement time in the ISM will be given by:     
\begin{align}   
	\tau &\simeq f_v \frac{L}{v_{hc}} + (1 - f_v) \frac{L}{v_{ha}}. \label{eq.timej}
\end{align}
Here $v_{hc}$ is the advance of the jet trough dense clouds with mean density $n_c$ and 
$v_{ha}$ is the advance speed through the gaps between the clouds with low-density ambient halo gas (e.g., $n_a \sim 0.1~\mbox{cm}^{-3}$).
\end{itemize}

\begin{figure}[H]

\includegraphics[width=0.55\linewidth,keepaspectratio]{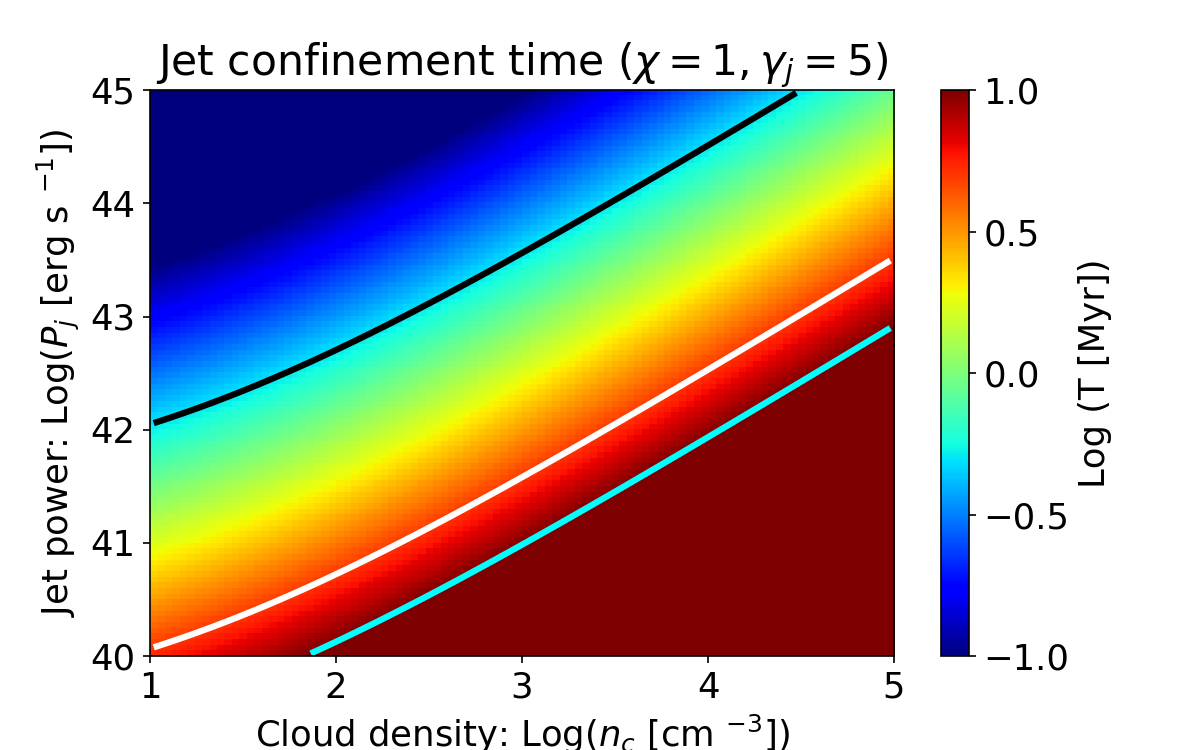}
\includegraphics[width=0.4\linewidth,keepaspectratio]{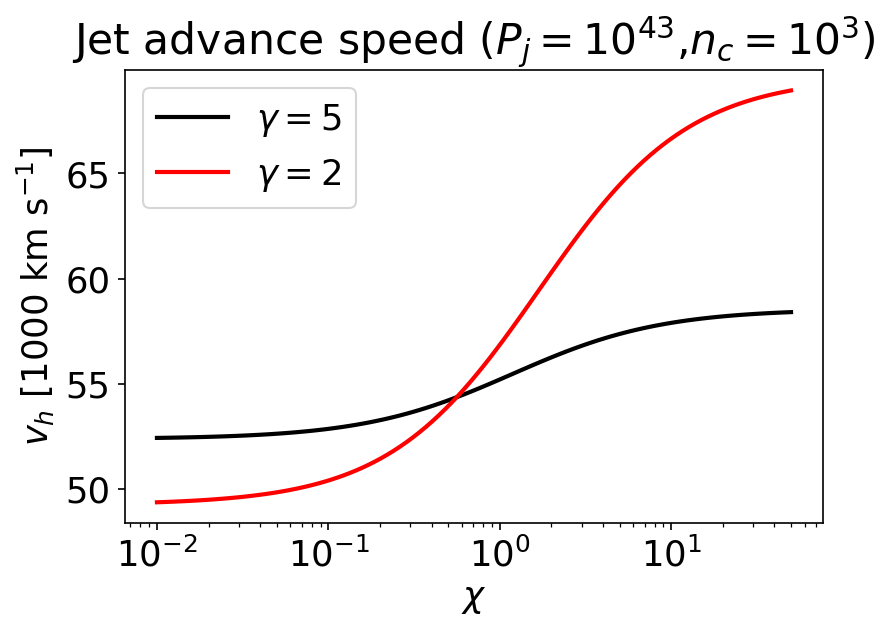}
\caption{{Left:} 
	Confinement timescales (see Equation~(\ref{eq.timej})) for different jet powers and 
	mean cloud densities, with $\chi=1$ and $\gamma_j=5$. The black, white and cyan 
	contours correspond to $\tau = (0.5,5,10)$ Myr. Right: Jet advance speed's variation with 
	$\chi$ for two different jet Lorentz factors. The jet power is $P_j = 10^{43}\ergs$ and 
	cloud density $n_c=10^3 \cc$.}
\label{fig:jetconf}
\end{figure}

The results for different jet powers and cloud density are presented in the left panel of 
Figure~\ref{fig:jetconf}, for jets with $\chi = 1$ and $\gamma_j=5$  {and a dense gas 
volume filling factor of $f_v = 0.1$}.  Typical confinement times are seen to range from a 
few hundred kilo-years for $P_j \sim 10^{43}$--$10^{45} \ergs$ and cloud densities $n_c 
\sim 10^2$--$10^4 \cc$ {(left of the black contour in left panel of Figure~\ref{fig:jetconf})} 
to $\tau \lesssim$5 Myr for higher densities and lower power {(white contour in 
Figure~\ref{fig:jetconf})}. Such approximate estimates align very well with the results from 
the various (relativistic) hydrodynamic simulations presented in this review. The results are also in 
agreement with more detailed semi-analytical dynamical models  \citep{carvalho98a}. The jet confinement times  only weakly  depend on  $\chi$, as shown in the right panel of 
Figure~\ref{fig:jetconf}. The advance speeds tend to asymptote to terminal values for 
$\chi \lesssim 0.1$ and $\chi \gtrsim 10$. It is interesting to note that lower-power jets 
$P_j \lesssim 10^{41} \ergs$ can remain trapped in the central regions of a galaxy for very 
long times ($\tau \gtrsim 10$ Myr, {shown by the cyan contour in 
Figure~\ref{fig:jetconf}}), failing to reach the breakout phase. Thus, even modest ISM 
parameters of $f_v \sim 0.1$ and $n_c \gtrsim 1000$ can trap low-power jets, restricting 
their large-scale growth.

\section{Observations of Jet--ISM Interaction}\label{sec.sourceList}
We list in Table~\ref{tab.sourceList1} a selection of galaxies or 
broad survey efforts highlighting the prominent detection of outflows and feedback processes 
induced by a relativistic jets. The list is not meant to be a complete census of jet--ISM 
interaction, but rather a representation of the diverse nature of  objects where observational 
studies of jet--ISM interactions have been reported. The third column denotes the gas 
phase where the more prominent impacts of jets have been observed. The recent 
references studying such jet--ISM interaction for each source have been listed in the fourth column. Many of 
these sources are well studied in the literature. The complete description of the existing studies 
on each source may be found from within the references cited.

\begin{table}[H]

\caption{A selection of cases with observations of jet--gas interaction. The second column 
mentions the major gas phase where outflows/feedback effects are observed. The label 
Ionized in general denotes standard emission lines in optical bands, with some having data 
at other frequencies such as IR and X-rays. WH2 in the molecular phase denotes warm 
H2.}\label{tab.sourceList1}

\begin{adjustwidth}{-\extralength}{0cm}
	\begin{tabularx}{\fulllength}{ABCD}
		\toprule
		& \textbf{Source/Survey}  & \textbf{Gas Phase} & \textbf{Comments} and 
		\textbf{References} \\
		\midrule
		1 & J1430 (Tea Cup), J1509, J1356, part of the
		QSOFEED survey & Molecular (CO, WH2, PAH), Ionized (Optical+Xrays) & Part of a 
		sample of 48  Type-2 Seyferts (44 detected in radio) with several examples of well 
		defined jetted system driving outflows. 
		\cite{lansbury18a,almeida22a,audibert23a,venturi23a,bessiere24a,zanchettin25a,almeida25a}
		 \\
		\midrule
		2 &  NGC 5929 & Molecular (WH2), Ionized (FeII) & Outflows perpendicular to the jet 
		axis.~\citep{riffel14a,riffel15a} \\
		\midrule
		3 & QFeedS survey  & Molecular (CO), Ionized  & A survey of 42 sources 
		\citep{jarvis21a},  many radio-quiet \citep{jarvis19a,jarvis21a} but showing jet-like 
		features ($
		\sim$88\% cases \citep{njeri25a}) and dense gas (for 17 sources 
		\cite{jarvis20a,molyneux24a}).  Multi-wavelength studies of feedback performed for a 
		subset of sources, including outflows perpendicular to the 
		jet.~\cite{girdhar22a,girdhar24a,ulivi24a}\\
		\midrule
		4 & NGC 5972 & Ionized  & Detection of jet-induced shocks. \cite{ali25a} \\
		\midrule
		5 & 3C 293 (UGC 8782) & Molecular (WH2), Atomic (HI absorption), Ionized  & 
		\cite{mahony13a,lanz15b,mahony16a,riffel23a,CostaSouza24a}
		\\
		\midrule
		
		6 & IC 5063 & Molecular (CO,WH2), Ionized (IR/optical+X-rays) & A very well studied 
		source with a jet strongly inclined into a kpc-scale disk. Shows outflow perpendicular 
		to the 
		jet.~\cite{morganti15a,dasyra16a,oosterloo17a,fonseca23a,travascio21a,dasyra24a} \\
		\bottomrule
	\end{tabularx}
\end{adjustwidth}
\end{table}
\begin{table}[H]\ContinuedFloat

\caption{\emph{Cont.}}

\begin{adjustwidth}{-\extralength}{0cm}
	\begin{tabularx}{\fulllength}{ABCD}
		\toprule
		& \textbf{Source/Survey}  & \textbf{Gas Phase} & \textbf{Comments} and 
		\textbf{References} \\
		
		\midrule
		7 & NGC 5643, NGC 1068, NGC 1386, NGC 1365 & Ionized & Part of the MAGNUM 
		survey, also including IC 5063. Several of these sources show outflow perpendicular 
		to the jet.  \cite{cresci15a,venturi21a,marconcini25a} \\
		\midrule 
		8 & NGC 3393 & Molecular (CO), Ionized (Optical+X-rays) & 
		\cite{finlez18a,maksym17a,maksym19a} \\
		\midrule
		9 & NGC 7319 in Stephan’s~quintet & Molecular (CO), Ionized & A well studied 
		group of 5 interacting galaxies with one showing prominent jet--ISM interaction. 
		\cite{santaella22a,emonts25a} \\
		\midrule
		10 & 3C 326 & Molecular (CO,WH2), Ionized & Early evidence of strong jet-induced 
		turbulence, refined with better spatial resolution (JWST) to uncover in situ outflows 
		\cite{nesvadba10a,nesvadba11b,villaVelez24a,leftley24a} \\
		\midrule
		11 & GATOS survey & Molecular (CO), Ionized & A survey of  dusty CND of 19+ 
		Seyferts \citep{garcia21a,arriba23a}. Several  show very prominent jet--ISM 
		interaction, reported as part of this survey and also  from other multi-wavelength 
		observations. \cite{garciaBernete21a,herrero23a,esposito24a,zhang24a} \\
		\midrule
		12 & WIDE-AEGIS-2018003848 & Ionized & Detection of strong shock from emission 
		line modelling, likely powered by the radio jet. \cite{Deugino25a} \\
		\midrule
		13 & B2 0258+35 (NGC 1167) & Molecular (CO), Ionized (X-ray) & A confirmed 
		detection of jet clearing the central kpc of dense gas. Tentative confirmation of thermal 
		\mbox{X-rays. \cite{murthy19a,murthy22a,fabbiano22a,murthy25a}} \\
		\midrule
		14 & NGC 3100, IC 1531, NGC~3557 & Molecular (CO, tentative HCO+) & A subset 
		from a survey of 11 LERGs, showing evidence of only mild jet--ISM interaction, in spite 
		of potential conditions available for more stronger effects observed elsewhere. 
		\cite{ruffa19a,ruffa19b,ruffa20a,ruffa22a}\\
		\midrule
		15 & NGC 1052 & Ionized & Prominent ionized bubble along the galaxy's minor axis, 
		blown by a jet inclined towards a nuclear gas disk, besides detection of large-scale 
		disturbed kinematics and shocks. \citep{dopita15a,cazzoli22a,molina18a,goold24a} \\
		\bottomrule
	\end{tabularx}
\end{adjustwidth}
\end{table}



\begin{table}[H]\ContinuedFloat

\caption{\emph{Cont.}}
\begin{adjustwidth}{-\extralength}{0cm}
	\begin{tabularx}{\linewidth}{ABCD}
		\toprule
		& \textbf{Source/Survey}  & \textbf{Gas Phase} & \textbf{Comments} and 
		\textbf{References} \\
		\midrule
		16 & NGC 3079 & Radio (deceleration of knots), Ionized & A well studied source with 
		prominent gas filaments from nuclear outflows \citep{cecil01a}. Observed pc-scale 
		jet--ISM interaction~\citep{middelberg07a,fernandez23a}, which may power the large-scale outflow~\citep{shafi15a,veilleux21a}. \\
		\midrule
		17 & XID2028 & Molecular (CO), Ionized & Co-spatial collimated molecular, ionized 
		jet-driven outflows outflow piercing gas shells ($\gtrsim$6 kpc) from the nucleus. 
		\citep{brusa18a,cresci23a} \\
		\midrule
		18 & 4C 31.04 & Ionized, Neutral & CSS source with $\sim 100$ pc jet but large-scale 
		($\sim$0.3--2 kpc) shocked gas. \citep{zovaro19a,murthy24a} \\
		\midrule
		19 & NGC 3998 & Radio & Indirect evidence of jet--medium interaction from radio 
		emission. \citep{sarrvesh20a} \\
		\midrule
		20 & NGC 4579 (Messier 58) & Molecular (CO,WH2,PAH), Ionized & \cite{Ogle24a} 
		\\
		\bottomrule
	\end{tabularx}
\end{adjustwidth}
\end{table}

\begin{table}[H]\ContinuedFloat

\caption{\emph{Cont.}}

\begin{adjustwidth}{-\extralength}{0cm}
	\begin{tabularx}{\linewidth}{ABCD}
		\toprule
		& \textbf{Source/Survey}  & \textbf{Gas Phase} & \textbf{Comments} and 
		\textbf{References} \\
		\midrule
		21 & IRAS 10565+2448 & Molecular (CO), Atomic (HI emission+absorption), Ionized & 
		\cite{renzhi23a} \\
		
		\midrule
		
		22 & 4C 41.17 & Molecular (CO), Ionized & A $z=3.792$ galaxy associated with 
		positive feedback~\cite{bicknell00a,nesvadba20a} \\
		\midrule
		23 & PKS 1549-79 & Molecular (CO), Ionized, Atomic (HI absorption) & Nuclear 
		molecular outflow, extended ionized outflow.~\cite{holt06a,oosterloo19a}      \\
		\midrule
		24 & Sub sample of 9 sources from the southern 2 Jy sample \citep{tadhunter98a} & 
		Ionized & Broad integrated outflowing emission lines ($v_{\rm out} \gtrsim 800 
		\kms$, FWHM $\gtrsim 700\kms$) driven by jets. \cite{santoro20a} \\
		\midrule
		25 & 3C 273 & Molecular (CO), Ionized & Expanding jet-driven cocoon impinging on a 
		gas disk.~\cite{husemann19b} \\
		\midrule
		26 & HE 1353-1917, HE0040-1105  & Ionized & Nuclear-scale jet-driven outflow. Part 
		of the CARS survey.~\cite{husemann19a,singha23a} \\
		\midrule
		27 & 4C 12.50 (F13451+1232) & Molecular (CO,WH2), Ionized & Strong jet-driven 
		nuclear ($\lesssim$100 pc) outflow~\citep{morganti13a,villarMartin23a,holden24a}, 
		but not on large scales\citep{holden25a}.\\
		\midrule
		28 & TNJ 1338-1942 & Ionized & Jet impact on extra-galactic gas cloud with extreme 
		kinematics. \cite{duncan23a,roy24a,saxena24a} \\
		\midrule
		29 & NGC 6328 (PKS 1718-649) & Molecular (CO) & GPS source with pc-scale jet 
		interacting with ambient gas.~\cite{papachristou23a} \\
		\midrule
		30 & PKS 0023-26 & Molecular (CO) & \cite{morganti21a} \\
        		\bottomrule
	\end{tabularx}
\end{adjustwidth}
\end{table}
\begin{table}[H]\ContinuedFloat

\caption{\emph{Cont.}}

\begin{adjustwidth}{-\extralength}{0cm}
	\begin{tabularx}{\fulllength}{ABCD}
		\toprule
		& \textbf{Source/Survey}  & \textbf{Gas Phase} & \textbf{Comments} and 
		\textbf{References} \\
		
		\midrule
		31 & HzRG-MRC 0152-209 (Dragonfly galaxy) & Molecular (CO) & Molecular outflow 
		(jet/AGN-driven) perpendicular to the jet, with indications of jet--ISM interaction at 
		small scales.~\cite{yuxing24a} \\
		\midrule
		32 & ESO 420-G13 & Molecular (CO), Ionized & \cite{ontiveros20a} \\
		\midrule
		33 & Jet-driven HI outflows  (including 3C 236, 3C 305, 3C 459, OQ 208) & Molecular 
		(CO), Atomic (HI absorption) & \cite{morganti05a,struve12a,schulz18a,guillard12a} 
		\\
		\midrule
		34 & NGC 4258 (Messier 106) & Molecular (WH2), Ionized, X-rays & Detection of 
		shocks and turbulence induced by \mbox{jets. \cite{cecil2000a,ogle14a,appleton18a}} 
		\\
		\midrule
		35 & Molecular Hydrogen Emission line Galaxies (\mbox{MOHEG}) & Molecular 
		(WH2) & A sample of 17 Radio-Loud galaxies with detections of warm H2 lines and 
		indications of jet-driven shocks. \cite{ogle10a} \\
		\midrule
		36 & PKS B1934-63 & Ionized, WH2 & Compact GPC source with ionized outflow  but 
		not in molecular phase. \cite{santoro18a} \\
		\midrule
		37 & Cen A (NGC 5128) & Molecular (CO) & Jet-induced inefficient star formation in 
		filaments along the jet. \citep{salome16a,salome17a} \\
		\midrule
		38 & Cygnus A & Molecular (WH2, PAH), Ionized & High velocity $\sim$6 kpc-scale 
		outflow driven by jet. \citep{ogle25a} \\
		\midrule
		39 & SINFONI survey of RLAGN & Molecular (WH2), Ionized & A survey of 33 
		powerful RLAGN, confirming widespread jet-driven extreme gas kinematics. 
		\cite{collet16a,nesvadba17a,nesvadba17b} \\



		
		\midrule
		40 & NGC 6951 & Ionized & \cite{may16a}\\  
		\midrule
		41 & PKS 2152-69 & Ionized (Optical+X-rays) & Highly ionized gas cloud 8 kpc from 
		the galaxy impacted by a jet. \cite{tadhunter87a,tadhunter88a,worrall12a}. \\
		\midrule
		42 & PKS PKS2250-41 & Ionized & Interaction of a jet with gas in a companion 
		\mbox{galaxy. \cite{clark97a,villarmartin99a,inskip08a}} \\
		\midrule
		43 & IRAS 00183-7111 & Molecular & Jet--ISM interaction in a ULIRG \cite{ruffa25a} \\
		\midrule
		44 & J165315.06+234943.0 (\mbox{Beetle}) & Ionized & Detection of extreme gas 
		kinematics up to 46 kpc from the galaxy due to shocks from a jet in a radio-quiet 
		quasar~\cite{villarMartin17a}. \\
		\bottomrule
	\end{tabularx}
\end{adjustwidth}
\end{table}


\begin{adjustwidth}{-\extralength}{0cm}
\printendnotes[custom] 


\reftitle{References}

\def\apj{ApJ}%
\def\mnras{MNRAS}%
\def\aap{A\&A}%
\def\apjl{ApJL}
\def\physrep{PhR}
\def\apjs{ApJS}
\def\pasa{PASA}
\def\pasj{PASJ}
\def\nat{Nature}
\def\memsai{MmSAI}
\def\aj{AJ}%
\def\aaps{A\&AS}%
\def\iaucirc{IAU~Circ.}%
\def\sovast{Soviet~Ast.}%
\def\apss{Ap\&SS}
\def\aplett{ApJL}
\def\aapr{Astron Astrophys Rev}
\def\pasp{PASP}
\def\araa{ARAA}%
\def\nar{New Astronomy Reviews}
\def\ssr{Space Science Reviews}


\begin{thebibliography}{999}
	
	\bibitem[{Fabian}(2012)]{fabian12a}
	{Fabian}, A.C.
	\newblock {Observational Evidence of Active Galactic Nuclei Feedback}.
	\newblock \emph{{Annu. Rev. Astron. Astrophys.} 
	} {\bf 2012}, {\em 50},~455--489.
	\newblock {\url{https://doi.org/10.1146/annurev-astro-081811-125521}}.
	
	\bibitem[{Harrison} and {Ramos Almeida}(2024)]{harrison24a}
	{Harrison}, C.M.; {Ramos Almeida}, C.
	\newblock {Observational Tests of Active Galactic Nuclei Feedback: An Overview
		of Approaches and Interpretation}.
	\newblock {\em Galaxies} {\bf 2024}, {\em 12}, {17.} 
	\newblock {\url{https://doi.org/10.3390/galaxies12020017}}.
	
	\bibitem[{Lea} et~al.(1973){Lea}, {Silk}, {Kellogg}, and {Murray}]{lea73a}
	{Lea}, S.M.; {Silk}, J.; {Kellogg}, E.; {Murray}, S.
	\newblock {Thermal-Bremsstrahlung Interpretation of Cluster X-Ray Sources}.
	\newblock {\em Astrophys. J.} {\bf 1973}, {\em 184},~L105.
	\newblock {\url{https://doi.org/10.1086/181300}}.
	
	\bibitem[{Cowie} and {Binney}(1977)]{cowie77a}
	{Cowie}, L.L.; {Binney}, J.
	\newblock {Radiative regulation of gas flow within clusters of galaxies: A
		model for cluster X-ray sources.}
	\newblock {\em Astrophys. J.} {\bf 1977}, {\em 215},~723--732.
	\newblock {\url{https://doi.org/10.1086/155406}}.
	
	\bibitem[{Fabian} and {Nulsen}(1977)]{fabian77a}
	{Fabian}, A.C.; {Nulsen}, P.E.J.
	\newblock {Subsonic accretion of cooling gas in clusters of galaxies.}
	\newblock {\em Mon. Not. R. Astron. Soc.} {\bf 1977}, {\em 180},~479--484.
	\newblock {\url{https://doi.org/10.1093/mnras/180.3.479}}.
	
	\bibitem[{Fabian} and {Nulsen}(1994)]{fabian94a}
	{Fabian}, A.C.; {Nulsen}, P.E.J.
	\newblock {Cooling flows, low-mass objects and the Galactic halo}.
	\newblock {\em Mon. Not. R. Astron. Soc.} {\bf 1994}, {\em 269},~L33.
	
	\bibitem[{Peterson} et~al.(2001){Peterson}, {Paerels}, {Kaastra}, {Arnaud},
	{Reiprich}, {Fabian}, {Mushotzky}, {Jernigan}, and {Sakelliou}]{peterson01a}
	{Peterson}, J.R.; {Paerels}, F.B.S.; {Kaastra}, J.S.; {Arnaud}, M.; {Reiprich},
	T.H.; {Fabian}, A.C.; {Mushotzky}, R.F.; {Jernigan}, J.G.; {Sakelliou}, I.
	\newblock {X-ray imaging-spectroscopy of Abell 1835}.
	\newblock {\em Astron. Astrophys.} {\bf 2001}, {\em 365},~L104--L109.
	\newblock {\url{https://doi.org/10.1051/0004-6361:20000021}}.
	
	\bibitem[{Tamura} et~al.(2001){Tamura}, {Kaastra}, {Peterson}, {Paerels},
	{Mittaz}, {Trudolyubov}, {Stewart}, {Fabian}, {Mushotzky}, {Lumb}, and
	{Ikebe}]{tamura01a}
	{Tamura}, T.; {Kaastra}, J.S.; {Peterson}, J.R.; {Paerels}, F.B.S.; {Mittaz},
	J.P.D.; {Trudolyubov}, S.P.; {Stewart}, G.; {Fabian}, A.C.; {Mushotzky},
	R.F.; {Lumb}, D.H.;  et~al.
	\newblock {X-ray spectroscopy of the cluster of galaxies Abell 1795 with
		XMM-Newton}.
	\newblock {\em Astron. Astrophys.} {\bf 2001}, {\em 365},~L87--L92.
	\newblock {\url{https://doi.org/10.1051/0004-6361:20000038}}.
	
	\bibitem[{Croton} et~al.(2006){Croton}, {Springel}, {White}, {De Lucia},
	{Frenk}, {Gao}, {Jenkins}, {Kauffmann}, {Navarro}, and {Yoshida}]{croton06a}
	{Croton}, D.J.; {Springel}, V.; {White}, S.D.M.; {De Lucia}, G.; {Frenk}, C.S.;
	{Gao}, L.; {Jenkins}, A.; {Kauffmann}, G.; {Navarro}, J.F.; {Yoshida}, N.
	\newblock {The many lives of active galactic nuclei: Cooling flows, black holes
		and the luminosities and colours of galaxies}.
	\newblock {\em Mon. Not. R. Astron. Soc.} {\bf 2006}, {\em 365},~11--28.
	\newblock {\url{https://doi.org/10.1111/j.1365-2966.2005.09675.x}}.
	
	\bibitem[{Silk} and {Rees}(1998)]{silk98a}
	{Silk}, J.; {Rees}, M.J.
	\newblock {Quasars and galaxy formation}.
	\newblock {\em Astron. Astrophys.} {\bf 1998}, {\em 331},~L1--L4.
	
	\bibitem[{Bower} et~al.(2006){Bower}, {Benson}, {Malbon}, {Helly}, {Frenk},
	{Baugh}, {Cole}, and {Lacey}]{bower06a}
	{Bower}, R.G.; {Benson}, A.J.; {Malbon}, R.; {Helly}, J.C.; {Frenk}, C.S.;
	{Baugh}, C.M.; {Cole}, S.; {Lacey}, C.G.
	\newblock {Breaking the hierarchy of galaxy formation}.
	\newblock {\em Mon. Not. R. Astron. Soc.} {\bf 2006}, {\em 370},~645--655.
	\newblock {\url{https://doi.org/10.1111/j.1365-2966.2006.10519.x}}.
	
	\bibitem[{Blandford} et~al.(2019){Blandford}, {Meier}, and
	{Readhead}]{blandford19a}
	{Blandford}, R.; {Meier}, D.; {Readhead}, A.
	\newblock {Relativistic Jets from Active Galactic Nuclei}.
	\newblock {\em Annu. Rev. Astron. Astrophys.} {\bf 2019}, {\em 57},~467--509.
	\newblock {\url{https://doi.org/10.1146/annurev-astro-081817-051948}}.
	
	\bibitem[{Mart{\'\i}}(2019)]{marti19a}
	{Mart{\'\i}}, J.M.
	\newblock {Numerical Simulations of Jets from Active Galactic Nuclei}.
	\newblock {\em Galaxies} {\bf 2019}, {\em 7},~24.
	\newblock {\url{https://doi.org/10.3390/galaxies7010024}}.
	
	\bibitem[{Komissarov} and {Porth}(2021)]{komissarov21a}
	{Komissarov}, S.; {Porth}, O.
	\newblock {Numerical simulations of jets}.
	\newblock {\em New Astron. Rev.} {\bf 2021}, {\em 92},~101610.
	\newblock {\url{https://doi.org/10.1016/j.newar.2021.101610}}.
	
	\bibitem[{Perucho} and {L{\'o}pez-Miralles}(2023)]{perucho23a}
	{Perucho, M.; L{\'o}pez-Miralles, J.} 
	\newblock {Numerical simulations of relativistic jets}.
	\newblock {\em J. Plasma Phys.} {\bf 2023}, {\em 89},~915890501.
	\newblock {\url{https://doi.org/10.1017/S0022377823000892}}.
	
	\bibitem[{Veilleux} et~al.(2020){Veilleux}, {Maiolino}, {Bolatto}, and
	{Aalto}]{veilleux20a}
	{Veilleux}, S.; {Maiolino}, R.; {Bolatto}, A.D.; {Aalto}, S.
	\newblock {Cool outflows in galaxies and their implications}.
	\newblock {\em Astron. Astrophys. Rev.} {\bf 2020}, {\em 28},~2.
	\newblock {\url{https://doi.org/10.1007/s00159-019-0121-9}}.
	
	\bibitem[{Laha} et~al.(2021){Laha}, {Reynolds}, {Reeves}, {Kriss}, {Guainazzi},
	{Smith}, {Veilleux}, and {Proga}]{laha21a}
	{Laha}, S.; {Reynolds}, C.S.; {Reeves}, J.; {Kriss}, G.; {Guainazzi}, M.;
	{Smith}, R.; {Veilleux}, S.; {Proga}, D.
	\newblock {Ionized outflows from active galactic nuclei as the essential
		elements of feedback}.
	\newblock {\em Nat. Astron.} {\bf 2021}, {\em 5},~13--24.
	\newblock {\url{https://doi.org/10.1038/s41550-020-01255-2}}.
	
	\bibitem[{Harrison}(2017)]{harrison17a}
	{Harrison}, C.M.
	\newblock {Impact of supermassive black hole growth on star formation}.
	\newblock {\em Nat. Astron.} {\bf 2017}, {\em 1},~0165.
	\newblock {\url{https://doi.org/10.1038/s41550-017-0165}}.
	
	\bibitem[{Morganti}(2017)]{morganti17a}
	{Morganti}, R.
	\newblock {The many routes to AGN feedback}.
	\newblock {\em Front. Astron. Space Sci.} {\bf 2017}, {\em 4},~42.  
	\newblock {\url{https://doi.org/10.3389/fspas.2017.00042}}.
	
	\bibitem[{Eckert} et~al.(2021){Eckert}, {Gaspari}, {Gastaldello}, {Le Brun},
	and {O'Sullivan}]{eckert21a}
	{Eckert}, D.; {Gaspari}, M.; {Gastaldello}, F.; {Le Brun}, A.M.C.;
	{O'Sullivan}, E.
	\newblock {Feedback from Active Galactic Nuclei in Galaxy Groups}.
	\newblock {\em Universe} {\bf 2021}, {\em 7},~142.
	\newblock {\url{https://doi.org/10.3390/universe7050142}}.
	
	\bibitem[{Combes}(2021)]{combes21a}
	{Combes}, F.
	\newblock {\em {Active Galactic Nuclei: Fueling and Feedback}}; {IoP Publishing: 
	Bristol, UK,} 
	2021.
	\newblock {\url{https://doi.org/10.1088/2514-3433/ac2a27}}.
	
	\bibitem[{Bourne} and {Yang}(2023)]{yang23a}
	{Bourne}, M.A.; {Yang}, H.Y.K.
	\newblock {Recent Progress in Modeling the Macro- and Micro-Physics of Radio
		Jet Feedback in Galaxy Clusters}.
	\newblock {\em Galaxies} {\bf 2023}, {\em 11},~73.
	\newblock {\url{https://doi.org/10.3390/galaxies11030073}}.
	
	\bibitem[{Tadhunter}(2016)]{tadhunter16a}
	{Tadhunter}, C.
	\newblock {Radio AGN in the local universe: Unification, triggering and
		evolution}.
	\newblock {\em Astron. Astrophys. Rev.} {\bf 2016}, {\em 24},~10.
	\newblock {\url{https://doi.org/10.1007/s00159-016-0094-x}}.
	
	\bibitem[{O'Dea} and {Saikia}(2021)]{odea21a}
	{O'Dea}, C.P.; {Saikia}, D.J.
	\newblock {Compact steep-spectrum and peaked-spectrum radio sources}.
	\newblock {\em Astron. Astrophys. Rev.} {\bf 2021}, {\em 29},~3.
	\newblock {\url{https://doi.org/10.1007/s00159-021-00131-w}}.
	
	\bibitem[{Hardcastle} and {Croston}(2020)]{hardcastle20a}
	{Hardcastle}, M.J.; {Croston}, J.H.
	\newblock {Radio galaxies and feedback from AGN jets}.
	\newblock {\em New Astron. Rev.} {\bf 2020}, {\em 88},~101539.
	\newblock {\url{https://doi.org/10.1016/j.newar.2020.101539}}.
	
	\bibitem[{Baldi}(2023)]{baldi23a}
	{Baldi}, R.D.
	\newblock {The nature of compact radio sources: The case of FR 0 radio
		galaxies}.
	\newblock {\em Astron. Astrophys. Rev.} {\bf 2023}, {\em 31},~3.
	\newblock {\url{https://doi.org/10.1007/s00159-023-00148-3}}.
	
	\bibitem[{Morganti} and {Oosterloo}(2018)]{morganti18a}
	{Morganti}, R.; {Oosterloo}, T.
	\newblock {The interstellar and circumnuclear medium of active nuclei traced by
		H i 21 cm absorption}.
	\newblock {\em Astron. Astrophys. Rev.} {\bf 2018}, {\em 26},~4.
	\newblock {\url{https://doi.org/10.1007/s00159-018-0109-x}}.
	
	\bibitem[{Storchi-Bergmann} and {Schnorr-M{\"u}ller}(2019)]{storchiBergmann19a}
	{Storchi-Bergmann}, T.; {Schnorr-M{\"u}ller}, A.
	\newblock {Observational constraints on the feeding of supermassive black
		holes}.
	\newblock {\em Nat. Astron.} {\bf 2019}, {\em 3},~48--61.
	\newblock {\url{https://doi.org/10.1038/s41550-018-0611-0}}.
	
	\bibitem[{Gaspari} et~al.(2020){Gaspari}, {Tombesi}, and {Cappi}]{gaspari20a}
	{Gaspari}, M.; {Tombesi}, F.; {Cappi}, M.
	\newblock {Linking macro-, meso- and microscales in multiphase AGN feeding and
		feedback}.
	\newblock {\em Nat. Astron.} {\bf 2020}, {\em 4},~10--13.
	\newblock {\url{https://doi.org/10.1038/s41550-019-0970-1}}.
	
	\bibitem[{Combes}(2023)]{combes23a}
	{Combes}, F.
	\newblock {Fueling Processes on (Sub-)kpc Scales}.
	\newblock {\em Galaxies} {\bf 2023}, {\em 11},~120.
	\newblock {\url{https://doi.org/10.3390/galaxies11060120}}.
	
	\bibitem[{Wagner} et~al.(2016){Wagner}, {Bicknell}, {Umemura}, {Sutherland},
	and {Silk}]{wagner16a}
	{Wagner}, A.Y.; {Bicknell}, G.V.; {Umemura}, M.; {Sutherland}, R.S.; {Silk}, J.
	\newblock {Galaxy-scale AGN feedback---Theory}.
	\newblock {\em Astron. Nachrichten} {\bf 2016}, {\em 337},~167.
	\newblock {\url{https://doi.org/10.1002/asna.201512287}}.
	
	\bibitem[{Mukherjee} et~al.(2021){Mukherjee}, {Bicknell}, and
	{Wagner}]{mukherjee21b}
	{Mukherjee}, D.; {Bicknell}, G.V.; {Wagner}, A.Y.
	\newblock {Resolved simulations of jet{\textendash}ISM interaction:
		Implications for gas dynamics and star formation}.
	\newblock {\em Astron. Nachrichten} {\bf 2021}, {\em 342},~1140--1145.
	\newblock {\url{https://doi.org/10.1002/asna.20210061}}.
	
	\bibitem[{Morganti} et~al.(2023){Morganti}, {Murthy}, {Guillard}, {Oosterloo},
	and {Garcia-Burillo}]{morganti23a}
	{Morganti}, R.; {Murthy}, S.; {Guillard}, P.; {Oosterloo}, T.;
	{Garcia-Burillo}, S.
	\newblock {Young Radio Sources Expanding in Gas-Rich ISM: Using Cold Molecular
		Gas to Trace Their Impact}.
	\newblock {\em Galaxies} {\bf 2023}, {\em 11},~24.
	\newblock {\url{https://doi.org/10.3390/galaxies11010024}}.
	
	\bibitem[{Krause}(2023)]{krause23a}
	{Krause}, M.G.H.
	\newblock {Jet Feedback in Star-Forming Galaxies}.
	\newblock {\em Galaxies} {\bf 2023}, {\em 11},~29.
	\newblock {\url{https://doi.org/10.3390/galaxies11010029}}.
	
	
	
	\bibitem[{Blandford} and {Rees}(1974)]{blandford74a}
	{Blandford}, R.D.; {Rees}, M.J.
	\newblock {A ``twin-exhaust'' model for double radio sources.}
	\newblock {\em Mon. Not. R. Astron. Soc.} {\bf 1974}, {\em 169},~395--415.
	\newblock {\url{https://doi.org/10.1093/mnras/169.3.395}}.
	
	\bibitem[Scheuer(1974)]{scheuer74a}
	Scheuer, P.A.G.
	\newblock Models of extragalactic radio sources with a continuous energy supply
	from a central object.
	\newblock {\em Mon. Not. R. Astron. Soc.} {\bf 1974}, {\em 166},~513.
	
	\bibitem[{Blandford} and {Znajek}(1977)]{blandford77a}
	{Blandford}, R.D.; {Znajek}, R.L.
	\newblock {Electromagnetic extraction of energy from Kerr black holes}.
	\newblock {\em Mon. Not. R. Astron. Soc.} {\bf 1977}, {\em 179},~433--456.
	
	\bibitem[{Blandford} and {Ostriker}(1978)]{blandford78a}
	{Blandford}, R.D.; {Ostriker}, J.P.
	\newblock {Particle acceleration by astrophysical shocks.}
	\newblock {\em Astrophys. J.} {\bf 1978}, {\em 221},~L29--L32.
	\newblock {\url{https://doi.org/10.1086/182658}}.
	
	\bibitem[{Rayburn}(1977)]{rayburn77a}
	{Rayburn}, D.R.
	\newblock {A numerical study of the continuous beam model of extragalactic
		radio sources.}
	\newblock {\em Mon. Not. R. Astron. Soc.} {\bf 1977}, {\em 179},~603--617.
	\newblock {\url{https://doi.org/10.1093/mnras/179.4.603}}.
	
	\bibitem[{Yokosawa} et~al.(1982){Yokosawa}, {Ikeuchi}, and
	{Sakashita}]{yokosawa82a}
	{Yokosawa}, M.; {Ikeuchi}, S.; {Sakashita}, S.
	\newblock {Structure and Expansion Law of a Hypersonic Beam}.
	\newblock {\em Publ. Astron. Soc. Jpn.} {\bf 1982}, {\em 34},~461.
	
	\bibitem[{Norman} et~al.(1982){Norman}, {Winkler}, {Smarr}, and
	{Smith}]{norman82a}
	{Norman}, M.L.; {Winkler}, K.H.A.; {Smarr}, L.; {Smith}, M.D.
	\newblock {Structure and dynamics of supersonic jets.}
	\newblock {\em Astron. Astrophys.} {\bf 1982}, {\em 113},~285--302.
	
	\bibitem[{Wilson} and {Scheuer}(1983)]{wilson83a}
	{Wilson}, M.J.; {Scheuer}, P.A.G.
	\newblock {The anisotropy of emission from hotspots in extragalactic radio
		sources}.
	\newblock {\em Mon. Not. R. Astron. Soc.} {\bf 1983}, {\em 205},~449--463.
	\newblock {\url{https://doi.org/10.1093/mnras/205.2.449}}.
	
	\bibitem[{Williams} and {Gull}(1984)]{williams84a}
	{Williams}, A.G.; {Gull}, S.F.
	\newblock {A three-dimensional model of the fluid dynamics of radio-trail
		sources}.
	\newblock {\em \nat} {\bf 1984}, {\em 310},~33--36.
	\newblock {\url{https://doi.org/10.1038/310033a0}}.
	
	\bibitem[{Arnold} and {Arnett}(1986)]{arnold86a}
	{Arnold}, C.N.; {Arnett}, W.D.
	\newblock {Three-dimensional Structure and Dynamics of a Supersonic Jet}.
	\newblock {\em Astrophys. J.} {\bf 1986}, {\em 305},~L57.
	\newblock {\url{https://doi.org/10.1086/184684}}.
	
	\bibitem[Hardee and Clarke(1992)]{hardee92}
	Hardee, P.E.; Clarke, D.A.
	\newblock The non-Linear Dynamics of a Three-Dimensional Jet.
	\newblock {\em Astrophys. J.} {\bf 1992}, {\em 400},~L9.
	
	\bibitem[{Norman} and {Balsara}(1993)]{norman93a}
	{Norman}, M.L.; {Balsara}, D.S.
	\newblock {3-D Hydrodynamical Simulations of Extragalactic Jets}. In {\em Jets
		in Extragalactic Radio Sources}; {R{\"o}ser}, H.J., {Meisenheimer}, K., Eds.; 
		{Springer: Berlin/Heidelberg, Germany,} 1993; Volume 421, p. 229. 
	\newblock {\url{https://doi.org/10.1007/3-540-57164-7_98}}.
	
	\bibitem[{Norman} and {Hardee}(1988)]{norman88a}
	{Norman}, M.L.; {Hardee}, P.E.
	\newblock {Spatial Stability of the Slab Jet. II. Numerical Simulations}.
	\newblock {\em Astrophys. J.} {\bf 1988}, {\em 334},~80.
	\newblock {\url{https://doi.org/10.1086/166819}}.
	
	\bibitem[{Clarke} et~al.(1986){Clarke}, {Norman}, and {Burns}]{clarke86a}
	{Clarke}, D.A.; {Norman}, M.L.; {Burns}, J.O.
	\newblock {Numerical Simulations of a Magnetically Confined Jet}.
	\newblock {\em Astrophys. J.} {\bf 1986}, {\em 311},~L63.
	\newblock {\url{https://doi.org/10.1086/184799}}.
	
	\bibitem[{Clarke} et~al.(1989){Clarke}, {Norman}, and {Burns}]{clarke89a}
	{Clarke}, D.A.; {Norman}, M.L.; {Burns}, J.O.
	\newblock {Numerical Observations of a Simulated Jet with a Passive Helical
		Magnetic Field}.
	\newblock {\em Astrophys. J.} {\bf 1989}, {\em 342},~700.
	\newblock {\url{https://doi.org/10.1086/167631}}.
	
	\bibitem[{Koessl} and {Mueller}(1988)]{koessl88a}
	{Koessl}, D.; {Mueller}, E.
	\newblock {Numerical simulations of astrophysical jets: The influence of
		boundary conditions and grid resolution.}
	\newblock {\em Astron. Astrophys.} {\bf 1988}, {\em 206},~204--218.
	
	\bibitem[{Mignone} et~al.(2010){Mignone}, {Rossi}, {Bodo}, {Ferrari}, and
	{Massaglia}]{mignone10a}
	{Mignone}, A.; {Rossi}, P.; {Bodo}, G.; {Ferrari}, A.; {Massaglia}, S.
	\newblock {High-resolution 3D relativistic MHD simulations of jets}.
	\newblock {\em Mon. Not. R. Astron. Soc.} {\bf 2010}, {\em 402},~7--12.
	\newblock {\url{https://doi.org/10.1111/j.1365-2966.2009.15642.x}}.
	
	
	\bibitem[{Sutherland} and {Bicknell}(2007)]{sutherland07a}
	{Sutherland}, R.S.; {Bicknell}, G.V.
	\newblock {Interactions of a Light Hypersonic Jet with a Nonuniform
		Interstellar Medium}.
	\newblock {\em Astrophys. J. Suppl. Ser.} {\bf 2007}, {\em 173},~37.
	\newblock {\url{https://doi.org/10.1086/520640}}.
	
	
	\bibitem[{Cohen} et~al.(1977){Cohen}, {Kellermann}, {Shaffer}, {Linfield},
	{Moffet}, {Romney}, {Seielstad}, {Pauliny-Toth}, {Preuss}, {Witzel},
	{Schilizzi}, and {Geldzahler}]{cohen77a}
	{Cohen}, M.H.; {Kellermann}, K.I.; {Shaffer}, D.B.; {Linfield}, R.P.; {Moffet},
	A.T.; {Romney}, J.D.; {Seielstad}, G.A.; {Pauliny-Toth}, I.I.K.; {Preuss},
	E.; {Witzel}, A.;  et~al.
	\newblock {Radio sources with superluminal velocities}.
	\newblock {\em \nat} {\bf 1977}, {\em 268},~405--409.
	\newblock {\url{https://doi.org/10.1038/268405a0}}.
	
	
	
	\bibitem[{Cohen} et~al.(1979){Cohen}, {Pearson}, {Readhead}, {Seielstad},
	{Simon}, and {Walker}]{cohen79a}
	{Cohen}, M.H.; {Pearson}, T.J.; {Readhead}, A.C.S.; {Seielstad}, G.A.; {Simon},
	R.S.; {Walker}, R.C.
	\newblock {Superluminal variations in 3C 120, 3C 273, and 3C 345.}
	\newblock {\em Astrophys. J.} {\bf 1979}, {\em 231},~293--298.
	\newblock {\url{https://doi.org/10.1086/157192}}.
	
	\bibitem[{O'Dell}(1978)]{O'dell78a}
	{O'Dell}, S.L.
	\newblock {The continuum radiation of compact extragalactic objects.}
	\newblock In Proceedings of the BL Lac Objects, {Pittsburgh, PA, USA, 24--26 April 
	1978;} 
	{Wolfe}, A.M., Ed.;  {University of Pittsburgh: Pittsburgh, PA, USA,}  1978; pp.  
	312--325.
	
	\bibitem[{Blandford} et~al.(1977){Blandford}, {McKee}, and
	{Rees}]{blandford77b}
	{Blandford}, R.D.; {McKee}, C.F.; {Rees}, M.J.
	\newblock {Super-luminal expansion in extragalactic radio sources}.
	\newblock {\em \nat} {\bf 1977}, {\em 267},~211--216.
	\newblock {\url{https://doi.org/10.1038/267211a0}}.
	
	\bibitem[{Blandford} and {K{\"o}nigl}(1979)]{blandford79b}
	{Blandford}, R.D.; {K{\"o}nigl}, A.
	\newblock {Relativistic jets as compact radio sources.}
	\newblock {\em Astrophys. J.} {\bf 1979}, {\em 232},~34--48.
	\newblock {\url{https://doi.org/10.1086/157262}}.
	
	\bibitem[{Wilson}(1987)]{wilson87b}
	{Wilson}, M.J.
	\newblock {Steady relativistic fluid jets}.
	\newblock {\em Mon. Not. R. Astron. Soc.} {\bf 1987}, {\em 226},~447--454.
	\newblock {\url{https://doi.org/10.1093/mnras/226.2.447}}.
	
	\bibitem[{van Putten}(1993)]{vanputten93a}
	{van Putten}, M.H.P.M.
	\newblock {A Two-dimensional Relativistic (Gamma = 3.25) Jet Simulation}.
	\newblock {\em Astrophys. J.} {\bf 1993}, {\em 408},~L21.
	\newblock {\url{https://doi.org/10.1086/186821}}.
	
	\bibitem[{Marti} et~al.(1994){Marti}, {Mueller}, and {Ibanez}]{marti94a}
	{Marti}, J.M.; {Mueller}, E.; {Ibanez}, J.M.
	\newblock {Hydrodynamical simulations of relativistic jets}.
	\newblock {\em Astron. Astrophys.} {\bf 1994}, {\em 281},~L9--L12.
	
	\bibitem[{Marti} et~al.(1995){Marti}, {Muller}, {Font}, and {Ibanez}]{marti95a}
	{Marti}, J.M.A.; {Muller}, E.; {Font}, J.A.; {Ibanez}, J.M.
	\newblock {Morphology and Dynamics of Highly Supersonic Relativistic Jets}.
	\newblock {\em Astrophys. J.} {\bf 1995}, {\em 448},~L105.
	\newblock {\url{https://doi.org/10.1086/309606}}.
	
	\bibitem[{Mart{\'\i}} et~al.(1997){Mart{\'\i}}, {M{\"u}ller}, {Font},
	{Ib{\'a}{\~n}ez}, and {Marquina}]{marti97a}
	{Mart{\'\i}}, J.M.; {M{\"u}ller}, E.; {Font}, J.A.; {Ib{\'a}{\~n}ez}, J.M.Z.;
	{Marquina}, A.
	\newblock {Morphology and Dynamics of Relativistic Jets}.
	\newblock {\em Astrophys. J.} {\bf 1997}, {\em 479},~151--163.
	\newblock {\url{https://doi.org/10.1086/303842}}.
	
	\bibitem[{Duncan} and {Hughes}(1994)]{duncan94a}
	{Duncan}, G.C.; {Hughes}, P.A.
	\newblock {Simulations of Relativistic Extragalactic Jets}.
	\newblock {\em Astrophys. J.} {\bf 1994}, {\em 436},~L119.
	\newblock {\url{https://doi.org/10.1086/187647}}.
	
	\bibitem[{Koide} et~al.(1996){Koide}, {Nishikawa}, and {Mutel}]{koide96a}
	{Koide}, S.; {Nishikawa}, K.I.; {Mutel}, R.L.
	\newblock {A Two-dimensional Simulation of Relativistic Magnetized Jet}.
	\newblock {\em Astrophys. J.} {\bf 1996}, {\em 463},~L71.
	\newblock {\url{https://doi.org/10.1086/310054}}.
	
	\bibitem[{Rosen} et~al.(1999){Rosen}, {Hughes}, {Duncan}, and
	{Hardee}]{rosen99a}
	{Rosen}, A.; {Hughes}, P.A.; {Duncan}, G.C.; {Hardee}, P.E.
	\newblock {A Comparison of the Morphology and Stability of Relativistic and
		Nonrelativistic Jets}.
	\newblock {\em Astrophys. J.} {\bf 1999}, {\em 516},~729--743.
	\newblock {\url{https://doi.org/10.1086/307143}}.
	
	\bibitem[{Bodo} et~al.(2013){Bodo}, {Mamatsashvili}, {Rossi}, and
	{Mignone}]{bodo13a}
	{Bodo}, G.; {Mamatsashvili}, G.; {Rossi}, P.; {Mignone}, A.
	\newblock {Linear stability analysis of magnetized relativistic jets: The
		non-rotating case}.
	\newblock {\em Mon. Not. R. Astron. Soc.} {\bf 2013}, {\em 434},~3030--3046.
	\newblock {\url{https://doi.org/10.1093/mnras/stt1225}}.
	
	\bibitem[{van Putten}(1996)]{vanputten96a}
	{van Putten}, M.H.P.M.
	\newblock {Knots in Simulations of Magnetized Relativistic Jets}.
	\newblock {\em Astrophys. J.} {\bf 1996}, {\em 467},~L57.
	\newblock {\url{https://doi.org/10.1086/310196}}.
	
	\bibitem[{Nishikawa} et~al.(1997){Nishikawa}, {Koide}, {Sakai},
	{Christodoulou}, {Sol}, and {Mutel}]{nishikawa97a}
	{Nishikawa}, K.I.; {Koide}, S.; {Sakai}, J.i.; {Christodoulou}, D.M.; {Sol},
	H.; {Mutel}, R.L.
	\newblock {Three-Dimensional Magnetohydrodynamic Simulations of Relativistic
		Jets Injected along a Magnetic Field}.
	\newblock {\em Astrophys. J.} {\bf 1997}, {\em 483},~L45--L48.
	\newblock {\url{https://doi.org/10.1086/310736}}.
	
	\bibitem[{Nishikawa} et~al.(1998){Nishikawa}, {Koide}, {Sakai},
	{Christodoulou}, {Sol}, and {Mutel}]{nishikawa98a}
	{Nishikawa}, K.I.; {Koide}, S.; {Sakai}, J.i.; {Christodoulou}, D.M.; {Sol},
	H.; {Mutel}, R.L.
	\newblock {Three-dimensional Magnetohydrodynamic Simulations of Relativistic
		Jets Injected into an Oblique Magnetic Field}.
	\newblock {\em Astrophys. J.} {\bf 1998}, {\em 498},~166--169.
	\newblock {\url{https://doi.org/10.1086/305556}}.
	
	\bibitem[{Komissarov}(1999{\natexlab{a}})]{komissarov99a}
	{Komissarov}, S.S.
	\newblock {Numerical simulations of relativistic magnetized jets}.
	\newblock {\em Mon. Not. R. Astron. Soc.} {\bf 1999}, {\em 308},~1069--1076.
	\newblock {\url{https://doi.org/10.1046/j.1365-8711.1999.02783.x}}.
	
	\bibitem[{Komissarov}(1999{\natexlab{b}})]{komissarov99b}
	{Komissarov}, S.S.
	\newblock {A Godunov-type scheme for relativistic magnetohydrodynamics}.
	\newblock {\em Mon. Not. R. Astron. Soc.} {\bf 1999}, {\em 303},~343--366.
	\newblock {\url{https://doi.org/10.1046/j.1365-8711.1999.02244.x}}.
	
	\bibitem[{Koldoba} et~al.(2002){Koldoba}, {Kuznetsov}, and
	{Ustyugova}]{koldoba02a}
	{Koldoba}, A.V.; {Kuznetsov}, O.A.; {Ustyugova}, G.V.
	\newblock {An approximate Riemann solver for relativistic
		magnetohydrodynamics}.
	\newblock {\em Mon. Not. R. Astron. Soc.} {\bf 2002}, {\em 333},~932--942.
	\newblock {\url{https://doi.org/10.1046/j.1365-8711.2002.05474.x}}.
	
	\bibitem[{Del Zanna} and {Bucciantini}(2002)]{delzanna02a}
	{Del Zanna}, L.; {Bucciantini}, N.
	\newblock {An efficient shock-capturing central-type scheme for
		multidimensional relativistic flows. I. Hydrodynamics}.
	\newblock {\em Astron. Astrophys.} {\bf 2002}, {\em 390},~1177--1186.
	\newblock {\url{https://doi.org/10.1051/0004-6361:20020776}}.
	
	\bibitem[{Del Zanna} et~al.(2003){Del Zanna}, {Bucciantini}, and
	{Londrillo}]{delzanna03a}
	{Del Zanna}, L.; {Bucciantini}, N.; {Londrillo}, P.
	\newblock {An efficient shock-capturing central-type scheme for
		multidimensional relativistic flows. II. Magnetohydrodynamics}.
	\newblock {\em Astron. Astrophys.} {\bf 2003}, {\em 400},~397--413.
	\newblock {\url{https://doi.org/10.1051/0004-6361:20021641}}.
	
	\bibitem[{Leismann} et~al.(2005){Leismann}, {Ant{\'o}n}, {Aloy}, {M{\"u}ller},
	{Mart{\'\i}}, {Miralles}, and {Ib{\'a}{\~n}ez}]{leismann05a}
	{Leismann}, T.; {Ant{\'o}n}, L.; {Aloy}, M.A.; {M{\"u}ller}, E.; {Mart{\'\i}},
	J.M.; {Miralles}, J.A.; {Ib{\'a}{\~n}ez}, J.M.
	\newblock {Relativistic MHD simulations of extragalactic jets}.
	\newblock {\em Astron. Astrophys.} {\bf 2005}, {\em 436},~503--526.
	\newblock {\url{https://doi.org/10.1051/0004-6361:20042520}}.
	
	\bibitem[{Mignone} et~al.(2005){Mignone}, {Plewa}, and {Bodo}]{mignone05a}
	{Mignone}, A.; {Plewa}, T.; {Bodo}, G.
	\newblock {The Piecewise Parabolic Method for Multidimensional Relativistic
		Fluid Dynamics}.
	\newblock {\em Astrophys. J. Suppl. Ser.} {\bf 2005}, {\em 160},~199--219.
	\newblock {\url{https://doi.org/10.1086/430905}}.
	
	\bibitem[{Mignone} and {Bodo}(2006)]{mignone06a}
	{Mignone}, A.; {Bodo}, G.
	\newblock {An HLLC Riemann solver for relativistic flows---II.
		Magnetohydrodynamics}.
	\newblock {\em Mon. Not. R. Astron. Soc.} {\bf 2006}, {\em 368},~1040--1054.
	\newblock {\url{https://doi.org/10.1111/j.1365-2966.2006.10162.x}}.
	
	\bibitem[{Mart{\'\i}} and {M{\"u}ller}(2003)]{marti03a}
	{Mart{\'\i}}, J.M.; {M{\"u}ller}, E.
	\newblock {Numerical Hydrodynamics in Special Relativity}.
	\newblock {\em Living Rev. Relativ.} {\bf 2003}, {\em 6},~7.
	\newblock {\url{https://doi.org/10.12942/lrr-2003-7}}.
	
	\bibitem[{Mukherjee} et~al.(2020){Mukherjee}, {Bodo}, {Mignone}, {Rossi}, and
	{Vaidya}]{mukherjee20a}
	{Mukherjee}, D.; {Bodo}, G.; {Mignone}, A.; {Rossi}, P.; {Vaidya}, B.
	\newblock {Simulating the dynamics and non-thermal emission of relativistic
		magnetized jets I. Dynamics}.
	\newblock {\em Mon. Not. R. Astron. Soc.} {\bf 2020}, {\em 499},~681--701.
	\newblock {\url{https://doi.org/10.1093/mnras/staa2934}}.
	
	\bibitem[{Meenakshi} et~al.(2023){Meenakshi}, {Mukherjee}, {Bodo}, and
	{Rossi}]{meenakshi23a}
	{Meenakshi}, M.; {Mukherjee}, D.; {Bodo}, G.; {Rossi}, P.
	\newblock {A polarization study of jets interacting with turbulent magnetic
		fields}.
	\newblock {\em Mon. Not. R. Astron. Soc.} {\bf 2023}, {\em 526},~5418--5440.
	\newblock {\url{https://doi.org/10.1093/mnras/stad3092}}.
	
	\bibitem[{Mattia} et~al.(2023){Mattia}, {Del Zanna}, {Bugli}, {Pavan},
	{Ciolfi}, {Bodo}, and {Mignone}]{mattia23a}
	{Mattia}, G.; {Del Zanna}, L.; {Bugli}, M.; {Pavan}, A.; {Ciolfi}, R.; {Bodo},
	G.; {Mignone}, A.
	\newblock {Resistive relativistic MHD simulations of astrophysical jets}.
	\newblock {\em Astron. Astrophys.} {\bf 2023}, {\em 679},~A49.
	\newblock {\url{https://doi.org/10.1051/0004-6361/202347126}}.

	
	\bibitem[{Rossi} et~al.(2024){Rossi}, {Bodo}, {Massaglia}, and
	{Capetti}]{rossi24a}
	{Rossi}, P.; {Bodo}, G.; {Massaglia}, S.; {Capetti}, A.
	\newblock {The different flavors of extragalactic jets: Magnetized relativistic
		flows}.
	\newblock {\em Astron. Astrophys.} {\bf 2024}, {\em 685},~A4.
	\newblock {\url{https://doi.org/10.1051/0004-6361/202348864}}.
	
	\bibitem[{Upreti} et~al.(2024){Upreti}, {Vaidya}, and {Shukla}]{upreti24a}
	{Upreti}, N.; {Vaidya}, B.; {Shukla}, A.
	\newblock {Bridging simulations of kink instability in relativistic magnetized
		jets with radio emission and polarisation}.
	\newblock {\em J. High Energy Astrophys.} {\bf 2024}, {\em
		44},~146--163.  
	\newblock {\url{https://doi.org/10.1016/j.jheap.2024.09.007}}.
	
	\bibitem[{Costa} et~al.(2024){Costa}, {Bodo}, {Tavecchio}, {Rossi}, {Capetti},
	{Massaglia}, {Sciaccaluga}, {Baldi}, and {Giovannini}]{costa24a}
	{Costa}, A.; {Bodo}, G.; {Tavecchio}, F.; {Rossi}, P.; {Capetti}, A.;
	{Massaglia}, S.; {Sciaccaluga}, A.; {Baldi}, R.D.; {Giovannini}, G.
	\newblock {FR0 jets and recollimation-induced instabilities}.
	\newblock {\em Astron. Astrophys.} {\bf 2024}, {\em 682},~L19.
	\newblock {\url{https://doi.org/10.1051/0004-6361/202348954}}.
	
	\bibitem[{Costa} et~al.(2025){Costa}, {Bodo}, {Tavecchio}, {Rossi}, {Coppi},
	{Sciaccaluga}, and {Boula}]{costa25a}
	{Costa}, A.; {Bodo}, G.; {Tavecchio}, F.; {Rossi}, P.; {Coppi}, P.;
	{Sciaccaluga}, A.; {Boula}, S.
	\newblock {How do recollimation-induced instabilities shape the propagation of
		hydrodynamic relativistic jets?}
	\newblock {\em arXiv} {\bf 2025}, arXiv:2503.18602.
	\newblock {\url{https://doi.org/10.48550/arXiv.2503.18602}}.
	
	\bibitem[{Perucho} and {Lobanov}(2007)]{perucho07a}
	{Perucho}, M.; {Lobanov}, A.P.
	\newblock {Physical properties of the jet in <ASTROBJ>0836+710</ASTROBJ>
		revealed by its transversal structure}.
	\newblock {\em Astron. Astrophys.} {\bf 2007}, {\em 469},~L23--L26.
	\newblock {\url{https://doi.org/10.1051/0004-6361:20077610}}.
	
	\bibitem[{Rossi} et~al.(2008){Rossi}, {Mignone}, {Bodo}, {Massaglia}, and
	{Ferrari}]{rossi08a}
	{Rossi}, P.; {Mignone}, A.; {Bodo}, G.; {Massaglia}, S.; {Ferrari}, A.
	\newblock {Formation of dynamical structures in relativistic jets: The FRI
		case}.
	\newblock {\em Astron. Astrophys.} {\bf 2008}, {\em 488},~795--806.
	\newblock {\url{https://doi.org/10.1051/0004-6361:200809687}}.
	
	\bibitem[{Perucho} et~al.(2014){Perucho}, {Mart{\'\i}}, {Laing}, and
	{Hardee}]{perucho14b}
	{Perucho}, M.; {Mart{\'\i}}, J.M.; {Laing}, R.A.; {Hardee}, P.E.
	\newblock {On the deceleration of Fanaroff-Riley Class I jets: Mass loading by
		stellar winds}.
	\newblock {\em Mon. Not. R. Astron. Soc.} {\bf 2014}, {\em 441},~1488--1503.
	\newblock {\url{https://doi.org/10.1093/mnras/stu676}}.
	
	\bibitem[{Massaglia} et~al.(2016){Massaglia}, {Bodo}, {Rossi}, {Capetti}, and
	{Mignone}]{massaglia16a}
	{Massaglia}, S.; {Bodo}, G.; {Rossi}, P.; {Capetti}, S.; {Mignone}, A.
	\newblock {Making Faranoff-Riley I radio sources. I. Numerical hydrodynamic 3D
		simulations of low-power jets}.
	\newblock {\em Astron. Astrophys.} {\bf 2016}, {\em 596},~A12.
	\newblock {\url{https://doi.org/10.1051/0004-6361/201629375}}.
	
	\bibitem[{Massaglia} et~al.(2019){Massaglia}, {Bodo}, {Rossi}, {Capetti}, and
	{Mignone}]{massaglia19a}
	{Massaglia}, S.; {Bodo}, G.; {Rossi}, P.; {Capetti}, S.; {Mignone}, A.
	\newblock {Making Faranoff-Riley I radio sources. II. The effects of jet
		magnetization}.
	\newblock {\em Astron. Astrophys.} {\bf 2019}, {\em 621},~A132.
	\newblock {\url{https://doi.org/10.1051/0004-6361/201834512}}.
	
	\bibitem[{Rossi} et~al.(2020){Rossi}, {Bodo}, {Massaglia}, and
	{Capetti}]{rossi20a}
	{Rossi}, P.; {Bodo}, G.; {Massaglia}, S.; {Capetti}, A.
	\newblock {The different flavors of extragalactic jets: The role of
		relativistic flow deceleration}.
	\newblock {\em Astron. Astrophys.} {\bf 2020}, {\em 642},~A69.
	\newblock {\url{https://doi.org/10.1051/0004-6361/202038725}}.
	
	\bibitem[{Massaglia} et~al.(2022){Massaglia}, {Bodo}, {Rossi}, {Capetti}, and
	{Mignone}]{massaglia22a}
	{Massaglia}, S.; {Bodo}, G.; {Rossi}, P.; {Capetti}, A.; {Mignone}, A.
	\newblock {Making Fanaroff-Riley I radio sources. III. The effects of the
		magnetic field on relativistic jets' propagation and source morphologies}.
	\newblock {\em Astron. Astrophys.} {\bf 2022}, {\em 659},~A139.
	\newblock {\url{https://doi.org/10.1051/0004-6361/202038724}}.
	
	\bibitem[{Bhattacharjee} et~al.(2024){Bhattacharjee}, {Seo}, {Ryu}, and
	{Kang}]{bhattacharjee24a}
	{Bhattacharjee}, A.; {Seo}, J.; {Ryu}, D.; {Kang}, H.
	\newblock {A Simulation Study of Low-power Relativistic Jets: Flow Dynamics and
		Radio Morphology of FR-I Jets}.
	\newblock {\em Astrophys. J.} {\bf 2024}, {\em 976},~91.
	\newblock {\url{https://doi.org/10.3847/1538-4357/ad83cc}}.
	
	\bibitem[{Perucho} et~al.(2019){Perucho}, {Mart{\'\i}}, and
	{Quilis}]{perucho19a}
	{Perucho}, M.; {Mart{\'\i}}, J.M.; {Quilis}, V.
	\newblock {Long-term FRII jet evolution: Clues from three-dimensional
		simulations}.
	\newblock {\em Mon. Not. R. Astron. Soc.} {\bf 2019}, {\em 482},~3718--3735.
	\newblock {\url{https://doi.org/10.1093/mnras/sty2912}}.
	
	\bibitem[{Seo} et~al.(2021){Seo}, {Kang}, and {Ryu}]{seo21a}
	{Seo}, J.; {Kang}, H.; {Ryu}, D.
	\newblock {A Simulation Study of Ultra-relativistic Jets. II. Structures and
		Dynamics of FR-II Jets}.
	\newblock {\em Astrophys. J.} {\bf 2021}, {\em 920},~144.
	\newblock {\url{https://doi.org/10.3847/1538-4357/ac19b4}}.
	
	\bibitem[{Perucho} et~al.(2022){Perucho}, {Mart{\'\i}}, and
	{Quilis}]{perucho22a}
	{Perucho}, M.; {Mart{\'\i}}, J.M.; {Quilis}, V.
	\newblock {Long-term FRII jet evolution in dense environments}.
	\newblock {\em Mon. Not. R. Astron. Soc.} {\bf 2022}, {\em 510},~2084--2096.
	\newblock {\url{https://doi.org/10.1093/mnras/stab3560}}.
	
	\bibitem[Begelman and Cioffi(1989)]{begelman89a}
	Begelman, M.C.; Cioffi, D.F.
	\newblock Overpressured Cocoons in Extragalactic Radio Sources.
	\newblock {\em Astrophys. J.} {\bf 1989}, {\em 345},~L21.
	
	\bibitem[{Kaiser} and {Alexander}(1997)]{kaiser97a}
	{Kaiser}, C.R.; {Alexander}, P.
	\newblock {A self-similar model for extragalactic radio sources}.
	\newblock {\em Mon. Not. R. Astron. Soc.} {\bf 1997}, {\em 286},~215--222.
	\newblock {\url{https://doi.org/10.1093/mnras/286.1.215}}.
	
	\bibitem[{Falle}(1991)]{falle91a}
	{Falle}, S.A.E.G.
	\newblock Self-similar jets.
	\newblock {\em Mon. Not. R. Astron. Soc.} {\bf 1991}, {\em 250},~581.
	
	\bibitem[{Carvalho} and {O'Dea}(2002)]{carvalho02a}
	{Carvalho}, J.C.; {O'Dea}, C.P.
	\newblock {Evolution of Global Properties of Powerful Radio Sources. I.
		Hydrodynamical Simulations in a Constant Density Atmosphere and Comparison
		with Self-similar Models}.
	\newblock {\em Astrophys. J. Suppl. Ser.} {\bf 2002}, {\em 141},~337--370.
	\newblock {\url{https://doi.org/10.1086/340645}}.
	
	\bibitem[{O'Neill} et~al.(2005){O'Neill}, {Tregillis}, {Jones}, and
	{Ryu}]{o'neill05a}
	{O'Neill}, S.M.; {Tregillis}, I.L.; {Jones}, T.W.; {Ryu}, D.
	\newblock {Three-dimensional Simulations of MHD Jet Propagation through Uniform
		and Stratified External Environments}.
	\newblock {\em Astrophys. J.} {\bf 2005}, {\em 633},~717--732.
	\newblock {\url{https://doi.org/10.1086/491618}}.
	
	\bibitem[{Perucho} et~al.(2011){Perucho}, {Quilis}, and
	{Mart{\'\i}}]{perucho11a}
	{Perucho}, M.; {Quilis}, V.; {Mart{\'\i}}, J.M.
	\newblock {Intracluster Medium Reheating by Relativistic Jets}.
	\newblock {\em Astrophys. J.} {\bf 2011}, {\em 743},~42.
	\newblock {\url{https://doi.org/10.1088/0004-637X/743/1/42}}.
	
	\bibitem[{Perucho} et~al.(2014){Perucho}, {Mart{\'\i}}, {Quilis}, and
	{Ricciardelli}]{perucho14a}
	{Perucho}, M.; {Mart{\'\i}}, J.M.; {Quilis}, V.; {Ricciardelli}, E.
	\newblock {Large-scale jets from active galactic nuclei as a source of
		intracluster medium heating: Cavities and shocks}.
	\newblock {\em Mon. Not. R. Astron. Soc.} {\bf 2014}, {\em 445},~1462--1481.
	\newblock {\url{https://doi.org/10.1093/mnras/stu1828}}.
	
	\bibitem[{Hardcastle} and {Krause}(2013)]{hardcastle13b}
	{Hardcastle}, M.J.; {Krause}, M.G.H.
	\newblock {Numerical modelling of the lobes of radio galaxies in cluster
		environments}.
	\newblock {\em Mon. Not. R. Astron. Soc.} {\bf 2013}, {\em 430},~174--196.
	\newblock {\url{https://doi.org/10.1093/mnras/sts564}}.
	
	\bibitem[{Hardcastle} and {Krause}(2014)]{hardcastle14a}
	{Hardcastle}, M.J.; {Krause}, M.G.H.
	\newblock {Numerical modelling of the lobes of radio galaxies in cluster
		environments---II. Magnetic field configuration and observability}.
	\newblock {\em Mon. Not. R. Astron. Soc.} {\bf 2014}, {\em 443},~1482--1499.
	\newblock {\url{https://doi.org/10.1093/mnras/stu1229}}.
	
	\bibitem[{English} et~al.(2016){English}, {Hardcastle}, and
	{Krause}]{english16a}
	{English}, W.; {Hardcastle}, M.J.; {Krause}, M.G.H.
	\newblock {Numerical modelling of the lobes of radio galaxies in cluster
		environments---III. Powerful relativistic and non-relativistic jets}.
	\newblock {\em Mon. Not. R. Astron. Soc.} {\bf 2016}, {\em 461},~2025--2043.
	\newblock {\url{https://doi.org/10.1093/mnras/stw1407}}.
	
	\bibitem[{English} et~al.(2019){English}, {Hardcastle}, and
	{Krause}]{english19a}
	{English}, W.; {Hardcastle}, M.J.; {Krause}, M.G.H.
	\newblock {Numerical modelling of the lobes of radio galaxies in cluster
		environments---IV. Remnant radio galaxies}.
	\newblock {\em Mon. Not. R. Astron. Soc.} {\bf 2019}, {\em 490},~5807--5819.
	\newblock {\url{https://doi.org/10.1093/mnras/stz2978}}.
	
	\bibitem[{Chen} et~al.(2019){Chen}, {Heinz}, and {En{\ss}lin}]{chen19a}
	{Chen}, Y.H.; {Heinz}, S.; {En{\ss}lin}, T.A.
	\newblock {Jets, bubbles, and heat pumps in galaxy clusters}.
	\newblock {\em Mon. Not. R. Astron. Soc.} {\bf 2019}, {\em 489},~1939--1949.
	\newblock {\url{https://doi.org/10.1093/mnras/stz2256}}.
	
	\bibitem[{Jerrim} et~al.(2024){Jerrim}, {Shabala}, {Yates-Jones}, {Krause},
	{Turner}, {Anderson}, {Stewart}, {Power}, and {Rodman}]{jerrim24a}
	{Jerrim}, L.A.; {Shabala}, S.S.; {Yates-Jones}, P.M.; {Krause}, M.G.H.;
	{Turner}, R.J.; {Anderson}, C.S.; {Stewart}, G.S.C.; {Power}, C.; {Rodman},
	P.E.
	\newblock {Faraday rotation as a probe of radio galaxy environment in RMHD AGN
		jet simulations}.
	\newblock {\em Mon. Not. R. Astron. Soc.} {\bf 2024}, {\em 531},~2532--2550.
	\newblock {\url{https://doi.org/10.1093/mnras/stae1317}}.
	
	\bibitem[{Giri} et~al.(2025){Giri}, {Bagchi}, {Thorat}, {Deane}, {Delhaize},
	and {Saikia}]{giri25a}
	{Giri}, G.; {Bagchi}, J.; {Thorat}, K.; {Deane}, R.P.; {Delhaize}, J.;
	{Saikia}, D.J.
	\newblock {Probing the formation of megaparsec-scale giant radio galaxies: I.
		Dynamical insights from magnetohydrodynamic simulations}.
	\newblock {\em Astron. Astrophys.} {\bf 2025}, {\em 693},~A77.
	\newblock {\url{https://doi.org/10.1051/0004-6361/202451812}}.
	
	\bibitem[{Hardcastle}(2018)]{hardcastle18a}
	{Hardcastle}, M.J.
	\newblock {A simulation-based analytic model of radio galaxies}.
	\newblock {\em Mon. Not. R. Astron. Soc.} {\bf 2018}, {\em 475},~2768--2786.
	\newblock {\url{https://doi.org/10.1093/mnras/stx3358}}.
	
	\bibitem[{Matthews} et~al.(2019){Matthews}, {Bell}, {Blundell}, and
	{Araudo}]{matthews19a}
	{Matthews}, J.H.; {Bell}, A.R.; {Blundell}, K.M.; {Araudo}, A.T.
	\newblock {Ultrahigh energy cosmic rays from shocks in the lobes of powerful
		radio galaxies}.
	\newblock {\em Mon. Not. R. Astron. Soc.} {\bf 2019}, {\em 482},~4303--4321.
	\newblock {\url{https://doi.org/10.1093/mnras/sty2936}}.
	
	\bibitem[{Seo} et~al.(2023){Seo}, {Ryu}, and {Kang}]{seo23a}
	{Seo}, J.; {Ryu}, D.; {Kang}, H.
	\newblock {A Simulation Study of Ultra-relativistic Jets. III. Particle
		Acceleration in FR-II Jets}.
	\newblock {\em Astrophys. J.} {\bf 2023}, {\em 944},~199.
	\newblock {\url{https://doi.org/10.3847/1538-4357/acb3ba}}.
	
	\bibitem[{Seo} et~al.(2024){Seo}, {Ryu}, and {Kang}]{seo24a}
	{Seo}, J.; {Ryu}, D.; {Kang}, H.
	\newblock {Model Spectrum of Ultrahigh-energy Cosmic Rays Accelerated in FR-I
		Radio Galaxy Jets}.
	\newblock {\em Astrophys. J.} {\bf 2024}, {\em 962},~46.
	\newblock {\url{https://doi.org/10.3847/1538-4357/ad182c}}.
	
	\bibitem[{Ohmura} and {Machida}(2023)]{ohmura23a}
	{Ohmura}, T.; {Machida}, M.
	\newblock {Simulations of two-temperature jets in galaxy clusters. I. Effect of
		jet magnetization on dynamics and electron heating}.
	\newblock {\em Astron. Astrophys.} {\bf 2023}, {\em 679},~A160.
	\newblock {\url{https://doi.org/10.1051/0004-6361/202244690}}.
	
	\bibitem[{Ohmura} et~al.(2023){Ohmura}, {Machida}, and {Akamatsu}]{ohmura23b}
	{Ohmura}, T.; {Machida}, M.; {Akamatsu}, H.
	\newblock {Simulations of two-temperature jets in galaxy clusters. II. X-ray
		properties of the forward shock}.
	\newblock {\em Astron. Astrophys.} {\bf 2023}, {\em 679},~A161.
	\newblock {\url{https://doi.org/10.1051/0004-6361/202244692}}.
	
	\bibitem[{Joshi} and {Chattopadhyay}(2023)]{joshi23a}
	{Joshi}, R.K.; {Chattopadhyay}, I.
	\newblock {The Morphology and Dynamics of Relativistic Jets with Relativistic
		Equation of State}.
	\newblock {\em Astrophys. J.} {\bf 2023}, {\em 948},~13.
	\newblock {\url{https://doi.org/10.3847/1538-4357/acc93d}}.
	
	\bibitem[{Rossi} et~al.(2017){Rossi}, {Bodo}, {Capetti}, and
	{Massaglia}]{rossi17a}
	{Rossi}, P.; {Bodo}, G.; {Capetti}, A.; {Massaglia}, S.
	\newblock {3D relativistic MHD numerical simulations of X-shaped radio
		sources}.
	\newblock {\em Astron. Astrophys.} {\bf 2017}, {\em 606},~A57.
	\newblock {\url{https://doi.org/10.1051/0004-6361/201730594}}.
	
	\bibitem[{Nawaz} et~al.(2014){Nawaz}, {Wagner}, {Bicknell}, {Sutherland}, and
	{McNamara}]{nawaz14a}
	{Nawaz}, M.A.; {Wagner}, A.Y.; {Bicknell}, G.V.; {Sutherland}, R.S.;
	{McNamara}, B.R.
	\newblock {Jet-intracluster medium interaction in Hydra A--I. Estimates of jet
		velocity from inner knots}.
	\newblock {\em Mon. Not. R. Astron. Soc.} {\bf 2014}, {\em 444},~1600--1614.
	\newblock {\url{https://doi.org/10.1093/mnras/stu1563}}.
	
	\bibitem[{Nawaz} et~al.(2016){Nawaz}, {Wagner}, {Bicknell}, {Sutherland}, and
	{McNamara}]{nawaz16a}
	{Nawaz}, M.A.; {Wagner}, A.Y.; {Bicknell}, G.V.; {Sutherland}, R.S.;
	{McNamara}, B.R.
	\newblock {Jet-intracluster medium interaction in Hydra A---II The effect of
		jet precession}.
	\newblock {\em Mon. Not. R. Astron. Soc.} {\bf 2016}, {\textit{458}, 802--815.} 
	
	
	\bibitem[{Horton} et~al.(2020){Horton}, {Krause}, and {Hardcastle}]{horton20a}
	{Horton}, M.A.; {Krause}, M.G.H.; {Hardcastle}, M.J.
	\newblock {3D hydrodynamic simulations of large-scale precessing jets: Radio
		morphology}.
	\newblock {\em Mon. Not. R. Astron. Soc.} {\bf 2020}, {\em 499},~5765--5781.
	\newblock {\url{https://doi.org/10.1093/mnras/staa3020}}.
	
	\bibitem[{Giri} et~al.(2022){Giri}, {Vaidya}, {Rossi}, {Bodo}, {Mukherjee}, and
	{Mignone}]{giri22a}
	{Giri}, G.; {Vaidya}, B.; {Rossi}, P.; {Bodo}, G.; {Mukherjee}, D.; {Mignone},
	A.
	\newblock {Modelling X-shaped radio galaxies: Dynamical and emission signatures
		from the Back-flow model}.
	\newblock {\em Astron. Astrophys.} {\bf 2022}, {\em 662},~A5.
	\newblock {\url{https://doi.org/10.1051/0004-6361/202142546}}.
	
	\bibitem[{Giri} et~al.(2023){Giri}, {Vaidya}, and {Fendt}]{giri23a}
	{Giri}, G.; {Vaidya}, B.; {Fendt}, C.
	\newblock {Deciphering the Morphological Origins of X-shaped Radio Galaxies:
		Numerical Modeling of Backflow versus Jet Reorientation}.
	\newblock {\em Astrophys. J. Suppl. Ser.} {\bf 2023}, {\em 268},~49.
	\newblock {\url{https://doi.org/10.3847/1538-4365/acebca}}.
	
	\bibitem[{Giri} et~al.(2024){Giri}, {Fendt}, {Thorat}, {Bodo}, and
	{Rossi}]{giri24a}
	{Giri}, G.; {Fendt}, C.; {Thorat}, K.; {Bodo}, G.; {Rossi}, P.
	\newblock {X-shaped radio galaxies: Probing jet evolution, ambient medium
		dynamics, and their intricate interconnection}.
	\newblock {\em Front. Astron. Space Sci.} {\bf 2024}, {\em
		11},~1371101.  
	\newblock {\url{https://doi.org/10.3389/fspas.2024.1371101}}.
	
	\bibitem[{Tregillis} et~al.(2001){Tregillis}, {Jones}, and {Ryu}]{tregillis01a}
	{Tregillis}, I.L.; {Jones}, T.W.; {Ryu}, D.
	\newblock {Simulating Electron Transport and Synchrotron Emission in Radio
		Galaxies: Shock Acceleration and Synchrotron Aging in Three-dimensional
		Flows}.
	\newblock {\em Astrophys. J.} {\bf 2001}, {\em 557},~475--491.
	\newblock {\url{https://doi.org/10.1086/321657}}.
	
	\bibitem[{Tregillis} et~al.(2004){Tregillis}, {Jones}, and {Ryu}]{tregillis04a}
	{Tregillis}, I.L.; {Jones}, T.W.; {Ryu}, D.
	\newblock {Synthetic Observations of Simulated Radio Galaxies. I. Radio and
		X-Ray Analysis}.
	\newblock {\em Astrophys. J.} {\bf 2004}, {\em 601},~778--797.
	\newblock {\url{https://doi.org/10.1086/380756}}.
	
	\bibitem[{Vaidya} et~al.(2018){Vaidya}, {Mignone}, {Bodo}, {Rossi}, and
	{Massaglia}]{vaidya18a}
	{Vaidya}, B.; {Mignone}, A.; {Bodo}, G.; {Rossi}, P.; {Massaglia}, S.
	\newblock {A Particle Module for the PLUTO Code. II. Hybrid Framework for
		Modeling Nonthermal Emission from Relativistic Magnetized Flows}.
	\newblock {\em Astrophys. J.} {\bf 2018}, {\em 865},~144.
	\newblock {\url{https://doi.org/10.3847/1538-4357/aadd17}}.
	
	\bibitem[{Mukherjee} et~al.(2021){Mukherjee}, {Bodo}, {Rossi}, {Mignone}, and
	{Vaidya}]{mukherjee21a}
	{Mukherjee}, D.; {Bodo}, G.; {Rossi}, P.; {Mignone}, A.; {Vaidya}, B.
	\newblock {Simulating the dynamics and synchrotron emission from relativistic
		jets---II. Evolution of non-thermal electrons}.
	\newblock {\em Mon. Not. R. Astron. Soc.} {\bf 2021}, {\em 505},~2267--2284.
	\newblock {\url{https://doi.org/10.1093/mnras/stab1327}}.
	
	\bibitem[{Meenakshi} et~al.(2024){Meenakshi}, {Mukherjee}, {Bodo}, {Rossi}, and
	{Harrison}]{meenakshi24a}
	{Meenakshi}, M.; {Mukherjee}, D.; {Bodo}, G.; {Rossi}, P.; {Harrison}, C.M.
	\newblock {A comparative study of radio signatures from winds and jets:
		modelling synchrotron emission and polarization}.
	\newblock {\em Mon. Not. R. Astron. Soc.} {\bf 2024}, {\em 533},~2213--2231.
	\newblock {\url{https://doi.org/10.1093/mnras/stae1890}}.
	
	\bibitem[{Chen} et~al.(2023){Chen}, {Heinz}, and {Hooper}]{chen23a}
	{Chen}, Y.H.; {Heinz}, S.; {Hooper}, E.
	\newblock {A numerical study of the impact of jet magnetic topology on radio
		galaxy evolution}.
	\newblock {\em Mon. Not. R. Astron. Soc.} {\bf 2023}, {\em 522},~2850--2868.
	\newblock {\url{https://doi.org/10.1093/mnras/stad1074}}.
	
	\bibitem[{Dubey} et~al.(2023){Dubey}, {Fendt}, and {Vaidya}]{dubey23a}
	{Dubey}, R.P.; {Fendt}, C.; {Vaidya}, B.
	\newblock {Particles in Relativistic MHD Jets. I. Role of Jet Dynamics in
		Particle Acceleration}.
	\newblock {\em Astrophys. J.} {\bf 2023}, {\em 952},~1.
	\newblock {\url{https://doi.org/10.3847/1538-4357/ace0bf}}.
	
	\bibitem[{Dubey} et~al.(2024){Dubey}, {Fendt}, and {Vaidya}]{dubey24a}
	{Dubey}, R.P.; {Fendt}, C.; {Vaidya}, B.
	\newblock {Particles in Relativistic Magnetohydrodynamic Jets. II. Bridging Jet
		Dynamics with Multi{\textendash}wave band Nonthermal Emission Signatures}.
	\newblock {\em Astrophys. J.} {\bf 2024}, {\em 976},~144.
	\newblock {\url{https://doi.org/10.3847/1538-4357/ad8135}}.
	
	\bibitem[{Blandford} and {Koenigl}(1979)]{blandford79a}
	{Blandford}, R.D.; {Koenigl}, A.
	\newblock {A Model for the Knots in the M87 Jet}.
	\newblock {\em Astrophys. Lett.} {\bf 1979}, {\em 20},~15.
	
	\bibitem[{Schreier} et~al.(1982){Schreier}, {Gorenstein}, and
	{Feigelson}]{schreier82a}
	{Schreier}, E.J.; {Gorenstein}, P.; {Feigelson}, E.D.
	\newblock {High-resolution X-ray observations of M87---Nucleus, jet and radio
		halo}.
	\newblock {\em Astrophys. J.} {\bf 1982}, {\em 261},~42--50.
	\newblock {\url{https://doi.org/10.1086/160316}}.
	
	\bibitem[{Biretta} et~al.(1983){Biretta}, {Owen}, and {Hardee}]{biretta83a}
	{Biretta}, J.A.; {Owen}, F.N.; {Hardee}, P.E.
	\newblock {Observations of the M 87 jet at 15 GHz with 0''.12 resolution.}
	\newblock {\em Astrophys. J.} {\bf 1983}, {\em 274},~L27--L30.
	\newblock {\url{https://doi.org/10.1086/184144}}.
	
	\bibitem[{Falle} and {Wilson}(1985)]{falle85a}
	{Falle}, S.A.E.G.; {Wilson}, M.J.
	\newblock {A theoretical model of the M 87 jet.}
	\newblock {\em Mon. Not. R. Astron. Soc.} {\bf 1985}, {\em 216},~79--84.
	\newblock {\url{https://doi.org/10.1093/mnras/216.1.79}}.
	
	\bibitem[{Biretta} et~al.(1995){Biretta}, {Zhou}, and {Owen}]{biretta95a}
	{Biretta}, J.A.; {Zhou}, F.; {Owen}, F.N.
	\newblock {Detection of Proper Motions in the M87 Jet}.
	\newblock {\em Astrophys. J.} {\bf 1995}, {\em 447},~582.
	\newblock {\url{https://doi.org/10.1086/175901}}.
	
	\bibitem[{Perlman} et~al.(1999){Perlman}, {Biretta}, {Zhou}, {Sparks}, and
	{Macchetto}]{perlman99a}
	{Perlman}, E.S.; {Biretta}, J.A.; {Zhou}, F.; {Sparks}, W.B.; {Macchetto}, F.D.
	\newblock {Optical and Radio Polarimetry of the M87 Jet at 0.2'' Resolution}.
	\newblock {\em Astron. J.} {\bf 1999}, {\em 117},~2185--2198.
	\newblock {\url{https://doi.org/10.1086/300844}}.
	
	\bibitem[{Kraft} et~al.(2002){Kraft}, {Forman}, {Jones}, {Murray},
	{Hardcastle}, and {Worrall}]{kraft02a}
	{Kraft}, R.P.; {Forman}, W.R.; {Jones}, C.; {Murray}, S.S.; {Hardcastle}, M.J.;
	{Worrall}, D.M.
	\newblock {Chandra Observations of the X-Ray Jet in Centaurus A}.
	\newblock {\em Astrophys. J.} {\bf 2002}, {\em 569},~54--71.
	\newblock {\url{https://doi.org/10.1086/339062}}.
	
	\bibitem[{Hardcastle} et~al.(2003){Hardcastle}, {Worrall}, {Kraft}, {Forman},
	{Jones}, and {Murray}]{hardcastle03b}
	{Hardcastle}, M.J.; {Worrall}, D.M.; {Kraft}, R.P.; {Forman}, W.R.; {Jones},
	C.; {Murray}, S.S.
	\newblock {Radio and X-Ray Observations of the Jet in Centaurus A}.
	\newblock {\em Astrophys. J.} {\bf 2003}, {\em 593},~169--183.
	\newblock {\url{https://doi.org/10.1086/376519}}.
	
	\bibitem[{Worrall}(2009)]{worrall09a}
	{Worrall}, D.M.
	\newblock {The X-ray jets of active galaxies}.
	\newblock {\em Astron. Astrophys. Rev.} {\bf 2009}, {\em 17},~1--46.
	\newblock {\url{https://doi.org/10.1007/s00159-009-0016-7}}.
	
	\bibitem[{Bogensberger} et~al.(2024){Bogensberger}, {Miller}, {Mushotzky},
	{Brandt}, {Kammoun}, {Zoghbi}, and {Behar}]{Bogensberger24a}
	{Bogensberger}, D.; {Miller}, J.M.; {Mushotzky}, R.; {Brandt}, W.N.; {Kammoun},
	E.; {Zoghbi}, A.; {Behar}, E.
	\newblock {Superluminal proper motion in the X-ray jet of Centaurus A}.
	\newblock {\em arXiv} {\bf 2024}, arXiv:2408.14078.
	\newblock {\url{https://doi.org/10.48550/arXiv.2408.14078}}.
	
	\bibitem[{Hardcastle}(2003)]{hardcastle03a}
	{Hardcastle}, M.J.
	\newblock {Physical conditions in hotspots-what the new data are telling us}.
	\newblock {\em New Astron. Rev.} {\bf 2003}, {\em 47},~649--652.
	
	\bibitem[Begelman(1996)]{begelman96a}
	Begelman, M.C.
	\newblock Baby Cygnus A's.
	\newblock In \textit{Cygnus A: Study of a Radio Galaxy}; Carilli,
	C.L., Harris, D.A., Eds.; {Cambridge University Press: Cambridge, UK,} 
	1996; p. 209.
	
	\bibitem[Bicknell et~al.(1997)Bicknell, Dopita, and O'Dea]{bicknell97a}
	Bicknell, G.V.; Dopita, M.A.; O'Dea, C.P.
	\newblock Unification of the Radio and Optical Properties of GPS and CSS Radio
	Sources.
	\newblock {\em Astrophys. J.} {\bf 1997}, {\em 485},~112.
	
	\bibitem[{O'Dea}(1998)]{odea98a}
	{O'Dea}, C.P.
	\newblock {The Compact Steep-Spectrum and Gigahertz Peaked-Spectrum Radio
		Sources}.
	\newblock {\em Publ. Astron. Soc. Pac.} {\bf 1998}, {\em 110},~493--532.
	\newblock {\url{https://doi.org/10.1086/316162}}.
	
	\bibitem[{Begelman}(1999)]{begelman99a}
	{Begelman}, M.C.
	\newblock {Young radio galaxies and their environments}.
	\newblock In Proceedings of the The Most Distant Radio Galaxies, {Amsterdam, The 
	Netherlands, 15--17 October 1997;}
	{R{\"o}ttgering}, H.J.A., {Best}, P.N., {Lehnert}, M.D., Eds.; {Royal Netherlands 
	Academy of Arts and Sciences: Amsterdam, The Netherlands,} 1999; p. 173.
	
	
	
	\bibitem[DeYoung(1993)]{deyoung93a}
	DeYoung, D.S.
	\newblock F-R I and F-R II Radio Galaxies.
	\newblock {\em Astrophys. J.} {\bf 1993}, {\em 405},~L13.
	
	\bibitem[{Steffen} et~al.(1997){Steffen}, {G{\'o}mez}, {Raga}, and
	{Williams}]{steffen97b}
	{Steffen}, W.; {G{\'o}mez}, J.L.; {Raga}, A.C.; {Williams}, R.J.R.
	\newblock {Jet-Cloud Interactions and the Brightening of the Narrow-Line Region
		in Seyfert Galaxies}.
	\newblock {\em Astrophys. J.} {\bf 1997}, {\em 491},~L73--L76.
	\newblock {\url{https://doi.org/10.1086/311066}}.
	
	\bibitem[{Hooda} and {Wiita}(1996)]{hooda96a}
	{Hooda}, J.S.; {Wiita}, P.J.
	\newblock {Three-dimensional Simulations of Extragalactic Jets Crossing
		Interstellar Medium/Intracluster Medium Interfaces}.
	\newblock {\em Astrophys. J.} {\bf 1996}, {\em 470},~211.
	\newblock {\url{https://doi.org/10.1086/177862}}.
	
	\bibitem[{Hooda} and {Wiita}(1998)]{hooda98a}
	{Hooda}, J.S.; {Wiita}, P.J.
	\newblock {Instabilities in Three-dimensional Simulations of Astrophysical Jets
		Crossing Tilted Interfaces}.
	\newblock {\em Astrophys. J.} {\bf 1998}, {\em 493},~81--90.
	\newblock {\url{https://doi.org/10.1086/305099}}.
	
	\bibitem[{Hughes} et~al.(2002){Hughes}, {Miller}, and {Duncan}]{hughes02a}
	{Hughes}, P.A.; {Miller}, M.A.; {Duncan}, G.C.
	\newblock {Three-dimensional Hydrodynamic Simulations of Relativistic
		Extragalactic Jets}.
	\newblock {\em Astrophys. J.} {\bf 2002}, {\em 572},~713--728.
	\newblock {\url{https://doi.org/10.1086/340382}}.
	
	\bibitem[{Higgins} et~al.(1999){Higgins}, {O'Brien}, and {Dunlop}]{higgins99a}
	{Higgins}, S.W.; {O'Brien}, T.J.; {Dunlop}, J.S.
	\newblock {Structures produced by the collision of extragalactic jets with
		dense clouds}.
	\newblock {\em Mon. Not. R. Astron. Soc.} {\bf 1999}, {\em 309},~273--286.
	\newblock {\url{https://doi.org/10.1046/j.1365-8711.1999.02779.x}}.
	
	\bibitem[{Wang} et~al.(2000){Wang}, {Wiita}, and {Hooda}]{wang2000a}
	{Wang}, Z.; {Wiita}, P.J.; {Hooda}, J.S.
	\newblock {Radio Jet Interactions with Massive Clouds}.
	\newblock {\em Astrophys. J.} {\bf 2000}, {\em 534},~201--212.
	\newblock {\url{https://doi.org/10.1086/308743}}.
	
	\bibitem[{Wiita}(2004)]{witta04a}
	{Wiita}, P.J.
	\newblock {Jet Propagation Through Irregular Media and the Impact of Lobes on
		Galaxy Formation}.
	\newblock {\em Astrophys. Space Sci.} {\bf 2004}, {\em 293},~235--245.
	\newblock {\url{https://doi.org/10.1023/B:ASTR.0000044672.94932.c5}}.
	
	\bibitem[{Choi} et~al.(2007){Choi}, {Wiita}, and {Ryu}]{choi07a}
	{Choi}, E.; {Wiita}, P.J.; {Ryu}, D.
	\newblock {Hydrodynamic Interactions of Relativistic Extragalactic Jets with
		Dense Clouds}.
	\newblock {\em Astrophys. J.} {\bf 2007}, {\em 655},~769--780.
	\newblock {\url{https://doi.org/10.1086/510120}}.


    
	\bibitem[{Antonuccio-Delogu} and {Silk}(2008{\natexlab{b}})]{antonuccio08a}
	{Antonuccio-Delogu, V.; Silk, J.}
	\newblock {Active galactic nuclei jet-induced feedback in galaxies---I.
		Suppression of star formation}.
	\newblock {\em Mon. Not. R. Astron. Soc.} {\bf 2008}, {\em 389},~1750--1762.
	\newblock {\url{https://doi.org/10.1111/j.1365-2966.2008.13663.x}}.
	
	\bibitem[{Tortora} et~al.(2009){Tortora}, {Antonuccio-Delogu}, {Kaviraj},
	{Silk}, {Romeo}, and {Becciani}]{tortora09a}
	{Tortora}, C.; {Antonuccio-Delogu}, V.; {Kaviraj}, S.; {Silk}, J.; {Romeo},
	A.D.; {Becciani}, U.
	\newblock {AGN jet-induced feedback in galaxies---II. Galaxy colours from a
		multicloud simulation}.
	\newblock {\em Mon. Not. R. Astron. Soc.} {\bf 2009}, {\em 396},~61--77.
	\newblock {\url{https://doi.org/10.1111/j.1365-2966.2009.14718.x}}.
	
	\bibitem[{Dutta} et~al.(2024){Dutta}, {Sharma}, {Sarkar}, and
	{Stone}]{dutta24a}
	{Dutta}, R.; {Sharma}, P.; {Sarkar}, K.C.; {Stone}, J.M.
	\newblock {Dissipation of AGN Jets in a Clumpy Interstellar Medium}.
	\newblock {\em Astrophys. J.} {\bf 2024}, {\em 973},~148.
	\newblock {\url{https://doi.org/10.3847/1538-4357/ad67d7}}.

    
	\bibitem[{Fragile} et~al.(2004){Fragile}, {Murray}, {Anninos}, and {van
		Breugel}]{fragile04a}
	{Fragile}, P.C.; {Murray}, S.D.; {Anninos}, P.; {van Breugel}, W.
	\newblock {Radiative Shock-induced Collapse of Intergalactic Clouds}.
	\newblock {\em Astrophys. J.} {\bf 2004}, {\em 604},~74--87.
	
	\bibitem[{Fragile} et~al.(2017){Fragile}, {Anninos}, {Croft}, {Lacy}, and
	{Witry}]{fragile17a}
	{Fragile}, P.C.; {Anninos}, P.; {Croft}, S.; {Lacy}, M.; {Witry}, J.W.L.
	\newblock {Numerical Simulations of a Jet-Cloud Collision and Starburst:
		Application to Minkowski{\textquoteright}s Object}.
	\newblock {\em Astrophys. J.} {\bf 2017}, {\em 850},~171.
	\newblock {\url{https://doi.org/10.3847/1538-4357/aa95c6}}.
	
	\bibitem[{Krause} and {Alexander}(2007)]{krause07a}
	{Krause}, M.; {Alexander}, P.
	\newblock {Simulations of multiphase turbulence in jet cocoons}.
	\newblock {\em Mon. Not. R. Astron. Soc.} {\bf 2007}, {\em 376},~465--478.
	\newblock {\url{https://doi.org/10.1111/j.1365-2966.2007.11480.x}}.
	
	\bibitem[{Dugan} et~al.(2017){Dugan}, {Gaibler}, and {Silk}]{dugan17a}
	{Dugan}, Z.; {Gaibler}, V.; {Silk}, J.
	\newblock {Feedback by AGN Jets and Wide-angle Winds on a Galactic Scale}.
	\newblock {\em Astrophys. J.} {\bf 2017}, {\em 844},~37.
	\newblock {\url{https://doi.org/10.3847/1538-4357/aa7566}}.
	
	\bibitem[{Gardner} et~al.(2017){Gardner}, {Jones}, {Scannapieco}, and
	{Windhorst}]{gardner17a}
	{Gardner}, C.L.; {Jones}, J.R.; {Scannapieco}, E.; {Windhorst}, R.A.
	\newblock {Numerical Simulation of Star Formation by the Bow Shock of the
		Centaurus A Jet}.
	\newblock {\em Astrophys. J.} {\bf 2017}, {\em 835},~232.
	\newblock {\url{https://doi.org/10.3847/1538-4357/835/2/232}}.
	
	\bibitem[{Lau{\v{z}}ikas} and {Zubovas}(2024)]{lauzikas24a}
	{Lau{\v{z}}ikas}, M.; {Zubovas}, K.
	\newblock {Slow and steady does the trick: Slow outflows enhance the
		fragmentation of molecular clouds}.
	\newblock {\em Astron. Astrophys.} {\bf 2024}, {\em 690},~A396.
	\newblock {\url{https://doi.org/10.1051/0004-6361/202450286}}.
	
	\bibitem[{Mandal} et~al.(2024){Mandal}, {Mukherjee}, {Federrath}, {Bicknell},
	{Nesvadba}, and {Mignone}]{mandal24a}
	{Mandal}, A.; {Mukherjee}, D.; {Federrath}, C.; {Bicknell}, G.V.; {Nesvadba},
	N.P.H.; {Mignone}, A.
	\newblock {Probing the role of self-gravity in clouds impacted by AGN-driven
		winds}.
	\newblock {\em Mon. Not. R. Astron. Soc.} {\bf 2024}, {\em 531},~2079--2110.
	\newblock {\url{https://doi.org/10.1093/mnras/stae1295}}.
	
	
	\bibitem[{Jeyakumar}(2009)]{jeyakumar09a}
	{Jeyakumar}, S.
	\newblock {Interaction of radio jets with clouds in the ambient medium:
		Numerical simulations}.
	\newblock {\em Astronomische Nachrichten} {\bf 2009}, {\em 330},~287.
	\newblock {\url{https://doi.org/10.1002/asna.200811177}}.
	
	\bibitem[{Nolting} et~al.(2022){Nolting}, {Lacy}, {Croft}, {Fragile}, {Linden},
	{Nyland}, and {Patil}]{nolting22a}
	{Nolting}, C.; {Lacy}, M.; {Croft}, S.; {Fragile}, P.C.; {Linden}, S.T.;
	{Nyland}, K.; {Patil}, P.
	\newblock {Observations and Simulations of Radio Emission and Magnetic Fields
		in Minkowski's Object}.
	\newblock {\em Astrophys. J.} {\bf 2022}, {\em 936},~130.
	\newblock {\url{https://doi.org/10.3847/1538-4357/ac874b}}.


    	\bibitem[Klein et~al.(1994)Klein, McKee, and Colella]{klein94a}
	Klein, R.I.; McKee, C.F.; Colella, P.
	\newblock On the Hydrodynamic Interaction of Shock Waves with Interstellar
	Clouds. I. Nonradiative Shocks in Small Clouds.
	\newblock {\em Astrophys. J.} {\bf 1994}, {\em 420},~213.

   	\bibitem[{Banda-Barrag{\'a}n} et~al.(2016){Banda-Barrag{\'a}n}, {Parkin},
	{Federrath}, {Crocker}, and {Bicknell}]{bandabarragan16a}
	{Banda-Barrag{\'a}n}, W.E.; {Parkin}, E.R.; {Federrath}, C.; {Crocker}, R.M.;
	{Bicknell}, G.V.
	\newblock {Filament formation in wind-cloud interactions---I. Spherical clouds
		in uniform magnetic fields}.
	\newblock {\em Mon. Not. R. Astron. Soc.} {\bf 2016}, {\em 455},~1309--1333.
	\newblock {\url{https://doi.org/10.1093/mnras/stv2405}}.
    
	
	\bibitem[{Wagner} and {Bicknell}(2011)]{wagner11a}
	{Wagner}, A.Y.; {Bicknell}, G.V.
	\newblock {Relativistic Jet Feedback in Evolving Galaxies}.
	\newblock {\em Astrophys. J.} {\bf 2011}, {\em 728},~29.
	\newblock {\url{https://doi.org/10.1088/0004-637X/728/1/29}}.
	
	\bibitem[{Mukherjee} et~al.(2018){Mukherjee}, {Bicknell}, {Wagner},
	{Sutherland}, and {Silk}]{mukherjee18b}
	{Mukherjee}, D.; {Bicknell}, G.V.; {Wagner}, A.e.Y.; {Sutherland}, R.S.;
	{Silk}, J.
	\newblock {Relativistic jet feedback---III. Feedback on gas discs}.
	\newblock {\em Mon. Not. R. Astron. Soc.} {\bf 2018}, {\em 479},~5544--5566.
	\newblock {\url{https://doi.org/10.1093/mnras/sty1776}}.
	
	\bibitem[{Bicknell} et~al.(2003){Bicknell}, {Saxton}, and
	{Sutherland}]{bicknell03a}
	{Bicknell}, G.V.; {Saxton}, C.J.; {Sutherland}, R.S.
	\newblock {GPS and CSS Sources---Theory and Modelling}.
	\newblock {\em Publ. Astron. Soc. Aust.} {\bf 2003}, {\em 20},~102--109.
	\newblock {\url{https://doi.org/10.1071/AS02042}}.
	
	\bibitem[{Saxton} et~al.(2005){Saxton}, {Bicknell}, {Sutherland}, and
	{Midgley}]{saxton05a}
	{Saxton}, C.J.; {Bicknell}, G.V.; {Sutherland}, R.S.; {Midgley}, S.
	\newblock {Interactions of jets with inhomogeneous cloudy media}.
	\newblock {\em Mon. Not. R. Astron. Soc.} {\bf 2005}, {\em 359},~781--800.
	\newblock {\url{https://doi.org/10.1111/j.1365-2966.2005.08962.x}}.
	
	\bibitem[{Gaibler} et~al.(2011){Gaibler}, {Khochfar}, and {Krause}]{gaibler11a}
	{Gaibler}, V.; {Khochfar}, S.; {Krause}, M.
	\newblock {Asymmetries in extragalactic double radio sources: Clues from 3D
		simulations of jet-disc interaction}.
	\newblock {\em Mon. Not. R. Astron. Soc.} {\bf 2011}, {\em 411},~155--161.
	\newblock {\url{https://doi.org/10.1111/j.1365-2966.2010.17674.x}}.
	
	\bibitem[{Gaibler} et~al.(2012){Gaibler}, {Khochfar}, {Krause}, and
	{Silk}]{gaibler12a}
	{Gaibler}, V.; {Khochfar}, S.; {Krause}, M.; {Silk}, J.
	\newblock {Jet-induced star formation in gas-rich galaxies}.
	\newblock {\em Mon. Not. R. Astron. Soc.} {\bf 2012}, {\em 425},~438--449.
	\newblock {\url{https://doi.org/10.1111/j.1365-2966.2012.21479.x}}.
	
	\bibitem[{Dugan} et~al.(2014){Dugan}, {Bryan}, {Gaibler}, {Silk}, and
	{Haas}]{dugan14a}
	{Dugan}, Z.; {Bryan}, S.; {Gaibler}, V.; {Silk}, J.; {Haas}, M.
	\newblock {Stellar Signatures of AGN-jet-triggered Star Formation}.
	\newblock {\em Astrophys. J.} {\bf 2014}, {\em 796},~113.
	\newblock {\url{https://doi.org/10.1088/0004-637X/796/2/113}}.
	
	\bibitem[{Mukherjee} et~al.(2016){Mukherjee}, {Bicknell}, {Sutherland }, and
	{Wagner}]{mukherjee16a}
	{Mukherjee}, D.; {Bicknell}, G.V.; {Sutherland }, R.; {Wagner}, A.
	\newblock {Relativistic jet feedback in high-redshift galaxies---I. Dynamics}.
	\newblock {\em Mon. Not. R. Astron. Soc.} {\bf 2016}, {\em 461},~967--983.
	\newblock {\url{https://doi.org/10.1093/mnras/stw1368}}.
	
	\bibitem[{Tanner} and {Weaver}(2022)]{tanner22a}
	{Tanner}, R.; {Weaver}, K.A.
	\newblock {Simulations of AGN-driven Galactic Outflow Morphology and Content}.
	\newblock {\em Astron. J.} {\bf 2022}, {\em 163},~134.
	\newblock {\url{https://doi.org/10.3847/1538-3881/ac4d23}}.
	
	\bibitem[{Clavijo-Boh{\'o}rquez} et~al.(2024){Clavijo-Boh{\'o}rquez}, {de
		Gouveia Dal Pino}, and {Melioli}]{clavijo24a}
	{Clavijo-Boh{\'o}rquez}, W.E.; {de Gouveia Dal Pino}, E.M.; {Melioli}, C.
	\newblock {Role of AGN and star formation feedback in the evolution of galaxy
		outflows}.
	\newblock {\em Mon. Not. R. Astron. Soc.} {\bf 2024}, {\em 535},~1696--1720.
	\newblock {\url{https://doi.org/10.1093/mnras/stae487}}.
	
	\bibitem[{Asahina} et~al.(2017){Asahina}, {Nomura}, and {Ohsuga}]{asahina17a}
	{Asahina}, Y.; {Nomura}, M.; {Ohsuga}, K.
	\newblock {Enhancement of Feedback Efficiency by Active Galactic Nucleus
		Outflows via the Magnetic Tension Force in the Inhomogeneous Interstellar
		Medium}.
	\newblock {\em Astrophys. J.} {\bf 2017}, {\em 840},~25.
	\newblock {\url{https://doi.org/10.3847/1538-4357/aa6c5f}}.
	
	\bibitem[{Fiacconi} et~al.(2018){Fiacconi}, {Sijacki}, and
	{Pringle}]{fiacconi18a}
	{Fiacconi}, D.; {Sijacki}, D.; {Pringle}, J.E.
	\newblock {Galactic nuclei evolution with spinning black holes: Method and
		implementation}.
	\newblock {\em Mon. Not. R. Astron. Soc.} {\bf 2018}, {\em 477},~3807--3835.
	\newblock {\url{https://doi.org/10.1093/mnras/sty893}}.
	
	\bibitem[{Talbot} et~al.(2021){Talbot}, {Bourne}, and {Sijacki}]{talbot21a}
	{Talbot}, R.Y.; {Bourne}, M.A.; {Sijacki}, D.
	\newblock {Blandford-Znajek jets in galaxy formation simulations: Method and
		implementation}.
	\newblock {\em Mon. Not. R. Astron. Soc.} {\bf 2021}, {\em 504},~3619--3650.
	\newblock {\url{https://doi.org/10.1093/mnras/stab804}}.
	
	\bibitem[{Talbot} et~al.(2022){Talbot}, {Sijacki}, and {Bourne}]{talbot22a}
	{Talbot}, R.Y.; {Sijacki}, D.; {Bourne}, M.A.
	\newblock {Blandford-Znajek jets in galaxy formation simulations: Exploring the
		diversity of outflows produced by spin-driven AGN jets in Seyfert galaxies}.
	\newblock {\em Mon. Not. R. Astron. Soc.} {\bf 2022}, {\em 514},~4535--4559.
	\newblock {\url{https://doi.org/10.1093/mnras/stac1566}}.
	
	\bibitem[{Talbot} et~al.(2024){Talbot}, {Sijacki}, and {Bourne}]{talbot24a}
	{Talbot}, R.Y.; {Sijacki}, D.; {Bourne}, M.A.
	\newblock {Simulations of spin-driven AGN jets in gas-rich galaxy mergers}.
	\newblock {\em Mon. Not. R. Astron. Soc.} {\bf 2024}, {\em 528},~5432--5451.
	\newblock {\url{https://doi.org/10.1093/mnras/stae392}}.
	
	\bibitem[{Garc{\'\i}a-Burillo} et~al.(2014){Garc{\'\i}a-Burillo}, {Combes},
	{Usero}, {Aalto}, {Krips}, {Viti}, {Alonso-Herrero}, {Hunt}, {Schinnerer},
	{Baker}, {Boone}, {Casasola}, {Colina}, {Costagliola}, {Eckart}, {Fuente},
	{Henkel}, {Labiano}, {Mart{\'\i}n}, {M{\'a}rquez}, {Muller}, {Planesas},
	{Ramos Almeida}, {Spaans}, {Tacconi}, and {van der Werf}]{garcia14a}
	{Garc{\'\i}a-Burillo}, S.; {Combes}, F.; {Usero}, A.; {Aalto}, S.; {Krips}, M.;
	{Viti}, S.; {Alonso-Herrero}, A.; {Hunt}, L.K.; {Schinnerer}, E.; {Baker},
	A.J.;  et~al.
	\newblock {Molecular line emission in NGC 1068 imaged with ALMA. I. An
		AGN-driven outflow in the dense molecular gas}.
	\newblock {\em Astron. Astrophys.} {\bf 2014}, {\em 567},~A125.
	\newblock {\url{https://doi.org/10.1051/0004-6361/201423843}}.
	
	\bibitem[{Garc{\'\i}a-Burillo} et~al.(2021){Garc{\'\i}a-Burillo},
	{Alonso-Herrero}, {Ramos Almeida}, {Gonz{\'a}lez-Mart{\'\i}n}, {Combes},
	{Usero}, {H{\"o}nig}, {Querejeta}, {Hicks}, {Hunt}, {Rosario}, {Davies},
	{Boorman}, {Bunker}, {Burtscher}, {Colina}, {D{\'\i}az-Santos}, {Gandhi},
	{Garc{\'\i}a-Bernete}, {Garc{\'\i}a-Lorenzo}, {Ichikawa}, {Imanishi},
	{Izumi}, {Labiano}, {Levenson}, {L{\'o}pez-Rodr{\'\i}guez}, {Packham},
	{Pereira-Santaella}, {Ricci}, {Rigopoulou}, {Rouan}, {Shimizu}, {Stalevski},
	{Wada}, and {Williamson}]{garcia21a}
	{Garc{\'\i}a-Burillo, S.;} 
	{Alonso-Herrero}, A.; {Ramos Almeida}, C.;
	{Gonz{\'a}lez-Mart{\'\i}n}, O.; {Combes}, F.; {Usero}, A.; {H{\"o}nig}, S.;
	{Querejeta}, M.; {Hicks}, E.K.S.; {Hunt}, L.K.;  et~al.
	\newblock {The Galaxy Activity, Torus, and Outflow Survey (GATOS). I. ALMA
		images of dusty molecular tori in Seyfert galaxies}.
	\newblock {\em Astron. Astrophys.} {\bf 2021}, {\em 652},~A98.
	\newblock {\url{https://doi.org/10.1051/0004-6361/202141075}}.
	
	\bibitem[{Komissarov} and {Falle}(1996)]{komissarov96a}
	{Komissarov}, S.S.; {Falle}, S.A.E.G.
	\newblock {LargeScale Structure of Relativistic Jets}.
	\newblock In Proceedings of the Energy Transport in Radio Galaxies and Quasars, 
	{Tuscaloosa, Alabama, 19--23 September 1995;} {Hardee}, P.E.; {Bridle}, A.H.; 
	{Zensus}, J.A., Eds., {Astronomical Society of the Pacific (ASP): San Francisco, CA, 
	USA,} 1996; Volume 100,  Astronomical Society of the Pacific Conference Series, 
	\mbox{p. 173.}
	
	\bibitem[{Perucho} et~al.(2017){Perucho}, {Mart{\'\i}}, {Quilis}, and
	{Borja-Lloret}]{perucho17a}
	{Perucho}, M.; {Mart{\'\i}}, J.M.; {Quilis}, V.; {Borja-Lloret}, M.
	\newblock {Radio mode feedback: Does relativity matter?}
	\newblock {\em Mon. Not. R. Astron. Soc.} {\bf 2017}, {\em 471},~L120--L124.
	\newblock {\url{https://doi.org/10.1093/mnrasl/slx115}}.
	
	\bibitem[{Wagner} et~al.(2012){Wagner}, {Bicknell}, and {Umemura}]{wagner12a}
	{Wagner}, A.Y.; {Bicknell}, G.V.; {Umemura}, M.
	\newblock {Driving Outflows with Relativistic Jets and the Dependence of Active
		Galactic Nucleus Feedback Efficiency on Interstellar Medium Inhomogeneity}.
	\newblock {\em Astrophys. J.} {\bf 2012}, {\em 757},~136.
	\newblock {\url{https://doi.org/10.1088/0004-637X/757/2/136}}.
	
	\bibitem[{Hughes} et~al.(2010){Hughes}, {Wong}, {Ott}, {Muller}, {Pineda},
	{Mizuno}, {Bernard}, {Paradis}, {Maddison}, {Reach}, {Staveley-Smith},
	{Kawamura}, {Meixner}, {Kim}, {Onishi}, {Mizuno}, and {Fukui}]{hughes10a}
	{Hughes}, A.; {Wong}, T.; {Ott}, J.; {Muller}, E.; {Pineda}, J.L.; {Mizuno},
	Y.; {Bernard}, J.P.; {Paradis}, D.; {Maddison}, S.; {Reach}, W.T.;  et~al.
	\newblock {Physical properties of giant molecular clouds in the Large
		Magellanic Cloud}.
	\newblock {\em Mon. Not. R. Astron. Soc.} {\bf 2010}, {\em 406},~2065--2086.
	\newblock {\url{https://doi.org/10.1111/j.1365-2966.2010.16829.x}}.
	
	\bibitem[{Hughes} et~al.(2013){Hughes}, {Meidt}, {Colombo}, {Schinnerer},
	{Pety}, {Leroy}, {Dobbs}, {Garc{\'\i}a-Burillo}, {Thompson}, {Dumas},
	{Schuster}, and {Kramer}]{hughes13a}
	{Hughes}, A.; {Meidt}, S.E.; {Colombo}, D.; {Schinnerer}, E.; {Pety}, J.;
	{Leroy}, A.K.; {Dobbs}, C.L.; {Garc{\'\i}a-Burillo}, S.; {Thompson}, T.A.;
	{Dumas}, G.;  et~al.
	\newblock {A Comparative Study of Giant Molecular Clouds in M51, M33, and the
		Large Magellanic Cloud}.
	\newblock {\em Astrophys. J.} {\bf 2013}, {\em 779},~46.
	\newblock {\url{https://doi.org/10.1088/0004-637X/779/1/46}}.
	
	\bibitem[{Faesi} et~al.(2018){Faesi}, {Lada}, and {Forbrich}]{faesi18a}
	{Faesi}, C.M.; {Lada}, C.J.; {Forbrich}, J.
	\newblock {The ALMA View of GMCs in NGC 300: Physical Properties and Scaling
		Relations at 10 pc Resolution}.
	\newblock {\em Astrophys. J.} {\bf 2018}, {\em 857},~19.
	\newblock {\url{https://doi.org/10.3847/1538-4357/aaad60}}.
	
	\bibitem[{Zubovas} and {King}(2012)]{zubovas12a}
	{Zubovas}, K.; {King}, A.
	\newblock {Clearing Out a Galaxy}.
	\newblock {\em Astrophys. J.} {\bf 2012}, {\em 745},~L34.
	\newblock {\url{https://doi.org/10.1088/2041-8205/745/2/L34}}.
	
	\bibitem[{Faucher-Gigu{\`e}re} and {Quataert}(2012)]{faucher12a}
	{Faucher-Gigu{\`e}re}, C.A.; {Quataert}, E.
	\newblock {The physics of galactic winds driven by active galactic nuclei}.
	\newblock {\em Mon. Not. R. Astron. Soc.} {\bf 2012}, {\em 425},~605--622.
	\newblock {\url{https://doi.org/10.1111/j.1365-2966.2012.21512.x}}.
	
	\bibitem[{Mukherjee} et~al.(2017){Mukherjee}, {Bicknell}, {Sutherland}, and
	{Wagner}]{mukherjee17a}
	{Mukherjee}, D.; {Bicknell}, G.V.; {Sutherland}, R.; {Wagner}, A.
	\newblock {Erratum: Relativistic jet feedback in high-redshift galaxies I.
		Dynamics}.
	\newblock {\em Mon. Not. R. Astron. Soc.} {\bf 2017}, {\em 471},~2790--2800.
	\newblock {\url{https://doi.org/10.1093/mnras/stx1749}}.
	
	\bibitem[{Bicknell} et~al.(2018){Bicknell}, {Mukherjee}, {Wagner},
	{Sutherland}, and {Nesvadba}]{bicknell18a}
	{Bicknell}, G.V.; {Mukherjee}, D.; {Wagner}, A.Y.; {Sutherland}, R.S.;
	{Nesvadba}, N.P.H.
	\newblock {Relativistic jet feedback---II. Relationship to gigahertz peak
		spectrum and compact steep spectrum radio galaxies}.
	\newblock {\em Mon. Not. R. Astron. Soc.} {\bf 2018}, {\em 475},~3493--3501.
	\newblock {\url{https://doi.org/10.1093/mnras/sty070}}.
	
	\bibitem[{Mukherjee} et~al.(2018){Mukherjee}, {Wagner}, {Bicknell}, {Morganti},
	{Oosterloo}, {Nesvadba}, and {Sutherland}]{mukherjee18a}
	{Mukherjee}, D.; {Wagner}, A.Y.; {Bicknell}, G.V.; {Morganti}, R.; {Oosterloo},
	T.; {Nesvadba}, N.; {Sutherland}, R.S.
	\newblock {The jet-ISM interactions in IC 5063}.
	\newblock {\em Mon. Not. R. Astron. Soc.} {\bf 2018}, {\em 476},~80--95.
	\newblock {\url{https://doi.org/10.1093/mnras/sty067}}.
	
	\bibitem[{Borodina} et~al.(2025){Borodina}, {Ni}, {Bennett}, {Weinberger},
	{Bryan}, {Hirschmann}, {Farcy}, {Hlavacek-Larrondo}, and
	{Hernquist}]{borodina25a}
	{Borodina}, O.; {Ni}, Y.; {Bennett}, J.S.; {Weinberger}, R.; {Bryan}, G.L.;
	{Hirschmann}, M.; {Farcy}, M.; {Hlavacek-Larrondo}, J.; {Hernquist}, L.
	\newblock {You Shall Not Pass! The Propagation of Low-/Moderate-powered Jets
		Through a Turbulent Interstellar Medium}.
	\newblock {\em Astrophys. J.} {\bf 2025}, {\em 981},~149.
	\newblock {\url{https://doi.org/10.3847/1538-4357/adb016}}.
	
	\bibitem[{Bieri} et~al.(2016){Bieri}, {Dubois}, {Silk}, {Mamon}, and
	{Gaibler}]{bieri16a}
	{Bieri}, R.; {Dubois}, Y.; {Silk}, J.; {Mamon}, G.A.; {Gaibler}, V.
	\newblock {External pressure-triggering of star formation in a disc galaxy: A
		template for positive feedback}.
	\newblock {\em Mon. Not. R. Astron. Soc.} {\bf 2016}, {\em 455},~4166--4182.
	\newblock {\url{https://doi.org/10.1093/mnras/stv2551}}.
	
	\bibitem[{Mandal} et~al.(2021){Mandal}, {Mukherjee}, {Federrath}, {Nesvadba},
	{Bicknell}, {Wagner}, and {Meenakshi}]{mandal21a}
	{Mandal}, A.; {Mukherjee}, D.; {Federrath}, C.; {Nesvadba}, N.P.H.; {Bicknell},
	G.V.; {Wagner}, A.Y.; {Meenakshi}, M.
	\newblock {Impact of relativistic jets on the star formation rate: A
		turbulence-regulated framework}.
	\newblock {\em Mon. Not. R. Astron. Soc.} {\bf 2021}, {\em 508},~4738--4757.
	\newblock {\url{https://doi.org/10.1093/mnras/stab2822}}.
	
	\bibitem[{Cielo} et~al.(2018){Cielo}, {Bieri}, {Volonteri}, {Wagner}, and
	{Dubois}]{cielo18a}
	{Cielo}, S.; {Bieri}, R.; {Volonteri}, M.; {Wagner}, A.Y.; {Dubois}, Y.
	\newblock {AGN feedback compared: Jets versus radiation}.
	\newblock {\em Mon. Not. R. Astron. Soc.} {\bf 2018}, {\em 477},~1336--1355.
	\newblock {\url{https://doi.org/10.1093/mnras/sty708}}.
	
	\bibitem[{Ruffa} et~al.(2019){Ruffa}, {Prandoni}, {Laing}, {Paladino}, {Parma},
	{de Ruiter}, {Mignano}, {Davis}, {Bureau}, and {Warren}]{ruffa19a}
	{Ruffa}, I.; {Prandoni}, I.; {Laing}, R.A.; {Paladino}, R.; {Parma}, P.; {de
		Ruiter}, H.; {Mignano}, A.; {Davis}, T.A.; {Bureau}, M.; {Warren}, J.
	\newblock {The AGN fuelling/feedback cycle in nearby radio galaxies I. ALMA
		observations and early results}.
	\newblock {\em Mon. Not. R. Astron. Soc.} {\bf 2019}, {\em 484},~4239--4259.
	\newblock {\url{https://doi.org/10.1093/mnras/stz255}}.
	
	\bibitem[{Ostriker} et~al.(2010){Ostriker}, {Choi}, {Ciotti}, {Novak}, and
	{Proga}]{ostriker10a}
	{Ostriker}, J.P.; {Choi}, E.; {Ciotti}, L.; {Novak}, G.S.; {Proga}, D.
	\newblock {Momentum Driving: Which Physical Processes Dominate Active Galactic
		Nucleus Feedback?}
	\newblock {\em Astrophys. J.} {\bf 2010}, {\em 722},~642--652.
	\newblock {\url{https://doi.org/10.1088/0004-637X/722/1/642}}.
	
	\bibitem[{Meenakshi} et~al.(2022{\natexlab{a}}){Meenakshi}, {Mukherjee},
	{Wagner}, {Nesvadba}, {Morganti}, {Janssen}, and {Bicknell}]{meenakshi22a}
	{Meenakshi}, M.; {Mukherjee}, D.; {Wagner}, A.Y.; {Nesvadba}, N.P.H.;
	{Morganti}, R.; {Janssen}, R.M.J.; {Bicknell}, G.V.
	\newblock {The extent of ionization in simulations of radio-loud AGNs impacting
		kpc gas discs}.
	\newblock {\em Mon. Not. R. Astron. Soc.} {\bf 2022}, {\em 511},~1622--1636.
	\newblock {\url{https://doi.org/10.1093/mnras/stac167}}.
	
	\bibitem[{Meenakshi} et~al.(2022{\natexlab{b}}){Meenakshi}, {Mukherjee},
	{Wagner}, {Nesvadba}, {Bicknell}, {Morganti}, {Janssen}, {Sutherland}, and
	{Mandal}]{meenakshi22b}
	{Meenakshi}, M.; {Mukherjee}, D.; {Wagner}, A.Y.; {Nesvadba}, N.P.H.;
	{Bicknell}, G.V.; {Morganti}, R.; {Janssen}, R.M.J.; {Sutherland}, R.S.;
	{Mandal}, A.
	\newblock {Modelling observable signatures of jet-ISM interaction: Thermal
		emission and gas kinematics}.
	\newblock {\em Mon. Not. R. Astron. Soc.} {\bf 2022}, {\em 516},~766--786.
	\newblock {\url{https://doi.org/10.1093/mnras/stac2251}}.
	
	\bibitem[{Sutherland} and {Dopita}(2017)]{sutherland17a}
	{Sutherland}, R.S.; {Dopita}, M.A.
	\newblock {Effects of Preionization in Radiative Shocks. I. Self-consistent
		Models}.
	\newblock {\em Astrophys. J. Suppl. Ser.} {\bf 2017}, {\em 229},~34.
	\newblock {\url{https://doi.org/10.3847/1538-4365/aa6541}}.
	
	\bibitem[{Nesvadba} et~al.(2011){Nesvadba}, {De Breuck}, {Lehnert}, {Best},
	{Binette}, and {Proga}]{nesvadba11a}
	{Nesvadba}, N.P.H.; {De Breuck}, C.; {Lehnert}, M.D.; {Best}, P.N.; {Binette},
	L.; {Proga}, D.
	\newblock {The black holes of radio galaxies during the ``Quasar Era'': Masses,
		accretion rates, and evolutionary stage}.
	\newblock {\em Astron. Astrophys.} {\bf 2011}, {\em 525},~A43.
	\newblock {\url{https://doi.org/10.1051/0004-6361/201014960}}.
	
	\bibitem[{Zovaro} et~al.(2019){Zovaro}, {Sharp}, {Nesvadba}, {Bicknell},
	{Mukherjee}, {Wagner}, {Groves}, and {Krishna}]{zovaro19a}
	{Zovaro}, H.R.M.; {Sharp}, R.; {Nesvadba}, N.P.H.; {Bicknell}, G.V.;
	{Mukherjee}, D.; {Wagner}, A.Y.; {Groves}, B.; {Krishna}, S.
	\newblock {Jets blowing bubbles in the young radio galaxy 4C 31.04}.
	\newblock {\em Mon. Not. R. Astron. Soc.} {\bf 2019}, {\em 484},~3393--3409.
	\newblock {\url{https://doi.org/10.1093/mnras/stz233}}.
	
	\bibitem[{Morganti} et~al.(2015){Morganti}, {Oosterloo}, {Oonk}, {Frieswijk},
	and {Tadhunter}]{morganti15a}
	{Morganti}, R.; {Oosterloo}, T.; {Oonk}, J.B.R.; {Frieswijk}, W.; {Tadhunter},
	C.
	\newblock {The fast molecular outflow in the Seyfert galaxy IC 5063 as seen by
		ALMA}.
	\newblock {\em Astron. Astrophys.} {\bf 2015}, {\em 580},~A1.
	\newblock {\url{https://doi.org/10.1051/0004-6361/201525860}}.
	
	\bibitem[{Salom{\'e}} et~al.(2023){Salom{\'e}}, {Krongold}, {Longinotti},
	{Bischetti}, {Garc{\'\i}a-Burillo}, {Vega}, {S{\'a}nchez-Portal}, {Feruglio},
	{Jim{\'e}nez-Donaire}, and {Zanchettin}]{salome23a}
	{Salom{\'e}}, Q.; {Krongold}, Y.; {Longinotti}, A.L.; {Bischetti}, M.;
	{Garc{\'\i}a-Burillo}, S.; {Vega}, O.; {S{\'a}nchez-Portal}, M.; {Feruglio},
	C.; {Jim{\'e}nez-Donaire}, M.J.; {Zanchettin}, M.V.
	\newblock {Star formation efficiency and AGN feedback in narrow-line Seyfert 1
		galaxies with fast X-ray nuclear winds}.
	\newblock {\em Mon. Not. R. Astron. Soc.} {\bf 2023}, {\em 524},~3130--3145.
	\newblock {\url{https://doi.org/10.1093/mnras/stad2116}}.
	
	\bibitem[{Perucho} et~al.(2021){Perucho}, {L{\'o}pez-Miralles}, {Reynaldi}, and
	{Labiano}]{perucho21a}
	{Perucho}, M.; {L{\'o}pez-Miralles}, J.; {Reynaldi}, V.; {Labiano}, {\'A}.
	\newblock {Jet propagation through inhomogeneous media and shock ionization}.
	\newblock {\em Astron. Nachrichten} {\bf 2021}, {\em 342},~1171--1175.
	\newblock {\url{https://doi.org/10.1002/asna.20210051}}.
	
	\bibitem[{Perucho}(2024)]{perucho24a}
	{Perucho}, M.
	\newblock {Shocks, clouds, and atomic outflows in active galactic nuclei
		hosting relativistic jets}.
	\newblock {\em Astron. Astrophys.} {\bf 2024}, {\em 684},~A45.
	\newblock {\url{https://doi.org/10.1051/0004-6361/202348624}}.
	
	\bibitem[{Murthy} et~al.(2022){Murthy}, {Morganti}, {Wagner}, {Oosterloo},
	{Guillard}, {Mukherjee}, and {Bicknell}]{murthy22a}
	{Murthy}, S.; {Morganti}, R.; {Wagner}, A.Y.; {Oosterloo}, T.; {Guillard}, P.;
	{Mukherjee}, D.; {Bicknell}, G.
	\newblock {Cold gas removal from the centre of a galaxy by a low-luminosity
		jet}.
	\newblock {\em Nat. Astron.} {\bf 2022}, {\em 6},~488--495.
	\newblock {\url{https://doi.org/10.1038/s41550-021-01596-6}}.
	
	\bibitem[{Nesvadba} et~al.(2010){Nesvadba}, {Boulanger}, {Salom{\'e}},
	{Guillard}, {Lehnert}, {Ogle}, {Appleton}, {Falgarone}, and {Pineau Des
		Forets}]{nesvadba10a}
	{Nesvadba}, N.P.H.; {Boulanger}, F.; {Salom{\'e}}, P.; {Guillard}, P.;
	{Lehnert}, M.D.; {Ogle}, P.; {Appleton}, P.; {Falgarone}, E.; {Pineau Des
		Forets}, G.
	\newblock {Energetics of the molecular gas in the H$_{2}$ luminous radio galaxy
		3C 326: Evidence for negative AGN feedback}.
	\newblock {\em Astron. Astrophys.} {\bf 2010}, {\em 521},~A65.
	\newblock {\url{https://doi.org/10.1051/0004-6361/200913333}}.
	
	\bibitem[{Collet} et~al.(2016){Collet}, {Nesvadba}, {De Breuck}, {Lehnert},
	{Best}, {Bryant}, {Hunstead}, {Dicken}, and {Johnston}]{collet16a}
	{Collet}, C.; {Nesvadba}, N.P.H.; {De Breuck}, C.; {Lehnert}, M.D.; {Best}, P.;
	{Bryant}, J.J.; {Hunstead}, R.; {Dicken}, D.; {Johnston}, H.
	\newblock {Kinematic signatures of AGN feedback in moderately powerful radio
		galaxies at z \raisebox{-0.5ex}\textasciitilde 2 observed with SINFONI}.
	\newblock {\em Astron. Astrophys.} {\bf 2016}, {\em 586},~A152.
	\newblock {\url{https://doi.org/10.1051/0004-6361/201526872}}.
	
	\bibitem[{Nesvadba} et~al.(2017){Nesvadba}, {De Breuck}, {Lehnert}, {Best}, and
	{Collet}]{nesvadba17a}
	{Nesvadba}, N.P.H.; {De Breuck}, C.; {Lehnert}, M.D.; {Best}, P.N.; {Collet},
	C.
	\newblock {The SINFONI survey of powerful radio galaxies at z 2: Jet-driven AGN
		feedback during the Quasar Era}.
	\newblock {\em Astron. Astrophys.} {\bf 2017}, {\em 599},~A123.
	\newblock {\url{https://doi.org/10.1051/0004-6361/201528040}}.
	
	\bibitem[{Murthy} et~al.(2024){Murthy}, {Morganti}, {Oosterloo}, {Schulz}, and
	{Paragi}]{murthy24a}
	{Murthy}, S.; {Morganti}, R.; {Oosterloo}, T.; {Schulz}, R.; {Paragi}, Z.
	\newblock {Turbulent circumnuclear disc and cold gas outflow in the newborn
		radio source 4C 31.04}.
	\newblock {\em Astron. Astrophys.} {\bf 2024}, {\em 688},~A84.
	\newblock {\url{https://doi.org/10.1051/0004-6361/202450233}}.
	
	\bibitem[{Hardcastle} et~al.(2010){Hardcastle}, {Massaro}, and
	{Harris}]{hardcastle10a}
	{Hardcastle}, M.J.; {Massaro}, F.; {Harris}, D.E.
	\newblock {X-ray emission from the extended emission-line region of the
		powerful radio galaxy 3C171}.
	\newblock {\em Mon. Not. R. Astron. Soc.} {\bf 2010}, {\em 401},~2697--2705.
	\newblock {\url{https://doi.org/10.1111/j.1365-2966.2009.15855.x}}.
	
	\bibitem[{Hardcastle} et~al.(2012){Hardcastle}, {Massaro}, {Harris}, {Baum},
	{Bianchi}, {Chiaberge}, {Morganti}, {O'Dea}, and
	{Siemiginowska}]{hardcastle12a}
	{Hardcastle}, M.J.; {Massaro}, F.; {Harris}, D.E.; {Baum}, S.A.; {Bianchi}, S.;
	{Chiaberge}, M.; {Morganti}, R.; {O'Dea}, C.P.; {Siemiginowska}, A.
	\newblock {The nature of the jet-driven outflow in the radio galaxy 3C 305}.
	\newblock {\em Mon. Not. R. Astron. Soc.} {\bf 2012}, {\em 424},~1774--1789.
	\newblock {\url{https://doi.org/10.1111/j.1365-2966.2012.21247.x}}.
	
	\bibitem[{Worrall} et~al.(2012){Worrall}, {Birkinshaw}, {Young}, {Momtahan},
	{Fosbury}, {Morganti}, {Tadhunter}, and {Verdoes Kleijn}]{worrall12a}
	{Worrall}, D.M.; {Birkinshaw}, M.; {Young}, A.J.; {Momtahan}, K.; {Fosbury},
	R.A.E.; {Morganti}, R.; {Tadhunter}, C.N.; {Verdoes Kleijn}, G.
	\newblock {The jet-cloud interacting radio galaxy PKS B2152-699---I. Structures
		revealed in new deep radio and X-ray observations}.
	\newblock {\em Mon. Not. R. Astron. Soc.} {\bf 2012}, {\em 424},~1346--1362.
	\newblock {\url{https://doi.org/10.1111/j.1365-2966.2012.21320.x}}.
	
	\bibitem[{Fabbiano} et~al.(2022){Fabbiano}, {Paggi}, {Morganti},
	{Balokovi{\'c}}, {Elvis}, {Mukherjee}, {Meenakshi}, {Siemiginowska},
	{Murthy}, {Oosterloo}, {Wagner}, and {Bicknell}]{fabbiano22a}
	{Fabbiano}, G.; {Paggi}, A.; {Morganti}, R.; {Balokovi{\'c}}, M.; {Elvis}, M.;
	{Mukherjee}, D.; {Meenakshi}, M.; {Siemiginowska}, A.; {Murthy}, S.M.;
	{Oosterloo}, T.A.;  et~al.
	\newblock {Jet-ISM Interaction in NGC 1167/B2 0258+35, an LINER with an AGN
		Past}.
	\newblock {\em Astrophys. J.} {\bf 2022}, {\em 938},~105.
	\newblock {\url{https://doi.org/10.3847/1538-4357/ac8ff8}}.
	
	\bibitem[{Fabbiano} and {Elvis}(2022)]{fabbiano22b}
	{Fabbiano}, G.; {Elvis}, M.
	\newblock {The Interaction of the Active Nucleus with the Host Galaxy
		Interstellar Medium}. In {\em Handbook of X-ray and Gamma-ray Astrophysics};
	{Bambi}, C., {Sangangelo}, A., Eds.;  {Springer: Singapore,}  2022; p.~92.
	\newblock {\url{https://doi.org/10.1007/978-981-16-4544-0_111-1}}.
	
	\bibitem[{Gabor} and {Bournaud}(2014)]{gabor14a}
	{Gabor}, J.M.; {Bournaud}, F.
	\newblock {Active galactic nuclei-driven outflows without immediate quenching
		in simulations of high-redshift disc galaxies}.
	\newblock {\em Mon. Not. R. Astron. Soc.} {\bf 2014}, {\em 441},~1615--1627.
	\newblock {\url{https://doi.org/10.1093/mnras/stu677}}.
	
	\bibitem[{Costa} et~al.(2020){Costa}, {Pakmor}, and {Springel}]{costa20a}
	{Costa}, T.; {Pakmor}, R.; {Springel}, V.
	\newblock {Powering galactic superwinds with small-scale AGN winds}.
	\newblock {\em Mon. Not. R. Astron. Soc.} {\bf 2020}, {\em 497},~5229--5255.
	\newblock {\url{https://doi.org/10.1093/mnras/staa2321}}.
	
	\bibitem[{Bourne} et~al.(2015){Bourne}, {Zubovas}, and {Nayakshin}]{bourne15a}
	{Bourne}, M.A.; {Zubovas}, K.; {Nayakshin}, S.
	\newblock {The resolution bias: Low-resolution feedback simulations are better
		at destroying galaxies}.
	\newblock {\em Mon. Not. R. Astron. Soc.} {\bf 2015}, {\em 453},~1829--1842.
	\newblock {\url{https://doi.org/10.1093/mnras/stv1730}}.
	
	\bibitem[{Murthy} et~al.(2025){Murthy}, {Morganti}, {Oosterloo}, {Mukherjee},
	{Bayram}, {Guillard}, {Wagner}, and {Bicknell}]{murthy25a}
	{Murthy}, S.; {Morganti}, R.; {Oosterloo}, T.; {Mukherjee}, D.; {Bayram}, S.;
	{Guillard}, P.; {Wagner}, A.Y.; {Bicknell}, G.
	\newblock {Cold gas bubble inflated by a low-luminosity radio jet}.
	\newblock {\em Astron. Astrophys.} {\bf 2025}, {\em 694},~A110.
	\newblock {\url{https://doi.org/10.1051/0004-6361/202453139}}.
	
	\bibitem[{Venturi} et~al.(2021){Venturi}, {Cresci}, {Marconi}, {Mingozzi},
	{Nardini}, {Carniani}, {Mannucci}, {Marasco}, {Maiolino}, {Perna},
	{Treister}, {Bland-Hawthorn}, and {Gallimore}]{venturi21a}
	{Venturi}, G.; {Cresci}, G.; {Marconi}, A.; {Mingozzi}, M.; {Nardini}, E.;
	{Carniani}, S.; {Mannucci}, F.; {Marasco}, A.; {Maiolino}, R.; {Perna}, M.;
	et~al.
	\newblock {MAGNUM survey: Compact jets causing large turmoil in galaxies.
		Enhanced line widths perpendicular to radio jets as tracers of jet-ISM
		interaction}.
	\newblock {\em Astron. Astrophys.} {\bf 2021}, {\em 648},~A17.
	\newblock {\url{https://doi.org/10.1051/0004-6361/202039869}}.
	
	\bibitem[{Riffel} et~al.(2014){Riffel}, {Storchi-Bergmann}, and
	{Riffel}]{riffel14a}
	{Riffel}, R.A.; {Storchi-Bergmann}, T.; {Riffel}, R.
	\newblock {An Outflow Perpendicular to the Radio Jet in the Seyfert Nucleus of
		NGC 5929}.
	\newblock {\em Astrophys. J.} {\bf 2014}, {\em 780},~L24.
	\newblock {\url{https://doi.org/10.1088/2041-8205/780/2/L24}}.
	
	\bibitem[{Girdhar} et~al.(2022){Girdhar}, {Harrison}, {Mainieri}, {Bittner},
	{Costa}, {Kharb}, {Mukherjee}, {Arrigoni Battaia}, {Alexander}, {Calistro
		Rivera}, {Circosta}, {De Breuck}, {Edge}, {Farina}, {Kakkad}, {Lansbury},
	{Molyneux}, {Mullaney}, {Silpa}, {Thomson}, and {Ward}]{girdhar22a}
	{Girdhar}, A.; {Harrison}, C.M.; {Mainieri}, V.; {Bittner}, A.; {Costa}, T.;
	{Kharb}, P.; {Mukherjee}, D.; {Arrigoni Battaia}, F.; {Alexander}, D.M.;
	{Calistro Rivera}, G.;  et~al.
	\newblock {Quasar feedback survey: Multiphase outflows, turbulence, and
		evidence for feedback caused by low power radio jets inclined into the galaxy
		disc}.
	\newblock {\em Mon. Not. R. Astron. Soc.} {\bf 2022}, {\em 512},~1608--1628.
	\newblock {\url{https://doi.org/10.1093/mnras/stac073}}.
	
	\bibitem[{Ulivi} et~al.(2024){Ulivi}, {Venturi}, {Cresci}, {Marconi},
	{Marconcini}, {Amiri}, {Belfiore}, {Bertola}, {Carniani}, {D'Amato}, {Di
		Teodoro}, {Ginolfi}, {Girdhar}, {Harrison}, {Maiolino}, {Mannucci},
	{Mingozzi}, {Perna}, {Scialpi}, {Tomicic}, {Tozzi}, and {Treister}]{ulivi24a}
	{Ulivi}, L.; {Venturi}, G.; {Cresci}, G.; {Marconi}, A.; {Marconcini}, C.;
	{Amiri}, A.; {Belfiore}, F.; {Bertola}, E.; {Carniani}, S.; {D'Amato}, Q.;
	et~al.
	\newblock {Feedback and ionized gas outflows in four low-radio power AGN at z
		{\ensuremath{\sim}} 0.15}.
	\newblock {\em Astron. Astrophys.} {\bf 2024}, {\em 685},~A122.
	\newblock {\url{https://doi.org/10.1051/0004-6361/202347436}}.
	
	\bibitem[{Ruschel-Dutra} et~al.(2021){Ruschel-Dutra}, {Storchi-Bergmann},
	{Schnorr-M{\"u}ller}, {Riffel}, {Dall'Agnol de Oliveira}, {Lena}, {Robinson},
	{Nagar}, and {Elvis}]{ruschelDutra21a}
	{Ruschel-Dutra}, D.; {Storchi-Bergmann}, T.; {Schnorr-M{\"u}ller}, A.;
	{Riffel}, R.A.; {Dall'Agnol de Oliveira}, B.; {Lena}, D.; {Robinson}, A.;
	{Nagar}, N.; {Elvis}, M.
	\newblock {AGNIFS survey of local AGN: GMOS-IFU data and outflows in 30
		sources}.
	\newblock {\em Mon. Not. R. Astron. Soc.} {\bf 2021}, {\em 507},~74--89.
	\newblock {\url{https://doi.org/10.1093/mnras/stab2058}}.
	
	\bibitem[{Audibert} et~al.(2023){Audibert}, {Ramos Almeida},
	{Garc{\'\i}a-Burillo}, {Combes}, {Bischetti}, {Meenakshi}, {Mukherjee},
	{Bicknell}, and {Wagner}]{audibert23a}
	{Audibert}, A.; {Ramos Almeida}, C.; {Garc{\'\i}a-Burillo}, S.; {Combes}, F.;
	{Bischetti}, M.; {Meenakshi}, M.; {Mukherjee}, D.; {Bicknell}, G.; {Wagner},
	A.Y.
	\newblock {Jet-induced molecular gas excitation and turbulence in the Teacup}.
	\newblock {\em Astron. Astrophys.} {\bf 2023}, {\em 671},~L12.
	\newblock {\url{https://doi.org/10.1051/0004-6361/202345964}}.
	
	\bibitem[{Su} et~al.(2021){Su}, {Hopkins}, {Bryan}, {Somerville}, {Hayward},
	{Angl{\'e}s-Alc{\'a}zar}, {Faucher-Gigu{\`e}re}, {Wellons}, {Stern},
	{Terrazas}, {Chan}, {Orr}, {Hummels}, {Feldmann}, and {Kere{\v{s}}}]{su21a}
	{Su}, K.Y.; {Hopkins}, P.F.; {Bryan}, G.L.; {Somerville}, R.S.; {Hayward},
	C.C.; {Angl{\'e}s-Alc{\'a}zar}, D.; {Faucher-Gigu{\`e}re}, C.A.; {Wellons},
	S.; {Stern}, J.; {Terrazas}, B.A.;  et~al.
	\newblock {Which AGN jets quench star formation in massive galaxies?}
	\newblock {\em Mon. Not. R. Astron. Soc.} {\bf 2021}, {\em 507},~175--204.
	\newblock {\url{https://doi.org/10.1093/mnras/stab2021}}.
	
	\bibitem[{Weinberger} et~al.(2023){Weinberger}, {Su}, {Ehlert}, {Pfrommer},
	{Hernquist}, {Bryan}, {Springel}, {Li}, {Burkhart}, {Choi}, and
	{Faucher-Gigu{\`e}re}]{weinberger23a}
	{Weinberger}, R.; {Su}, K.Y.; {Ehlert}, K.; {Pfrommer}, C.; {Hernquist}, L.;
	{Bryan}, G.L.; {Springel}, V.; {Li}, Y.; {Burkhart}, B.; {Choi}, E.;  et~al.
	\newblock {Active galactic nucleus jet feedback in hydrostatic haloes}.
	\newblock {\em Mon. Not. R. Astron. Soc.} {\bf 2023}, {\em 523},~1104--1125.
	\newblock {\url{https://doi.org/10.1093/mnras/stad1396}}.
	
	\bibitem[{Salom{\'e}} et~al.(2015){Salom{\'e}}, {Salom{\'e}}, and
	{Combes}]{salome15a}
	{Salom{\'e}}, Q.; {Salom{\'e}}, P.; {Combes}, F.
	\newblock {Jet-induced star formation in 3C 285 and Minkowski's Object}.
	\newblock {\em Astron. Astrophys.} {\bf 2015}, {\em 574},~A34.
	\newblock {\url{https://doi.org/10.1051/0004-6361/201424932}}.
	
	\bibitem[{Lacy} et~al.(2017){Lacy}, {Croft}, {Fragile}, {Wood}, and
	{Nyland}]{lacy17a}
	{Lacy}, M.; {Croft}, S.; {Fragile}, C.; {Wood}, S.; {Nyland}, K.
	\newblock {ALMA Observations of the Interaction of a Radio Jet with Molecular
		Gas in Minkowski's Object}.
	\newblock {\em Astrophys. J.} {\bf 2017}, {\em 838},~146.
	\newblock {\url{https://doi.org/10.3847/1538-4357/aa65d7}}.
	
	\bibitem[{Nesvadba} et~al.(2020){Nesvadba}, {Bicknell}, {Mukherjee}, and
	{Wagner}]{nesvadba20a}
	{Nesvadba}, N.P.H.; {Bicknell}, G.V.; {Mukherjee}, D.; {Wagner}, A.Y.
	\newblock {Gas, dust, and star formation in the positive AGN feedback candidate
		4C 41.17 at z = 3.8}.
	\newblock {\em Astron. Astrophys.} {\bf 2020}, {\em 639},~L13.
	\newblock {\url{https://doi.org/10.1051/0004-6361/202038269}}.
	
	\bibitem[{Duggal} et~al.(2024){Duggal}, {O'Dea}, {Baum}, {Labiano},
	{Tadhunter}, {Worrall}, {Morganti}, {Tremblay}, and {Dicken}]{duggal24a}
	{Duggal}, C.; {O'Dea}, C.P.; {Baum}, S.A.; {Labiano}, A.; {Tadhunter}, C.;
	{Worrall}, D.M.; {Morganti}, R.; {Tremblay}, G.R.; {Dicken}, D.
	\newblock {Optical- and UV-continuum Morphologies of Compact Radio Source
		Hosts}.
	\newblock {\em Astrophys. J.} {\bf 2024}, {\em 965},~17.
	\newblock {\url{https://doi.org/10.3847/1538-4357/ad2513}}.
	
	\bibitem[{Nesvadba} et~al.(2021){Nesvadba}, {Wagner}, {Mukherjee}, {Mandal},
	{Janssen}, {Zovaro}, {Neumayer}, {Bagchi}, and {Bicknell}]{nesvadba21a}
	{Nesvadba}, N.P.H.; {Wagner}, A.Y.; {Mukherjee}, D.; {Mandal}, A.; {Janssen},
	R.M.J.; {Zovaro}, H.; {Neumayer}, N.; {Bagchi}, J.; {Bicknell}, G.
	\newblock {Jet-driven AGN feedback on molecular gas and low star-formation
		efficiency in a massive local spiral galaxy with a bright X-ray halo}.
	\newblock {\em Astron. Astrophys.} {\bf 2021}, {\em 654},~A8.
	\newblock {\url{https://doi.org/10.1051/0004-6361/202140544}}.
	
	\bibitem[{Krumholz} and {McKee}(2005)]{krumholz05a}
	{Krumholz}, M.R.; {McKee}, C.F.
	\newblock {A General Theory of Turbulence-regulated Star Formation, from
		Spirals to Ultraluminous Infrared Galaxies}.
	\newblock {\em Astrophys. J.} {\bf 2005}, {\em 630},~250--268.
	\newblock {\url{https://doi.org/10.1086/431734}}.
	
	\bibitem[{Federrath} and {Klessen}(2012)]{federrath12a}
	{Federrath}, C.; {Klessen}, R.S.
	\newblock {The Star Formation Rate of Turbulent Magnetized Clouds: Comparing
		Theory, Simulations, and Observations}.
	\newblock {\em Astrophys. J.} {\bf 2012}, {\em 761},~156.
	\newblock {\url{https://doi.org/10.1088/0004-637X/761/2/156}}.
	
	\bibitem[{Roos} et~al.(2015){Roos}, {Juneau}, {Bournaud}, and {Gabor}]{roos15a}
	{Roos}, O.; {Juneau}, S.; {Bournaud}, F.; {Gabor}, J.M.
	\newblock {Thermal and Radiative Active Galactic Nucleus Feedback have a
		Limited Impact on Star Formation in High-redshift Galaxies}.
	\newblock {\em Astrophys. J.} {\bf 2015}, {\em 800},~19.
	\newblock {\url{https://doi.org/10.1088/0004-637X/800/1/19}}.
	
	\bibitem[{Richings} and {Faucher-Gigu{\`e}re}(2018)]{richings17a}
	{Richings}, A.J.; {Faucher-Gigu{\`e}re}, C.A.
	\newblock {The origin of fast molecular outflows in quasars: Molecule formation
		in AGN-driven galactic winds}.
	\newblock {\em Mon. Not. R. Astron. Soc.} {\bf 2018}, {\em 474},~3673--3699,
	\newblock {\url{https://doi.org/10.1093/mnras/stx3014}}.
	
	\bibitem[{van der Kruit} et~al.(1972){van der Kruit}, {Oort}, and
	{Mathewson}]{kruit72a}
	{van der Kruit}, P.C.; {Oort}, J.H.; {Mathewson}, D.S.
	\newblock {The Radio Emission of NGC 4258 and the Possible Origin of Spiral
		Structure}.
	\newblock {\em Astron. Astrophys.} {\bf 1972}, {\em 21},~169.
	
	\bibitem[{Cecil} et~al.(2000){Cecil}, {Greenhill}, {DePree}, {Nagar}, {Wilson},
	{Dopita}, {P{\'e}rez-Fournon}, {Argon}, and {Moran}]{cecil2000a}
	{Cecil}, G.; {Greenhill}, L.J.; {DePree}, C.G.; {Nagar}, N.; {Wilson}, A.S.;
	{Dopita}, M.A.; {P{\'e}rez-Fournon}, I.; {Argon}, A.L.; {Moran}, J.M.
	\newblock {The Active Jet in NGC 4258 and Its Associated Shocks}.
	\newblock {\em Astrophys. J.} {\bf 2000}, {\em 536},~675--696.
	\newblock {\url{https://doi.org/10.1086/308959}}.
	
	\bibitem[{Ogle} et~al.(2014){Ogle}, {Lanz}, and {Appleton}]{ogle14a}
	{Ogle}, P.M.; {Lanz}, L.; {Appleton}, P.N.
	\newblock {Jet-shocked H$_{2}$ and CO in the Anomalous Arms of Molecular
		Hydrogen Emission Galaxy NGC 4258}.
	\newblock {\em Astrophys. J.} {\bf 2014}, {\em 788},~L33.
	\newblock {\url{https://doi.org/10.1088/2041-8205/788/2/L33}}.
	
	\bibitem[{Appleton} et~al.(2018){Appleton}, {Diaz-Santos}, {Fadda}, {Ogle},
	{Togi}, {Lanz}, {Alatalo}, {Fischer}, {Rich}, and {Guillard}]{appleton18a}
	{Appleton}, P.N.; {Diaz-Santos}, T.; {Fadda}, D.; {Ogle}, P.; {Togi}, A.;
	{Lanz}, L.; {Alatalo}, K.; {Fischer}, C.; {Rich}, J.; {Guillard}, P.
	\newblock {Jet-related Excitation of the [C II] Emission in the Active Galaxy
		NGC 4258 with SOFIA}.
	\newblock {\em Astrophys. J.} {\bf 2018}, {\em 869},~61.
	\newblock {\url{https://doi.org/10.3847/1538-4357/aaed2a}}.
	
	\bibitem[{Butcher} et~al.(1980){Butcher}, {van Breugel}, and
	{Miley}]{butcher80a}
	{Butcher}, H.R.; {van Breugel}, W.; {Miley}, G.K.
	\newblock {Optical observations of radio jets.}
	\newblock {\em Astrophys. J.} {\bf 1980}, {\em 235},~749--754.
	\newblock {\url{https://doi.org/10.1086/157677}}.
	
	\bibitem[{Miley} et~al.(1981){Miley}, {Heckman}, {Butcher}, and {van
		Breugel}]{miley81a}
	{Miley}, G.K.; {Heckman}, T.M.; {Butcher}, H.R.; {van Breugel}, W.J.M.
	\newblock {Optical emission from the extended radio source 3C 277.3 (Coma A).}
	\newblock {\em Astrophys. J.} {\bf 1981}, {\em 247},~L5--L9.
	\newblock {\url{https://doi.org/10.1086/183578}}.
	
	\bibitem[{Heckman} et~al.(1982){Heckman}, {Miley}, {Balick}, {van Breugel}, and
	{Butcher}]{heckman82a}
	{Heckman}, T.M.; {Miley}, G.K.; {Balick}, B.; {van Breugel}, W.J.M.; {Butcher},
	H.R.
	\newblock {An optical and radio investigation of the radio galaxy 3C 305.}
	\newblock {\em Astrophys. J.} {\bf 1982}, {\em 262},~529--553.
	\newblock {\url{https://doi.org/10.1086/160445}}.
	
	\bibitem[{Heckman} et~al.(1984){Heckman}, {van Breugel}, and
	{Miley}]{heckman84a}
	{Heckman}, T.M.; {van Breugel}, W.J.M.; {Miley}, G.K.
	\newblock {Emission-line gas associated with the radio lobes of the
		high-luminosity radiosource 3C 171.}
	\newblock {\em Astrophys. J.} {\bf 1984}, {\em 286},~509--516.
	\newblock {\url{https://doi.org/10.1086/162626}}.
	
	\bibitem[{van Breugel} et~al.(1984){van Breugel}, {Heckman}, {Butcher}, and
	{Miley}]{VanBreugel84b}
	{van Breugel}, W.; {Heckman}, T.; {Butcher}, H.; {Miley}, G.
	\newblock {Extended optical line emission from 3C 293 : Radio jets propagating
		through a rotating gaseous disk.}
	\newblock {\em Astrophys. J.} {\bf 1984}, {\em 277},~82--91.
	\newblock {\url{https://doi.org/10.1086/161673}}.
	
	\bibitem[{van Breugel} and {Fomalont}(1984)]{vanBreugel84c}
	{van Breugel}, W.; {Fomalont}, E.B.
	\newblock {Is 3C 310 blowing bubbles ?}
	\newblock {\em Astrophys. J.} {\bf 1984}, {\em 282},~L55--L58.
	\newblock {\url{https://doi.org/10.1086/184304}}.
	
	\bibitem[{Chambers} et~al.(1987){Chambers}, {Miley}, and {van
		Breugel}]{chambers87a}
	{Chambers}, K.C.; {Miley}, G.K.; {van Breugel}, W.
	\newblock {Alignment of radio and optical orientations in high-redshift radio
		galaxies}.
	\newblock {\em \nat} {\bf 1987}, {\em 329},~604--606.
	\newblock {\url{https://doi.org/10.1038/329604a0}}.
	
	\bibitem[{McCarthy} et~al.(1987){McCarthy}, {van Breugel}, {Spinrad}, and
	{Djorgovski}]{mccarthy87a}
	{McCarthy}, P.J.; {van Breugel}, W.; {Spinrad}, H.; {Djorgovski}, S.
	\newblock {A Correlation between the Radio and Optical Morphologies of Distant
		3 CR Radio Galaxies}.
	\newblock {\em Astrophys. J.} {\bf 1987}, {\em 321},~L29.
	\newblock {\url{https://doi.org/10.1086/185000}}.
	
	\bibitem[{de Vries} et~al.(1997){de Vries}, {O'Dea}, {Baum}, {Sparks},
	{Biretta}, {de Koff}, {Golombek}, {Lehnert}, {Macchetto}, {McCarthy}, and
	{Miley}]{devries97a}
	{de Vries}, W.H.; {O'Dea}, C.P.; {Baum}, S.A.; {Sparks}, W.B.; {Biretta}, J.;
	{de Koff}, S.; {Golombek}, D.; {Lehnert}, M.D.; {Macchetto}, F.; {McCarthy},
	P.;  et~al.
	\newblock {Hubble Space Telescope Imaging of Compact Steep-Spectrum Radio
		Sources}.
	\newblock {\em Astrophys. J. Suppl. Ser.} {\bf 1997}, {\em 110},~191--211.
	\newblock {\url{https://doi.org/10.1086/313001}}.
	
	\bibitem[Vries et~al.(1999)Vries, O'Dea, Baum, and Barthel]{devries99a}
	Vries, W.D.; O'Dea, C.P.; Baum, S.A.; Barthel, P.D.
	\newblock Optical-Radio Alignment in Compact Steep-Spectrum Radio Sources.
	\newblock {\em Astrophys. J.} {\bf 1999}, {\em 526},~27--39.
	
	\bibitem[{McCarthy}(1993)]{mccarthy93a}
	{McCarthy}, P.J.
	\newblock {High redshift radio galaxies.}
	\newblock {\em Annu. Rev. Astron. Astrophys.} {\bf 1993}, {\em 31},~639--688.
	\newblock {\url{https://doi.org/10.1146/annurev.aa.31.090193.003231}}.
	
	\bibitem[{Fabian}(1989)]{fabian89a}
	{Fabian}, A.C.
	\newblock {The alignment of the optical continuum and radio axes of
		high-redshift radio galaxies : Electron scattering in intracluster gas ?}
	\newblock {\em Mon. Not. R. Astron. Soc.} {\bf 1989}, {\em 238},~41P--44.
	\newblock {\url{https://doi.org/10.1093/mnras/238.1.41P}}.
	
	\bibitem[{Tadhunter} et~al.(1992){Tadhunter}, {Scarrott}, {Draper}, and
	{Rolph}]{tadhunter92a}
	{Tadhunter}, C.N.; {Scarrott}, S.M.; {Draper}, P.; {Rolph}, C.
	\newblock {The optical polarizations of high- and intermediate-redshift radio
		galaxies}.
	\newblock {\em Mon. Not. R. Astron. Soc.} {\bf 1992}, {\em 256},~53P--58P.
	\newblock {\url{https://doi.org/10.1093/mnras/256.1.53P}}.
	
	\bibitem[{Dickson} et~al.(1995){Dickson}, {Tadhunter}, {Shaw}, {Clark}, and
	{Morganti}]{dickson95a}
	{Dickson}, R.; {Tadhunter}, C.; {Shaw}, M.; {Clark}, N.; {Morganti}, R.
	\newblock {The nebular contribution to the extended UV continua of powerful
		radio galaxies}.
	\newblock {\em Mon. Not. R. Astron. Soc.} {\bf 1995}, {\em 273},~L29--L33.
	\newblock {\url{https://doi.org/10.1093/mnras/273.1.L29}}.
	
	\bibitem[{Tadhunter} et~al.(2002){Tadhunter}, {Dickson}, {Morganti},
	{Robinson}, {Wills}, {Villar-Martin}, and {Hughes}]{tadhunter02a}
	{Tadhunter}, C.; {Dickson}, R.; {Morganti}, R.; {Robinson}, T.G.; {Wills}, K.;
	{Villar-Martin}, M.; {Hughes}, M.
	\newblock {The origin of the UV excess in powerful radio galaxies: Spectroscopy
		and polarimetry of a complete sample of intermediate-redshift radio
		galaxies}.
	\newblock {\em Mon. Not. R. Astron. Soc.} {\bf 2002}, {\em 330},~977--996.
	\newblock {\url{https://doi.org/10.1046/j.1365-8711.2002.05153.x}}.
	
	\bibitem[{Kukreti} et~al.(2023){Kukreti}, {Morganti}, {Tadhunter}, and
	{Santoro}]{kukreti23a}
	{Kukreti}, P.; {Morganti}, R.; {Tadhunter}, C.; {Santoro}, F.
	\newblock {Ionised gas outflows over the radio AGN life cycle}.
	\newblock {\em Astron. Astrophys.} {\bf 2023}, {\em 674},~A198.
	\newblock {\url{https://doi.org/10.1051/0004-6361/202245691}}.
	
	\bibitem[{Kukreti} et~al.(2025){Kukreti}, {Wylezalek}, {Alb\textbackslash'an},
	and {DallAgnol de Oliveira}]{kukreti25a}
	{Kukreti}, P.; {Wylezalek}, D.; {Alb\textbackslash'an}, M.; {DallAgnol de
		Oliveira}, B.
	\newblock {Feedback from low-to-moderate luminosity radio-AGN with MaNGA}.
	\newblock {\em arXiv} {\bf 2025}, arXiv:2503.20889.
	\newblock {\url{https://doi.org/10.48550/arXiv.2503.20889}}.
	
	\bibitem[{Calistro Rivera} et~al.(2024){Calistro Rivera}, {Alexander},
	{Harrison}, {Fawcett}, {Best}, {Williams}, {Hardcastle}, {Rosario}, {Smith},
	{Arnaudova}, {Escott}, {G{\"u}rkan}, {Kondapally}, {Miley}, {Morabito},
	{Petley}, {Prandoni}, {R{\"o}ttgering}, and {Yue}]{rivera24a}
	{Calistro Rivera}, G.; {Alexander}, D.M.; {Harrison}, C.M.; {Fawcett}, V.A.;
	{Best}, P.N.; {Williams}, W.L.; {Hardcastle}, M.J.; {Rosario}, D.J.; {Smith},
	D.J.B.; {Arnaudova}, M.I.;  et~al.
	\newblock {Ubiquitous radio emission in quasars: Predominant AGN origin and a
		connection to jets, dust, and winds}.
	\newblock {\em Astron. Astrophys.} {\bf 2024}, {\em 691},~A191.
	\newblock {\url{https://doi.org/10.1051/0004-6361/202348982}}.
	
	\bibitem[{Nandi} et~al.(2025){Nandi}, {Stalin}, and {Saikia}]{nandi25a}
	{Nandi}, P.; {Stalin}, C.S.; {Saikia}, D.J.
	\newblock {Warm Ionized Gas Outflows in Active Galactic Nuclei: What Causes
		Them?}
	\newblock {\em Astrophys. J.} {\bf 2025}, {\em 984},~20.
	\newblock {\url{https://doi.org/10.3847/1538-4357/adc110}}.
	
	\bibitem[{Cheung} et~al.(2016){Cheung}, {Bundy}, {Cappellari}, {Peirani},
	{Rujopakarn}, {Westfall}, {Yan}, {Bershady}, {Greene}, {Heckman}, {Drory},
	{Law}, {Masters}, {Thomas}, {Wake}, {Weijmans}, {Rubin}, {Belfiore},
	{Vulcani}, {Chen}, {Zhang}, {Gelfand}, {Bizyaev}, {Roman-Lopes}, and
	{Schneider}]{cheung16a}
	{Cheung}, E.; {Bundy}, K.; {Cappellari}, M.; {Peirani}, S.; {Rujopakarn}, W.;
	{Westfall}, K.; {Yan}, R.; {Bershady}, M.; {Greene}, J.E.; {Heckman}, T.M.;
	et~al.
	\newblock {Suppressing star formation in quiescent galaxies with supermassive
		black hole winds}.
	\newblock {\em \nat} {\bf 2016}, {\em 533},~504--508.
	\newblock {\url{https://doi.org/10.1038/nature18006}}.
	
	\bibitem[{Roy} et~al.(2018){Roy}, {Bundy}, {Cheung}, {Rujopakarn},
	{Cappellari}, {Belfiore}, {Yan}, {Heckman}, {Bershady}, {Greene}, {Westfall},
	{Drory}, {Rubin}, {Law}, {Zhang}, {Gelfand}, {Bizyaev}, {Wake}, {Masters},
	{Thomas}, {Li}, and {Riffel}]{roy18a}
	{Roy}, N.; {Bundy}, K.; {Cheung}, E.; {Rujopakarn}, W.; {Cappellari}, M.;
	{Belfiore}, F.; {Yan}, R.; {Heckman}, T.; {Bershady}, M.; {Greene}, J.;
	et~al.
	\newblock {Detecting Radio AGN Signatures in Red Geysers}.
	\newblock {\em Astrophys. J.} {\bf 2018}, {\em 869},~117.
	\newblock {\url{https://doi.org/10.3847/1538-4357/aaee72}}.
	
	\bibitem[{Roy} et~al.(2021){Roy}, {Moravec}, {Bundy}, {Hardcastle},
	{G{\"u}rkan}, {Diego Baldi}, {Leslie}, {Masters}, {Gelfand}, {Riffel},
	{Riffel}, {Mingo Fernandez}, and {Drabent}]{roy21a}
	{Roy}, N.; {Moravec}, E.; {Bundy}, K.; {Hardcastle}, M.J.; {G{\"u}rkan}, G.;
	{Diego Baldi}, R.; {Leslie}, S.K.; {Masters}, K.; {Gelfand}, J.; {Riffel},
	R.;  et~al.
	\newblock {Radio Morphology of Red Geysers}.
	\newblock {\em Astrophys. J.} {\bf 2021}, {\em 922},~230.
	\newblock {\url{https://doi.org/10.3847/1538-4357/ac24a0}}.
	
	\bibitem[{Murthy} et~al.(2019){Murthy}, {Morganti}, {Oosterloo}, {Schulz},
	{Mukherjee}, {Wagner}, {Bicknell}, {Prandoni}, and {Shulevski}]{murthy19a}
	{Murthy}, S.; {Morganti}, R.; {Oosterloo}, T.; {Schulz}, R.; {Mukherjee}, D.;
	{Wagner}, A.Y.; {Bicknell}, G.; {Prandoni}, I.; {Shulevski}, A.
	\newblock {Feedback from low-luminosity radio galaxies: B2 0258+35}.
	\newblock {\em Astron. Astrophys.} {\bf 2019}, {\em 629},~A58.
	\newblock {\url{https://doi.org/10.1051/0004-6361/201935931}}.

    
	
	\bibitem[{Drevet Mulard} et~al.(2023){Drevet Mulard}, {Nesvadba}, {Meenakshi},
	{Mukherjee}, {Wagner}, {Bicknell}, {Neumayer}, {Combes}, {Zovaro}, {Janssen},
	{Bagchi}, {Dabhade}, and {Prunet}]{mulard23a}
	{Drevet Mulard}, M.; {Nesvadba}, N.P.H.; {Meenakshi}, M.; {Mukherjee}, D.;
	{Wagner}, A.; {Bicknell}, G.; {Neumayer}, N.; {Combes}, F.; {Zovaro}, H.;
	{Janssen}, R.M.J.;  et~al.
	\newblock {Star formation in a massive spiral galaxy with a radio-AGN}.
	\newblock {\em Astron. Astrophys.} {\bf 2023}, {\em 676},~A35.
	\newblock {\url{https://doi.org/10.1051/0004-6361/202245173}}.

        \bibitem[{Orienti} et~al.(2025){Orienti}, {D'Ammando}, {Dallacasa}, {Migliori}, {Rossi}, and {Bodo}]{Orienti25a}
    {Orienti}, M.; {D'Ammando}, F.; {Dallacasa}, D.; {Migliori}, G.; {Rossi}, P.; {Bodo}, G.
    \newblock {VLBA observations of a sample of low-power compact symmetric objects}.
    \newblock {\em \aap} {\bf 2025}, {\em 698},~A157,  
    \newblock {\url{https://doi.org/10.1051/0004-6361/202553798}}.

        \bibitem[{Dallacasa} et~al.(2000){Dallacasa}, {Stanghellini}, {Centonza}, and {Fanti}]{dallacasa2000a}
    {Dallacasa}, D.; {Stanghellini}, C.; {Centonza}, M.; {Fanti}, R.
    \newblock {High frequency peakers. I. The bright sample}.
    \newblock {\em \aap} {\bf 2000}, {\em 363},~887--900,  
    \newblock {\url{https://doi.org/10.48550/arXiv.astro-ph/0012428}}.

    
	\bibitem[{Murgia}(2003)]{murgia03a}
	{Murgia}, M.
	\newblock {Spectral Ages of CSOs and CSS Sources}.
	\newblock {\em Publ. Astron. Soc. Aust.} {\bf 2003}, {\em 20},~19--24.
	\newblock {\url{https://doi.org/10.1071/AS02033}}.
	
	\bibitem[{An} and {Baan}(2012)]{an12a}
	{An}, T.; {Baan}, W.A.
	\newblock {The Dynamic Evolution of Young Extragalactic Radio Sources}.
	\newblock {\em Astrophys. J.} {\bf 2012}, {\em 760},~77.
	\newblock {\url{https://doi.org/10.1088/0004-637X/760/1/77}}.
	
	\bibitem[{Patil} et~al.(2020){Patil}, {Nyland}, {Whittle}, {Lonsdale}, {Lacy},
	{Lonsdale}, {Mukherjee}, {Trapp}, {Kimball}, {Lanz}, {Wilkes}, {Blain},
	{Harwood}, {Efstathiou}, and {Vlahakis}]{patil20a}
	{Patil}, P.; {Nyland}, K.; {Whittle}, M.; {Lonsdale}, C.; {Lacy}, M.;
	{Lonsdale}, C.; {Mukherjee}, D.; {Trapp}, A.C.; {Kimball}, A.E.; {Lanz}, L.;
	et~al.
	\newblock {High-resolution VLA Imaging of Obscured Quasars: Young Radio Jets
		Caught in a Dense ISM}.
	\newblock {\em Astrophys. J.} {\bf 2020}, {\em 896},~18.
	\newblock {\url{https://doi.org/10.3847/1538-4357/ab9011}}.
	
	\bibitem[{Rossetti} et~al.(2008){Rossetti}, {Dallacasa}, {Fanti}, {Fanti}, and
	{Mack}]{rossetti08a}
	{Rossetti}, A.; {Dallacasa}, D.; {Fanti}, C.; {Fanti}, R.; {Mack}, K.H.
	\newblock {The B3-VLA CSS sample. VII. WSRT polarisation observations and the
		ambient Faraday medium properties revisited}.
	\newblock {\em Astron. Astrophys.} {\bf 2008}, {\em 487},~865--883.
	\newblock {\url{https://doi.org/10.1051/0004-6361:20079047}}.
	
	\bibitem[{Mantovani} et~al.(2013){Mantovani}, {Rossetti}, {Junor}, {Saikia},
	and {Salter}]{mantovani13a}
	{Mantovani}, F.; {Rossetti}, A.; {Junor}, W.; {Saikia}, D.J.; {Salter}, C.J.
	\newblock {Radio polarimetry of compact steep spectrum sources at sub-arcsecond
		resolution}.
	\newblock {\em Astron. Astrophys.} {\bf 2013}, {\em 555},~A4.
	\newblock {\url{https://doi.org/10.1051/0004-6361/201220769}}.
	
	\bibitem[{Orienti}(2016)]{orienti16a}
	{Orienti}, M.
	\newblock {Radio properties of Compact Steep Spectrum and GHz-Peaked Spectrum
		radio sources}.
	\newblock {\em Astron. Nachrichten} {\bf 2016}, {\em 337},~9.
	\newblock {\url{https://doi.org/10.1002/asna.201512257}}.
	
	\bibitem[{Guainazzi} et~al.(2006){Guainazzi}, {Siemiginowska}, {Stanghellini},
	{Grandi}, {Piconcelli}, and {Azubike Ugwoke}]{guiainazzi06a}
	{Guainazzi}, M.; {Siemiginowska}, A.; {Stanghellini}, C.; {Grandi}, P.;
	{Piconcelli}, E.; {Azubike Ugwoke}, C.
	\newblock {A hard X-ray view of giga-hertz peaked spectrum radio galaxies}.
	\newblock {\em Astron. Astrophys.} {\bf 2006}, {\em 446},~87--96.
	\newblock {\url{https://doi.org/10.1051/0004-6361:20053374}}.
	
	\bibitem[{Siemiginowska} et~al.(2008){Siemiginowska}, {LaMassa}, {Aldcroft},
	{Bechtold}, and {Elvis}]{Siemiginowska08a}
	{Siemiginowska}, A.; {LaMassa}, S.; {Aldcroft}, T.L.; {Bechtold}, J.; {Elvis},
	M.
	\newblock {X-Ray Properties of the Gigahertz Peaked and Compact Steep Spectrum
		Sources}.
	\newblock {\em Astrophys. J.} {\bf 2008}, {\em 684},~811--821.
	\newblock {\url{https://doi.org/10.1086/589437}}.
	
	\bibitem[{Siemiginowska} et~al.(2016){Siemiginowska}, {Sobolewska}, {Migliori},
	{Guainazzi}, {Hardcastle}, {Ostorero}, and {Stawarz}]{Siemiginowska16a}
	{Siemiginowska}, A.; {Sobolewska}, M.; {Migliori}, G.; {Guainazzi}, M.;
	{Hardcastle}, M.; {Ostorero}, L.; {Stawarz}, {\L}.
	\newblock {X-Ray Properties of the Youngest Radio Sources and Their
		Environments}.
	\newblock {\em Astrophys. J.} {\bf 2016}, {\em 823},~57.
	\newblock {\url{https://doi.org/10.3847/0004-637X/823/1/57}}.
	
	\bibitem[{Ostorero} et~al.(2010){Ostorero}, {Moderski}, {Stawarz}, {Diaferio},
	{Kowalska}, {Cheung}, {Kataoka}, {Begelman}, and {Wagner}]{ostorero10a}
	{Ostorero}, L.; {Moderski}, R.; {Stawarz}, {\L}.; {Diaferio}, A.; {Kowalska},
	I.; {Cheung}, C.C.; {Kataoka}, J.; {Begelman}, M.C.; {Wagner}, S.J.
	\newblock {X-ray-emitting GHz-peaked-spectrum Galaxies: Testing a
		Dynamical-Radiative Model with Broadband Spectra}.
	\newblock {\em Astrophys. J.} {\bf 2010}, {\em 715},~1071--1093.
	\newblock {\url{https://doi.org/10.1088/0004-637X/715/2/1071}}.
	
	\bibitem[{Ostorero} et~al.(2017){Ostorero}, {Morganti}, {Diaferio},
	{Siemiginowska}, {Stawarz}, {Moderski}, and {Labiano}]{ostorero17a}
	{Ostorero}, L.; {Morganti}, R.; {Diaferio}, A.; {Siemiginowska}, A.; {Stawarz},
	{\L}.; {Moderski}, R.; {Labiano}, A.
	\newblock {Correlation between X-Ray and Radio Absorption in Compact Radio
		Galaxies}.
	\newblock {\em Astrophys. J.} {\bf 2017}, {\em 849},~34.
	\newblock {\url{https://doi.org/10.3847/1538-4357/aa8ef6}}.
	
	\bibitem[{Patil} et~al.(2022){Patil}, {Whittle}, {Nyland}, {Lonsdale}, {Lacy},
	{Kimball}, {Lonsdale}, {Peters}, {Clarke}, {Efstathiou}, {Giacintucci},
	{Kim}, {Lanz}, {Mukherjee}, and {Polisensky}]{patil22a}
	{Patil}, P.; {Whittle}, M.; {Nyland}, K.; {Lonsdale}, C.; {Lacy}, M.;
	{Kimball}, A.E.; {Lonsdale}, C.; {Peters}, W.; {Clarke}, T.E.; {Efstathiou},
	A.;  et~al.
	\newblock {Radio Spectra of Luminous, Heavily Obscured WISE-NVSS Selected
		Quasars}.
	\newblock {\em Astrophys. J.} {\bf 2022}, {\em 934},~26.
	\newblock {\url{https://doi.org/10.3847/1538-4357/ac71b0}}.
	
	\bibitem[{Nascimento} et~al.(2022){Nascimento}, {Rodr{\'\i}guez-Ardila},
	{Dahmer-Hahn}, {Fonseca-Faria}, {Riffel}, {Marinello}, {Beuchert}, and
	{Callingham}]{nascimento22a}
	{Nascimento}, R.S.; {Rodr{\'\i}guez-Ardila}, A.; {Dahmer-Hahn}, L.;
	{Fonseca-Faria}, M.A.; {Riffel}, R.; {Marinello}, M.; {Beuchert}, T.;
	{Callingham}, J.R.
	\newblock {Optical properties of Peaked Spectrum radio sources}.
	\newblock {\em Mon. Not. R. Astron. Soc.} {\bf 2022}, {\em 511},~214--230.
	\newblock {\url{https://doi.org/10.1093/mnras/stab3791}}.
	
	\bibitem[{Fanti}(2009)]{fanti09a}
	{Fanti}, C.
	\newblock {Radio properties of CSSs and GPSs}.
	\newblock {\em Astron. Nachrichten} {\bf 2009}, {\em 330},~120--127.
	\newblock {\url{https://doi.org/10.1002/asna.200811137}}.
	
	\bibitem[{Miranda Marques} et~al.(2025){Miranda Marques},
	{Rodr{\'\i}guez-Ardila}, {Fonseca-Faria}, and {Panda}]{miranda25a}
	{Miranda Marques}, B.L.; {Rodr{\'\i}guez-Ardila}, A.; {Fonseca-Faria}, M.A.;
	{Panda}, S.
	\newblock {Powerful Outflows of Compact Radio Galaxies}.
	\newblock {\em Astrophys. J.} {\bf 2025}, {\em 978},~16.
	\newblock {\url{https://doi.org/10.3847/1538-4357/ad8f40}}.
	
	\bibitem[{Baldi} et~al.(2018){Baldi}, {Capetti}, and {Massaro}]{baldi18a}
	{Baldi}, R.D.; {Capetti}, A.; {Massaro}, F.
	\newblock {FR0CAT: A FIRST catalog of FR 0 radio galaxies}.
	\newblock {\em Astron. Astrophys.} {\bf 2018}, {\em 609},~A1.
	\newblock {\url{https://doi.org/10.1051/0004-6361/201731333}}.
	
	\bibitem[{Jarvis} et~al.(2021){Jarvis}, {Harrison}, {Mainieri}, {Alexander},
	{Arrigoni Battaia}, {Calistro Rivera}, {Circosta}, {Costa}, {De Breuck},
	{Edge}, {Girdhar}, {Kakkad}, {Kharb}, {Lansbury}, {Molyneux}, {Mukherjee},
	{Mullaney}, {Farina}, {Silpa}, {Thomson}, and {Ward}]{jarvis21a}
	{Jarvis}, M.E.; {Harrison}, C.M.; {Mainieri}, V.; {Alexander}, D.M.; {Arrigoni
		Battaia}, F.; {Calistro Rivera}, G.; {Circosta}, C.; {Costa}, T.; {De
		Breuck}, C.; {Edge}, A.C.;  et~al.
	\newblock {The quasar feedback survey: Discovering hidden Radio-AGN and their
		connection to the host galaxy ionized gas}.
	\newblock {\em Mon. Not. R. Astron. Soc.} {\bf 2021}, {\em 503},~1780--1797.
	\newblock {\url{https://doi.org/10.1093/mnras/stab549}}.
	
	\bibitem[{Njeri} et~al.(2025){Njeri}, {Harrison}, {Kharb}, {Beswick},
	{Calistro-Rivera}, {Circosta}, {Mainieri}, {Molyneux}, {Mullaney}, and
	{Sasikumar}]{njeri25a}
	{Njeri}, A.; {Harrison}, C.M.; {Kharb}, P.; {Beswick}, R.; {Calistro-Rivera},
	G.; {Circosta}, C.; {Mainieri}, V.; {Molyneux}, S.; {Mullaney}, J.;
	{Sasikumar}, S.
	\newblock {The quasar feedback survey: Zooming into the origin of radio
		emission with e-MERLIN}.
	\newblock {\em Mon. Not. R. Astron. Soc.} {\bf 2025}, {\textit{537}, 705--722.}{} 
	\newblock {\url{https://doi.org/10.1093/mnras/staf020}}.
	
	\bibitem[{Jarvis} et~al.(2020){Jarvis}, {Harrison}, {Mainieri}, {Calistro
		Rivera}, {Jethwa}, {Zhang}, {Alexander}, {Circosta}, {Costa}, {De Breuck},
	{Kakkad}, {Kharb}, {Lansbury}, and {Thomson}]{jarvis20a}
	{Jarvis}, M.E.; {Harrison}, C.M.; {Mainieri}, V.; {Calistro Rivera}, G.;
	{Jethwa}, P.; {Zhang}, Z.Y.; {Alexander}, D.M.; {Circosta}, C.; {Costa}, T.;
	{De Breuck}, C.;  et~al.
	\newblock {High molecular gas content and star formation rates in local
		galaxies that host quasars, outflows, and jets}.
	\newblock {\em Mon. Not. R. Astron. Soc.} {\bf 2020}, {\em 498},~1560--1575.
	\newblock {\url{https://doi.org/10.1093/mnras/staa2196}}.
	
	\bibitem[{Molyneux} et~al.(2024){Molyneux}, {Calistro Rivera}, {De Breuck},
	{Harrison}, {Mainieri}, {Lundgren}, {Kakkad}, {Circosta}, {Girdhar}, {Costa},
	{Mullaney}, {Kharb}, {Arrigoni Battaia}, {Farina}, {Alexander}, {Ward},
	{Silpa}, and {Smit}]{molyneux24a}
	{Molyneux}, S.J.; {Calistro Rivera}, G.; {De Breuck}, C.; {Harrison}, C.M.;
	{Mainieri}, V.; {Lundgren}, A.; {Kakkad}, D.; {Circosta}, C.; {Girdhar}, A.;
	{Costa}, T.;  et~al.
	\newblock {The Quasar Feedback Survey: Characterizing CO excitation in quasar
		host galaxies}.
	\newblock {\em Mon. Not. R. Astron. Soc.} {\bf 2024}, {\em 527},~4420--4439.
	\newblock {\url{https://doi.org/10.1093/mnras/stad3133}}.
	
	\bibitem[{Davis} et~al.(2019){Davis}, {Greene}, {Ma}, {Blakeslee}, {Dawson},
	{Pandya}, {Veale}, and {Zabel}]{davis19a}
	{Davis}, T.A.; {Greene}, J.E.; {Ma}, C.P.; {Blakeslee}, J.P.; {Dawson}, J.M.;
	{Pandya}, V.; {Veale}, M.; {Zabel}, N.
	\newblock {The MASSIVE survey---XI. What drives the molecular gas properties of
		early-type galaxies}.
	\newblock {\em Mon. Not. R. Astron. Soc.} {\bf 2019}, {\em 486},~1404--1423.
	\newblock {\url{https://doi.org/10.1093/mnras/stz871}}.
	
	\bibitem[{Tadhunter} et~al.(2024){Tadhunter}, {Oosterloo}, {Morganti}, {Ramos
		Almeida}, {Mart{\'\i}n}, {Emonts}, and {Dicken}]{tadhunter24a}
	{Tadhunter}, C.; {Oosterloo}, T.; {Morganti}, R.; {Ramos Almeida}, C.;
	{Mart{\'\i}n}, M.V.; {Emonts}, B.; {Dicken}, D.
	\newblock {An ALMA CO(1-0) survey of the 2Jy sample: Large and massive
		molecular discs in radio AGN host galaxies}.
	\newblock {\em Mon. Not. R. Astron. Soc.} {\bf 2024}, {\em 532},~4463--4485.
	\newblock {\url{https://doi.org/10.1093/mnras/stae1745}}.
	
	\bibitem[{Ruffa} and {Davis}(2024)]{ruffa24a}
	{Ruffa}, I.; {Davis}, T.A.
	\newblock {Molecular Gas Kinematics in Local Early-Type Galaxies with ALMA}.
	\newblock {\em Galaxies} {\bf 2024}, {\em 12},~36.
	\newblock {\url{https://doi.org/10.3390/galaxies12040036}}.
	
	\bibitem[{Audibert} et~al.(2022){Audibert}, {Dasyra}, {Papachristou},
	{Fern{\'a}ndez-Ontiveros}, {Ruffa}, {Bisigello}, {Combes}, {Salom{\'e}}, and
	{Gruppioni}]{audibert22a}
	{Audibert}, A.; {Dasyra}, K.M.; {Papachristou}, M.; {Fern{\'a}ndez-Ontiveros},
	J.A.; {Ruffa}, I.; {Bisigello}, L.; {Combes}, F.; {Salom{\'e}}, P.;
	{Gruppioni}, C.
	\newblock {CO in the ALMA Radio-source Catalogue (ARC): The molecular gas
		content of radio galaxies as a function of redshift}.
	\newblock {\em Astron. Astrophys.} {\bf 2022}, {\em 668},~A67.
	\newblock {\url{https://doi.org/10.1051/0004-6361/202243666}}.
	
	\bibitem[{Best} et~al.(2005){Best}, {Kauffmann}, {Heckman}, {Brinchmann},
	{Charlot}, {Ivezi{\'c}}, and {White}]{best05a}
	{Best}, P.N.; {Kauffmann}, G.; {Heckman}, T.M.; {Brinchmann}, J.; {Charlot},
	S.; {Ivezi{\'c}}, {\v{Z}}.; {White}, S.D.M.
	\newblock {The host galaxies of radio-loud active galactic nuclei: Mass
		dependences, gas cooling and active galactic nuclei feedback}.
	\newblock {\em Mon. Not. R. Astron. Soc.} {\bf 2005}, {\em 362},~25--40.
	\newblock {\url{https://doi.org/10.1111/j.1365-2966.2005.09192.x}}.
	
	\bibitem[{Mauch} and {Sadler}(2007)]{mauch07a}
	{Mauch}, T.; {Sadler}, E.M.
	\newblock {Radio sources in the 6dFGS: Local luminosity functions at 1.4GHz for
		star-forming galaxies and radio-loud AGN}.
	\newblock {\em Mon. Not. R. Astron. Soc.} {\bf 2007}, {\em 375},~931--950.
	\newblock {\url{https://doi.org/10.1111/j.1365-2966.2006.11353.x}}.
	
	\bibitem[{Sabater} et~al.(2019){Sabater}, {Best}, {Hardcastle}, {Shimwell},
	{Tasse}, {Williams}, {Br{\"u}ggen}, {Cochrane}, {Croston}, {de Gasperin},
	{Duncan}, {G{\"u}rkan}, {Mechev}, {Morabito}, {Prandoni}, {R{\"o}ttgering},
	{Smith}, {Harwood}, {Mingo}, {Mooney}, and {Saxena}]{sabater19a}
	{Sabater}, J.; {Best}, P.N.; {Hardcastle}, M.J.; {Shimwell}, T.W.; {Tasse}, C.;
	{Williams}, W.L.; {Br{\"u}ggen}, M.; {Cochrane}, R.K.; {Croston}, J.H.; {de
		Gasperin}, F.;  et~al.
	\newblock {The LoTSS view of radio AGN in the local Universe. The most massive
		galaxies are always switched on}.
	\newblock {\em Astron. Astrophys.} {\bf 2019}, {\em 622},~A17.
	\newblock {\url{https://doi.org/10.1051/0004-6361/201833883}}.
	
	\bibitem[{Weymann} et~al.(1982){Weymann}, {Scott}, {Schiano}, and
	{Christiansen}]{weymann82a}
	{Weymann}, R.J.; {Scott}, J.S.; {Schiano}, A.V.R.; {Christiansen}, W.A.
	\newblock {A thermal wind model for the broad emission line region of quasars}.
	\newblock {\em Astrophys. J.} {\bf 1982}, {\em 262},~497--510.
	\newblock {\url{https://doi.org/10.1086/160443}}.
	
	\bibitem[{Schiano}(1985)]{schiano85a}
	{Schiano}, A.V.R.
	\newblock {The hydrodynamic effects of nuclear active galaxy winds on host
		galaxies}.
	\newblock {\em Astrophys. J.} {\bf 1985}, {\em 299},~24--40.
	\newblock {\url{https://doi.org/10.1086/163680}}.
	
	\bibitem[{Schiano}(1986)]{schiano86a}
	{Schiano}, A.V.R.
	\newblock {The Interaction of Interstellar Clouds with a ``Quasar Wind'' and
		Its Relation to the Narrow Emmision Line Regions of Seyfert Galaxies}.
	\newblock {\em Astrophys. J.} {\bf 1986}, {\em 302},~81.
	\newblock {\url{https://doi.org/10.1086/163975}}.
	
	\bibitem[{King}(2003)]{king03a}
	{King}, A.
	\newblock {Black Holes, Galaxy Formation, and the M$_{BH}$-{$\sigma$}
		Relation}.
	\newblock {\em Astrophys. J.} {\bf 2003}, {\em 596},~L27--L29.
	\newblock {\url{https://doi.org/10.1086/379143}}.
	
	\bibitem[{King}(2005)]{king05a}
	{King}, A.
	\newblock {The AGN-Starburst Connection, Galactic Superwinds, and
		$M_{BH}$-{$\sigma$}}.
	\newblock {\em Astrophys. J. Lett.} {\bf 2005}, {\em 635},~L121--L123.
	\newblock {\url{https://doi.org/10.1086/499430}}.
	
	\bibitem[{Zubovas} and {Nayakshin}(2014)]{zubovas14a}
	{Zubovas}, K.; {Nayakshin}, S.
	\newblock {Energy- and momentum-conserving AGN feedback outflows}.
	\newblock {\em Mon. Not. R. Astron. Soc.} {\bf 2014}, {\em 440},~2625--2635.
	\newblock {\url{https://doi.org/10.1093/mnras/stu431}}.
	
	\bibitem[{Wagner} et~al.(2013){Wagner}, {Umemura}, and {Bicknell}]{wagner13a}
	{Wagner}, A.Y.; {Umemura}, M.; {Bicknell}, G.V.
	\newblock {Ultrafast Outflows: Galaxy-scale Active Galactic Nucleus Feedback}.
	\newblock {\em Astrophys. J. Lett.} {\bf 2013}, {\em 763},~L18.
	\newblock {\url{https://doi.org/10.1088/2041-8205/763/1/L18}}.
	
	\bibitem[{Hopkins} et~al.(2016){Hopkins}, {Torrey}, {Faucher-Gigu{\`e}re},
	{Quataert}, and {Murray}]{hopkins16a}
	{Hopkins}, P.F.; {Torrey}, P.; {Faucher-Gigu{\`e}re}, C.A.; {Quataert}, E.;
	{Murray}, N.
	\newblock {Stellar and quasar feedback in concert: Effects on AGN accretion,
		obscuration, and outflows}.
	\newblock {\em Mon. Not. R. Astron. Soc.} {\bf 2016}, {\em 458},~816--831.
	\newblock {\url{https://doi.org/10.1093/mnras/stw289}}.
	
	\bibitem[{Bieri} et~al.(2017){Bieri}, {Dubois}, {Rosdahl}, {Wagner}, {Silk},
	and {Mamon}]{bieri17b}
	{Bieri}, R.; {Dubois}, Y.; {Rosdahl}, J.; {Wagner}, A.; {Silk}, J.; {Mamon},
	G.A.
	\newblock {Outflows driven by quasars in high-redshift galaxies with radiation
		hydrodynamics}.
	\newblock {\em Mon. Not. R. Astron. Soc.} {\bf 2017}, {\em 464},~1854--1873.
	\newblock {\url{https://doi.org/10.1093/mnras/stw2380}}.
	
	\bibitem[{Costa} et~al.(2018){Costa}, {Rosdahl}, {Sijacki}, and
	{Haehnelt}]{costa18a}
	{Costa}, T.; {Rosdahl}, J.; {Sijacki}, D.; {Haehnelt}, M.G.
	\newblock {Quenching star formation with quasar outflows launched by trapped IR
		radiation}.
	\newblock {\em Mon. Not. R. Astron. Soc.} {\bf 2018}, {\em 479},~2079--2111.
	\newblock {\url{https://doi.org/10.1093/mnras/sty1514}}.
	
	\bibitem[{Ward} et~al.(2024){Ward}, {Costa}, {Harrison}, and
	{Mainieri}]{ward24a}
	{Ward}, S.R.; {Costa}, T.; {Harrison}, C.M.; {Mainieri}, V.
	\newblock {AGN-driven outflows in clumpy media: Multiphase structure and
		scaling relations}.
	\newblock {\em Mon. Not. R. Astron. Soc.} {\bf 2024}, {\em 533},~1733--1755.
	\newblock {\url{https://doi.org/10.1093/mnras/stae1816}}.
	
	\bibitem[{Tombesi} et~al.(2010){Tombesi}, {Cappi}, {Reeves}, {Palumbo},
	{Yaqoob}, {Braito}, and {Dadina}]{tombesi10a}
	{Tombesi}, F.; {Cappi}, M.; {Reeves}, J.N.; {Palumbo}, G.G.C.; {Yaqoob}, T.;
	{Braito}, V.; {Dadina}, M.
	\newblock {Evidence for ultra-fast outflows in radio-quiet AGNs. I. Detection
		and statistical incidence of Fe K-shell absorption lines}.
	\newblock {\em Astron. Astrophys.} {\bf 2010}, {\em 521},~A57.
	\newblock {\url{https://doi.org/10.1051/0004-6361/200913440}}.
	
	\bibitem[{Tombesi} et~al.(2014){Tombesi}, {Tazaki}, {Mushotzky}, {Ueda},
	{Cappi}, {Gofford}, {Reeves}, and {Guainazzi}]{tombesi14a}
	{Tombesi}, F.; {Tazaki}, F.; {Mushotzky}, R.F.; {Ueda}, Y.; {Cappi}, M.;
	{Gofford}, J.; {Reeves}, J.N.; {Guainazzi}, M.
	\newblock {Ultrafast outflows in radio-loud active galactic nuclei}.
	\newblock {\em Mon. Not. R. Astron. Soc.} {\bf 2014}, {\em 443},~2154--2182.
	\newblock {\url{https://doi.org/10.1093/mnras/stu1297}}.
	
	\bibitem[{Rankine} et~al.(2020){Rankine}, {Hewett}, {Banerji}, and
	{Richards}]{rankine20a}
	{Rankine}, A.L.; {Hewett}, P.C.; {Banerji}, M.; {Richards}, G.T.
	\newblock {BAL and non-BAL quasars: Continuum, emission, and absorption
		properties establish a common parent sample}.
	\newblock {\em Mon. Not. R. Astron. Soc.} {\bf 2020}, {\em 492},~4553--4575.
	\newblock {\url{https://doi.org/10.1093/mnras/staa130}}.
	
	\bibitem[{Zakamska} and {Greene}(2014)]{zakamska14a}
	{Zakamska}, N.L.; {Greene}, J.E.
	\newblock {Quasar feedback and the origin of radio emission in radio-quiet
		quasars}.
	\newblock {\em Mon. Not. R. Astron. Soc.} {\bf 2014}, {\em 442},~784--804.
	\newblock {\url{https://doi.org/10.1093/mnras/stu842}}.
	
	\bibitem[{Nims} et~al.(2015){Nims}, {Quataert}, and
	{Faucher-Gigu{\`e}re}]{nims15a}
	{Nims}, J.; {Quataert}, E.; {Faucher-Gigu{\`e}re}, C.A.
	\newblock {Observational signatures of galactic winds powered by active
		galactic nuclei}.
	\newblock {\em Mon. Not. R. Astron. Soc.} {\bf 2015}, {\em 447},~3612--3622.
	\newblock {\url{https://doi.org/10.1093/mnras/stu2648}}.
	
	\bibitem[{Panessa} et~al.(2019){Panessa}, {Baldi}, {Laor}, {Padovani}, {Behar},
	and {McHardy}]{panessa19a}
	{Panessa}, F.; {Baldi}, R.D.; {Laor}, A.; {Padovani}, P.; {Behar}, E.;
	{McHardy}, I.
	\newblock {The origin of radio emission from radio-quiet active galactic
		nuclei}.
	\newblock {\em Nature Astronomy} {\bf 2019}, {\em 3},~387--396.
	\newblock {\url{https://doi.org/10.1038/s41550-019-0765-4}}.
	
	\bibitem[{Wylezalek} and {Zakamska}(2016)]{wylezalek16a}
	{Wylezalek}, D.; {Zakamska}, N.L.
	\newblock {Evidence of suppression of star formation by quasar-driven winds in
		gas-rich host galaxies at z < 1?}
	\newblock {\em Mon. Not. R. Astron. Soc.} {\bf 2016}, {\em 461},~3724--3739.
	\newblock {\url{https://doi.org/10.1093/mnras/stw1557}}.
	
	\bibitem[{Morabito} et~al.(2019){Morabito}, {Matthews}, {Best}, {G{\"u}rkan},
	{Jarvis}, {Prandoni}, {Duncan}, {Hardcastle}, {Kunert-Bajraszewska},
	{Mechev}, {Mooney}, {Sabater}, {R{\"o}ttgering}, {Shimwell}, {Smith},
	{Tasse}, and {Williams}]{morabito19a}
	{Morabito}, L.K.; {Matthews}, J.H.; {Best}, P.N.; {G{\"u}rkan}, G.; {Jarvis},
	M.J.; {Prandoni}, I.; {Duncan}, K.J.; {Hardcastle}, M.J.;
	{Kunert-Bajraszewska}, M.; {Mechev}, A.P.;  et~al.
	\newblock {The origin of radio emission in broad absorption line quasars:
		Results from the LOFAR Two-metre Sky Survey}.
	\newblock {\em Astron. Astrophys.} {\bf 2019}, {\em 622},~A15.
	\newblock {\url{https://doi.org/10.1051/0004-6361/201833821}}.
	
	\bibitem[{Alb{\'a}n} et~al.(2024){Alb{\'a}n}, {Wylezalek}, {Comerford},
	{Greene}, and {Riffel}]{alban24a}
	{Alb{\'a}n}, M.; {Wylezalek}, D.; {Comerford}, J.M.; {Greene}, J.E.; {Riffel},
	R.A.
	\newblock {Mapping AGN winds: A connection between radio-mode AGNs and the 
	AGN
		feedback cycle}.
	\newblock {\em Astron. Astrophys.} {\bf 2024}, {\em 691},~A124.
	\newblock {\url{https://doi.org/10.1051/0004-6361/202451738}}.
	
	\bibitem[{Fawcett} et~al.(2025){Fawcett}, {Harrison}, {Alexander}, {Morabito},
	{Kharb}, {Rosario}, {Baghel}, {Ghosh}, {Silpa}, {Petley}, {Sargent}, and
	{Calistro Rivera}]{fawcett25a}
	{Fawcett}, V.A.; {Harrison}, C.M.; {Alexander}, D.M.; {Morabito}, L.K.;
	{Kharb}, P.; {Rosario}, D.J.; {Baghel}, J.; {Ghosh}, S.; {Silpa}, S.;
	{Petley}, J.;  et~al.
	\newblock {Connection between steep radio spectral slopes and dust extinction
		in QSOs: Evidence for outflow-driven shocks in dusty QSOs}.
	\newblock {\em Mon. Not. R. Astron. Soc.} {\bf 2025}, {\em 537},~2003--2023.
	\newblock {\url{https://doi.org/10.1093/mnras/staf105}}.
	
	\bibitem[{Gaspari} et~al.(2011){Gaspari}, {Melioli}, {Brighenti}, and
	{D'Ercole}]{gaspari11a}
	{Gaspari}, M.; {Melioli}, C.; {Brighenti}, F.; {D'Ercole}, A.
	\newblock {The dance of heating and cooling in galaxy clusters:
		three-dimensional simulations of self-regulated active galactic nuclei
		outflows}.
	\newblock {\em Mon. Not. R. Astron. Soc.} {\bf 2011}, {\em 411},~349--372.
	\newblock {\url{https://doi.org/10.1111/j.1365-2966.2010.17688.x}}.
	
	\bibitem[{Gaspari} et~al.(2012){Gaspari}, {Ruszkowski}, and
	{Sharma}]{gaspari12a}
	{Gaspari}, M.; {Ruszkowski}, M.; {Sharma}, P.
	\newblock {Cause and Effect of Feedback: Multiphase Gas in Cluster Cores Heated
		by AGN Jets}.
	\newblock {\em Astrophys. J.} {\bf 2012}, {\em 746},~94.
	\newblock {\url{https://doi.org/10.1088/0004-637X/746/1/94}}.
	
	\bibitem[{Yang} and {Reynolds}(2016)]{yang16a}
	{Yang}, H.Y.K.; {Reynolds}, C.S.
	\newblock {How AGN Jets Heat the Intracluster Medium--Insights from
		Hydrodynamic Simulations}.
	\newblock {\em Astrophys. J.} {\bf 2016}, {\em 829},~90.
	\newblock {\url{https://doi.org/10.3847/0004-637X/829/2/90}}.
	
	\bibitem[{Proga} and {Kallman}(2004)]{proga04a}
	{Proga}, D.; {Kallman}, T.R.
	\newblock {Dynamics of Line-driven Disk Winds in Active Galactic Nuclei. II.
		Effects of Disk Radiation}.
	\newblock {\em Astrophys. J.} {\bf 2004}, {\em 616},~688--695.
	\newblock {\url{https://doi.org/10.1086/425117}}.
	
	\bibitem[{Proga} et~al.(2014){Proga}, {Jiang}, {Davis}, {Stone}, and
	{Smith}]{proga14a}
	{Proga}, D.; {Jiang}, Y.F.; {Davis}, S.W.; {Stone}, J.M.; {Smith}, D.
	\newblock {The Effects of Irradiation on Cloud Evolution in Active Galactic
		Nuclei}.
	\newblock {\em Astrophys. J.} {\bf 2014}, {\em 780},~51.
	\newblock {\url{https://doi.org/10.1088/0004-637X/780/1/51}}.
	
	\bibitem[{Dyda} et~al.(2025){Dyda}, {Dannen}, {Kallman}, {Davis}, and
	{Proga}]{dyda25a}
	{Dyda}, S.; {Dannen}, R.C.; {Kallman}, T.R.; {Davis}, S.W.; {Proga}, D.
	\newblock {Time-Dependent AGN Disc Winds II -- Effects of Photoionization}.
	\newblock {\em arXiv} {\bf 2025}, arXiv:2504.00117.
	\newblock {\url{https://doi.org/10.48550/arXiv.2504.00117}}.
	
	\bibitem[{Dasyra} et~al.(2022){Dasyra}, {Paraschos}, {Bisbas}, {Combes}, and
	{Fern{\'a}ndez-Ontiveros}]{dasyra22a}
	{Dasyra}, K.M.; {Paraschos}, G.F.; {Bisbas}, T.G.; {Combes}, F.;
	{Fern{\'a}ndez-Ontiveros}, J.A.
	\newblock {Insights into the collapse and expansion of molecular clouds in
		outflows from observable pressure gradients}.
	\newblock {\em Nat. Astron.} {\bf 2022}, {\em 6},~1077--1084.
	\newblock {\url{https://doi.org/10.1038/s41550-022-01725-9}}.
	
	\bibitem[{Snodin} et~al.(2006){Snodin}, {Brandenburg}, {Mee}, and
	{Shukurov}]{snodin06a}
	{Snodin}, A.P.; {Brandenburg}, A.; {Mee}, A.J.; {Shukurov}, A.
	\newblock {Simulating field-aligned diffusion of a cosmic ray gas}.
	\newblock {\em Mon. Not. R. Astron. Soc.} {\bf 2006}, {\em 373},~643--652.
	\newblock {\url{https://doi.org/10.1111/j.1365-2966.2006.11034.x}}.
	
	\bibitem[{Dubois} and {Commer{\c{c}}on}(2016)]{dubois16a}
	{Dubois}, Y.; {Commer{\c{c}}on}, B.
	\newblock {An implicit scheme for solving the anisotropic diffusion of heat and
		cosmic rays in the RAMSES code}.
	\newblock {\em Astron. Astrophys.} {\bf 2016}, {\em 585},~A138.
	\newblock {\url{https://doi.org/10.1051/0004-6361/201527126}}.
	
	\bibitem[{Thomas} and {Pfrommer}(2019)]{thomas19a}
	{Thomas}, T.; {Pfrommer}, C.
	\newblock {Cosmic-ray hydrodynamics: Alfv{\'e}n-wave regulated transport of
		cosmic rays}.
	\newblock {\em Mon. Not. R. Astron. Soc.} {\bf 2019}, {\em 485},~2977--3008.
	\newblock {\url{https://doi.org/10.1093/mnras/stz263}}.
	
	\bibitem[{Thomas} et~al.(2021){Thomas}, {Pfrommer}, and {Pakmor}]{thomas21a}
	{Thomas}, T.; {Pfrommer}, C.; {Pakmor}, R.
	\newblock {A finite volume method for two-moment cosmic ray hydrodynamics on a
		moving mesh}.
	\newblock {\em Mon. Not. R. Astron. Soc.} {\bf 2021}, {\em 503},~2242--2264.
	\newblock {\url{https://doi.org/10.1093/mnras/stab397}}.
	
	\bibitem[{Chan} et~al.(2019){Chan}, {Kere{\v{s}}}, {Hopkins}, {Quataert}, {Su},
	{Hayward}, and {Faucher-Gigu{\`e}re}]{chan19a}
	{Chan}, T.K.; {Kere{\v{s}}}, D.; {Hopkins}, P.F.; {Quataert}, E.; {Su}, K.Y.;
	{Hayward}, C.C.; {Faucher-Gigu{\`e}re}, C.A.
	\newblock {Cosmic ray feedback in the FIRE simulations: Constraining cosmic ray
		propagation with GeV {\ensuremath{\gamma}}-ray emission}.
	\newblock {\em Mon. Not. R. Astron. Soc.} {\bf 2019}, {\em 488},~3716--3744.
	\newblock {\url{https://doi.org/10.1093/mnras/stz1895}}.
	
	\bibitem[{Guo} and {Mathews}(2011)]{fulai11a}
	{Guo}, F.; {Mathews}, W.G.
	\newblock {Cosmic-ray-dominated AGN Jets and the Formation of X-ray Cavities in
		Galaxy Clusters}.
	\newblock {\em Astrophys. J.} {\bf 2011}, {\em 728},~121.
	\newblock {\url{https://doi.org/10.1088/0004-637X/728/2/121}}.
	
	\bibitem[{Yang} et~al.(2012){Yang}, {Ruszkowski}, {Ricker}, {Zweibel}, and
	{Lee}]{yang12a}
	{Yang}, H.Y.K.; {Ruszkowski}, M.; {Ricker}, P.M.; {Zweibel}, E.; {Lee}, D.
	\newblock {The Fermi Bubbles: Supersonic Active Galactic Nucleus Jets with
		Anisotropic Cosmic-Ray Diffusion}.
	\newblock {\em Astrophys. J.} {\bf 2012}, {\em 761},~185.
	\newblock {\url{https://doi.org/10.1088/0004-637X/761/2/185}}.
	
	\bibitem[{Ruszkowski} et~al.(2017){Ruszkowski}, {Yang}, and
	{Reynolds}]{ruszkowski17a}
	{Ruszkowski}, M.; {Yang}, H.Y.K.; {Reynolds}, C.S.
	\newblock {Cosmic-Ray Feedback Heating of the Intracluster Medium}.
	\newblock {\em Astrophys. J.} {\bf 2017}, {\em 844},~13.
	\newblock {\url{https://doi.org/10.3847/1538-4357/aa79f8}}.
	
	\bibitem[{Ehlert} et~al.(2018){Ehlert}, {Weinberger}, {Pfrommer}, {Pakmor}, and
	{Springel}]{ehlert18a}
	{Ehlert}, K.; {Weinberger}, R.; {Pfrommer}, C.; {Pakmor}, R.; {Springel}, V.
	\newblock {Simulations of the dynamics of magnetized jets and cosmic rays in
		galaxy clusters}.
	\newblock {\em Mon. Not. R. Astron. Soc.} {\bf 2018}, {\em 481},~2878--2900.
	\newblock {\url{https://doi.org/10.1093/mnras/sty2397}}.
	
	\bibitem[{Farcy} et~al.(2022){Farcy}, {Rosdahl}, {Dubois}, {Blaizot}, and
	{Martin-Alvarez}]{farcy22a}
	{Farcy}, M.; {Rosdahl}, J.; {Dubois}, Y.; {Blaizot}, J.; {Martin-Alvarez}, S.
	\newblock {Radiation-magnetohydrodynamics simulations of cosmic ray feedback in
		disc galaxies}.
	\newblock {\em Mon. Not. R. Astron. Soc.} {\bf 2022}, {\em 513},~5000--5019.
	\newblock {\url{https://doi.org/10.1093/mnras/stac1196}}.
	
	\bibitem[{Su} et~al.(2024){Su}, {Bryan}, {Hayward}, {Somerville}, {Hopkins},
	{Emami}, {Faucher-Gigu{\`e}re}, {Quataert}, {Ponnada}, {Fielding}, and
	{Kere{\v{s}}}]{su24a}
	{Su}, K.Y.; {Bryan}, G.L.; {Hayward}, C.C.; {Somerville}, R.S.; {Hopkins},
	P.F.; {Emami}, R.; {Faucher-Gigu{\`e}re}, C.A.; {Quataert}, E.; {Ponnada},
	S.B.; {Fielding}, D.;  et~al.
	\newblock {Unravelling jet quenching criteria across L* galaxies and massive
		cluster ellipticals}.
	\newblock {\em Mon. Not. R. Astron. Soc.} {\bf 2024}, {\em 532},~2724--2740.
	\newblock {\url{https://doi.org/10.1093/mnras/stae1629}}.
	
	\bibitem[{Scheck} et~al.(2002){Scheck}, {Aloy}, {Mart{\'\i}}, {G{\'o}mez}, and
	{M{\"u}ller}]{scheck02a}
	{Scheck}, L.; {Aloy}, M.A.; {Mart{\'\i}}, J.M.; {G{\'o}mez}, J.L.;
	{M{\"u}ller}, E.
	\newblock {Does the plasma composition affect the long-term evolution of
		relativistic jets?}
	\newblock {\em Mon. Not. R. Astron. Soc.} {\bf 2002}, {\em 331},~615--634.
	\newblock {\url{https://doi.org/10.1046/j.1365-8711.2002.05210.x}}.
	
	\bibitem[{Chattopadhyay} and {Ryu}(2009)]{chattopadhyay2000a}
	{Chattopadhyay}, I.; {Ryu}, D.
	\newblock {Effects of Fluid Composition on Spherical Flows Around Black Holes}.
	\newblock {\em Astrophys. J.} {\bf 2009}, {\em 694},~492--501.
	\newblock {\url{https://doi.org/10.1088/0004-637X/694/1/492}}.
	
	\bibitem[{Chael}(2025)]{chael25a}
	{Chael}, A.
	\newblock {Survey of radiative, two-temperature magnetically arrested
		simulations of the black hole M87* I: Turbulent electron heating}.
	\newblock {\em Mon. Not. R. Astron. Soc.} {\bf 2025}, {\em 537},~2496--2515.
	\newblock {\url{https://doi.org/10.1093/mnras/staf200}}.
	
	\bibitem[{Salas} et~al.(2025){Salas}, {Liska}, {Markoff}, {Chatterjee},
	{Musoke}, {Porth}, {Ripperda}, {Yoon}, and {Mulaudzi}]{salas25a}
	{Salas}, L.D.S.; {Liska}, M.T.P.; {Markoff}, S.B.; {Chatterjee}, K.; {Musoke},
	G.; {Porth}, O.; {Ripperda}, B.; {Yoon}, D.; {Mulaudzi}, W.
	\newblock {Two-temperature treatments in magnetically arrested disc GRMHD
		simulations more accurately predict light curves of Sagittarius A*}.
	\newblock {\em Mon. Not. R. Astron. Soc.} {\bf 2025}, {\em 538},~698--710.
	\newblock {\url{https://doi.org/10.1093/mnras/staf240}}.
	
	\bibitem[{Nishikawa} et~al.(2003){Nishikawa}, {Hardee}, {Richardson}, {Preece},
	{Sol}, and {Fishman}]{nishikawa03a}
	{Nishikawa}, K.I.; {Hardee}, P.; {Richardson}, G.; {Preece}, R.; {Sol}, H.;
	{Fishman}, G.J.
	\newblock {Particle Acceleration in Relativistic Jets Due to Weibel
		Instability}.
	\newblock {\em Astrophys. J.} {\bf 2003}, {\em 595},~555--563.
	\newblock {\url{https://doi.org/10.1086/377260}}.
	
	\bibitem[{Liang} et~al.(2013){Liang}, {Boettcher}, and {Smith}]{liang13a}
	{Liang}, E.; {Boettcher}, M.; {Smith}, I.
	\newblock {Magnetic Field Generation and Particle Energization at Relativistic
		Shear Boundaries in Collisionless Electron-Positron Plasmas}.
	\newblock {\em Astrophys. J.} {\bf 2013}, {\em 766},~L19.
	\newblock {\url{https://doi.org/10.1088/2041-8205/766/2/L19}}.
	
	\bibitem[{Nishikawa} et~al.(2016){Nishikawa}, {Frederiksen}, {Nordlund},
	{Mizuno}, {Hardee}, {Niemiec}, {G{\'o}mez}, {Pe'er}, {Du{\c{t}}an}, {Meli},
	{Sol}, {Pohl}, and {Hartmann}]{nishikawa16a}
	{Nishikawa}, K.I.; {Frederiksen}, J.T.; {Nordlund}, {\r{A}}.; {Mizuno}, Y.;
	{Hardee}, P.E.; {Niemiec}, J.; {G{\'o}mez}, J.L.; {Pe'er}, A.; {Du{\c{t}}an},
	I.; {Meli}, A.;  et~al.
	\newblock {Evolution of Global Relativistic Jets: Collimations and Expansion
		with kKHI and the Weibel Instability}.
	\newblock {\em Astrophys. J.} {\bf 2016}, {\em 820},~94.
	\newblock {\url{https://doi.org/10.3847/0004-637X/820/2/94}}.
	
	\bibitem[{Rieger} and {Duffy}(2019)]{rieger19a}
	{Rieger}, F.M.; {Duffy}, P.
	\newblock {Particle Acceleration in Shearing Flows: Efficiencies and Limits}.
	\newblock {\em Astrophys. J.} {\bf 2019}, {\em 886},~L26.
	\newblock {\url{https://doi.org/10.3847/2041-8213/ab563f}}.
	
	\bibitem[{Chand} and {B{\"o}ttcher}(2024)]{chand24a}
	{Chand}, T.; {B{\"o}ttcher}, M.
	\newblock {Inverse Compton Emission and Cooling of Relativistic Particles
		Accelerated at Shear Boundary Layers in Relativistic Jets}.
	\newblock {\em Astrophys. J.} {\bf 2024}, {\em 962},~31.
	\newblock {\url{https://doi.org/10.3847/1538-4357/ad0a63}}.
	
	\bibitem[{Du{\c{t}}an} et~al.(2025){Du{\c{t}}an}, {Nishikawa}, {Meli},
	{Kobzar}, {K{\"o}hn}, {Mizuno}, {MacDonald}, {G{\'o}mez}, and
	{Hirotani}]{dutan25a}
	{Du{\c{t}}an}, I.; {Nishikawa}, K.; {Meli}, A.; {Kobzar}, O.; {K{\"o}hn}, C.;
	{Mizuno}, Y.; {MacDonald}, N.; {G{\'o}mez}, J.L.; {Hirotani}, K.
	\newblock {Synthetic spectra from particle-in-cell simulations of relativistic
		jets containing an initial toroidal magnetic field}.
	\newblock {\em Mon. Not. R. Astron. Soc.} {\bf 2025}, {\em 540},~1043--1054.
	\newblock {\url{https://doi.org/10.1093/mnras/staf626}}.
	
	\bibitem[{Meli} and {Nishikawa}(2021)]{meli21a}
	{Meli}, A.; {Nishikawa}, K.i.
	\newblock {Particle-in-Cell Simulations of Astrophysical Relativistic Jets}.
	\newblock {\em Universe} {\bf 2021}, {\em 7},~450.
	\newblock {\url{https://doi.org/10.3390/universe7110450}}.
	
	\bibitem[{Sironi} et~al.(2015){Sironi}, {Keshet}, and {Lemoine}]{sironi15a}
	{Sironi}, L.; {Keshet}, U.; {Lemoine}, M.
	\newblock {Relativistic Shocks: Particle Acceleration and Magnetization}.
	\newblock {\em Space Sci. Rev.} {\bf 2015}, {\em 191},~519--544.
	\newblock {\url{https://doi.org/10.1007/s11214-015-0181-8}}.
	
	\bibitem[{Sironi} et~al.(2025){Sironi}, {Uzdensky}, and {Giannios}]{sironi25a}
	{Sironi}, L.; {Uzdensky}, D.A.; {Giannios}, D.
	\newblock {Relativistic Magnetic Reconnection in Astrophysical Plasmas: A
		Powerful Mechanism of Nonthermal Emission}.
	\newblock {\em arXiv} {\bf 2025}, arXiv:2506.02101.
	\newblock {\url{https://doi.org/10.48550/arXiv.2506.02101}}.
	
	\bibitem[{Bodo} et~al.(1994){Bodo}, {Massaglia}, {Ferrari}, and
	{Trussoni}]{bodo94a}
	{Bodo}, G.; {Massaglia}, S.; {Ferrari}, A.; {Trussoni}, E.
	\newblock {Kelvin-Helmholtz instability of hydrodynamic supersonic jets.}
	\newblock {\em Astron. Astrophys.} {\bf 1994}, {\em 283},~655--676.
	
	\bibitem[{Hardee}(2000)]{hardee2000a}
	{Hardee}, P.E.
	\newblock {On Three-dimensional Structures in Relativistic Hydrodynamic Jets}.
	\newblock {\em Astrophys. J.} {\bf 2000}, {\em 533},~176--193.
	\newblock {\url{https://doi.org/10.1086/308656}}.
	
	\bibitem[{Perucho} et~al.(2004){Perucho}, {Hanasz}, {Mart{\'\i}}, and
	{Sol}]{perucho04a}
	{Perucho}, M.; {Hanasz}, M.; {Mart{\'\i}}, J.M.; {Sol}, H.
	\newblock {Stability of hydrodynamical relativistic planar jets. I. Linear
		evolution and saturation of Kelvin-Helmholtz modes}.
	\newblock {\em Astron. Astrophys.} {\bf 2004}, {\em 427},~415--429.
	\newblock {\url{https://doi.org/10.1051/0004-6361:20040349}}.
	
	\bibitem[{Perucho} et~al.(2005){Perucho}, {Mart{\'\i}}, and
	{Hanasz}]{perucho05a}
	{Perucho}, M.; {Mart{\'\i}}, J.M.; {Hanasz}, M.
	\newblock {Nonlinear stability of relativistic sheared planar jets}.
	\newblock {\em Astron. Astrophys.} {\bf 2005}, {\em 443},~863--881.
	\newblock {\url{https://doi.org/10.1051/0004-6361:20053115}}.
	
	\bibitem[{Giannios} and {Spruit}(2006)]{giannios06a}
	{Giannios}, D.; {Spruit}, H.C.
	\newblock {The role of kink instability in Poynting-flux dominated jets}.
	\newblock {\em Astron. Astrophys.} {\bf 2006}, {\em 450},~887--898.
	\newblock {\url{https://doi.org/10.1051/0004-6361:20054107}}.
	
	\bibitem[{Moll} et~al.(2008){Moll}, {Spruit}, and {Obergaulinger}]{moll08a}
	{Moll}, R.; {Spruit}, H.C.; {Obergaulinger}, M.
	\newblock {Kink instabilities in jets from rotating magnetic fields}.
	\newblock {\em Astron. Astrophys.} {\bf 2008}, {\em 492},~621--630.
	\newblock {\url{https://doi.org/10.1051/0004-6361:200810523}}.
	
	\bibitem[{Mizuno} et~al.(2014){Mizuno}, {Hardee}, and {Nishikawa}]{mizuno14a}
	{Mizuno}, Y.; {Hardee}, P.E.; {Nishikawa}, K.I.
	\newblock {Spatial Growth of the Current-driven Instability in Relativistic
		Jets}.
	\newblock {\em Astrophys. J.} {\bf 2014}, {\em 784},~167.
	\newblock {\url{https://doi.org/10.1088/0004-637X/784/2/167}}.
	
	\bibitem[{Mizuno} et~al.(2009){Mizuno}, {Lyubarsky}, {Nishikawa}, and
	{Hardee}]{mizuno09a}
	{Mizuno}, Y.; {Lyubarsky}, Y.; {Nishikawa}, K.I.; {Hardee}, P.E.
	\newblock {Three-Dimensional Relativistic Magnetohydrodynamic Simulations of
		Current-Driven Instability. I. Instability of a Static Column}.
	\newblock {\em Astrophys. J.} {\bf 2009}, {\em 700},~684--693.
	\newblock {\url{https://doi.org/10.1088/0004-637X/700/1/684}}.
	
	\bibitem[{Bromberg} et~al.(2019){Bromberg}, {Singh}, {Davelaar}, and
	{Philippov}]{bromberg19a}
	{Bromberg}, O.; {Singh}, C.B.; {Davelaar}, J.; {Philippov}, A.A.
	\newblock {Kink Instability: Evolution and Energy Dissipation in Relativistic
		Force-free Nonrotating Jets}.
	\newblock {\em Astrophys. J.} {\bf 2019}, {\em 884},~39.
	\newblock {\url{https://doi.org/10.3847/1538-4357/ab3fa5}}.
	
	\bibitem[{Mizuno} et~al.(2007){Mizuno}, {Hardee}, and {Nishikawa}]{mizuno07a}
	{Mizuno}, Y.; {Hardee}, P.; {Nishikawa}, K.I.
	\newblock {Three-dimensional Relativistic Magnetohydrodynamic Simulations of
		Magnetized Spine-Sheath Relativistic Jets}.
	\newblock {\em Astrophys. J.} {\bf 2007}, {\em 662},~835--850.
	\newblock {\url{https://doi.org/10.1086/518106}}.
	
	\bibitem[{McKinney} and {Blandford}(2009)]{mckinney09a}
	{McKinney}, J.C.; {Blandford}, R.D.
	\newblock {Stability of relativistic jets from rotating, accreting black holes
		via fully three-dimensional magnetohydrodynamic simulations}.
	\newblock {\em Mon. Not. R. Astron. Soc.} {\bf 2009}, {\em 394},~L126--L130.
	\newblock {\url{https://doi.org/10.1111/j.1745-3933.2009.00625.x}}.
	
	\bibitem[{Musso} et~al.(2024){Musso}, {Bodo}, {Mamatsashvili}, {Rossi}, and
	{Mignone}]{musso24a}
	{Musso}, M.; {Bodo}, G.; {Mamatsashvili}, G.; {Rossi}, P.; {Mignone}, A.
	\newblock {Evolution of current- and pressure-driven instabilities in
		relativistic jets}.
	\newblock {\em Mon. Not. R. Astron. Soc.} {\bf 2024}, {\em 532},~4810--4825.
	\newblock {\url{https://doi.org/10.1093/mnras/stae1788}}.
	
	\bibitem[{Perucho} et~al.(2010){Perucho}, {Mart{\'\i}}, {Cela}, {Hanasz}, {de
		La Cruz}, and {Rubio}]{perucho10a}
	{Perucho}, M.; {Mart{\'\i}}, J.M.; {Cela}, J.M.; {Hanasz}, M.; {de La Cruz},
	R.; {Rubio}, F.
	\newblock {Stability of three-dimensional relativistic jets: Implications for
		jet collimation}.
	\newblock {\em Astron. Astrophys.} {\bf 2010}, {\em 519},~A41.
	\newblock {\url{https://doi.org/10.1051/0004-6361/200913012}}.
	
	\bibitem[{Chow} et~al.(2023){Chow}, {Davelaar}, {Rowan}, and {Sironi}]{chow23a}
	{Chow}, A.; {Davelaar}, J.; {Rowan}, M.E.; {Sironi}, L.
	\newblock {The Kelvin-Helmholtz Instability at the Boundary of Relativistic
		Magnetized Jets}.
	\newblock {\em Astrophys. J.} {\bf 2023}, {\em 951},~L23.
	\newblock {\url{https://doi.org/10.3847/2041-8213/acdfcf}}.
	
	\bibitem[{Bodo} et~al.(2016){Bodo}, {Mamatsashvili}, {Rossi}, and
	{Mignone}]{bodo16a}
	{Bodo}, G.; {Mamatsashvili}, G.; {Rossi}, P.; {Mignone}, A.
	\newblock {Linear stability analysis of magnetized jets: The rotating case}.
	\newblock {\em Mon. Not. R. Astron. Soc.} {\bf 2016}, {\em 462},~3031--3052.
	\newblock {\url{https://doi.org/10.1093/mnras/stw1650}}.
	
	\bibitem[{Bodo} et~al.(2019){Bodo}, {Mamatsashvili}, {Rossi}, and
	{Mignone}]{bodo19a}
	{Bodo}, G.; {Mamatsashvili}, G.; {Rossi}, P.; {Mignone}, A.
	\newblock {Linear stability analysis of magnetized relativistic rotating jets}.
	\newblock {\em Mon. Not. R. Astron. Soc.} {\bf 2019}, {\em 485},~2909--2921.
	\newblock {\url{https://doi.org/10.1093/mnras/stz591}}.
	
	\bibitem[{Bromberg} and {Tchekhovskoy}(2016)]{bromberg16a}
	{Bromberg}, O.; {Tchekhovskoy}, A.
	\newblock {Relativistic MHD simulations of core-collapse GRB jets: 3D
		instabilities and magnetic dissipation}.
	\newblock {\em Mon. Not. R. Astron. Soc.} {\bf 2016}, {\em 456},~1739--1760.
	\newblock {\url{https://doi.org/10.1093/mnras/stv2591}}.
	
	\bibitem[{Tchekhovskoy} and {Bromberg}(2016)]{tchekhovskoy16a}
	{Tchekhovskoy}, A.; {Bromberg}, O.
	\newblock {Three-dimensional relativistic MHD simulations of active galactic
		nuclei jets: Magnetic kink instability and Fanaroff-Riley dichotomy}.
	\newblock {\em Mon. Not. R. Astron. Soc.} {\bf 2016}, {\em 461},~L46--L50.
	\newblock {\url{https://doi.org/10.1093/mnrasl/slw064}}.
	
	\bibitem[{Lobanov} and {Zensus}(2001)]{lobanov01a}
	{Lobanov}, A.P.; {Zensus}, J.A.
	\newblock {A Cosmic Double Helix in the Archetypical Quasar 3C273}.
	\newblock {\em Science} {\bf 2001}, {\em 294},~128--131.
	\newblock {\url{https://doi.org/10.1126/science.1063239}}.
	
	\bibitem[{Perucho} et~al.(2006){Perucho}, {Lobanov}, {Mart{\'\i}}, and
	{Hardee}]{perucho06a}
	{Perucho}, M.; {Lobanov}, A.P.; {Mart{\'\i}}, J.M.; {Hardee}, P.E.
	\newblock {The role of Kelvin-Helmholtz instability in the internal structure
		of relativistic outflows. The case of the jet in 3C 273}.
	\newblock {\em Astron. Astrophys.} {\bf 2006}, {\em 456},~493--504.
	\newblock {\url{https://doi.org/10.1051/0004-6361:20065310}}.
	
	\bibitem[{Nikonov} et~al.(2023){Nikonov}, {Kovalev}, {Kravchenko},
	{Pashchenko}, and {Lobanov}]{nikonov23a}
	{Nikonov}, A.S.; {Kovalev}, Y.Y.; {Kravchenko}, E.V.; {Pashchenko}, I.N.;
	{Lobanov}, A.P.
	\newblock {Properties of the jet in M87 revealed by its helical structure
		imaged with the VLBA at 8 and 15 GHz}.
	\newblock {\em Mon. Not. R. Astron. Soc.} {\bf 2023}, {\em 526},~5949--5963.
	\newblock {\url{https://doi.org/10.1093/mnras/stad3061}}.
	
	\bibitem[{Lobanov} et~al.(2003){Lobanov}, {Hardee}, and {Eilek}]{lobanov03a}
	{Lobanov}, A.; {Hardee}, P.; {Eilek}, J.
	\newblock {Internal structure and dynamics of the kiloparsec-scale jet in M87}.
	\newblock {\em New Astron. Rev.} {\bf 2003}, {\em 47},~629--632.
	\newblock {\url{https://doi.org/10.1016/S1387-6473(03)00109-X}}.
	
	\bibitem[{Hardee} and {Eilek}(2011)]{hardee11a}
	{Hardee}, P.E.; {Eilek}, J.A.
	\newblock {Using Twisted Filaments to Model the Inner Jet in M 87}.
	\newblock {\em Astrophys. J.} {\bf 2011}, {\em 735},~61.
	\newblock {\url{https://doi.org/10.1088/0004-637X/735/1/61}}.
	
	\bibitem[{Mertens} et~al.(2016){Mertens}, {Lobanov}, {Walker}, and
	{Hardee}]{martens16a}
	{Mertens}, F.; {Lobanov}, A.P.; {Walker}, R.C.; {Hardee}, P.E.
	\newblock {Kinematics of the jet in M 87 on scales of 100-1000 Schwarzschild
		radii}.
	\newblock {\em Astron. Astrophys.} {\bf 2016}, {\em 595},~A54.
	\newblock {\url{https://doi.org/10.1051/0004-6361/201628829}}.
	
	\bibitem[{Perucho} et~al.(2012){Perucho}, {Kovalev}, {Lobanov}, {Hardee}, and
	{Agudo}]{perucho12a}
	{Perucho}, M.; {Kovalev}, Y.Y.; {Lobanov}, A.P.; {Hardee}, P.E.; {Agudo}, I.
	\newblock {Anatomy of Helical Extragalactic Jets: The Case of S5 0836+710}.
	\newblock {\em Astrophys. J.} {\bf 2012}, {\em 749},~55.
	\newblock {\url{https://doi.org/10.1088/0004-637X/749/1/55}}.
	
	\bibitem[{Vega-Garc{\'\i}a} et~al.(2019){Vega-Garc{\'\i}a}, {Perucho}, and
	{Lobanov}]{VegaGarcia19a}
	{Vega-Garc{\'\i}a}, L.; {Perucho}, M.; {Lobanov}, A.P.
	\newblock {Derivation of the physical parameters of the jet in S5 0836+710 from
		stability analysis}.
	\newblock {\em Astron. Astrophys.} {\bf 2019}, {\em 627},~A79.
	\newblock {\url{https://doi.org/10.1051/0004-6361/201935119}}.
	
	\bibitem[{Worrall} et~al.(2007){Worrall}, {Birkinshaw}, {Laing}, {Cotton}, and
	{Bridle}]{worrall07a}
	{Worrall}, D.M.; {Birkinshaw}, M.; {Laing}, R.A.; {Cotton}, W.D.; {Bridle},
	A.H.
	\newblock {The inner jet of radio galaxy NGC 315 as observed with Chandra and
		the Very Large Array}.
	\newblock {\em Mon. Not. R. Astron. Soc.} {\bf 2007}, {\em 380},~2--14.
	\newblock {\url{https://doi.org/10.1111/j.1365-2966.2007.11998.x}}.
	
	\bibitem[{Park} et~al.(2024){Park}, {Zhao}, {Nakamura}, {Mizuno}, {Pu},
	{Asada}, {Takahashi}, {Toma}, {Kino}, {Cho}, {Hada}, {Edwards}, {Ro}, {Kam},
	{Yi}, {Lee}, {Koyama}, {Byun}, {Phillips}, {Reynolds}, {Hodgson}, and
	{Lee}]{park24a}
	{Park}, J.; {Zhao}, G.Y.; {Nakamura}, M.; {Mizuno}, Y.; {Pu}, H.Y.; {Asada},
	K.; {Takahashi}, K.; {Toma}, K.; {Kino}, M.; {Cho}, I.;  et~al.
	\newblock {Discovery of Limb Brightening in the Parsec-scale Jet of NGC 315
		through Global Very Long Baseline Interferometry Observations and Its
		Implications for Jet Models}.
	\newblock {\em Astrophys. J.} {\bf 2024}, {\em 973},~L45.
	\newblock {\url{https://doi.org/10.3847/2041-8213/ad7137}}.
	
	\bibitem[{Fuentes} et~al.(2023){Fuentes}, {G{\'o}mez}, {Mart{\'\i}}, {Perucho},
	{Zhao}, {Lico}, {Lobanov}, {Bruni}, {Kovalev}, {Chael}, {Akiyama}, {Bouman},
	{Sun}, {Cho}, {Traianou}, {Toscano}, {Dahale}, {Foschi}, {Gurvits},
	{Jorstad}, {Kim}, {Marscher}, {Mizuno}, {Ros}, and {Savolainen}]{fuentes23a}
	{Fuentes}, A.; {G{\'o}mez}, J.L.; {Mart{\'\i}}, J.M.; {Perucho}, M.; {Zhao},
	G.Y.; {Lico}, R.; {Lobanov}, A.P.; {Bruni}, G.; {Kovalev}, Y.Y.; {Chael}, A.;
	et~al.
	\newblock {Filamentary structures as the origin of blazar jet radio
		variability}.
	\newblock {\em Nat. Astron.} {\bf 2023}, {\em 7},~1359--1367.
	\newblock {\url{https://doi.org/10.1038/s41550-023-02105-7}}.
	
	\bibitem[{Paraschos} and {Mpisketzis}(2025)]{paraschos25a}
	{Paraschos}, G.F.; {Mpisketzis}, V.
	\newblock {Unravelling the dynamics of cosmic vortices: Probing a
		Kelvin-Helmholtz instability in the jet of 3C 84}.
	\newblock {\em Astron. Astrophys.} {\bf 2025}, {\em 696},~L7.
	\newblock {\url{https://doi.org/10.1051/0004-6361/202554201}}.
	
	\bibitem[{Perucho}(2020)]{perucho20a}
	{Perucho}, M.
	\newblock {Triggering mixing and deceleration in FRI jets: A solution}.
	\newblock {\em Mon. Not. R. Astron. Soc.} {\bf 2020}, {\em 494},~L22--L26.
	\newblock {\url{https://doi.org/10.1093/mnrasl/slaa031}}.
	
	\bibitem[Bicknell(1994)]{bicknell94a}
	Bicknell, G.V.
	\newblock On the Relationship between BL Lacertae Objects and Fanaroff-Riley I
	Radio Galaxies.
	\newblock {\em Astrophys. J.} {\bf 1994}, {\em 422},~542.
	
	\bibitem[{Carvalho}(1998)]{carvalho98a}
	{Carvalho}, J.C.
	\newblock {The evolution of GHz-peaked-spectrum radio sources}.
	\newblock {\em Astron. Astrophys.} {\bf 1998}, {\em 329},~845--852.
	
	\bibitem[{Lansbury} et~al.(2018){Lansbury}, {Jarvis}, {Harrison}, {Alexander},
	{Del Moro}, {Edge}, {Mullaney}, and {Thomson}]{lansbury18a}
	{Lansbury}, G.B.; {Jarvis}, M.E.; {Harrison}, C.M.; {Alexander}, D.M.; {Del
		Moro}, A.; {Edge}, A.C.; {Mullaney}, J.R.; {Thomson}, A.P.
	\newblock {Storm in a Teacup: X-Ray View of an Obscured Quasar and
		Superbubble}.
	\newblock {\em Astrophys. J.} {\bf 2018}, {\em 856},~L1.
	\newblock {\url{https://doi.org/10.3847/2041-8213/aab357}}.
	
	\bibitem[{Ramos Almeida} et~al.(2022){Ramos Almeida}, {Bischetti},
	{Garc{\'\i}a-Burillo}, {Alonso-Herrero}, {Audibert}, {Cicone}, {Feruglio},
	{Tadhunter}, {Pierce}, {Pereira-Santaella}, and {Bessiere}]{almeida22a}
	{Ramos Almeida}, C.; {Bischetti}, M.; {Garc{\'\i}a-Burillo}, S.;
	{Alonso-Herrero}, A.; {Audibert}, A.; {Cicone}, C.; {Feruglio}, C.;
	{Tadhunter}, C.N.; {Pierce}, J.C.S.; {Pereira-Santaella}, M.;  et~al.
	\newblock {The diverse cold molecular gas contents, morphologies, and
		kinematics of type-2 quasars as seen by ALMA}.
	\newblock {\em Astron. Astrophys.} {\bf 2022}, {\em 658},~A155.
	\newblock {\url{https://doi.org/10.1051/0004-6361/202141906}}.
	
	\bibitem[{Venturi} et~al.(2023){Venturi}, {Treister}, {Finlez}, {D'Ago},
	{Bauer}, {Harrison}, {Ramos Almeida}, {Revalski}, {Ricci}, {Sartori},
	{Girdhar}, {Keel}, and {Tub{\'\i}n}]{venturi23a}
	{Venturi}, G.; {Treister}, E.; {Finlez}, C.; {D'Ago}, G.; {Bauer}, F.;
	{Harrison}, C.M.; {Ramos Almeida}, C.; {Revalski}, M.; {Ricci}, F.;
	{Sartori}, L.F.;  et~al.
	\newblock {Complex AGN feedback in the Teacup galaxy. A powerful ionised
		galactic outflow, jet-ISM interaction, and evidence for AGN-triggered star
		formation in a giant bubble}.
	\newblock {\em Astron. Astrophys.} {\bf 2023}, {\em 678},~A127.
	\newblock {\url{https://doi.org/10.1051/0004-6361/202347375}}.
	
	\bibitem[{Bessiere} et~al.(2024){Bessiere}, {Ramos Almeida}, {Holden},
	{Tadhunter}, and {Canalizo}]{bessiere24a}
	{Bessiere}, P.S.; {Ramos Almeida}, C.; {Holden}, L.R.; {Tadhunter}, C.N.;
	{Canalizo}, G.
	\newblock {QSOFEED: Relationship between star formation and active galactic
		nuclei feedback}.
	\newblock {\em Astron. Astrophys.} {\bf 2024}, {\em 689},~A271.
	\newblock {\url{https://doi.org/10.1051/0004-6361/202348795}}.
	
	\bibitem[{Zanchettin} et~al.(2025){Zanchettin}, {Ramos Almeida}, {Audibert},
	{Acosta-Pulido}, {Cezar}, {Hicks}, {Lapi}, and {Mullaney}]{zanchettin25a}
	{Zanchettin}, M.V.; {Ramos Almeida}, C.; {Audibert}, A.; {Acosta-Pulido}, J.A.;
	{Cezar}, P.H.; {Hicks}, E.; {Lapi}, A.; {Mullaney}, J.
	\newblock {Unveiling the warm molecular outflow component of type-2 quasars
		with SINFONI}.
	\newblock {\em Astron. Astrophys.} {\bf 2025}, {\em 695},~A185.
	\newblock {\url{https://doi.org/10.1051/0004-6361/202453224}}.
	
	\bibitem[{Ramos Almeida} et~al.(2025){Ramos Almeida}, {Garcia-Bernete},
	{Pereira-Santaella}, {Speranza}, {Maiolino}, {Ji}, {Audibert}, {Cezar},
	{Acosta-Pulido}, {Alonso-Herrero}, {Garcia-Burillo}, {Gonzalez-Martin},
	{Rigopoulou}, {Tadhunter}, {Labiano}, {Levenson}, and {Donnan}]{almeida25a}
	{Ramos Almeida}, C.; {Garcia-Bernete}, I.; {Pereira-Santaella}, M.; {Speranza},
	G.; {Maiolino}, R.; {Ji}, X.; {Audibert}, A.; {Cezar}, P.H.; {Acosta-Pulido},
	J.A.; {Alonso-Herrero}, A.;  et~al.
	\newblock {JWST MIRI reveals the diversity of nuclear mid-infrared spectra of
		nearby type-2 quasars}.
	\newblock {\em arXiv} {\bf 2025}, arXiv:2504.01595.
	\newblock {\url{https://doi.org/10.48550/arXiv.2504.01595}}.
	
	\bibitem[{Riffel} et~al.(2015){Riffel}, {Storchi-Bergmann}, and
	{Riffel}]{riffel15a}
	{Riffel}, R.A.; {Storchi-Bergmann}, T.; {Riffel}, R.
	\newblock {Feeding versus feedback in active galactic nuclei from near-infrared
		integral field spectroscopy---X. NGC 5929}.
	\newblock {\em Mon. Not. R. Astron. Soc.} {\bf 2015}, {\em 451},~3587--3605.
	\newblock {\url{https://doi.org/10.1093/mnras/stv1129}}.
	
	\bibitem[{Jarvis} et~al.(2019){Jarvis}, {Harrison}, {Thomson}, {Circosta},
	{Mainieri}, {Alexander}, {Edge}, {Lansbury}, {Molyneux}, and
	{Mullaney}]{jarvis19a}
	{Jarvis}, M.E.; {Harrison}, C.M.; {Thomson}, A.P.; {Circosta}, C.; {Mainieri},
	V.; {Alexander}, D.M.; {Edge}, A.C.; {Lansbury}, G.B.; {Molyneux}, S.J.;
	{Mullaney}, J.R.
	\newblock {Prevalence of radio jets associated with galactic outflows and
		feedback from quasars}.
	\newblock {\em Mon. Not. R. Astron. Soc.} {\bf 2019}, {\em 485},~2710--2730.
	\newblock {\url{https://doi.org/10.1093/mnras/stz556}}.
	
	\bibitem[{Girdhar} et~al.(2024){Girdhar}, {Harrison}, {Mainieri},
	{Fern{\'a}ndez Aranda}, {Alexander}, {Arrigoni Battaia}, {Bianchin},
	{Calistro Rivera}, {Circosta}, {Costa}, {Edge}, {Farina}, {Kakkad}, {Kharb},
	{Molyneux}, {Mukherjee}, {Njeri}, {Silpa}, {Venturi}, and {Ward}]{girdhar24a}
	{Girdhar}, A.; {Harrison}, C.M.; {Mainieri}, V.; {Fern{\'a}ndez Aranda}, R.;
	{Alexander}, D.M.; {Arrigoni Battaia}, F.; {Bianchin}, M.; {Calistro Rivera},
	G.; {Circosta}, C.; {Costa}, T.;  et~al.
	\newblock {Quasar feedback survey: Molecular gas affected by central outflows
		and by 10-kpc radio lobes reveal dual feedback effects in 'radio quiet'
		quasars}.
	\newblock {\em Mon. Not. R. Astron. Soc.} {\bf 2024}, {\em 527},~9322--9342.
	\newblock {\url{https://doi.org/10.1093/mnras/stad3453}}.
	
	\bibitem[{Ali} et~al.(2025){Ali}, {Sebastian}, {Kakkad}, {Silpa}, {Kharb},
	{O'Dea}, {Singha}, {K}, {Rubinur}, {Baum}, {Bait}, {Vaddi}, and
	{Kurapati}]{ali25a}
	{Ali}, A.; {Sebastian}, B.; {Kakkad}, D.; {Silpa}, S.; {Kharb}, P.; {O'Dea},
	C.P.; {Singha}, M.; {K}.; {Rubinur}.; {Baum}, S.A.;  et~al.
	\newblock {Jet-mode feedback in NGC 5972: Insights from resolved MUSE, GMRT and
		VLA observations}.
	\newblock {\em arXiv} {\bf 2025},  arXiv:2503.19031.
	\newblock {\url{https://doi.org/10.48550/arXiv.2503.19031}}.
	
	\bibitem[{Mahony} et~al.(2013){Mahony}, {Morganti}, {Emonts}, {Oosterloo}, and
	{Tadhunter}]{mahony13a}
	{Mahony}, E.K.; {Morganti}, R.; {Emonts}, B.H.C.; {Oosterloo}, T.A.;
	{Tadhunter}, C.
	\newblock {The location and impact of jet-driven outflows of cold gas: The case
		of 3C 293}.
	\newblock {\em Mon. Not. R. Astron. Soc.} {\bf 2013}, {\em 435},~L58--L62.
	\newblock {\url{https://doi.org/10.1093/mnrasl/slt094}}.
	
	\bibitem[{Lanz} et~al.(2015){Lanz}, {Ogle}, {Evans}, {Appleton}, {Guillard},
	and {Emonts}]{lanz15b}
	{Lanz}, L.; {Ogle}, P.M.; {Evans}, D.; {Appleton}, P.N.; {Guillard}, P.;
	{Emonts}, B.
	\newblock {Jet-ISM Interaction in the Radio Galaxy 3C 293: Jet-driven Shocks
		Heat ISM to Power X-Ray and Molecular H$_{2}$ Emission}.
	\newblock {\em Astrophys. J.} {\bf 2015}, {\em 801},~17.
	\newblock {\url{https://doi.org/10.1088/0004-637X/801/1/17}}.
	
	\bibitem[{Mahony} et~al.(2016){Mahony}, {Oonk}, {Morganti}, {Tadhunter},
	{Bessiere}, {Short}, {Emonts}, and {Oosterloo}]{mahony16a}
	{Mahony}, E.K.; {Oonk}, J.B.R.; {Morganti}, R.; {Tadhunter}, C.; {Bessiere},
	P.; {Short}, P.; {Emonts}, B.H.C.; {Oosterloo}, T.A.
	\newblock {Jet-driven outflows of ionized gas in the nearby radio galaxy 3C
		293}.
	\newblock {\em Mon. Not. R. Astron. Soc.} {\bf 2016}, {\em 455},~2453--2460.
	\newblock {\url{https://doi.org/10.1093/mnras/stv2456}}.
	
	\bibitem[{Riffel} et~al.(2023){Riffel}, {Riffel}, {Bianchin},
	{Storchi-Bergmann}, {Souza-Oliveira}, and {Zakamska}]{riffel23a}
	{Riffel}, R.A.; {Riffel}, R.; {Bianchin}, M.; {Storchi-Bergmann}, T.;
	{Souza-Oliveira}, G.L.; {Zakamska}, N.L.
	\newblock {Spatially resolved observations of outflows in the radio loud AGN of
		UGC 8782}.
	\newblock {\em Mon. Not. R. Astron. Soc.} {\bf 2023}, {\em 521},~3260--3272.
	\newblock {\url{https://doi.org/10.1093/mnras/stad776}}.
	
	\bibitem[{Costa-Souza} et~al.(2024){Costa-Souza}, {Riffel}, {Souza-Oliveira},
	{Zakamska}, {Bianchin}, {Storchi-Bergmann}, and {Riffel}]{CostaSouza24a}
	{Costa-Souza}, J.H.; {Riffel}, R.A.; {Souza-Oliveira}, G.L.; {Zakamska}, N.L.;
	{Bianchin}, M.; {Storchi-Bergmann}, T.; {Riffel}, R.
	\newblock {Blowing Star Formation Away in Active Galactic Nuclei Hosts. I.
		Observation of Warm Molecular Outflows with JWST MIRI}.
	\newblock {\em Astrophys. J.} {\bf 2024}, {\em 974},~127.
	\newblock {\url{https://doi.org/10.3847/1538-4357/ad702a}}.
	
	\bibitem[{Dasyra} et~al.(2016){Dasyra}, {Combes}, {Oosterloo}, {Oonk},
	{Morganti}, {Salom{\'e}}, and {Vlahakis}]{dasyra16a}
	{Dasyra}, K.M.; {Combes}, F.; {Oosterloo}, T.; {Oonk}, J.B.R.; {Morganti}, R.;
	{Salom{\'e}}, P.; {Vlahakis}, N.
	\newblock {ALMA reveals optically thin, highly excited CO gas in the jet-driven
		winds of the galaxy IC 5063}.
	\newblock {\em Astron. Astrophys.} {\bf 2016}, {\em 595},~L7.
	\newblock {\url{https://doi.org/10.1051/0004-6361/201629689}}.
	
	\bibitem[{Oosterloo} et~al.(2017){Oosterloo}, {Raymond Oonk}, {Morganti},
	{Combes}, {Dasyra}, {Salom{\'e}}, {Vlahakis}, and {Tadhunter}]{oosterloo17a}
	{Oosterloo}, T.; {Raymond Oonk}, J.B.; {Morganti}, R.; {Combes}, F.; {Dasyra},
	K.; {Salom{\'e}}, P.; {Vlahakis}, N.; {Tadhunter}, C.
	\newblock {Properties of the molecular gas in the fast outflow in the Seyfert
		galaxy IC 5063}.
	\newblock {\em Astron. Astrophys.} {\bf 2017}, {\em 608},~A38.
	\newblock {\url{https://doi.org/10.1051/0004-6361/201731781}}.
	
	\bibitem[{Fonseca-Faria} et~al.(2023){Fonseca-Faria}, {Rodr{\'\i}guez-Ardila},
	{Contini}, {Dahmer-Hahn}, and {Morganti}]{fonseca23a}
	{Fonseca-Faria}, M.A.; {Rodr{\'\i}guez-Ardila}, A.; {Contini}, M.;
	{Dahmer-Hahn}, L.G.; {Morganti}, R.
	\newblock {Physical conditions and extension of the coronal line region in IC
		5063}.
	\newblock {\em Mon. Not. R. Astron. Soc.} {\bf 2023}, {\em 524},~143--160.
	\newblock {\url{https://doi.org/10.1093/mnras/stad1871}}.
	
	\bibitem[{Travascio} et~al.(2021){Travascio}, {Fabbiano}, {Paggi}, {Elvis},
	{Maksym}, {Morganti}, {Oosterloo}, and {Fiore}]{travascio21a}
	{Travascio}, A.; {Fabbiano}, G.; {Paggi}, A.; {Elvis}, M.; {Maksym}, W.P.;
	{Morganti}, R.; {Oosterloo}, T.; {Fiore}, F.
	\newblock {AGN-Host Interaction in IC 5063. I. Large-scale X-Ray Morphology and
		Spectral Analysis}.
	\newblock {\em Astrophys. J.} {\bf 2021}, {\em 921},~129.
	\newblock {\url{https://doi.org/10.3847/1538-4357/ac18c7}}.
	
	\bibitem[{Dasyra} et~al.(2024){Dasyra}, {Paraschos}, {Combes}, {Patapis},
	{Helou}, {Papachristou}, {Fernandez-Ontiveros}, {Bisbas}, {Spinoglio},
	{Armus}, and {Malkan}]{dasyra24a}
	{Dasyra}, K.M.; {Paraschos}, G.F.; {Combes}, F.; {Patapis}, P.; {Helou}, G.;
	{Papachristou}, M.; {Fernandez-Ontiveros}, J.A.; {Bisbas}, T.G.; {Spinoglio},
	L.; {Armus}, L.;  et~al.
	\newblock {A Case Study of Gas Impacted by Black-hole Jets with the JWST:
		Outflows, Bow Shocks, and High Excitation of the Gas in the Galaxy IC 5063}.
	\newblock {\em Astrophys. J.} {\bf 2024}, {\em 977},~156.
	\newblock {\url{https://doi.org/10.3847/1538-4357/ad89ba}}.
	
	\bibitem[{Cresci} et~al.(2015){Cresci}, {Marconi}, {Zibetti}, {Risaliti},
	{Carniani}, {Mannucci}, {Gallazzi}, {Maiolino}, {Balmaverde}, {Brusa},
	{Capetti}, {Cicone}, {Feruglio}, {Bland-Hawthorn}, {Nagao}, {Oliva},
	{Salvato}, {Sani}, {Tozzi}, {Urrutia}, and {Venturi}]{cresci15a}
	{Cresci}, G.; {Marconi}, A.; {Zibetti}, S.; {Risaliti}, G.; {Carniani}, S.;
	{Mannucci}, F.; {Gallazzi}, A.; {Maiolino}, R.; {Balmaverde}, B.; {Brusa},
	M.;  et~al.
	\newblock {The MAGNUM survey: Positive feedback in the nuclear region of NGC
		5643 suggested by MUSE}.
	\newblock {\em Astron. Astrophys.} {\bf 2015}, {\em 582},~A63.
	\newblock {\url{https://doi.org/10.1051/0004-6361/201526581}}.
	
	\bibitem[{Marconcini} et~al.(2025){Marconcini}, {Marconi}, {Cresci},
	{Mannucci}, {Ulivi}, {Venturi}, {Scialpi}, {Tozzi}, {Belfiore}, {Bertola},
	{Carniani}, {Cataldi}, {Chakraborty}, {D'Amato}, {Di Teodoro}, {Feltre},
	{Ginolfi}, {Moreschini}, {Orientale}, {Trefoloni}, and {King}]{marconcini25a}
	{Marconcini}, C.; {Marconi}, A.; {Cresci}, G.; {Mannucci}, F.; {Ulivi}, L.;
	{Venturi}, G.; {Scialpi}, M.; {Tozzi}, G.; {Belfiore}, F.; {Bertola}, E.;
	et~al.
	\newblock {Evidence of the fast acceleration of AGN-driven winds at kiloparsec
		scales}.
	\newblock {\em Nat. Astron.} {\textbf{2025}.}, {\em 9}, ~907--915. 
	\newblock {\url{https://doi.org/10.1038/s41550-025-02518-6}}.
	
	\bibitem[{Finlez} et~al.(2018){Finlez}, {Nagar}, {Storchi-Bergmann},
	{Schnorr-M{\"u}ller}, {Riffel}, {Lena}, {Mundell}, and {Elvis}]{finlez18a}
	{Finlez}, C.; {Nagar}, N.M.; {Storchi-Bergmann}, T.; {Schnorr-M{\"u}ller}, A.;
	{Riffel}, R.A.; {Lena}, D.; {Mundell}, C.G.; {Elvis}, M.S.
	\newblock {The complex jet- and bar-perturbed kinematics in NGC 3393 as
		revealed with ALMA and GEMINI-GMOS/IFU}.
	\newblock {\em Mon. Not. R. Astron. Soc.} {\bf 2018}, {\em 479},~3892--3908.
	\newblock {\url{https://doi.org/10.1093/mnras/sty1555}}.
	
	\bibitem[{Maksym} et~al.(2017){Maksym}, {Fabbiano}, {Elvis}, {Karovska},
	{Paggi}, {Raymond}, {Wang}, and {Storchi-Bergmann}]{maksym17a}
	{Maksym}, W.P.; {Fabbiano}, G.; {Elvis}, M.; {Karovska}, M.; {Paggi}, A.;
	{Raymond}, J.; {Wang}, J.; {Storchi-Bergmann}, T.
	\newblock {CHEERS Results from NGC 3393. II. Investigating the Extended
		Narrow-line Region Using Deep Chandra Observations and Hubble Space Telescope
		Narrow-line Imaging}.
	\newblock {\em Astrophys. J.} {\bf 2017}, {\em 844},~69.
	\newblock {\url{https://doi.org/10.3847/1538-4357/aa78a4}}.
	
	\bibitem[{Maksym} et~al.(2019){Maksym}, {Fabbiano}, {Elvis}, {Karovska},
	{Paggi}, {Raymond}, {Wang}, {Storchi-Bergmann}, and {Risaliti}]{maksym19a}
	{Maksym}, W.P.; {Fabbiano}, G.; {Elvis}, M.; {Karovska}, M.; {Paggi}, A.;
	{Raymond}, J.; {Wang}, J.; {Storchi-Bergmann}, T.; {Risaliti}, G.
	\newblock {CHEERS Results from NGC 3393. III. Chandra X-Ray Spectroscopy of the
		Narrow Line Region}.
	\newblock {\em Astrophys. J.} {\bf 2019}, {\em 872},~94.
	\newblock {\url{https://doi.org/10.3847/1538-4357/aaf4f5}}.
	
	\bibitem[{Pereira-Santaella} et~al.(2022){Pereira-Santaella},
	{{\'A}lvarez-M{\'a}rquez}, {Garc{\'\i}a-Bernete}, {Labiano}, {Colina},
	{Alonso-Herrero}, {Bellocchi}, {Garc{\'\i}a-Burillo}, {H{\"o}nig}, {Ramos
		Almeida}, and {Rosario}]{santaella22a}
	{Pereira-Santaella}, M.; {{\'A}lvarez-M{\'a}rquez}, J.; {Garc{\'\i}a-Bernete},
	I.; {Labiano}, A.; {Colina}, L.; {Alonso-Herrero}, A.; {Bellocchi}, E.;
	{Garc{\'\i}a-Burillo}, S.; {H{\"o}nig}, S.F.; {Ramos Almeida}, C.;  et~al.
	\newblock {Low-power jet-interstellar medium interaction in NGC 7319 revealed
		by JWST/MIRI MRS}.
	\newblock {\em Astron. Astrophys.} {\bf 2022}, {\em 665},~L11.
	\newblock {\url{https://doi.org/10.1051/0004-6361/202244725}}.
	
	\bibitem[{Emonts} et~al.(2025){Emonts}, {Appleton}, {Lisenfeld}, {Guillard},
	{Xu}, {Reach}, {Barcos-Mu{\~n}oz}, {Labiano}, {Ogle}, {O'Sullivan}, {Togi},
	{Gallagher}, {Aromal}, {Duc}, {Alatalo}, {Boulanger}, {D{\'\i}az-Santos}, and
	{Helou}]{emonts25a}
	{Emonts}, B.H.C.; {Appleton}, P.N.; {Lisenfeld}, U.; {Guillard}, P.; {Xu},
	C.K.; {Reach}, W.T.; {Barcos-Mu{\~n}oz}, L.; {Labiano}, A.; {Ogle}, P.M.;
	{O'Sullivan}, E.;  et~al.
	\newblock {Bird's-eye View of Molecular Gas across Stephan's Quintet Galaxy
		Group and Intragroup Medium}.
	\newblock {\em Astrophys. J.} {\bf 2025}, {\em 978},~111.
	\newblock {\url{https://doi.org/10.3847/1538-4357/ad957c}}.
	
	\bibitem[{Nesvadba} et~al.(2011){Nesvadba}, {Boulanger}, {Lehnert}, {Guillard},
	and {Salome}]{nesvadba11b}
	{Nesvadba}, N.P.H.; {Boulanger}, F.; {Lehnert}, M.D.; {Guillard}, P.; {Salome},
	P.
	\newblock {Dense gas without star formation: The kpc-sized turbulent molecular
		disk in 3C 326 N}.
	\newblock {\em Astron. Astrophys.} {\bf 2011}, {\em 536},~L5.
	\newblock {\url{https://doi.org/10.1051/0004-6361/201118018}}.


        \bibitem[{Villa-V{\'e}lez} et~al.(2024){Villa-V{\'e}lez}, {Godard}, {Guillard}, and {Pineau des For{\^e}ts}]{villaVelez24a}
        {Villa-V{\'e}lez}, J.A.; {Godard}, B.; {Guillard}, P.; {Pineau des For{\^e}ts}, G.
        \newblock {Radiative and mechanical energies in galaxies. I. Contributions of molecular shocks and PDRs in 3C 326 N}.
     \newblock {\em Astron. Astrophys} {\bf 2024}, {\em 688},~A96,  
    \newblock {\url{https://doi.org/10.1051/0004-6361/202449212}}. 
	
	\bibitem[{Leftley} et~al.(2024){Leftley}, {Nesvadba}, {Bicknell}, {Janssen},
	{Mukherjee}, {Petrov}, {Shende}, and {Zovaro}]{leftley24a}
	{Leftley}, J.H.; {Nesvadba}, N.P.H.; {Bicknell}, G.V.; {Janssen}, R.M.J.;
	{Mukherjee}, D.; {Petrov}, R.; {Shende}, M.B.; {Zovaro}, H.R.M.
	\newblock {JWST/NIRSpec and MIRI observations of an expanding, jet-driven
		bubble of warm H$_{2}$ in the radio galaxy 3C 326 N}.
	\newblock {\em Astron. Astrophys.} {\bf 2024}, {\em 689},~A314.
	\newblock {\url{https://doi.org/10.1051/0004-6361/202449848}}.

	
	\bibitem[{Peralta de Arriba} et~al.(2023){Peralta de Arriba}, {Alonso-Herrero},
	{Garc{\'\i}a-Burillo}, {Garc{\'\i}a-Bernete}, {Villar-Mart{\'\i}n},
	{Garc{\'\i}a-Lorenzo}, {Davies}, {Rosario}, {H{\"o}nig}, {Levenson},
	{Packham}, {Ramos Almeida}, {Pereira-Santaella}, {Audibert}, {Bellocchi},
	{Hicks}, {Labiano}, {Ricci}, and {Rigopoulou}]{arriba23a}
	{Peralta de Arriba}, L.; {Alonso-Herrero}, A.; {Garc{\'\i}a-Burillo}, S.;
	{Garc{\'\i}a-Bernete}, I.; {Villar-Mart{\'\i}n}, M.; {Garc{\'\i}a-Lorenzo},
	B.; {Davies}, R.; {Rosario}, D.J.; {H{\"o}nig}, S.F.; {Levenson}, N.A.;
	et~al.
	\newblock {A radio-jet-driven outflow in the Seyfert 2 galaxy NGC 2110?}
	\newblock {\em Astron. Astrophys.} {\bf 2023}, {\em 675},~A58.
	\newblock {\url{https://doi.org/10.1051/0004-6361/202245408}}.
	
	\bibitem[{Garc{\'\i}a-Bernete} et~al.(2021{\natexlab{a}}){Garc{\'\i}a-Bernete},
	{Alonso-Herrero}, {Garc{\'\i}a-Burillo}, {Pereira-Santaella},
	{Garc{\'\i}a-Lorenzo}, {Carrera}, {Rigopoulou}, {Ramos Almeida}, {Villar
		Mart{\'\i}n}, {Gonz{\'a}lez-Mart{\'\i}n}, {Hicks}, {Labiano}, {Ricci}, and
	{Mateos}]{garciaBernete21a}
	{Garc{\'\i}a-Bernete, I.;} 
	{Alonso-Herrero}, A.; {Garc{\'\i}a-Burillo}, S.;
	{Pereira-Santaella}, M.; {Garc{\'\i}a-Lorenzo}, B.; {Carrera}, F.J.;
	{Rigopoulou}, D.; {Ramos Almeida}, C.; {Villar Mart{\'\i}n}, M.;
	{Gonz{\'a}lez-Mart{\'\i}n}, O.;  et~al.
	\newblock {Multiphase feedback processes in the Sy2 galaxy NGC 5643}.
	\newblock {\em Astron. Astrophys.} {\bf 2021}, {\em 645},~A21.
	\newblock {\url{https://doi.org/10.1051/0004-6361/202038256}}.

	
	\bibitem[{Alonso Herrero} et~al.(2023){Alonso Herrero}, {Garc{\'\i}a-Burillo},
	{Pereira-Santaella}, {Shimizu}, {Combes}, {Hicks}, {Davies}, {Ramos Almeida},
	{Garc{\'\i}a-Bernete}, {H{\"o}nig}, {Levenson}, {Packham}, {Bellocchi},
	{Hunt}, {Imanishi}, {Ricci}, and {Roche}]{herrero23a}
	{Alonso Herrero}, A.; {Garc{\'\i}a-Burillo}, S.; {Pereira-Santaella}, M.;
	{Shimizu}, T.; {Combes}, F.; {Hicks}, E.K.S.; {Davies}, R.; {Ramos Almeida},
	C.; {Garc{\'\i}a-Bernete}, I.; {H{\"o}nig}, S.F.;  et~al.
	\newblock {AGN feedback in action in the molecular gas ring of the Seyfert
		galaxy NGC 7172}.
	\newblock {\em Astron. Astrophys.} {\bf 2023}, {\em 675},~A88.
	\newblock {\url{https://doi.org/10.1051/0004-6361/202346074}}.
	
	\bibitem[{Esposito} et~al.(2024){Esposito}, {Alonso-Herrero},
	{Garc{\'\i}a-Burillo}, {Casasola}, {Combes}, {Dallacasa}, {Davies},
	{Garc{\'\i}a-Bernete}, {Garc{\'\i}a-Lorenzo}, {Hermosa Mu{\~n}oz}, {de
		Arriba}, {Pereira-Santaella}, {Pozzi}, {Ramos Almeida}, {Shimizu}, {Vallini},
	{Bellocchi}, {Gonz{\'a}lez-Mart{\'\i}n}, {Hicks}, {H{\"o}nig}, {Labiano},
	{Levenson}, {Ricci}, and {Rosario}]{esposito24a}
	{Esposito, F.;}{Alonso-Herrero}, A.; {Garc{\'\i}a-Burillo}, S.; {Casasola},
	V.; {Combes}, F.; {Dallacasa}, D.; {Davies}, R.; {Garc{\'\i}a-Bernete}, I.;
	{Garc{\'\i}a-Lorenzo}, B.; {Hermosa Mu{\~n}oz}, L.;  et~al.
	\newblock {AGN feedback in the Local Universe: Multiphase outflow of the
		Seyfert galaxy NGC 5506}.
	\newblock {\em Astron. Astrophys.} {\bf 2024}, {\em 686},~A46.
	\newblock {\url{https://doi.org/10.1051/0004-6361/202449245}}.
	
	\bibitem[{Zhang} et~al.(2024){Zhang}, {Packham}, {Hicks}, {Davies}, {Shimizu},
	{Alonso-Herrero}, {Hermosa Mu{\~n}oz}, {Garc{\'\i}a-Bernete},
	{Pereira-Santaella}, {Audibert}, {L{\'o}pez-Rodr{\'\i}guez}, {Bellocchi},
	{Bunker}, {Combes}, {D{\'\i}az-Santos}, {Gandhi}, {Garc{\'\i}a-Burillo},
	{Garc{\'\i}a-Lorenzo}, {Gonz{\'a}lez-Mart{\'\i}n}, {Imanishi}, {Labiano},
	{Leist}, {Levenson}, {Ramos Almeida}, {Ricci}, {Rigopoulou}, {Rosario},
	{Stalevski}, {Ward}, {Esparza-Arredondo}, {Delaney}, {Fuller}, {Haidar},
	{H{\"o}nig}, {Izumi}, and {Rouan}]{zhang24a}
	{Zhang}, L.; {Packham}, C.; {Hicks}, E.K.S.; {Davies}, R.I.; {Shimizu}, T.T.;
	{Alonso-Herrero}, A.; {Hermosa Mu{\~n}oz}, L.; {Garc{\'\i}a-Bernete}, I.;
	{Pereira-Santaella}, M.; {Audibert}, A.;  et~al.
	\newblock {The Galaxy Activity, Torus, and Outflow Survey (GATOS). IV.
		Exploring Ionized Gas Outflows in Central Kiloparsec Regions of GATOS
		Seyferts}.
	\newblock {\em Astrophys. J.} {\bf 2024}, {\em 974},~195.
	\newblock {\url{https://doi.org/10.3847/1538-4357/ad6a4b}}.
	
	\bibitem[{D'Eugenio} et~al.(2025){D'Eugenio}, {Maiolino}, {Mahatma},
	{Mazzolari}, {Carniani}, {de Graaff}, {Maseda}, {Parlanti}, {Bunker}, {Ji},
	{Jones}, {Helton}, {Morganti}, {Scholtz}, {Tacchella}, {Tadhunter},
	{{\"U}bler}, and {Venturi}]{Deugino25a}
	{D'Eugenio}, F.; {Maiolino}, R.; {Mahatma}, V.H.; {Mazzolari}, G.; {Carniani},
	S.; {de Graaff}, A.; {Maseda}, M.V.; {Parlanti}, E.; {Bunker}, A.J.; {Ji},
	X.;  et~al.
	\newblock {JWST/NIRSpec WIDE survey: A z = 4.6 low-mass star-forming galaxy
		hosting a jet-driven shock with low ionization and solar metallicity}.
	\newblock {\em Mon. Not. R. Astron. Soc.} {\bf 2025}, {\em 536},~51--71.
	\newblock {\url{https://doi.org/10.1093/mnras/stae2545}}.
	
	\bibitem[{Ruffa} et~al.(2019){Ruffa}, {Davis}, {Prandoni}, {Laing}, {Paladino},
	{Parma}, {de Ruiter}, {Casasola}, {Bureau}, and {Warren}]{ruffa19b}
	{Ruffa}, I.; {Davis}, T.A.; {Prandoni}, I.; {Laing}, R.A.; {Paladino}, R.;
	{Parma}, P.; {de Ruiter}, H.; {Casasola}, V.; {Bureau}, M.; {Warren}, J.
	\newblock {The AGN fuelling/feedback cycle in nearby radio galaxies---II.
		Kinematics of the molecular gas}.
	\newblock {\em Mon. Not. R. Astron. Soc.} {\bf 2019}, {\em 489},~3739--3757.
	\newblock {\url{https://doi.org/10.1093/mnras/stz2368}}.
	
	\bibitem[{Ruffa} et~al.(2020){Ruffa}, {Laing}, {Prandoni}, {Paladino}, {Parma},
	{Davis}, and {Bureau}]{ruffa20a}
	{Ruffa}, I.; {Laing}, R.A.; {Prandoni}, I.; {Paladino}, R.; {Parma}, P.;
	{Davis}, T.A.; {Bureau}, M.
	\newblock {The AGN fuelling/feedback cycle in nearby radio galaxies---III. 3D
		relative orientations of radio jets and CO discs and their interaction}.
	\newblock {\em Mon. Not. R. Astron. Soc.} {\bf 2020}, {\em 499},~5719--5731.
	\newblock {\url{https://doi.org/10.1093/mnras/staa3166}}.
	
	\bibitem[{Ruffa} et~al.(2022){Ruffa}, {Prandoni}, {Davis}, {Laing}, {Paladino},
	{Casasola}, {Parma}, and {Bureau}]{ruffa22a}
	{Ruffa}, I.; {Prandoni}, I.; {Davis}, T.A.; {Laing}, R.A.; {Paladino}, R.;
	{Casasola}, V.; {Parma}, P.; {Bureau}, M.
	\newblock {The AGN fuelling/feedback cycle in nearby radio galaxies---IV.
		Molecular gas conditions and jet-ISM interaction in NGC 3100}.
	\newblock {\em Mon. Not. R. Astron. Soc.} {\bf 2022}, {\em 510},~4485--4503.
	\newblock {\url{https://doi.org/10.1093/mnras/stab3541}}.
	
	\bibitem[{Dopita} et~al.(2015){Dopita}, {Shastri}, {Davies}, {Kewley},
	{Hampton}, {Scharw{\"a}chter}, {Sutherland}, {Kharb}, {Jose}, {Bhatt},
	{Ramya}, {Jin}, {Banfield}, {Zaw}, {Juneau}, {James}, and
	{Srivastava}]{dopita15a}
	{Dopita}, M.A.; {Shastri}, P.; {Davies}, R.; {Kewley}, L.; {Hampton}, E.;
	{Scharw{\"a}chter}, J.; {Sutherland}, R.; {Kharb}, P.; {Jose}, J.; {Bhatt},
	H.;  et~al.
	\newblock {Probing the Physics of Narrow Line Regions in Active Galaxies. II.
		The Siding Spring Southern Seyfert Spectroscopic Snapshot Survey (S7)}.
	\newblock {\em Astrophys. J. Suppl. Ser.} {\bf 2015}, {\em 217},~12.
	\newblock {\url{https://doi.org/10.1088/0067-0049/217/1/12}}.
	
	\bibitem[{Cazzoli} et~al.(2022){Cazzoli}, {Hermosa Mu{\~n}oz}, {M{\'a}rquez},
	{Masegosa}, {Castillo-Morales}, {Gil de Paz}, {Hern{\'a}ndez-Garc{\'\i}a},
	{La Franca}, and {Ramos Almeida}]{cazzoli22a}
	{Cazzoli}, S.; {Hermosa Mu{\~n}oz}, L.; {M{\'a}rquez}, I.; {Masegosa}, J.;
	{Castillo-Morales}, {\'A}.; {Gil de Paz}, A.; {Hern{\'a}ndez-Garc{\'\i}a},
	L.; {La Franca}, F.; {Ramos Almeida}, C.
	\newblock {Unexplored outflows in nearby low luminosity AGNs. The case of NGC
		1052}.
	\newblock {\em Astron. Astrophys.} {\bf 2022}, {\em 664},~A135.
	\newblock {\url{https://doi.org/10.1051/0004-6361/202142695}}.
	
	\bibitem[{Molina} et~al.(2018){Molina}, {Eracleous}, {Barth}, {Maoz}, {Runnoe},
	{Ho}, {Shields}, and {Walsh}]{molina18a}
	{Molina}, M.; {Eracleous}, M.; {Barth}, A.J.; {Maoz}, D.; {Runnoe}, J.C.; {Ho},
	L.C.; {Shields}, J.C.; {Walsh}, J.L.
	\newblock {The Shocking Power Sources of LINERs}.
	\newblock {\em Astrophys. J.} {\bf 2018}, {\em 864},~90.
	\newblock {\url{https://doi.org/10.3847/1538-4357/aad5ed}}.
	
	\bibitem[{Goold} et~al.(2024){Goold}, {Seth}, {Molina}, {Ohlson}, {Runnoe},
	{B{\"o}ker}, {Davis}, {Dumont}, {Eracleous}, {Fern{\'a}ndez-Ontiveros},
	{Gallo}, {Goulding}, {Greene}, {Ho}, {Markoff}, {Neumayer}, {Plotkin},
	{Prieto}, {Satyapal}, {van de Ven}, {Walsh}, {Yuan}, {Feldmeier-Krause},
	{G{\"u}ltekin}, {H{\"o}nig}, {Kirkpatrick}, {L{\"u}tzgendorf}, {Reines},
	{Strader}, {Trump}, and {Voggel}]{goold24a}
	{Goold}, K.; {Seth}, A.; {Molina}, M.; {Ohlson}, D.; {Runnoe}, J.C.;
	{B{\"o}ker}, T.; {Davis}, T.A.; {Dumont}, A.; {Eracleous}, M.;
	{Fern{\'a}ndez-Ontiveros}, J.A.;  et~al.
	\newblock {ReveaLLAGN 0: First Look at JWST MIRI Data of Sombrero and NGC
		1052}.
	\newblock {\em Astrophys. J.} {\bf 2024}, {\em 966},~204.
	\newblock {\url{https://doi.org/10.3847/1538-4357/ad3065}}.
	
	\bibitem[{Cecil} et~al.(2001){Cecil}, {Bland-Hawthorn}, {Veilleux}, and
	{Filippenko}]{cecil01a}
	{Cecil}, G.; {Bland-Hawthorn}, J.; {Veilleux}, S.; {Filippenko}, A.V.
	\newblock {Jet- and Wind-driven Ionized Outflows in the Superbubble and
		Star-forming Disk of NGC 3079}.
	\newblock {\em Astrophys. J.} {\bf 2001}, {\em 555},~338--355.
	\newblock {\url{https://doi.org/10.1086/321481}}.
	
	\bibitem[{Middelberg} et~al.(2007){Middelberg}, {Agudo}, {Roy}, and
	{Krichbaum}]{middelberg07a}
	{Middelberg}, E.; {Agudo}, I.; {Roy}, A.L.; {Krichbaum}, T.P.
	\newblock {Jet-cloud collisions in the jet of the Seyfert galaxy NGC3079}.
	\newblock {\em Mon. Not. R. Astron. Soc.} {\bf 2007}, {\em 377},~731--740.
	\newblock {\url{https://doi.org/10.1111/j.1365-2966.2007.11639.x}}.
	
	\bibitem[{Fernandez} et~al.(2023){Fernandez}, {Secrest}, {Johnson}, and
	{Fischer}]{fernandez23a}
	{Fernandez}, L.C.; {Secrest}, N.J.; {Johnson}, M.C.; {Fischer}, T.C.
	\newblock {FRAMEx. IV. Mechanical Feedback from the Active Galactic Nucleus in
		NGC 3079}.
	\newblock {\em Astrophys. J.} {\bf 2023}, {\em 958},~61.
	\newblock {\url{https://doi.org/10.3847/1538-4357/acfeda}}.
	
	\bibitem[{Shafi} et~al.(2015){Shafi}, {Oosterloo}, {Morganti}, {Colafrancesco},
	and {Booth}]{shafi15a}
	{Shafi}, N.; {Oosterloo}, T.A.; {Morganti}, R.; {Colafrancesco}, S.; {Booth},
	R.
	\newblock {The `shook up' galaxy NGC 3079: The complex interplay between H I,
		activity and environment}.
	\newblock {\em Mon. Not. R. Astron. Soc.} {\bf 2015}, {\em 454},~1404--1415.
	\newblock {\url{https://doi.org/10.1093/mnras/stv2034}}.
	
	\bibitem[{Veilleux} et~al.(2021){Veilleux}, {Mel{\'e}ndez}, {Stone}, {Cecil},
	{Hodges-Kluck}, {Bland-Hawthorn}, {Bregman}, {Heitsch}, {Martin}, {Mueller},
	{Rupke}, {Sturm}, {Tanner}, and {Engelbracht}]{veilleux21a}
	{Veilleux}, S.; {Mel{\'e}ndez}, M.; {Stone}, M.; {Cecil}, G.; {Hodges-Kluck},
	E.; {Bland-Hawthorn}, J.; {Bregman}, J.; {Heitsch}, F.; {Martin}, C.L.;
	{Mueller}, T.;  et~al.
	\newblock {Exploring the dust content of galactic haloes with Herschel---IV.
		NGC 3079}.
	\newblock {\em Mon. Not. R. Astron. Soc.} {\bf 2021}, {\em 508},~4902--4918.
	\newblock {\url{https://doi.org/10.1093/mnras/stab2881}}.
	
	\bibitem[{Brusa} et~al.(2018){Brusa}, {Cresci}, {Daddi}, {Paladino}, {Perna},
	{Bongiorno}, {Lusso}, {Sargent}, {Casasola}, {Feruglio}, {Fraternali},
	{Georgiev}, {Mainieri}, {Carniani}, {Comastri}, {Duras}, {Fiore}, {Mannucci},
	{Marconi}, {Piconcelli}, {Zamorani}, {Gilli}, {La Franca}, {Lanzuisi},
	{Lutz}, {Santini}, {Scoville}, {Vignali}, {Vito}, {Rabien}, {Busoni}, and
	{Bonaglia}]{brusa18a}
	{Brusa}, M.; {Cresci}, G.; {Daddi}, E.; {Paladino}, R.; {Perna}, M.;
	{Bongiorno}, A.; {Lusso}, E.; {Sargent}, M.T.; {Casasola}, V.; {Feruglio},
	C.;  et~al.
	\newblock {Molecular outflow and feedback in the obscured quasar XID2028
		revealed by ALMA}.
	\newblock {\em Astron. Astrophys.} {\bf 2018}, {\em 612},~A29.
	\newblock {\url{https://doi.org/10.1051/0004-6361/201731641}}.
	
	\bibitem[{Cresci} et~al.(2023){Cresci}, {Tozzi}, {Perna}, {Brusa},
	{Marconcini}, {Marconi}, {Carniani}, {Brienza}, {Giroletti}, {Belfiore},
	{Ginolfi}, {Mannucci}, {Ulivi}, {Scholtz}, {Venturi}, {Arribas}, {{\"U}bler},
	{D'Eugenio}, {Mingozzi}, {Balmaverde}, {Capetti}, {Parlanti}, and
	{Zana}]{cresci23a}
	{Cresci}, G.; {Tozzi}, G.; {Perna}, M.; {Brusa}, M.; {Marconcini}, C.;
	{Marconi}, A.; {Carniani}, S.; {Brienza}, M.; {Giroletti}, M.; {Belfiore},
	F.;  et~al.
	\newblock {Bubbles and outflows: The novel JWST/NIRSpec view of the z = 1.59
		obscured quasar XID2028}.
	\newblock {\em Astron. Astrophys.} {\bf 2023}, {\em 672},~A128.
	\newblock {\url{https://doi.org/10.1051/0004-6361/202346001}}.
	
	\bibitem[{Sridhar} et~al.(2020){Sridhar}, {Morganti}, {Nyland}, {Frank},
	{Harwood}, and {Oosterloo}]{sarrvesh20a}
	{Sridhar}, S.S.; {Morganti}, R.; {Nyland}, K.; {Frank}, B.S.; {Harwood}, J.;
	{Oosterloo}, T.
	\newblock {LOFAR view of NGC 3998, a sputtering AGN}.
	\newblock {\em Astron. Astrophys.} {\bf 2020}, {\em 634},~A108.
	\newblock {\url{https://doi.org/10.1051/0004-6361/201936796}}.
	
	\bibitem[{Ogle} et~al.(2024){Ogle}, {L{\'o}pez}, {Reynaldi}, {Togi}, {Rich},
	{Rom{\'a}n}, {Caceres}, {Li}, {Donnelly}, {Smith}, {Appleton}, and
	{Lanz}]{Ogle24a}
	{Ogle}, P.M.; {L{\'o}pez}, I.E.; {Reynaldi}, V.; {Togi}, A.; {Rich}, R.M.;
	{Rom{\'a}n}, J.; {Caceres}, O.; {Li}, Z.C.; {Donnelly}, G.; {Smith}, J.D.T.;
	et~al.
	\newblock {Radio Jet Feedback on the Inner Disk of Virgo Spiral Galaxy Messier
		58}.
	\newblock {\em Astrophys. J.} {\bf 2024}, {\em 962},~196.
	\newblock {\url{https://doi.org/10.3847/1538-4357/ad1242}}.
	
	\bibitem[{Su} et~al.(2023){Su}, {Mahony}, {Gu}, {Sadler}, {Curran}, {Allison},
	{Yoon}, {Aditya}, {Chandola}, {Chen}, {Moss}, {Wu}, {Shao}, {Liu},
	{Glowacki}, {Whiting}, and {Weng}]{renzhi23a}
	{Su}, R.; {Mahony}, E.K.; {Gu}, M.; {Sadler}, E.M.; {Curran}, S.J.; {Allison},
	J.R.; {Yoon}, H.; {Aditya}, J.N.H.S.; {Chandola}, Y.; {Chen}, Y.;  et~al.
	\newblock {Does a radio jet drive the massive multiphase outflow in the
		ultra-luminous infrared galaxy IRAS 10565 + 2448?}
	\newblock {\em Mon. Not. R. Astron. Soc.} {\bf 2023}, {\em 520},~5712--5723.
	\newblock {\url{https://doi.org/10.1093/mnras/stad370}}.
	
	\bibitem[{Bicknell} et~al.(2000){Bicknell}, {Sutherland}, {van Breugel},
	{Dopita}, {Dey}, and {Miley}]{bicknell00a}
	{Bicknell}, G.; {Sutherland}, R.; {van Breugel}, W.; {Dopita}, M.; {Dey}, A.;
	{Miley}, G.
	\newblock Jet-induced Emission-Line Nebulosity and Star Formation in the
	High-Redshift Radio Galaxy 4C 41.17.
	\newblock {\em Astrophys. J. Lett.} {\bf 2000}, {\em 540},~678.
	
	\bibitem[{Holt} et~al.(2006){Holt}, {Tadhunter}, {Morganti}, {Bellamy},
	{Gonz{\'a}lez Delgado}, {Tzioumis}, and {Inskip}]{holt06a}
	{Holt}, J.; {Tadhunter}, C.; {Morganti}, R.; {Bellamy}, M.; {Gonz{\'a}lez
		Delgado}, R.M.; {Tzioumis}, A.; {Inskip}, K.J.
	\newblock {The co-evolution of the obscured quasar PKS 1549-79 and its host
		galaxy: Evidence for a high accretion rate and warm outflow}.
	\newblock {\em Mon. Not. R. Astron. Soc.} {\bf 2006}, {\em 370},~1633--1650.
	\newblock {\url{https://doi.org/10.1111/j.1365-2966.2006.10604.x}}.
	
	\bibitem[{Oosterloo} et~al.(2019){Oosterloo}, {Morganti}, {Tadhunter}, {Raymond
		Oonk}, {Bignall}, {Tzioumis}, and {Reynolds}]{oosterloo19a}
	{Oosterloo}, T.; {Morganti}, R.; {Tadhunter}, C.; {Raymond Oonk}, J.B.;
	{Bignall}, H.E.; {Tzioumis}, T.; {Reynolds}, C.
	\newblock {ALMA observations of PKS 1549-79: A case of feeding and feedback in
		a young radio quasar}.
	\newblock {\em Astron. Astrophys.} {\bf 2019}, {\em 632},~A66.
	\href{http://arxiv.org/abs/1910.07865}{{\normalfont
			[arXiv:astro-ph.GA/1910.07865]}}.
	\newblock {\url{https://doi.org/10.1051/0004-6361/201936248}}.
	
	\bibitem[{Tadhunter} et~al.(1998){Tadhunter}, {Morganti}, {Robinson},
	{Dickson}, {Villar-Martin}, and {Fosbury}]{tadhunter98a}
	{Tadhunter}, C.N.; {Morganti}, R.; {Robinson}, A.; {Dickson}, R.;
	{Villar-Martin}, M.; {Fosbury}, R.A.E.
	\newblock {The nature of the optical-radio correlations for powerful radio
		galaxies}.
	\newblock {\em Mon. Not. R. Astron. Soc.} {\bf 1998}, {\em 298},~1035--1047.
	\newblock {\url{https://doi.org/10.1046/j.1365-8711.1998.01706.x}}.
	
	\bibitem[{Santoro} et~al.(2020){Santoro}, {Tadhunter}, {Baron}, {Morganti}, and
	{Holt}]{santoro20a}
	{Santoro}, F.; {Tadhunter}, C.; {Baron}, D.; {Morganti}, R.; {Holt}, J.
	\newblock {AGN-driven outflows and the AGN feedback efficiency in young radio
		galaxies}.
	\newblock {\em Astron. Astrophys.} {\bf 2020}, {\em 644},~A54.
	\newblock {\url{https://doi.org/10.1051/0004-6361/202039077}}.
	
	\bibitem[{Husemann} et~al.(2019{\natexlab{a}}){Husemann}, {Bennert}, {Jahnke},
	{Davis}, {Woo}, {Scharw{\"a}chter}, {Schulze}, {Gaspari}, and
	{Zwaan}]{husemann19b}
	{Husemann}, B.; {Bennert}, V.N.; {Jahnke}, K.; {Davis}, T.A.; {Woo}, J.H.;
	{Scharw{\"a}chter}, J.; {Schulze}, A.; {Gaspari}, M.; {Zwaan}, M.A.
	\newblock {Jet-driven Galaxy-scale Gas Outflows in the Hyperluminous Quasar 3C
		273}.
	\newblock {\em Astrophys. J.} {\bf 2019}, {\em 879},~75.
	\href{http://arxiv.org/abs/1905.10387}{{\normalfont
			[arXiv:astro-ph.GA/1905.10387]}}.
	\newblock {\url{https://doi.org/10.3847/1538-4357/ab24bc}}.
	
	\bibitem[{Husemann} et~al.(2019{\natexlab{b}}){Husemann}, {Scharw{\"a}chter},
	{Davis}, {P{\'e}rez-Torres}, {Smirnova-Pinchukova}, {Tremblay}, {Krumpe},
	{Combes}, {Baum}, {Busch}, {Connor}, {Croom}, {Gaspari}, {Kraft}, {O'Dea},
	{Powell}, {Singha}, and {Urrutia}]{husemann19a}
	{Husemann}, B.; {Scharw{\"a}chter}, J.; {Davis}, T.A.; {P{\'e}rez-Torres}, M.;
	{Smirnova-Pinchukova}, I.; {Tremblay}, G.R.; {Krumpe}, M.; {Combes}, F.;
	{Baum}, S.A.; {Busch}, G.;  et~al.
	\newblock {The Close AGN Reference Survey (CARS). A massive multi-phase outflow
		impacting the edge-on galaxy HE 1353-1917}.
	\newblock {\em Astron. Astrophys.} {\bf 2019}, {\em 627},~A53.
	\newblock {\url{https://doi.org/10.1051/0004-6361/201935283}}.
	
	\bibitem[{Singha} et~al.(2023){Singha}, {Winkel}, {Vaddi}, {Perez Torres},
	{Gaspari}, {Smirnova-Pinchukova}, {O'Dea}, {Combes}, {Omoruyi}, {Rose},
	{McElroy}, {Husemann}, {Davis}, {Baum}, {Lawlor-Forsyth}, {Neumann}, and
	{Tremblay}]{singha23a}
	{Singha}, M.; {Winkel}, N.; {Vaddi}, S.; {Perez Torres}, M.; {Gaspari}, M.;
	{Smirnova-Pinchukova}, I.; {O'Dea}, C.P.; {Combes}, F.; {Omoruyi}, O.;
	{Rose}, T.;  et~al.
	\newblock {The Close AGN Reference Survey (CARS): An Interplay between Radio
		Jets and AGN Radiation in the Radio-quiet AGN HE0040-1105}.
	\newblock {\em Astrophys. J.} {\bf 2023}, {\em 959},~107.
	\newblock {\url{https://doi.org/10.3847/1538-4357/ad004d}}.
	
	\bibitem[{Morganti} et~al.(2013){Morganti}, {Fogasy}, {Paragi}, {Oosterloo},
	and {Orienti}]{morganti13a}
	{Morganti}, R.; {Fogasy}, J.; {Paragi}, Z.; {Oosterloo}, T.; {Orienti}, M.
	\newblock {Radio Jets Clearing the Way Through a Galaxy: Watching Feedback in
		Action}.
	\newblock {\em Science} {\bf 2013}, {\em 341},~1082--1085.
	\newblock {\url{https://doi.org/10.1126/science.1240436}}.
	
	\bibitem[{Villar Mart{\'\i}n} et~al.(2023){Villar Mart{\'\i}n},
	{Castro-Rodr{\'\i}guez}, {Pereira Santaella}, {Lamperti}, {Tadhunter},
	{Emonts}, {Colina}, {Alonso Herrero}, {Cabrera-Lavers}, and
	{Bellocchi}]{villarMartin23a}
	{Villar Mart{\'\i}n}, M.; {Castro-Rodr{\'\i}guez}, N.; {Pereira Santaella}, M.;
	{Lamperti}, I.; {Tadhunter}, C.; {Emonts}, B.; {Colina}, L.; {Alonso
		Herrero}, A.; {Cabrera-Lavers}, A.; {Bellocchi}, E.
	\newblock {Limited impact of jet-induced feedback in the multi-phase nuclear
		interstellar medium of 4C12.50}.
	\newblock {\em Astron. Astrophys.} {\bf 2023}, {\em 673},~A25.
	\newblock {\url{https://doi.org/10.1051/0004-6361/202245418}}.
	
	\bibitem[{Holden} et~al.(2024){Holden}, {Tadhunter}, {Audibert}, {Oosterloo},
	{Ramos Almeida}, {Morganti}, {Pereira-Santaella}, and {Lamperti}]{holden24a}
	{Holden}, L.R.; {Tadhunter}, C.; {Audibert}, A.; {Oosterloo}, T.; {Ramos
		Almeida}, C.; {Morganti}, R.; {Pereira-Santaella}, M.; {Lamperti}, I.
	\newblock {ALMA reveals a compact and massive molecular outflow driven by the
		young AGN in a nearby ULIRG}.
	\newblock {\em Mon. Not. R. Astron. Soc.} {\bf 2024}, {\em 530},~446--456.
	\newblock {\url{https://doi.org/10.1093/mnras/stae810}}.
	
	\bibitem[{Holden} and {Tadhunter}(2025)]{holden25a}
	{Holden}, L.R.; {Tadhunter}, C.N.
	\newblock {No evidence for fast, galaxy-wide ionized outflows in a nearby
		quasar---The importance of accounting for beam smearing}.
	\newblock {\em Mon. Not. R. Astron. Soc.} {\bf 2025}, {\em 536},~1857--1877.
	\newblock {\url{https://doi.org/10.1093/mnras/stae2661}}.
	
	\bibitem[{Duncan} et~al.(2023){Duncan}, {Windhorst}, {Koekemoer},
	{R{\"o}ttgering}, {Cohen}, {Jansen}, {Summers}, {Tompkins}, {Hutchison},
	{Conselice}, {Driver}, {Yan}, {Adams}, {Cheng}, {Coe}, {Diego}, {Dole},
	{Frye}, {Gim}, {Grogin}, {Holwerda}, {Lim}, {Marshall}, {Nonino}, {Pirzkal},
	{Robotham}, {Ryan}, and {Willmer}]{duncan23a}
	{Duncan}, K.J.; {Windhorst}, R.A.; {Koekemoer}, A.M.; {R{\"o}ttgering}, H.J.A.;
	{Cohen}, S.H.; {Jansen}, R.A.; {Summers}, J.; {Tompkins}, S.; {Hutchison},
	T.A.; {Conselice}, C.J.;  et~al.
	\newblock {JWST's PEARLS: TN J1338-1942---I. Extreme jet-triggered star
		formation in a z = 4.11 luminous radio galaxy}.
	\newblock {\em Mon. Not. R. Astron. Soc.} {\bf 2023}, {\em 522},~4548--4564.
	\newblock {\url{https://doi.org/10.1093/mnras/stad1267}}.
	
	\bibitem[{Roy} et~al.(2024){Roy}, {Heckman}, {Overzier}, {Saxena}, {Duncan},
	{Miley}, {Villar Mart{\'\i}n}, {Gab{\'a}nyi}, {Aydar}, {Bosman},
	{Rottgering}, {Pentericci}, {Onoue}, and {Reynaldi}]{roy24a}
	{Roy}, N.; {Heckman}, T.; {Overzier}, R.; {Saxena}, A.; {Duncan}, K.; {Miley},
	G.; {Villar Mart{\'\i}n}, M.; {Gab{\'a}nyi}, K.{\'E}.; {Aydar}, C.; {Bosman},
	S.E.I.;  et~al.
	\newblock {JWST Reveals Powerful Feedback from Radio Jets in a Massive Galaxy
		at z = 4.1}.
	\newblock {\em Astrophys. J.} {\bf 2024}, {\em 970},~69.
	\newblock {\url{https://doi.org/10.3847/1538-4357/ad4bda}}.
	
	\bibitem[{Saxena} et~al.(2024){Saxena}, {Overzier}, {Villar-Mart{\'\i}n},
	{Heckman}, {Roy}, {Duncan}, {R{\"o}ttgering}, {Miley}, {Aydar}, {Best},
	{Bosman}, {Cameron}, {Gab{\'a}nyi}, {Humphrey}, {Morais}, {Onoue},
	{Pentericci}, {Reynaldi}, and {Venemans}]{saxena24a}
	{Saxena}, A.; {Overzier}, R.A.; {Villar-Mart{\'\i}n}, M.; {Heckman}, T.; {Roy},
	N.; {Duncan}, K.J.; {R{\"o}ttgering}, H.; {Miley}, G.; {Aydar}, C.; {Best},
	P.;  et~al.
	\newblock {Widespread AGN feedback in a forming brightest cluster galaxy at z =
		4.1, unveiled by JWST}.
	\newblock {\em Mon. Not. R. Astron. Soc.} {\bf 2024}, {\em 531},~4391--4407.
	\newblock {\url{https://doi.org/10.1093/mnras/stae1406}}.
	
	\bibitem[{Papachristou} et~al.(2023){Papachristou}, {Dasyra},
	{Fern{\'a}ndez-Ontiveros}, {Audibert}, {Ruffa}, {Combes}, {Polkas}, and
	{Gkogkou}]{papachristou23a}
	{Papachristou}, M.; {Dasyra}, K.M.; {Fern{\'a}ndez-Ontiveros}, J.A.;
	{Audibert}, A.; {Ruffa}, I.; {Combes}, F.; {Polkas}, M.; {Gkogkou}, A.
	\newblock {A plausible link between dynamically unsettled molecular gas and the
		radio jet in NGC 6328}.
	\newblock {\em Astron. Astrophys.} {\bf 2023}, {\em 679},~A115.
	\newblock {\url{https://doi.org/10.1051/0004-6361/202346464}}.
	
	\bibitem[{Morganti} et~al.(2021){Morganti}, {Oosterloo}, {Tadhunter},
	{Bernhard}, and {Raymond Oonk}]{morganti21a}
	{Morganti}, R.; {Oosterloo}, T.; {Tadhunter}, C.; {Bernhard}, E.P.; {Raymond
		Oonk}, J.B.
	\newblock {Taking snapshots of the jet-ISM interplay: The case of PKS 0023-26}.
	\newblock {\em Astron. Astrophys.} {\bf 2021}, {\em 656},~A55.
	\newblock {\url{https://doi.org/10.1051/0004-6361/202141766}}.
	
	\bibitem[{Zhong} et~al.(2024){Zhong}, {Inoue}, {Sugahara}, {Morokuma-Matsui},
	{Komugi}, {Kaneko}, and {Fudamoto}]{yuxing24a}
	{Zhong}, Y.; {Inoue}, A.K.; {Sugahara}, Y.; {Morokuma-Matsui}, K.; {Komugi},
	S.; {Kaneko}, H.; {Fudamoto}, Y.
	\newblock {Revisiting the Dragonfly galaxy II. Young, radiatively efficient
		radio-loud AGN drives massive molecular outflow in a starburst merger at z =
		1.92}.
	\newblock {\em Mon. Not. R. Astron. Soc.} {\bf 2024}, {\em 529},~4531--4553.
	\newblock {\url{https://doi.org/10.1093/mnras/stae798}}.
	
	\bibitem[{Fern{\'a}ndez-Ontiveros} et~al.(2020){Fern{\'a}ndez-Ontiveros},
	{Dasyra}, {Hatziminaoglou}, {Malkan}, {Pereira-Santaella}, {Papachristou},
	{Spinoglio}, {Combes}, {Aalto}, {Nagar}, {Imanishi}, {Andreani}, {Ricci}, and
	{Slater}]{ontiveros20a}
	{Fern{\'a}ndez-Ontiveros}, J.A.; {Dasyra}, K.M.; {Hatziminaoglou}, E.;
	{Malkan}, M.A.; {Pereira-Santaella}, M.; {Papachristou}, M.; {Spinoglio}, L.;
	{Combes}, F.; {Aalto}, S.; {Nagar}, N.;  et~al.
	\newblock {A CO molecular gas wind 340 pc away from the Seyfert 2 nucleus in
		ESO 420-G13 probes an elusive radio jet}.
	\newblock {\em Astron. Astrophys.} {\bf 2020}, {\em 633},~A127.
	\newblock {\url{https://doi.org/10.1051/0004-6361/201936552}}.
	
	\bibitem[{Morganti} et~al.(2005){Morganti}, {Tadhunter}, and
	{Oosterloo}]{morganti05a}
	{Morganti}, R.; {Tadhunter}, C.N.; {Oosterloo}, T.A.
	\newblock {Fast neutral outflows in powerful radio galaxies: A major source of
		feedback in massive galaxies}.
	\newblock {\em Astron. Astrophys.} {\bf 2005}, {\em 444},~L9--L13.
	\newblock {\url{https://doi.org/10.1051/0004-6361:200500197}}.
	
	\bibitem[{Struve} and {Conway}(2012)]{struve12a}
	{Struve}, C.; {Conway}, J.E.
	\newblock {The circumnuclear cold gas environments of the powerful radio
		galaxies 3C 236 and 4C 31.04}.
	\newblock {\em Astron. Astrophys.} {\bf 2012}, {\em 546},~A22.
	\newblock {\url{https://doi.org/10.1051/0004-6361/201218768}}.
	
	\bibitem[{Schulz} et~al.(2018){Schulz}, {Morganti}, {Nyland}, {Paragi},
	{Mahony}, and {Oosterloo}]{schulz18a}
	{Schulz}, R.; {Morganti}, R.; {Nyland}, K.; {Paragi}, Z.; {Mahony}, E.K.;
	{Oosterloo}, T.
	\newblock {Mapping the neutral atomic hydrogen gas outflow in the restarted
		radio galaxy 3C 236}.
	\newblock {\em Astron. Astrophys.} {\bf 2018}, {\em 617},~A38.
	\newblock {\url{https://doi.org/10.1051/0004-6361/201833108}}.
	
	\bibitem[{Guillard} et~al.(2012){Guillard}, {Ogle}, {Emonts}, {Appleton},
	{Morganti}, {Tadhunter}, {Oosterloo}, {Evans}, and {Evans}]{guillard12a}
	{Guillard}, P.; {Ogle}, P.M.; {Emonts}, B.H.C.; {Appleton}, P.N.; {Morganti},
	R.; {Tadhunter}, C.; {Oosterloo}, T.; {Evans}, D.A.; {Evans}, A.S.
	\newblock {Strong Molecular Hydrogen Emission and Kinematics of the Multiphase
		Gas in Radio Galaxies with Fast Jet-driven Outflows}.
	\newblock {\em Astrophys. J.} {\bf 2012}, {\em 747},~95.
	\newblock {\url{https://doi.org/10.1088/0004-637X/747/2/95}}.
	
	\bibitem[{Ogle} et~al.(2010){Ogle}, {Boulanger}, {Guillard}, {Evans},
	{Antonucci}, {Appleton}, {Nesvadba}, and {Leipski}]{ogle10a}
	{Ogle}, P.; {Boulanger}, F.; {Guillard}, P.; {Evans}, D.A.; {Antonucci}, R.;
	{Appleton}, P.N.; {Nesvadba}, N.; {Leipski}, C.
	\newblock {Jet-powered Molecular Hydrogen Emission from Radio Galaxies}.
	\newblock {\em Astrophys. J.} {\bf 2010}, {\em 724},~1193--1217.
	\newblock {\url{https://doi.org/10.1088/0004-637X/724/2/1193}}.
	
	\bibitem[{Santoro} et~al.(2018){Santoro}, {Rose}, {Morganti}, {Tadhunter},
	{Oosterloo}, and {Holt}]{santoro18a}
	{Santoro}, F.; {Rose}, M.; {Morganti}, R.; {Tadhunter}, C.; {Oosterloo}, T.A.;
	{Holt}, J.
	\newblock {Probing multi-phase outflows and AGN feedback in compact radio
		galaxies: The case of PKS B1934-63}.
	\newblock {\em Astron. Astrophys.} {\bf 2018}, {\em 617},~A139.
	\newblock {\url{https://doi.org/10.1051/0004-6361/201833248}}.
	
	\bibitem[{Salom{\'e}} et~al.(2016){Salom{\'e}}, {Salom{\'e}}, {Combes},
	{Hamer}, and {Heywood}]{salome16a}
	{Salom{\'e}}, Q.; {Salom{\'e}}, P.; {Combes}, F.; {Hamer}, S.; {Heywood}, I.
	\newblock {Star formation efficiency along the radio jet in Centaurus A}.
	\newblock {\em Astron. Astrophys.} {\bf 2016}, {\em 586},~A45.
	\newblock {\url{https://doi.org/10.1051/0004-6361/201526409}}.
	
	\bibitem[{Salom{\'e}} et~al.(2017){Salom{\'e}}, {Salom{\'e}},
	{Miville-Desch{\^e}nes}, {Combes}, and {Hamer}]{salome17a}
	{Salom{\'e}}, Q.; {Salom{\'e}}, P.; {Miville-Desch{\^e}nes}, M.A.; {Combes},
	F.; {Hamer}, S.
	\newblock {Inefficient jet-induced star formation in Centaurus A. High
		resolution ALMA observations of the northern filaments}.
	\newblock {\em Astron. Astrophys.} {\bf 2017}, {\em 608},~A98.
	\newblock {\url{https://doi.org/10.1051/0004-6361/201731429}}.
	
	\bibitem[{Ogle} et~al.(2025){Ogle}, {Sebastian}, {Aravindan}, {McDonald},
	{Canalizo}, {Ashby}, {Azadi}, {Antonucci}, {Barthel}, {Baum}, {Birkinshaw},
	{Carilli}, {Chiaberge}, {Duggal}, {Gebhardt}, {Hyman}, {Kuraszkiewicz},
	{Lopez-Rodriguez}, {Medling}, {Miley}, {Omoruyi}, {O'Dea}, {Perley},
	{Perley}, {Perlman}, {Reynaldi}, {Singha}, {Sparks}, {Tremblay}, {Wilkes},
	{Willner}, and {Worrall}]{ogle25a}
	{Ogle}, P.M.; {Sebastian}, B.; {Aravindan}, A.; {McDonald}, M.; {Canalizo}, G.;
	{Ashby}, M.L.N.; {Azadi}, M.; {Antonucci}, R.; {Barthel}, P.; {Baum}, S.;
	et~al.
	\newblock {The JWST View of Cygnus A: Jet-driven Coronal Outflow with a Twist}.
	\newblock {\em Astrophys. J.} {\bf 2025}, {\em 983},~98.
	\newblock {\url{https://doi.org/10.3847/1538-4357/adb71a}}.
	
	\bibitem[{Nesvadba} et~al.(2017){Nesvadba}, {Drouart}, {De Breuck}, {Best},
	{Seymour}, and {Vernet}]{nesvadba17b}
	{Nesvadba}, N.P.H.; {Drouart}, G.; {De Breuck}, C.; {Best}, P.; {Seymour}, N.;
	{Vernet}, J.
	\newblock {Gas kinematics in powerful radio galaxies at z 2: Energy supply from
		star formation, AGN, and radio jets}.
	\newblock {\em Astron. Astrophys.} {\bf 2017}, {\em 600},~A121.
	\newblock {\url{https://doi.org/10.1051/0004-6361/201629357}}.
	
	\bibitem[{May} et~al.(2016){May}, {Steiner}, {Ricci}, {Menezes}, and
	{Andrade}]{may16a}
	{May}, D.; {Steiner}, J.E.; {Ricci}, T.V.; {Menezes}, R.B.; {Andrade}, I.S.
	\newblock {Digging process in NGC 6951: The molecular disc bumped by the jet}.
	\newblock {\em Mon. Not. R. Astron. Soc.} {\bf 2016}, {\em 457},~949--970.
	\newblock {\url{https://doi.org/10.1093/mnras/stv2929}}.
	
	\bibitem[{Tadhunter} et~al.(1987){Tadhunter}, {Fosbury}, {Binette}, {Danziger},
	and {Robinson}]{tadhunter87a}
	{Tadhunter}, C.N.; {Fosbury}, R.A.E.; {Binette}, L.; {Danziger}, I.J.;
	{Robinson}, A.
	\newblock {Detached nuclear-like activity in the radio galaxy PKS 2152-69}.
	\newblock {\em \nat} {\bf 1987}, {\em 325},~504--507.
	\newblock {\url{https://doi.org/10.1038/325504a0}}.
	
	\bibitem[{Tadhunter} et~al.(1988){Tadhunter}, {Fosbury}, {di Serego Alighieri},
	{Bland}, {Danziger}, {Goss}, {McAdam}, and {Snijders}]{tadhunter88a}
	{Tadhunter}, C.N.; {Fosbury}, R.A.E.; {di Serego Alighieri}, S.; {Bland}, J.;
	{Danziger}, I.J.; {Goss}, W.M.; {McAdam}, W.B.; {Snijders}, M.A.J.
	\newblock {Very extended ionized gas in radio galaxies---IV. PKS 2152-69.}
	\newblock {\em Mon. Not. R. Astron. Soc.} {\bf 1988}, {\em 235},~403--423.
	\newblock {\url{https://doi.org/10.1093/mnras/235.2.403}}.
	
	\bibitem[{Clark} et~al.(1997){Clark}, {Tadhunter}, {Morganti}, {Killeen},
	{Fosbury}, {Hook}, {Siebert}, and {Shaw}]{clark97a}
	{Clark}, N.E.; {Tadhunter}, C.N.; {Morganti}, R.; {Killeen}, N.E.B.; {Fosbury},
	R.A.E.; {Hook}, R.N.; {Siebert}, J.; {Shaw}, M.A.
	\newblock {Radio, optical and X-ray observations of PKS 2250-41: A jet/galaxy
		collision?}
	\newblock {\em Mon. Not. R. Astron. Soc.} {\bf 1997}, {\em 286},~558--582.
	\newblock {\url{https://doi.org/10.1093/mnras/286.3.558}}.
	
	\bibitem[{Villar-Mart{\'\i}n} et~al.(1999){Villar-Mart{\'\i}n}, {Tadhunter},
	{Morganti}, {Axon}, and {Koekemoer}]{villarmartin99a}
	{Villar-Mart{\'\i}n}, M.; {Tadhunter}, C.; {Morganti}, R.; {Axon}, D.;
	{Koekemoer}, A.
	\newblock {PKS 2250-41 and the role of jet-cloud interactions in powerful radio
		galaxies}.
	\newblock {\em Mon. Not. R. Astron. Soc.} {\bf 1999}, {\em 307},~24--40.
	\newblock {\url{https://doi.org/10.1046/j.1365-8711.1999.02603.x}}.
	
	\bibitem[{Inskip} et~al.(2008){Inskip}, {Villar-Mart{\'\i}n}, {Tadhunter},
	{Morganti}, {Holt}, and {Dicken}]{inskip08a}
	{Inskip}, K.J.; {Villar-Mart{\'\i}n}, M.; {Tadhunter}, C.N.; {Morganti}, R.;
	{Holt}, J.; {Dicken}, D.
	\newblock {PKS2250-41: A case study for triggering}.
	\newblock {\em Mon. Not. R. Astron. Soc.} {\bf 2008}, {\em 386},~1797--1810.
	\newblock {\url{https://doi.org/10.1111/j.1365-2966.2008.13171.x}}.
	
	\bibitem[{Ruffa} et~al.(2025){Ruffa}, {Spavone}, {Iodice}, {Garcia-Burillo},
	{Davis}, {Iwasawa}, {Spoon}, {Paladino}, {Perna}, and {Vignali}]{ruffa25a}
	{Ruffa}, I.; {Spavone}, M.; {Iodice}, E.; {Garcia-Burillo}, S.; {Davis}, T.A.;
	{Iwasawa}, K.; {Spoon}, H.W.W.; {Paladino}, R.; {Perna}, M.; {Vignali}, C.
	\newblock {The link between galaxy merger, radio jet expansion and molecular
		outflow in the ULIRG IRAS 00183-7111}.
	\newblock {\em arXiv} {\bf 2025},  arXiv:2506.07852.
	\newblock {\url{https://doi.org/10.48550/arXiv.2506.07852}}.
	
	\bibitem[{Villar-Mart{\'\i}n} et~al.(2017){Villar-Mart{\'\i}n}, {Emonts},
	{Cabrera Lavers}, {Tadhunter}, {Mukherjee}, {Humphrey}, {Rodr{\'\i}guez
		Zaur{\'\i}n}, {Ramos Almeida}, {P{\'e}rez Torres}, and
	{Bessiere}]{villarMartin17a}
	{Villar-Mart{\'\i}n}, M.; {Emonts}, B.; {Cabrera Lavers}, A.; {Tadhunter}, C.;
	{Mukherjee}, D.; {Humphrey}, A.; {Rodr{\'\i}guez Zaur{\'\i}n}, J.; {Ramos
		Almeida}, C.; {P{\'e}rez Torres}, M.; {Bessiere}, P.
	\newblock {Galaxy-wide radio-induced feedback in a radio-quiet quasar}.
	\newblock {\em Mon. Not. R. Astron. Soc.} {\bf 2017}, {\em 472},~4659--4678.
	\newblock {\url{https://doi.org/10.1093/mnras/stx2209}}.
	
	
\bibitem[{Seidl} et~al.(2025){Seidl}, {Gronke}, {Farber}, and {Dolag}]{seidl25a}
{Seidl}, B.S.; {Gronke}, M.; {Farber}, R.J.; {Dolag}, K.
\newblock {Multi-cloud crushing -- the collective survival of cold clouds in galactic outflows}.
\newblock {\em arXiv e-prints} {\bf 2025}, p. arXiv:2506.05448,  
\newblock {\url{https://doi.org/10.48550/arXiv.2506.05448}}.
	
	\bibitem[{Cooper} et~al.(2009){Cooper}, {Bicknell}, {Sutherland}, and
	{Bland-Hawthorn}]{cooper09a}
	{Cooper}, J.L.; {Bicknell}, G.V.; {Sutherland}, R.S.; {Bland-Hawthorn}, J.
	\newblock {Starburst-Driven Galactic Winds: Filament Formation and Emission
		Processes}.
	\newblock {\em Astrophys. J.} {\bf 2009}, {\em 703},~330.
	\newblock {\url{https://doi.org/10.1088/0004-637X/703/1/330}}.
	
	\bibitem[{Pittard} et~al.(2010){Pittard}, {Hartquist}, and {Falle}]{pittard10a}
	{Pittard}, J.M.; {Hartquist}, T.W.; {Falle}, S.A.E.G.
	\newblock {The turbulent destruction of clouds---II. Mach number dependence,
		mass-loss rates and tail formation}.
	\newblock {\em Mon. Not. R. Astron. Soc.} {\bf 2010}, {\em 405},~821--838.
	\newblock {\url{https://doi.org/10.1111/j.1365-2966.2010.16504.x}}.
	
	\bibitem[{Scannapieco} and {Br{\"u}ggen}(2015)]{scannapieco15a}
	{Scannapieco}, E.; {Br{\"u}ggen}, M.
	\newblock {The Launching of Cold Clouds by Galaxy Outflows. I. Hydrodynamic
		Interactions with Radiative Cooling}.
	\newblock {\em Astrophys. J.} {\bf 2015}, {\em 805},~158.
	\newblock {\url{https://doi.org/10.1088/0004-637X/805/2/158}}.
	
	
	\bibitem[{Pittard} and {Parkin}(2016)]{pittard16a}
	{Pittard}, J.M.; {Parkin}, E.R.
	\newblock {The turbulent destruction of clouds---III. Three-dimensional
		adiabatic shock-cloud simulations}.
	\newblock {\em Mon. Not. R. Astron. Soc.} {\bf 2016}, {\em 457},~4470--4498.
	\newblock {\url{https://doi.org/10.1093/mnras/stw025}}.
	
	\bibitem[{Banda-Barrag{\'a}n} et~al.(2018){Banda-Barrag{\'a}n}, {Federrath},
	{Crocker}, and {Bicknell}]{bandabarragan18a}
	{Banda-Barrag{\'a}n}, W.E.; {Federrath}, C.; {Crocker}, R.M.; {Bicknell}, G.V.
	\newblock {Filament formation in wind-cloud interactions- II. Clouds with
		turbulent density, velocity, and magnetic fields}.
	\newblock {\em Mon. Not. R. Astron. Soc.} {\bf 2018}, {\em 473},~3454--3489.
	\newblock {\url{https://doi.org/10.1093/mnras/stx2541}}.
	
	\bibitem[{Gronke} and {Oh}(2018)]{gronke18a}
	{Gronke}, M.; {Oh}, S.P.
	\newblock {The growth and entrainment of cold gas in a hot wind}.
	\newblock {\em Mon. Not. R. Astron. Soc.} {\bf 2018}, {\em 480},~L111--L115.
	\newblock {\url{https://doi.org/10.1093/mnrasl/sly131}}.
	
	\bibitem[{Cottle} et~al.(2020){Cottle}, {Scannapieco}, {Br{\"u}ggen},
	{Banda-Barrag{\'a}n}, and {Federrath}]{cottle20a}
	{Cottle}, J.; {Scannapieco}, E.; {Br{\"u}ggen}, M.; {Banda-Barrag{\'a}n}, W.;
	{Federrath}, C.
	\newblock {The Launching of Cold Clouds by Galaxy Outflows. III. The Influence
		of Magnetic Fields}.
	\newblock {\em Astrophys. J.} {\bf 2020}, {\em 892},~59.
	\newblock {\url{https://doi.org/10.3847/1538-4357/ab76d1}}.
	
	\bibitem[Sutherland and Dopita(1993)]{sutherland93c}
Sutherland, R.S.; Dopita, M.A.
\newblock Cooling functions for low-density astrophysical plasmas.
\newblock {\em ApJS} {\bf 1993}, {\em 88},~253.  

\bibitem[{Sutherland} and {Dopita}(2017)]{sutherland17a}
{Sutherland}, R.S.; {Dopita}, M.A.
\newblock {Effects of Preionization in Radiative Shocks. I. Self-consistent Models}.
\newblock {\em \apjs} {\bf 2017}, {\em 229},~34,  
\newblock {\url{https://doi.org/10.3847/1538-4365/aa6541}}.
	
	
	
	
	
	
	
\end{thebibliography}

\PublishersNote{}
\end{adjustwidth}
\end{document}